\documentclass[longauth]{aa}
\newcommand{\lt}{<}

\newcommand{\ml}{$\alpha_\mathrm{ML}$}

\newcommand{\teff}{T$_{\rm eff}$}

\usepackage{graphicx}
\usepackage{txfonts}
\begin{document} 

\title{The Gaia-ESO Survey: open clusters in {\it Gaia}-DR1 -  
a way forward to stellar age calibration
\thanks{Based on observations collected with the FLAMES instrument
at VLT/UT2 telescope (Paranal Observatory, ESO, Chile), for the Gaia-ESO Large Public Spectroscopic Survey (188.B-3002, 193.B-0936).}}

\author{S. Randich\inst{1}\and
          E. Tognelli\inst{2,3}\and
          R. Jackson\inst{4} \and
          R.~D. Jeffries\inst{4} \and
          S. Degl'Innocenti\inst{2,3}\and
          E. Pancino\inst{1} \and
          P. Re Fiorentin\inst{5} \and 
          A. Spagna\inst{5} \and
          G. Sacco\inst{1} \and
          A. Bragaglia\inst{6} \and
          L. Magrini\inst{1} \and
          P.G. Prada Moroni\inst{2,3}\and
          E. Alfaro\inst{7} \and
          E. Franciosini\inst{1}\and
          L. Morbidelli\inst{1}\and
          V. Roccatagliata\inst{1}\and
          H. Bouy\inst{8}\and
          L. Bravi\inst{1,9}\and
          F. M. Jim\'{e}nez-Esteban\inst{10}\and
          C. Jordi\inst{11}\and
          E. Zari\inst{12}\and  
          G. Tautvai\v{s}iene\inst{13}\and
          A. Drazdauskas\inst{13}\and
          S. Mikolaitis\inst{13}\and
          G. Gilmore\inst{14}\and
          S. Feltzing\inst{15}\and 
          A. Vallenari\inst{16}\and
          T. Bensby\inst{15}\and
          S. Koposov\inst{14}\and
          A. Korn\inst{17}\and 
          A. Lanzafame\inst{18}\and 
          R. Smiljanic\inst{19}\and
          A. Bayo\inst{20}\and
          G. Carraro\inst{21}\and
          M. T. Costado\inst{7}\and
          U. Heiter\inst{17}\and
          A. Hourihane\inst{14}\and
          P. Jofr\'e\inst{22}\and
          J. Lewis\inst{14}\and 
          L. Monaco\inst{23}\and 
          L. Prisinzano\inst{24}\and 
          L. Sbordone\inst{25}\and
          S.~G. Sousa\inst{26}\and              
          C.~C. Worley\inst{14}\and 
          S. Zaggia\inst{16}}
\institute{INAF-Osservatorio Astrofisico di Arcetri, Largo E. Fermi 5, 
50125 Firenze, Italy\\
\email{randich@arcetri.astro.it}
\and
Department of Physics `E.Fermi', University of Pisa, Largo Bruno Pontecorvo 3, I-56127 Pisa, Italy
\and
INFN, Section of Pisa, Largo Bruno Pontecorvo 3, I-56127 Pisa, Italy
\and
Astrophysics Group, University of Keele, Keele, Staffordshire ST5 5BG, 
United Kingdom   
\and
INAF-Osservatorio Astrofisico di Torino, Via Osservatorio 20, 10025 
Pino Torinese, Italy
\and
INAF-Osservatorio Astronomico di Bologna, Via Gobetti 93/3, 40129 Bologna, Italy
\and
Instituto de Astrof\`{i}sica de Andaluc\`{i}a (IAA-CSIC), Glorieta de la Astronom\`ia, E-18008 Granada, Spain 
\and
Laboratoire d'Astrophysique de Bordeaux, Universit\'e Bordeaux, CNRS, B18N,
Alle\'ee Geoffroy Saint/Hilaire, F-33615 Pessac, France
\and
Dipartimento di Fisica e Astronomia, Universit\`a degli Studi di Firenze, Via G. Sansone 1, 50019 Sesto Fiorentino (Firenze), Italy
\and
Centro de Astrobiolog\'{i}a (INTA-CSIC), Departamento de Astrof\'{i}sica, PO Box 78, E-28691, Villanueva de la Ca\~nada, Madrid, Spain
\and
Departament de F\`{i}s\`{i}ca Quant\`{i}ca i Astrof\`{i}sica, Institut C\`{i}encies Cosmos (ICCUB), Universitat de Barcelona, Barcelona, Spain
\and
Leiden Observatory, Niels Bohrweg 2, 2333 CA Leiden, The Netherlands
\and
Institute of Theoretical Physics and Astronomy, Vilnius University, Saul\.{e}tekio al. 3, LT-10222 Vilnius, Lithuania
\and
Institute of Astronomy, Madingley Road, University of Cambridge, CB3 0HA, UK
\and
Lund Observatory, Department of Astronomy and Theoretical Physics, Box 43, SE-221 00 Lund, Sweden
\and
INAF - Osservatorio Astronomico di Padova, Vicolo dell'Osservatorio 5, 35122 Padova, Italy
\and
Department of Physics and Astronomy, Uppsala University, Box 516, SE-751 20 Uppsala, Sweden
\and
Dipartimento di Fisica e Astronomia, Sezione Astrofisica, Universit\`a di Catania, via S. Sofia 78, 95123, Catania, Italy
\and
Nicolaus Copernicus Astronomical Center, Polish Academy of Sciences, ul. Bartycka 18, 00-716, Warsaw, Poland
\and
Instituto de F\'isica y Astronomi\'ia, Universidad de Valpar\'iso, Chile
\and
Dipartimento di Fisica e Astronomia, Universit\`a di Padova, Vicolo dell'Osservatorio 3, 35122 Padova, Italy
\and
N\'ucleo de Astronom\'a, Universidad Diego Portales, Av. Ej\'exrcito 441, Santiago, Chile
\and
Departamento de Ciencias Fisicas, Universidad Andres Bello, Fernandez
Concha 700, Las Condes, Santiago, Chile
\and
INAF - Osservatorio Astronomico di Palermo, Piazza del Parlamento 1, 90134, Palermo, Italy
\and
European Southern Observatory, Alonso de Cordova 3107 Vitacura, Santiago de Chile, Chile
\and
Instituto de Astrof\'isica e Ci\^encias do Espa\c{c}o, Universidade do Porto, CAUP, Rua das Estrelas, 4150-762 Porto, Portugal
}
\titlerunning{The Gaia-ESO Survey: open clusters in {\it Gaia}-DR1}
   \date{received date ---; accepted date ---}

  \abstract
{Determination and calibration of the ages of stars, 
which heavily rely on stellar evolutionary models,
are very challenging, while
representing a crucial aspect in many astrophysical areas.}
{We describe the methodologies that, taking advantage of {\it Gaia}-DR1 and the Gaia-ESO Survey data, enable the comparison of observed open star cluster sequences with stellar evolutionary models. The final, long-term goal is the exploitation of open clusters as age calibrators.}
{We perform a homogeneous analysis of eight open clusters using the {\it Gaia}-DR1 
TGAS catalogue for bright members and 
information from the Gaia-ESO Survey for fainter stars. 
Cluster membership probabilities for the Gaia-ESO Survey targets are derived based on several spectroscopic tracers. The
Gaia-ESO Survey also provides the cluster chemical composition. 
We obtain cluster
parallaxes using two methods. The first one relies on the astrometric selection of a sample of {\it bona fide} members, while 
the other one fits the parallax distribution of a larger sample of TGAS sources. Ages and reddening values are recovered through a Bayesian analysis using the 2MASS magnitudes and three sets of standard models.
Lithium depletion boundary (LDB)
ages are also determined
using literature observations and the same models employed for the Bayesian
analysis.}
{For all but one cluster, parallaxes derived by us agree
with those presented in {\it Gaia} Collaboration et al. (2017), while a 
discrepancy is found for NGC~2516; we provide evidence supporting our own 
determination. Our age determinations are robust against models and are generally
consistent with literature values. 
}
{The systematic parallax errors inherent in the Gaia DR1 data presently limit the precision of our results. Nevertheless, we have been able to place these eight clusters onto the same age scale for the first time, 
with good agreement between isochronal and LDB ages where there is overlap. Our approach appears promising and demonstrates the potential of combining {\it Gaia} and ground-based spectroscopic datasets.}
   \maketitle
\section{Introduction}\label{intro}
Determination of stellar ages is vital in addressing 
virtually all branches of stellar and Galactic astrophysics,
from the duration of the star formation process and the timescales for the 
formation and evolution of planetary systems, to the larger-scale formation and evolution of the Milky Way galaxy, and to the age of the Universe.
The age of a star cannot be measured directly and it is widely acknowledged 
that age estimation is one of the most challenging problems in astrophysics 
(e.g. Palla  et al. 2005; Soderblom 2010; Soderblom et al. 2014
and references therein).
The most commonly employed method for inferring the ages of stars
is the comparison of different quantities which can be derived
from observations (colours,
temperature, surface gravity, luminosity) with outputs from stellar evolutionary codes, possibly using  some appropriate (statistical) inference scheme. 
This approach is widely used for field stars as well as for star
clusters; however,
the precision of those ages and, even more importantly, the absolute ages, are set by the fidelity of the evolutionary models, the physical uncertainties inherent to them, and uncertainties in the transformations between theoretical and observational quantities. 

Indeed, in spite of considerable progress in understanding stellar physics and evolution 
in the last decades, several issues are not completely settled, 
among which we mention the lack of a precise treatment of convection and overshooting;
(e.g. Salaris \& Cassisi 2015; Viallet et al. 2015);  
the input physics adopted in the calculations
(e.g. opacities, equation of state -EOS-, nuclear reactions etc. --Naylor 2009; Valle et al. 2013; Tognelli et al. 2015b); 
the effects of internal rotation and magnetic fields 
(e.g. Eggenberger et al. 2010;
Feiden \& Chaboyer 2012; Charbonnel et al. 2013;
Feiden et al.  2014, 2015); and
the outer boundary conditions and atmosphere adopted for the star (e.g. Baraffe et al. 2002). The latter issue is especially true 
at low masses where the formation of molecules or even dust is a difficult 
problem and there may be further 
issues associated with the usual assumption of homogeneous, plane-parallel atmospheres when many stars, particularly those with high levels of 
magnetic activity, have demonstrable surface inhomogeneities 
(Jackson \& Jeffries 2014; Somers \& Pinsonneault 2015; Jeffries et al. 2017).
Comparison between theory and observations clearly constitutes a benchmark 
for stellar evolutionary theory. 

Asteroseismology represents a relatively new, very powerful tool for determining stellar ages, in particular for evolved giants (see Miglio et al. 2017
and references therein). However, star clusters remain key age calibrators for stars in all evolutionary phases, and continue to serve as a critical tool to put constraints on evolutionary models. The distribution of non-binary members of a cluster in the colour-magnitude or
Hertzsprung-Russell diagrams (CMDs, HRDs) is expected to be very narrow, 
reflecting the homogeneity in age and chemical composition. At the same time, members of a given cluster cover different masses and 
evolutionary stages, at the cluster age and metallicity. Each cluster thus represents a snapshot of stellar evolution; 
linking together observations of many clusters at different 
ages and chemical compositions empirically
reveals the story of stellar evolution, to be compared with the predictions of theoretical models. 
Much of stellar, and ultimately Galactic, astrophysics hinges on these crucial comparisons between cluster observations and the predictions of the models.
Only once the models are appropriately adjusted to represent 
the data as obtained for a wide range of clusters of different ages and compositions, reliable stellar masses and ages can be determined. Focusing on open clusters (OCs), we note that
they play a key role as stellar age calibrators also because the ages of their stars can be estimated using a variety of independent techniques that work accurately (or at least more precisely) when applied to ensembles of stars. 
These different techniques have different sensitivities to uncertainties in model physics, and in some cases are weakly sensitive to these uncertainties, allowing (almost) model-independent  absolute age calibration (e.g. the lithium depletion boundary technique, Burke et al. 2004; Tognelli et al. 2015b). 

Precise distances to the clusters are clearly the key to making  a big step forward and the {\it Gaia} mission (Gaia Collaboration et al. 2016a) will enable a revolution in this area. In the last decade or so a variety of surveys allowed significant progress in deriving empirical CMDs for clusters; however, {\it Gaia} will provide precise parallaxes and photometric measurements  for several tens or hundreds
of nearby OCs and for large samples of members, finally making fully accessible the unique potential of OCs to empirically resolve the CMDs and HRDs and to put tighter constraints on stellar evolutionary models for Population~I stars, from the pre-main sequence (PMS) phase to the latest evolutionary stages.  
In order to best exploit the exquisite information from {\it Gaia}, spectroscopy from the ground is also necessary, mainly 
to firmly derive the cluster chemical composition
and to confirm cluster membership using radial velocities 
and other membership tracers. Ground-based spectroscopy
also allows the determination of rotation,
effective temperature, surface gravity, and, in general, 
a full characterisation of the stellar properties down to faint stars.\footnote{Radial velocities and astrophysical parameters will also be available from {\it Gaia}; however, they are expected to be of lower precision and/or limited to brighter magnitudes than achievable from ground-based spectroscopy.}

The Gaia-ESO Large Public Spectroscopic Survey 
(GES --Gilmore et al. 2012; Randich et al. 2013)
is a high-precision, high-resolution survey 
specifically designed  to cover the {\it Gaia} range of stellar populations. GES is using the multi-object system 
FLAMES at ESO VLT (Pasquini et al. 2002), 
with its high-resolution UVES and its 
medium-resolution GIRAFFE spectrographs, to target about 10$^5$ stars, systematically covering all the major components of the Milky Way, and 
providing a homogeneous overview of the distributions of 
kinematics and elemental abundances in stars as faint as $V\sim 19$ and 
$V\sim 16.5$, for GIRAFFE and UVES targets, respectively.
GES complements {\it Gaia} and vice-versa: in particular,
the GES OC dataset, 
combined with {\it Gaia} astrometry (and eventually, spectrophotometry),
represents an unrivalled resource to attack the problem of 
stellar model and age calibration. GES is providing homogeneous and precise measurements of radial velocities, astrophysical parameters, metallicity, detailed abundances for several chemical elements, and assessments of 
magnetic/accretion activity from H$\alpha$ emission for thousands of stars,
over a wide range in masses, 
in about 60--70 OC with ages from a few Myr to several Gyr. 
As mentioned, in the last decade or so, significant progress has been made both on testing evolutionary models and on star cluster age determination (e.g. Lyra et al. 2006; Silaj \& Landstreet 2014; Kopytova et al. 2016).
The combination of {\it Gaia} and GES cluster datasets will offer 
clean CMDs and HRDs of well-defined samples of cluster members, allowing further
tests of evolutionary models and stellar atmospheres, and making
it possible to choose the best models to estimate ages. 

The first intermediate {\it Gaia} data release 
({\it Gaia}-DR1 -Gaia Collaboration et al. 2016b) 
includes positions and {\it Gaia} G band magnitudes for about 1 billion stars,
as well as the five-parameter astrometric solution - positions, parallaxes, and proper motions - for about two million sources in common between the Tycho-2 Catalogue (H\o g et al. 2000) and {\it Gaia}. This part of the
{\it Gaia} DR1 dataset is based on the Tycho-Gaia Astrometric Solution (TGAS -Michalik et al. 2015; Lindegren et al. 2016), 
obtained by combining the Tycho 2 and {\it Gaia} data together, which provides
a long enough baseline to break the degeneracy between motion and parallax.
The limiting magnitude of the TGAS catalogue is  $V\sim 12$ mag;
bright stars in several nearby cluster fields are thus contained in it, allowing not only the validation of {\it Gaia} astrometry, 
but also the determination of average cluster parallaxes ({\it Gaia} Collaboration et al. 2017 -hereafter VL17). 
A fraction of the clusters whose members are included in TGAS
have been observed by GES during the first 30 months of data 
taking and the parameters of the target stars have been
derived during the latest analysis cycle (iDR4) and released internally to the consortium in the GEiDR4Final catalogue.\footnote{The GEiDR4Final catalogue is 
available for the members of the GES consortium at http:\//ges.roe.ac.uk/. Parameters for a large fraction of the stars
have also been released to ESO during the Phase 3 delivery and are publicly available
at the ESO archive facility (http://www.eso.org/qi).}  
Within each cluster, the GES target samples are at this stage generally complementary to the TGAS ones, mostly including stars much fainter than the TGAS limit,
although a few stars in common are present in some of the clusters.
The two datasets, the TGAS one for the bright members and the GES one for the fainter stars down to the 19 magnitude in the $V$ band, will hence be used mainly independently. Namely, the GES data will allow us to define the cluster sequences in a way that is unbiased with respect to the CMD, over a broad magnitude range 
(equivalent to a range of masses). GES data will also yield the 
cluster chemical composition. Members identified in TGAS will instead mainly be employed to determine the cluster parallaxes. 
The two datasets together will critically allow us to match evolutionary models, with metallicity and distance as fixed parameters, and
to put the clusters into a model-dependent age sequence.

{\it Gaia}-DR2 will yield more precise and accurate parallaxes for 
stars down to about the 20th magnitude, allowing one to reach more distant 
clusters and/or lower mass members in nearby clusters. At the same time, GES will complete observations and analysis of all its sample clusters.
Along with the GES data, {\it Gaia}-DR2 will thus make possible a more detailed analysis
based on both a larger number of clusters covering a larger interval of ages and metallicity, and on more secure astrometry. With the present analysis and 
paper, however, we aim to set the methodology, to identify major sources of uncertainty, and
to show, exploiting {\it Gaia}-DR1 already, how the combination 
of {\it Gaia} and GES cluster data can be used
to calibrate stellar evolution and ages. The methods that we outline here will then be exploited using {\it Gaia}-DR2 and the full sample of GES OCs.

The paper is structured as follows: in Sect.~\ref{sample} we present the sample OCs, GES target selection, available spectroscopic information, and cluster chemical composition. The
cluster membership determination using GES data is presented in Sect.~\ref{member_ges}, 
while the astrometric analysis and determination
of the cluster parallaxes are discussed in Sect.~\ref{TGAS}. 
Section~\ref{sec:comparison} is dedicated to the comparison with different
evolutionary models and age determination; the presentation and discussion of the results,
and conclusions are summarised in Sects.~\ref{sec:discussion} and \ref{sec:conclusions}, respectively.
\section{The sample clusters and Gaia-ESO Survey information}\label{sample}
We considered clusters that have bright stars included in TGAS, have been observed by GES, and have been analysed in iDR4. We
further selected OCs close enough that their parallaxes are not greatly affected by {\it Gaia}-DR1 systematic errors in parallax (see Sect.~\ref{TGAS}), and old enough not to be characterised by possible age dispersions.
The final sample includes eight clusters, listed 
in Table~\ref{cluster_sample}, along with their main
properties from the literature.
\begin{table*}
\begin{center}
\caption{Sample clusters.}             
\label{cluster_sample}      
\tiny
\begin{tabular}{lccllllcl}        
\hline\hline                 
Cluster & RA & DEC & age & ref. & E($B-V$) & \multicolumn{1}{c}{ref.} & pre-{\it Gaia} &\multicolumn{1}{c}{ref.} \\
 & \multicolumn{2}{c}{J2000}& (Myr) & & & &  distance modulus & \\
 \hline
 &  &  & &  & & & & \\
NGC~2451A  & 07~44~27.00 & $-37$~40~00.00 & 50-80 & (1), (2)& 0.01 & (2) & $6.32\pm 0.04$ & van Leeuwen (2009)\\
NGC~2451B  & 07~44~27.00 & $-37$~40~00.00 & 50    & (1) &    0.05-0.12 & (3), (4) & $7.83\pm 0.35$ & Carrier et al. (1999) \\
NGC~2516   & 07~58~04.00 & $-60$~45~12.00 & 70-150 & (5), (6), (7), (8) & 0.09-0.15 & (9), (10), (11) &7.68$\pm 0.07$ &  van Leeuwen (2009) \\
NGC~2547   & 08~10~25.70 & $-49$~10~03.00 & 35-45 & (12), (13), (14) & 0.038-0.12 & (9), (13), (15) &
8.38$\pm 0.17$ & van Leeuwen (2009) \\
IC~2391    & 08~40~32.00 & $-53$~02~00.00 & 30-50 & (16), (17), (18) &$\leq 0.05$ & (9) & 5.80$\pm 0.04$ &  van Leeuwen (2009) \\
IC~2602    & 10~42~58.00 & $-64$~24~00.00 & 30-46 & (18), (19)& 0.02-0.04 & (9), (20) & 5.86$\pm 0.03$ &  van Leeuwen (2009) \\
IC~4665    & 17~46~18.00 & $+05$~43~00.00 & 28-40 & (21), (22)& 0.16-0.19 & (9), (22) & 7.75$\pm 0.21$ & 
van Leeuwen (2009) \\
NGC~6633   & 18~27~15.00 & $+06$~30~30.00 & 425-575 & (24), (25), (26) & 0.17-0.18 & (9), (23) & 7.87$\pm 0.26$ & 
van Leeuwen (2009) \\ \hline 
\end{tabular}
\end{center}
(1) H\"unsch et al. (2003); (2) Platais et al. (2001); (3) Carrier et al. (1999); (4) Balog et al. (2009); (5) Jeffries et al. (1998); (6) Lyra et al. (2006); (7) Silaj \& Landstreet (2014); (8) Tadross et al. (2002); (9) Nicolet (1981), (10) Sung et al. (2002); (11) Terndrup et al. (2002); (12) Jeffries \& Oliveira (2005); (13) Naylor \& Jeffries (2006); (14) Paunzen et al. (2014); (15) Claria (1982); (16) Barrado y Navascu\' es (1999); (17) Barrado y Navascu\' es (2004);
(18) Stauffer et al. (1997); (19) Dobbie et al. (2010); (20) Hill \& Perry (1969); (21) Manzi et al. (2008) ; (22) Cargile \& James (2010); (23) Gurklyte \& Strayzis (1981); (24) van Leeuwen (2009); (25) Dias et al. (2002); (26) Williams \& Bolte (2007).
\end{table*}
\subsection{Target selection and characteristics}\label{selection}
The selection of the target stars within each cluster is described in detail in Bragaglia et al. (in preparation). We summarise here the relevant features. Briefly, the selection was performed following two main criteria: 
a) obtaining a large and unbiased sample of cluster candidates
to be observed with the 
GIRAFFE fibres; and b) obtaining a smaller sample of brighter stars to be observed with the
UVES fibres. More specifically, GIRAFFE targets were selected using 2MASS near-IR photometry \citep{2mass}, along with many 
public sources of optical photometry.
The cluster sequences were identified in optical and near-IR CMDs, with the help of known members from the literature and/or isochrones. We then inclusively
selected candidates, considering stars covering the same spatial extension as known members
and lying in a generous band around the identified cluster sequences. When information was available, UVES targets were instead chosen among the most likely previously known cluster members. 
By observing stars in all evolutionary phases and by combining the two approaches, we aim to derive a comprehensive picture of each cluster and of the complete sample of clusters.  
The number of candidates observed in the field of each cluster varies from about 100-200
to almost 2000, but significant contamination is expected (even among the UVES targets) due to our inclusive selection of candidates. 
Most of the OCs in the present paper are relatively young (age $<150$ Myr, but older than $\sim 10$~Myr) and 
the targets (both UVES and GIRAFFE ones)
are stars on the [pre/zero-age] main sequence. 
The only cluster which is slightly
older is NGC~6633 (age about 600 Myr), where we observed stars on 
the main sequence and in the red clump phase.

We refer to Pancino et al. (2017) for complete information on the GIRAFFE gratings employed in GES. We just mention here that the targets in the eight sample clusters are generally of late spectral type (F to M), with only a few earlier-type stars. We mainly observed them with the GIRAFFE setup HR15N, where H$\alpha$ and the Li {\sc i} 6708\AA\ line are located.  About 5-10 \% of the targets were observed with UVES and the 520 and
580 setups for stars earlier and later than F-type, respectively.
\subsection{Spectrum analysis and products}\label{spectro}
GES data reduction and spectrum analysis, including homogenisation have been described in a number of papers (Sacco et al. 2014; 
Smiljanic et al. 2014; Jackson et al. 2015; Lanzafame et al. 2015). In summary, pipeline data reduction, as well as radial and rotational velocity determinations are
centralised and performed at the Cambridge Astronomy Survey Unit (CASU) for GIRAFFE and at Arcetri for UVES spectra, respectively.
Spectrum analysis is instead distributed among five working groups (WGs), 
depending on the stellar-type and/or instrument and/or setup. From at least two up to several nodes contribute to the analysis within each WG; node astrophysical parameters and abundances are first homogenised within each WG and subsequently homogenised across WGs, to put them on a common scale. Homogenisation is performed 
using several calibrators, including benchmark stars and calibration
open/globular clusters, selected as described in Pancino et al. (2017). Radial velocities from the different instruments and settings are also homogenised.
The homogenised values constitute the recommended set and are included in 
the GEiDR4Final catalogue: unless otherwise mentioned, this paper makes use of these recommended values. 
The dataset released in the GEiDR4Final catalogue includes radial
and projected equatorial rotational velocities (RV and v$\sin i$), stellar parameters (\teff, $\log$g, and/or $\gamma$ index -see Damiani et al. 2014), global metallicity [Fe/H], and equivalent widths of the Li~{\sc i}~6708\AA\ feature (EWLi); for UVES targets 
individual abundances for a variety of species, such as light, $\alpha$, Fe--peak, and neutron capture elements are also provided. 
For each of the eight clusters and all the candidates considered in this study, the Gaia-ESO products are provided in 
the online tables.
\subsection{Metallicity and $\alpha$-element abundances}\label{metal}
Overall metallicities for the eight clusters included in this analysis were derived, based on the GEiDR4Final catalogue, by Jacobson et al. (2016 --NGC~2516 and NGC~6633) and by Spina et al. (2017 --the remaining clusters) and are listed in Table~\ref{tab_clu_abufe}. 

$\alpha-$element abundances were available for two of the sample clusters only (NGC~2516 and NGC~6633, also based on GEiDR4Final --Magrini et al. 2017). Thus, we consistently derived them here for the other clusters.
Specifically, average [El/Fe] ratios were obtained based on 
the secure cluster members observed with UVES (see following section) and considering only stars 
with rotational velocities below 15 km\,s$^{-1}$. [El/Fe] values are normalised to the GES recommended abundances for the Sun (see Magrini et al. 2017).
The results are listed in Table~\ref{tab_clu_abufe}, 
where for each cluster we present the metallicity and the abundance ratios with their 1-$\sigma$ dispersion. 
In the second to the last column we report the mean [$\alpha$/Fe], computed by averaging the available $\alpha$ elements. As discussed already by Spina et al. (2017), all clusters have close to solar metallicities with a relatively small dispersion. On the other hand, the table shows some scatter in elemental abundances, both within clusters and across clusters, with some elements being systematically somewhat below or above solar. 
Given the uncertainties, however, we can safely conclude (at least for the purposes of the present work) that all the clusters have [$\alpha$/Fe] consistent with the solar ratio, that is,  [$\alpha$/Fe]=0~dex.  Whilst in the following we will assume [Fe/H]$=0$ for all of the clusters, the effect of metallicity on the recovered ages and reddening values will be briefly discussed in Sect.~\ref{sec:discussion}. 
\begin{table*}
\begin{center}
\caption{Cluster metallicities and abundance ratios.} 
\tiny
\begin{tabular}{lrrrrrcc}
\hline \hline
  \multicolumn{1}{c}{Cluster} &
  \multicolumn{1}{c}{[Fe/H]} &
  \multicolumn{1}{c}{[MgI/Fe]} &
  \multicolumn{1}{c}{[SiI/Fe]} &
  \multicolumn{1}{c}{[CaI/Fe]} &
  \multicolumn{1}{c}{[TiI/Fe]} &
   \multicolumn{1}{c}{[$\alpha$/Fe]} &
  \multicolumn{1}{c}{[$\alpha$/Fe] provenance} \\ 
\hline
NGC~2451A & $-0.05\pm$0.02 & $-0.08$:: & $-0.17$:: & 0.02:: & $-0.17$::& $-0.1\pm 0.07$ & this paper, 1 star \\
NGC~2451B &$-0.01\pm$0.01 & $-0.02$:: & $-0.08$:: & 0.01:: & 0.05::& $-0.01\pm 0.07$ & this paper, 1 star \\
NGC~2516 & $-0.08\pm 0.02$ & $0.04\pm 0.06$ & $-0.05\pm 0.07$ &  $0.03\pm 0.05$ & $0.10\pm 0.09$ & $0.03\pm 0.08$ & Magrini et al. (2017)\\ 
NGC~2547        & $-0.01\pm 0.01$ & $0.00\pm 0.02$ & $-0.06\pm 0.06$ & $0.07\pm 0.01$ & $0.07\pm 0.03$ & $0.02\pm 0.06$ & this paper, 2 stars\\
IC~2391 & $-0.03\pm 0.02$ & $-0.10\pm 0.20$ & $-0.08\pm 0.02$ & $0.06\pm 0.05$ & $0.03\pm 0.04$ & $-0.02\pm 0.08$ & this paper 2 stars \\
IC~2602 &$-0.02\pm 0.02$ & $0.06\pm 0.08$ & $-0.08\pm 0.08$ & $0.08\pm 0.11$ & $0.0\pm 0.12$ & $0.01\pm 0.09$ & this paper, 7 stars \\
IC~4665 & $0.0\pm 0.02$ & $0.05\pm 0.09$ & $-0.10\pm 0.08$ & $0.08\pm 0.09$ & $0.01\pm 0.13$ & $0.01\pm 0.11$ & this paper, 6 stars \\
 NGC~6633 & $-0.06\pm 0.02$  & $-0.01\pm 0.01$ & $-0.02\pm 0.04$ &  $0.07\pm 0.06$ & $0.01\pm 0.07$ & $0.01\pm 0.04$  & Magrini et al. (2017)\\ 
\hline
\end{tabular}
\label{tab_clu_abufe}
\end{center}
\tablefoot{[Fe/H] values have been retrieved from Jacobson et al. (2016) and Spina et al. (2017) for all clusters, while $\alpha$-element abundances
have been taken from Magrini et al. (2017) or determined in this paper, as indicated.
Abundance ratios derived from one star only are marked with "::". Note that for the same clusters, [Fe/H] values were instead estimated using more stars (Spina et al.~2017).}
\end{table*}
\section{Gaia-ESO Survey cluster membership}\label{member_ges}
As described in Sect.~\ref{selection}, 
GES cluster target selection was designed to
be inclusive; as a result, many of the observed targets are not
cluster members. A key aim of this paper is to compare clean cluster sequences with theoretical predictions in the CM
diagrams and so these interlopers need to be removed, but without
relying on information from these two diagrams, to avoid biasing
the final results. In this Subsection we
describe a homogeneous approach to membership selection that was
applied to all the clusters.

To perform the membership analysis, which was applied to both GIRAFFE
and UVES targets,
we used the available photometry, together with information from GES
products and summarised at the end of Sect.~\ref{spectro}.
All of these data are
available for the majority of stars, but in what follows, we required a star to have a T$_{\rm eff}$, $\log$g or $\gamma$, and a RV in order to be considered as a candidate member. As mentioned, most of the data come from the 
GEiDR4Final internal release except v$\sin i$, where calibration issues meant that the previous release (iDR2/3) values were used in preference and are thus only available for three of the considered clusters (NGC~2547, IC~4665 and NGC~2516).
Also, for a few stars EWLi were 
not available in the GEiDR4Final catalogue; in those cases, EWLi determined
by one of the nodes were used. 

RV uncertainties are important to derive the RV distribution and membership 
probabilities.
For the clusters with available v$\sin i$ data (see above) RV uncertainties 
were calculated according to the detailed prescription described by Jackson et 
al. (2015), which takes into account stellar rotation. For the remaining clusters, the uncertainties were instead adopted from the GEiDR4Final catalogue. The comparison of the calculated uncertainties 
with those from the GEiDR4Final catalogue for candidate members of NGC~2547, IC~4665 and NGC~2516 showed that 
using the GEiDR4Final catalogue
RV uncertainty (where necessary) has only a marginal effect on the average membership probability as a function of RV. At worst, the effect of using the GEiDR4Final catalogue RV uncertainty is to change the membership classifications of a few targets in each cluster where the measured RV is close to the boundary of plausible cluster members.
\subsection{The selection sequence}
Whilst we initially retained all targets observed with the UVES 520 setup and available
parameters, the samples of stars observed with HR15N and UVES 580 
were cleaned of almost certain non-members in the following way.
\begin{enumerate}
\item For all clusters, likely 
giant contaminants were removed using a modified version of the
 $\gamma$ index ($\gamma^{\prime} = \gamma + \tau/6$, where
$\tau$ is the temperature-sensitive index -- see Damiani et
  al. 2014), which removes the temperature sensitivity of
  the index. 
  Targets with $4000 < \rm T_{\rm eff} <7000$~K and
    $\gamma^{\prime} > 1.335$ were thus rejected as giants. By comparing
    $\gamma^{\prime}$ with $\log g$ for the subset of stars for which 
    the gravity is also available, we find that this threshold
    corresponds to $\log g < 3.4 \pm 0.1$. We note that for NGC~6633 UVES sample giants were not discarded. 
\item The presence of lithium in the photosphere is an excellent
  empirical age indicator for low-mass stars within given 
   T$_{\rm eff}$ ranges (e.g. Soderblom et al. 2014). By
  examining compiled data from the literature for the sample clusters and other
  clusters in the same age interval (e.g. Soderblom et al. 1993; Jeffries et al. 1998; Randich et al. 2001; Sestito \& Randich 2005; Jeffries et al. 2014), we designed simple filters in the EWLi versus T$_{\rm eff}$ plane. These filters allowed the inclusion of almost all previously known members of the clusters at their assumed ages, whilst excluding a significant fraction of older contaminant stars at the same T$_{\rm eff}$. More specifically, if a valid EWLi measurement (or upper limit) was present for a star in the T$_{\rm eff}$ range indicated in Table~\ref{lifilter}, 
but was below the T$_{\rm eff}$-dependent threshold, 
then the star was rejected.
The separation between field stars and cluster members becomes smaller at older ages and Li selection was not applied for NGC~6633.
  \item Targets with $[{\rm Fe}/{\rm H}]<-0.5$ were excluded, since all of the clusters have approximately a solar metallicity 
  (see Sect.~\ref{spectro}). A very conservative threshold was adopted here.
\item For targets with valid v$\sin i$ estimates, if v$
  \sin i < 3$ km\,s$^{-1}$ (unresolved broadening at the GES spectral resolution) the target was rejected as unlikely to be a cluster member. This threshold was  only applied to those stars outside the T$_{\rm eff}$ range for which the lithium test was valid. The rationale here is that such stars are either relatively hot, in which case they do not spin down with age in the same way as cooler stars; or, they are so cool that their spin down timescales are extremely long, such that no slow rotators are
expected at the ages of the clusters considered here. Of course the unknown inclination angle might mean that a small fraction of rapidly rotating stars have a small v$\sin i$, but we were prepared to accept this loss in favour of a more efficient rejection of non-members. 
\end{enumerate}
The first of the above steps (gravity) identified the majority
of contaminants.
Lithium selection was very effective
in the youngest clusters, becoming less so in older clusters, yet still rejecting $\sim 10$ \% of the remaining candidates. 
Very few candidates were
removed because of their low metallicity or slow rotation.
\begin{figure*}
\centering
\includegraphics[width=13.0cm]{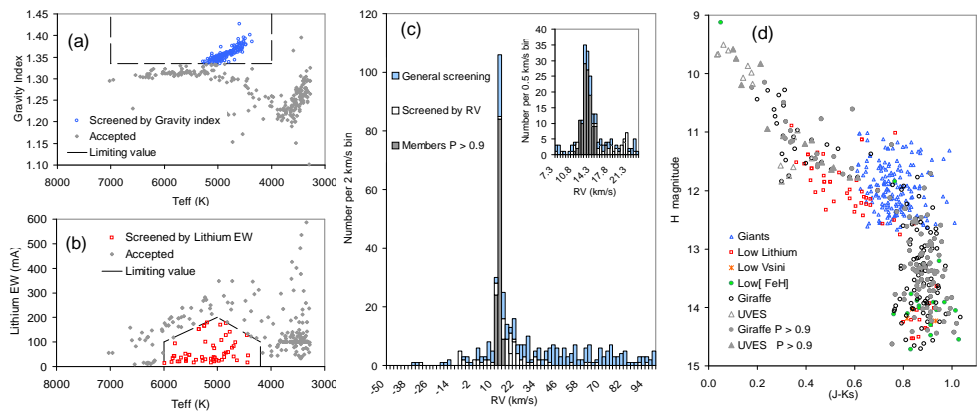}
\caption{Selection of likely cluster members of the open cluster NGC~2547. Plot (a) shows the initial selection by gravity index; grey diamonds being retained and blue circles being rejected as non-members. Plot (b) shows selection from the remaining targets according to lithium EW; red squares being rejected as 
non-members. Plot (c) shows a histogram of measured RVs. Blue targets were identified as non members by the general screening (gravity index, lithium 
EW, vsini and [FeH]). Grey targets have a probability larger than 90\% of membership based on their measured RVs. White targets have a lower probability of membership and are discarded from the sample. Plot (d) shows the $H$ versus $J-K_{\rm S}$ diagram colour coded 
according to the target selection status. Solid grey circles and triangles identify targets classed as likely cluster members retained for subsequent analysis.
}
\label{fig_sel}
\end{figure*}
\subsection{Radial velocity distributions and membership probabilities}
The radial velocity distributions of the filtered candidate members for
each cluster were modelled using the maximum likelihood technique
described in Jeffries et al. (2014), originally proposed by Pryor \& Meylan (1993) and updated by Cottar et al. (2012) to include the contribution of binaries. This technique assumes that the
observed radial velocities are taken from an intrinsic model broadened by 
the observational RV uncertainties and the effects of unresolved binaries; 
the latter have 
a distribution of RV offsets expected from a set of randomly oriented 
SB1 binary systems with a specified distribution of orbital periods, 
eccentricities, and mass ratios. 
The total likelihood of a star's observed
RV is then given by the sum of its likelihood if it were a single star and its likelihood if it were in an unresolved binary. 
\begin{equation}
\mathcal{L} = (1-f_\mathrm{bin}) \times \mathcal{L}_\mathrm{sing} + f_\mathrm{bin} \times \mathcal{L}_\mathrm{bin}
\label{eq:like_tot}
,\end{equation}
where $\mathcal{L}_\mathrm{sing}$ and $\mathcal{L}_\mathrm{bin}$ are the likelihood of single and binary stars and $f_\mathrm{bin}$ is the adopted
binary fraction. 
Given a model specified by a number of
free parameters (see below), the best-fitting model is found by
calculating the likelihood for each star and then maximising the summed logarithmic likelihood for all stars by varying the model parameters over a
grid of possibilities.

For the clusters considered here we constructed models that consisted
of multiple Gaussian components. One broad Gaussian was used to
represent the background of contaminating stars that are still likely
to be in the sample after filtering. The clusters themselves were
modelled as single Gaussians, except for
NGC~2547 and NGC~2451, where two Gaussians were required to adequately model the RV distribution. In the former case, the second kinematic population
was originally discovered by Sacco et al. (2015) and will not
be considered further
in the present paper. In the case of NGC~2451 the two populations
correspond to clusters A and B, which will be considered separately.
The characteristics of the binary population
were fixed in each case to be those of the solar-type field stars
estimated by Raghavan et al. (2010), with a binary fraction of 0.46.
The exact details of the assumed binary parameters make little difference to
the membership probabilities and ultimate selection of candidates to compare with isochrones.

Once the best-fitting parameters of the intrinsic distributions were found, the probability $p$ of individual stars belonging to the cluster component(s) was calculated, assuming
that each star must be a member of either the cluster component(s) or the background. We caution that these samples will not be cleaned of all contamination. For example, a star with $p=0.9$ still has a 10\%\ chance of being an unrelated contaminating star. For the comparison with theoretical models
we conservatively
retained all the objects with $p>0.9$, except for NGC~2415A and 
NGC~6633 where we considered $p>0.8$, as there are no stars with $p>0.9$, 
due to the less well defined RV peak and broader distribution.
\begin{table*}
\caption{Lithium selection criteria.}             
\label{lifilter}    
\centering
\begin{tabular}{lcccccc}
\hline\hline           
Cluster  & T$_1$ & T$_2$ & T$_3$ & EWLi$_1$ & EWLi$_2$ & EWLi$_3$ \\ 
         &  \multicolumn{3}{c}{(K)} & \multicolumn{3}{c}{(m\AA)}\\ 
\hline                       
All but NGC~2516&  4200 & 5000 & 6000 & 100 & 200 & 100\\
NGC~2516 &  4600 & 5000 & 6000 & 50  & 100 & 50  \\
NGC~6633 &  4600 & 5000 & 6000 &  0  &  0  & 0   \\
\hline                                   
\end{tabular}
\tablefoot{Targets were rejected if EWLi (or its upper limit) were below the lines defined by the coordinates (T$_1$,EW$_1$), (T$_2$,EW$_2$) and (T$_3$,EW$_3$). Targets with T$_{\rm eff}<\rm T_1$ or T$_{\rm eff}>\rm T_3$ were not rejected on the basis of EWLi.}
\end{table*}

As an example, the selection sequence and the results for the cluster NGC~2547 are summarised in Fig.~\ref{fig_sel}.
\section{TGAS analysis: Cluster membership and parallaxes}\label{TGAS}
TGAS-based parallaxes for all clusters included in this paper except NGC~2451B have been published in VL17. 
Nevertheless, we re-derived them here
to independently test our methodologies and because we aim
to perform a uniform analysis which would also include NGC~2451B.

Our approach is based on the trigonometric parallaxes and proper motions of the bright stars included in the TGAS catalogue.
We first compiled samples of previously known
cluster candidates brighter than $V\sim 12$ mag for each cluster, using catalogues available in the literature. These lists are used as initial guesses
for the TGAS membership analysis; hence,
we adopted a conservative approach and considered only high-probability 
members from the literature. Also, these samples
do not include GES targets, which, as noted,  are typically fainter than the 
TGAS limit.
We then cross-matched these samples of likely cluster members with 
the TGAS catalogue, finding at least ten members for each cluster with available
astrometric parameters. 

As mentioned, the samples of literature members were used to compute 
the initial estimates of the astrometric parameters
(i.e. the basic statistics of the proper motions and parallaxes) needed to 
refine the selection of the candidate
cluster members from the whole TGAS catalogue.
After that, we applied two complementary methods to analyse the TGAS subsets and to estimate the cluster trigonometric parallaxes. 
The first procedure relies on a conservative selection of a sample of {\it bona fide} astrometric members, based on tight confidence levels that minimise the contamination of the false positives.
Conversely, the second procedure adopts larger thresholds and fits the parallax distribution of a larger sample of TGAS sources with a model representing both cluster and field stars.
In the following, we describe in detail these two procedures and compare the results in Table~\ref{table:1}.
\subsection{Analysis of the `bona fide' cluster members}
\label{sect:method1}
We selected TGAS sources centred on the nominal cluster position listed in Table~\ref{cluster_sample} and 
within a search radius between $0.7^\circ$ and $2^\circ$; the radii correspond to 4-6 pc at the expected distances of the clusters. 
The smaller values were usually adopted for the clusters with proper motions that were not well separated from the field stars.
Also,
since the search radii are of the order or smaller than the scale-length, $\sim 2^\circ$, of the astrometric systematic error in TGAS (Lindegren et al. 2016),
we can reasonably assume that a global zero point uncertainty 
will possibly affect our parallax and distance estimations.

After identifying all TGAS sources within the search area,
we checked their astrometric solution and
rejected stars with formal errors 
on parallaxes and proper motions larger than 0.6 mas and 3 mas~yr$^{-1}$, respectively. The adopted thresholds were chosen after visual inspection of the corresponding distributions, in order to remove outliers corresponding to anomalous astrometric solutions.
In addition, we selected only TGAS astrometric solutions based on a minimum of 50 CCD observations and affected by an {\it excess of noise}\footnote{This parameter represents the residual error of the single star fitting. See Lindegren et al. (2016) for further details.} $\epsilon< 1$ mas 
in order to discard astrometric binaries and other anomalous cases. 

Candidate cluster members were then
selected in the 3D space of parallaxes and proper motions, $(\varpi, \mu_\alpha^*, \mu_\delta)$,  by means of the following procedure. 
We first estimated the mean values and dispersions of the cluster members selected from the literature. Secondly, in order to remove false positives, 
we further rejected catalogue sources by applying a $2\sigma$ threshold on both proper motions and parallaxes. 
Then, we recomputed the first- and second-order
moments of the three dimensional (3D) distribution, including the 
correlation between proper motions and parallaxes, and for the final selection we assumed a probability distribution in the form of a 3D Gaussian ellipsoid:
\begin{equation}
\begin{array}{lll}
f(\varpi,\mu_{\alpha^*},\mu_\delta)&=& {\rm const}\, \cdot  e^{-\frac{1}{2}E(\varpi,\mu_{\alpha^*},\mu_\delta)}
\end{array}
\label{e1}
,\end{equation}
where $E$ is the function defined by: 
\begin{equation}
\begin{array}{lll}
&&\lefteqn{E(\varpi,\mu_{\alpha^*},\mu_\delta)= }\\
&&\\
&&\frac{R_{\varpi\varpi}}{R}\left(\frac{\varpi-\langle \varpi\rangle}{\sigma_\varpi}\right)^2+\frac{R_{\mu_{\alpha^*}\mu_{\alpha^*}}}{R}\left(\frac{\mu_{\alpha^*}-\langle \mu_{\alpha^*}\rangle}{\sigma_{\mu_{\alpha^*}}}\right)^2+\frac{R_{\mu_\delta \mu_\delta}}{R}\left(\frac{\mu_\delta-\langle \mu_\delta\rangle}{\sigma_{\mu_\delta}}\right)^2+\\
&&\\
&&2\frac{R_{\varpi \mu_{\alpha^*}}}{R}\left(\frac{\varpi-\langle \varpi\rangle}{\sigma_\varpi}\right)\left(\frac{\mu_{\alpha^*}-\langle \mu_{\alpha^*}\rangle}{\sigma_{\mu_{\alpha^*}}}\right)+\\
&&\\
&&2\frac{R_{\mu_{\alpha^*}\mu_\delta}}{R}\left(\frac{\mu_{\alpha^*}-\langle \mu_{\alpha^*}\rangle}{\sigma_{\mu_{\alpha^*}}}\right)\left(\frac{\mu_\delta-\langle \mu_\delta\rangle}{\sigma_{\mu_\delta}}\right)+\\
&&\\
&&2\frac{R_{\varpi \mu_\delta}}{R}\left(\frac{\varpi-\langle \varpi\rangle}{\sigma_\varpi}\right)\left(\frac{\mu_\delta-\langle \mu_\delta\rangle}{\sigma_{\mu_\delta}}\right).
\end{array}
\label{e2}
\end{equation}
Here, $R$ represents the determinant of the symmetrical matrix $\mathcal{R}$ of the empirical correlation coefficients $\rho_{ij}=R_{ij} / R$ (for $i,j=\varpi,\mu_{\alpha^*},\mu_{\delta}$),
and $R_{ij}$ denotes the cofactor of the corresponding correlation element in $\mathcal{R}$ 
(Trumpler \& Weaver 1953).
A 2$\sigma$ threshold, $ E(\varpi, \mu_{\alpha^*}, \mu_\delta) < \zeta^2$ with $\zeta=2$,  corresponding to a confidence level (C.L.) of 73.8\% in 3D, was usually applied to select the {\it bona fide} cluster members, although a more conservative threshold of 1.5$\sigma$ (C.L. 48\%) was adopted for some
clusters (see Table~\ref{table:1}), 
to further minimise the contamination of field stars \footnote
{
The function in Eq.~2 represents a 3D Gaussian distribution, 
where $ E(\varpi, \mu_{\alpha^*}, \mu_\delta)$ is the sum of the squared and normalised residuals. }.

\begin{figure*}
\centering
\includegraphics[width=6.0cm]{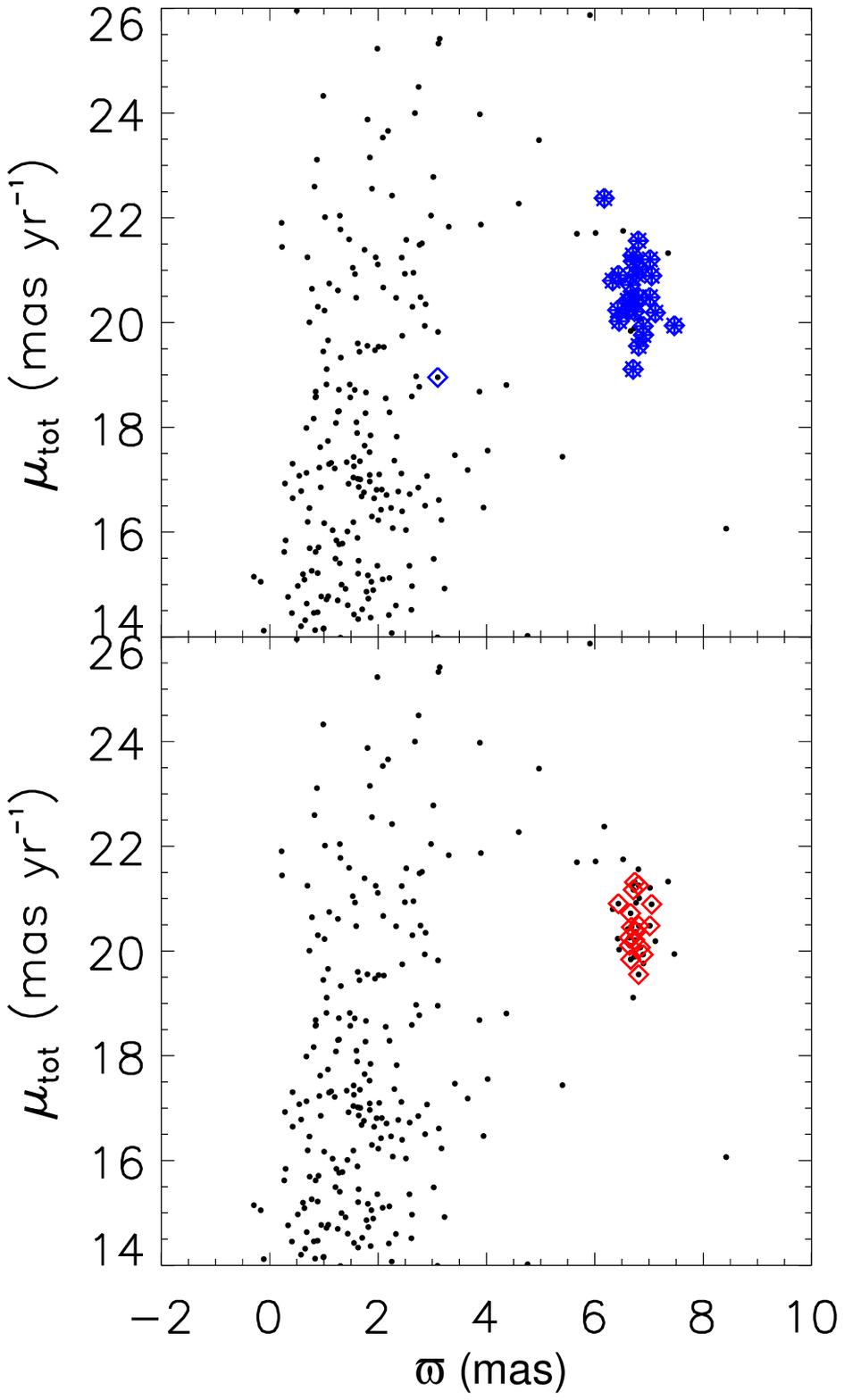}
\includegraphics[width=6.0cm]{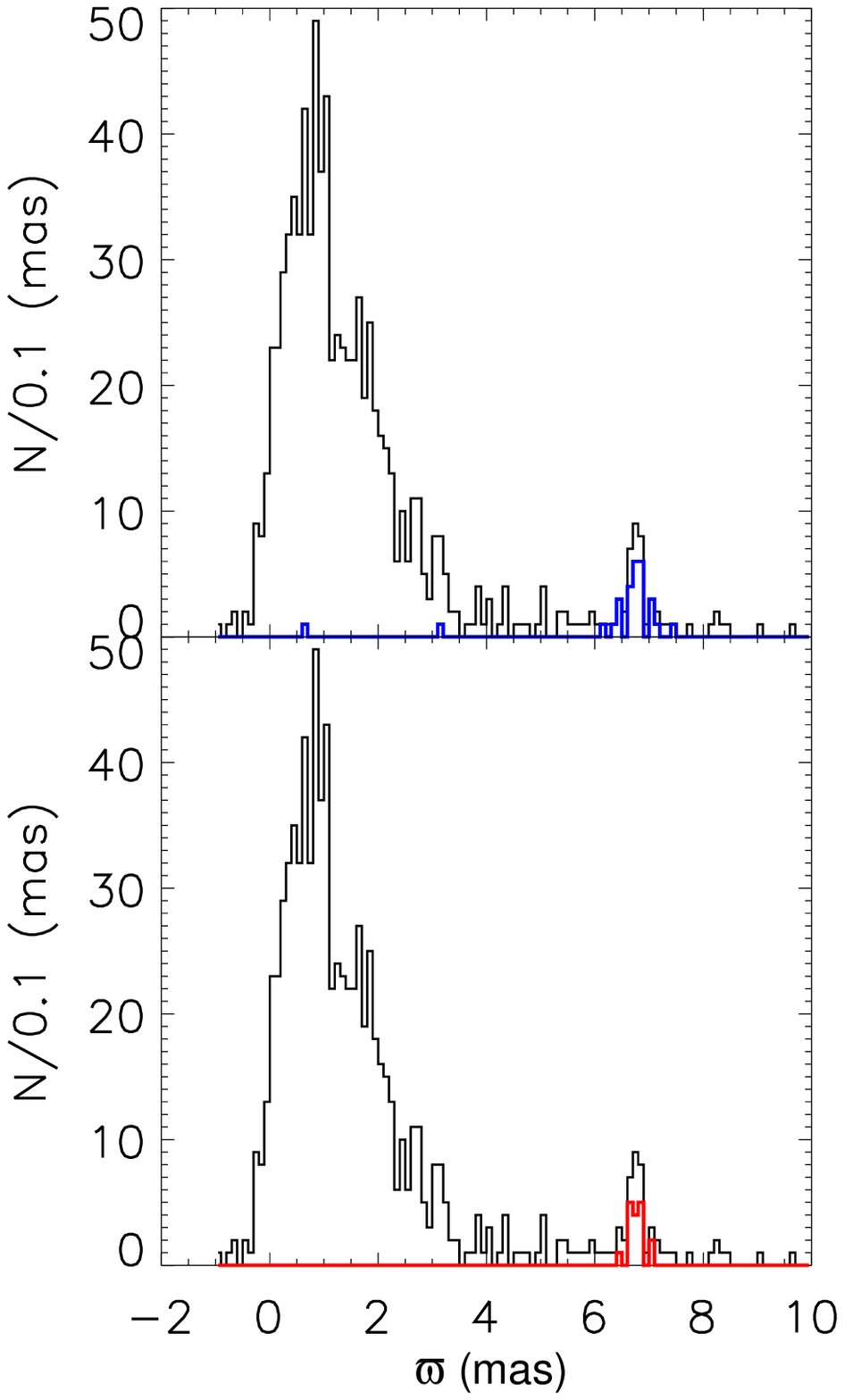}
\caption{Distribution of $\mu_{\rm tot} = \sqrt{\mu_{\alpha^*}^2 +\mu_\delta^2}$ vs.  $\varpi$ (left panel) and histogram of $\varpi$  (right panel) for IC~2602. The black colour represents the initial TGAS subset, while blue and red symbols mark the candidate members from the literature and those classified by the method described in Sect.~\ref{sect:method1}, respectively.}
         \label{fig:cluster1}
\medskip
   \includegraphics[width=6.0cm]{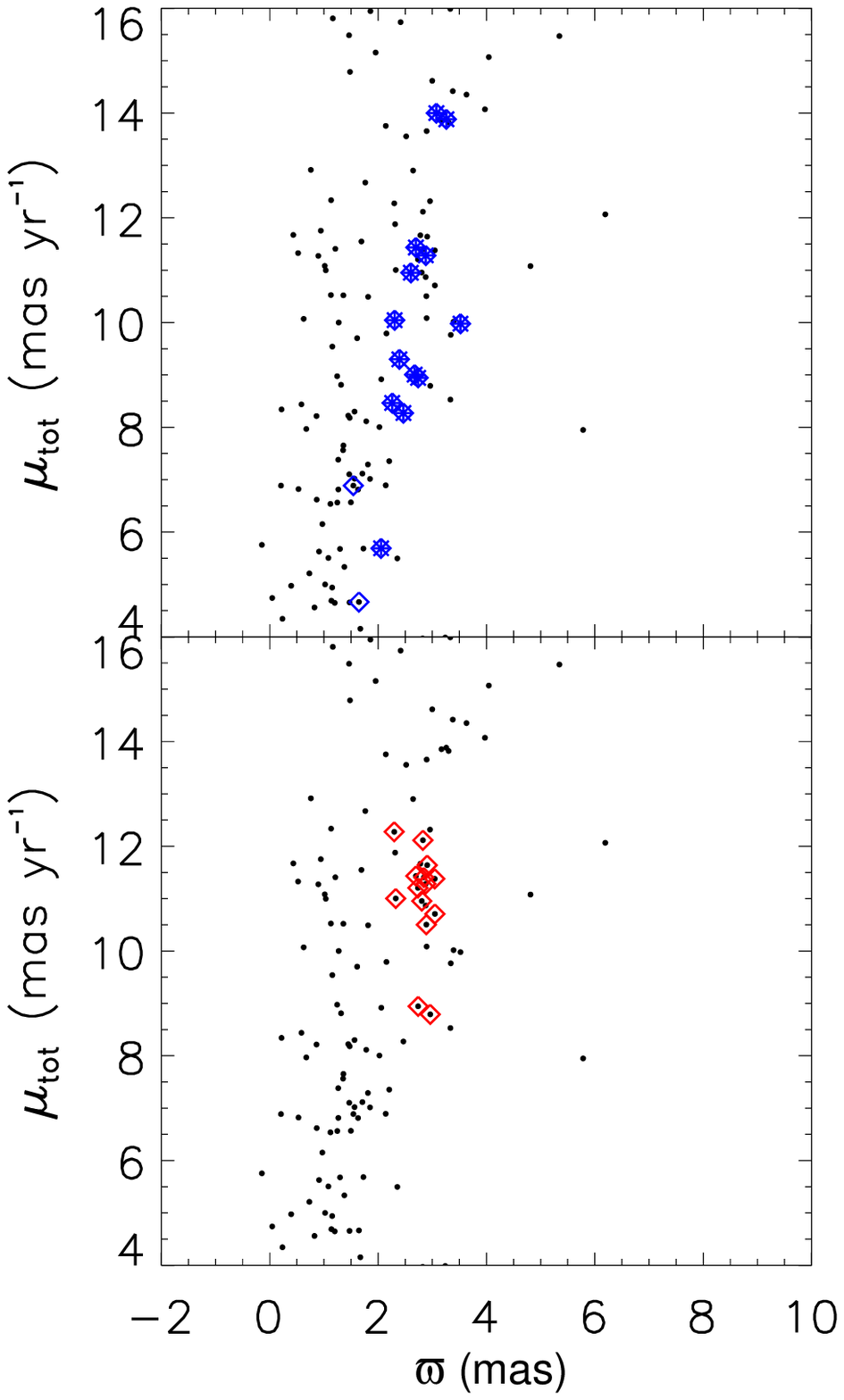}
  \includegraphics[width=6.0cm]{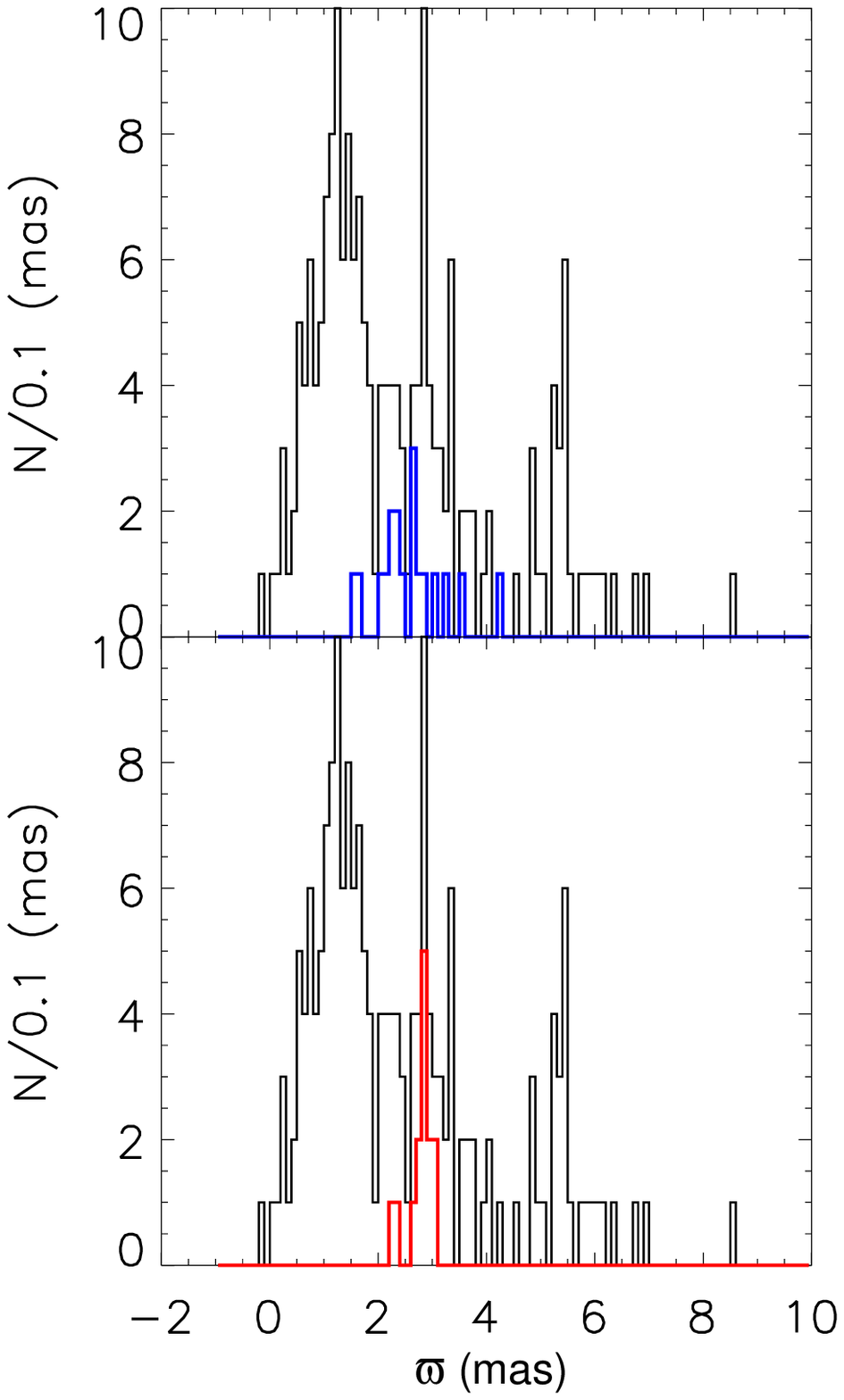}
      \caption{Same as Fig.~\ref{fig:cluster1}, but for NGC~2451B. 
     In this case, the efficiency of the astrometric selection in reducing the contamination of false positives classified as cluster members by previous studies 
      (blue symbols) is apparent. Notice also a third peak at $\varpi \sim 5.3 $ mas in the parallax distribution of the complete sample (black histogram), 
      which corresponds to members belonging to NGC~2541~A in the foreground.   }
         \label{fig:cluster2}
   \end{figure*}

The results of IC~2602 and NGC~2451~B are shown as examples in Figs.~\ref{fig:cluster1} and \ref{fig:cluster2}.
For IC~2602, the mean proper motions are well separated from the field stars, while for NGC~2451B, the mean proper motions overlap with the field. We also notice that, particularly for NGC~2451B, the candidate members collected from the literature (blue symbols) appear contaminated by several false members, 
in spite of our initial conservative selection;
nevertheless, the left panels of Figs.~\ref{fig:cluster1}-\ref{fig:cluster2} clearly show that the resulting astrometrically selected members (red symbols) 
match properly the cluster peak in the parallax distribution of the complete set of TGAS sources.

Finally, the cluster parallax was derived as a weighted mean of the 
parallaxes of the $n$ selected {\it members}: 
$ \langle\varpi\rangle = \left( \sum_{i=1}^{n} \varpi_{i}/\sigma_{\varpi,i}^2 \right) / \left(  \sum_{i=1}^{n} 1/\sigma_{\varpi,i}^2   \right)$.
 The corresponding error was estimated using the standard formula, 
$\epsilon_{\langle\varpi\rangle}^2 = \left( \sum_{i=1}^{n} 1/\sigma_{\varpi,i}^2 \right)^{-1}$,
which does not depend on the residuals and thus
is not biased due to
the cuts applied to the tail of the distribution by the procedure described 
above. 
We point out that the formal error on the weighted mean, $\epsilon_\varpi$, 
does not include the uncertainty due to the possible systematic errors 
affecting the TGAS catalogue. The latter was considered when converting 
parallaxes to distances.

The results are reported in Table~\ref{table:1} 
and compared to the independent results provided by the method based on the Gaussian maximum likelihood fitting that is described in more detail in the following section.
\subsection{Analysis of the loosely selected members}
In a different approach, for reference purposes, we pre-selected all TGAS stars
in the clusters of interest with positions, proper motions, and parallaxes
loosely compatible with those of the selected literature cluster members
described above. The selection was performed as follows:
$$
\frac{(\alpha - \alpha_0)^2 (\cos\delta)^2 }{r^2} +
\frac{(\delta-\delta_0)^2} {r^2} +
\frac{ (\mu_{\alpha^*}-\bar{\mu}_{\alpha^*})^2 } {(3\sigma_{\mu_{\alpha^*}})^2} +
$$
$$
+\frac{(\mu_\delta-\bar{\mu_\delta})^2}{(3\sigma_{\mu_\delta})^2} +
\frac{(\varpi-\bar\varpi)^2}{(5\sigma_{\varpi})^2} < 1,
\label{eq:loosely}
$$
where $r$ is a generous radius for membership selection (1 deg), 
$\alpha_0$ and $\delta_0$ are the centre
coordinates, $\bar{\mu}_{\alpha^*}$ and $\bar{\mu_\delta}$ are the median proper motions of the literature members with their dispersions, $\sigma_{\mu_{\alpha^*}}$ and $\sigma_{\mu_\delta}$, while $\bar{\varpi}$ is the median parallax with its dispersion, $\sigma_{\varpi}$.
The selection in parallax was less restrictive
than that on proper motions in order to have a large enough range of parallaxes to characterise the distribution of field stars.
%
\begin{figure}
\centering
\includegraphics[width=\columnwidth]{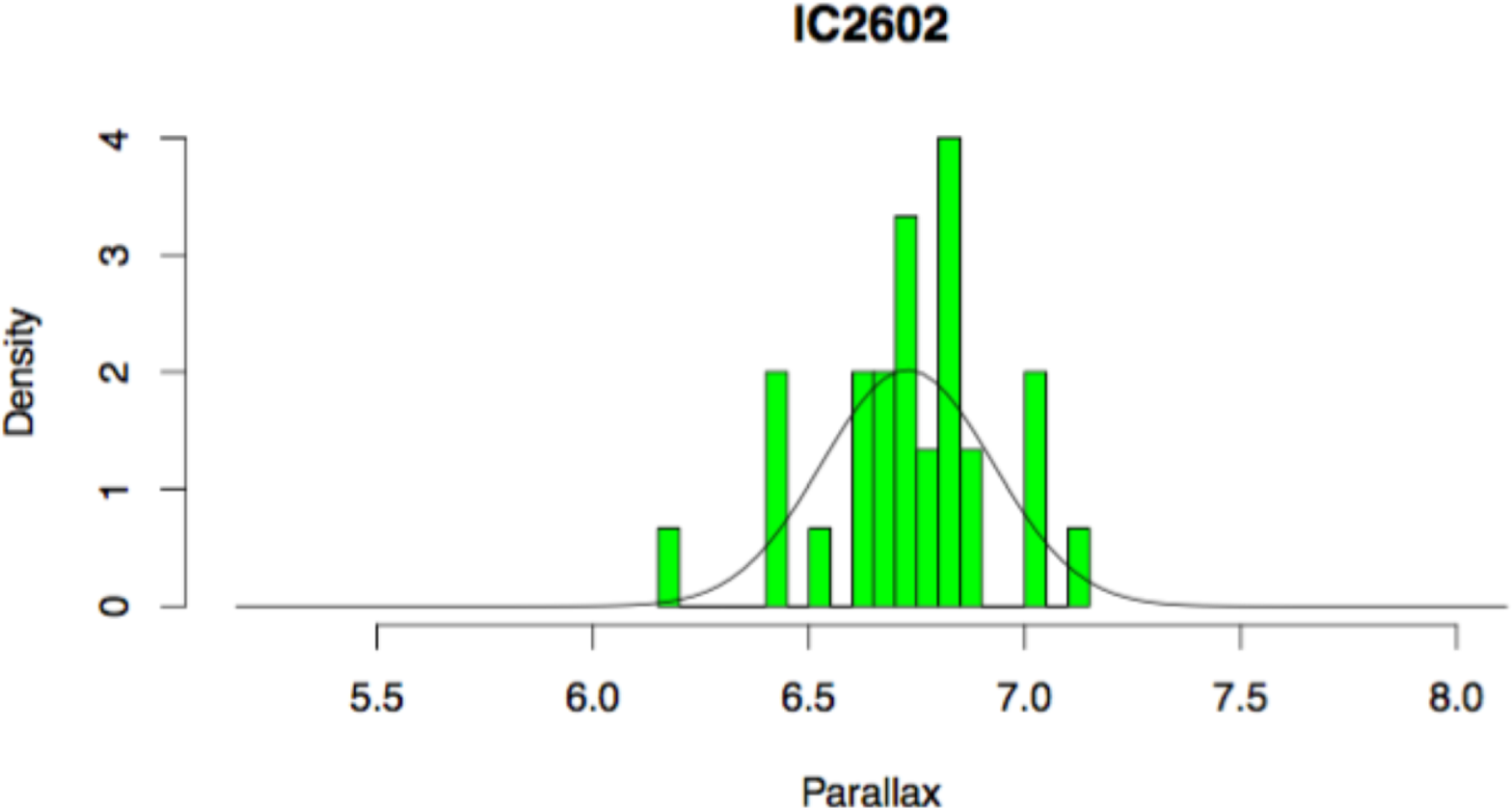}
\caption{Example of a one-Gaussian MLE fit for IC~2602 with the loosely selected
members method.}
\label{fig:2602}
\centering
\includegraphics[width=\columnwidth]{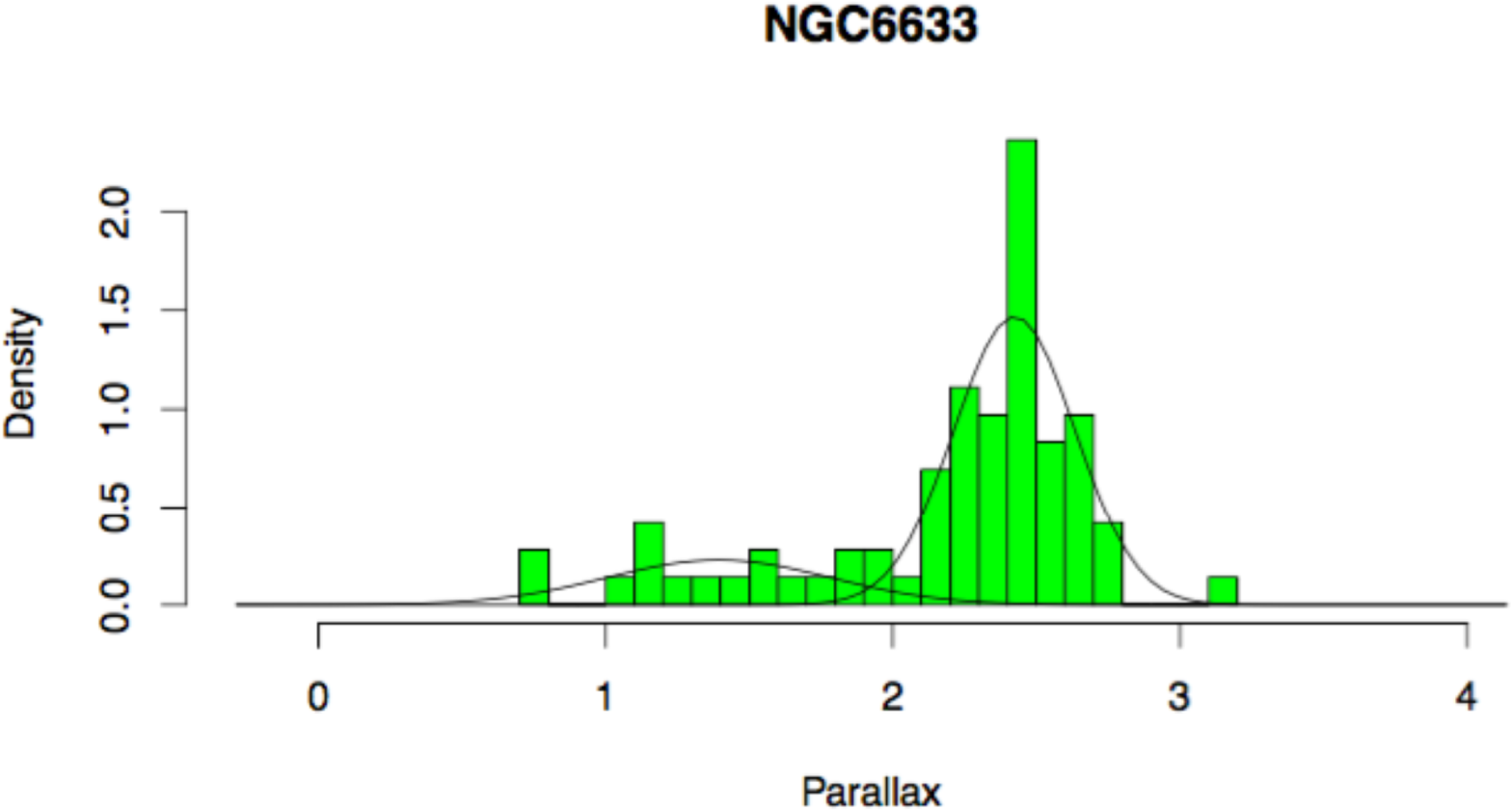}
\caption{Example of a two-Gaussians MLE fit for NGC~6633 with the loosely selected
members method.}
\label{fig:6633}
\end{figure}
We then fitted the distribution
of parallaxes of the selected stars
with both one Gaussian and the sum of two Gaussians with different mean 
and sigma, and maximum likelihood estimation (MLE). Specifically, 
we employed the likelihood estimator defined by Pryor \& Meylan (1993), 
Walker et al. (2006), and Martin et al. (2007), which takes into account the errors on measurements as well. Namely, the estimated uncertainty in the measurements was included in the model separately from the intrinsic dispersion of the cluster and background. We mention in passing that
this approach is very similar to the one adopted in Sect.~\ref{member_ges}
for the GES RV analysis, 
except that it does not take into account the presence of binaries.

In cases with a negligible field contamination
(IC~2602 being the best case, Fig.~\ref{fig:2602}), the one or two Gaussians fits gave the same result for the cluster parallax, within the uncertainties. In more difficult cases
(NGC~2451~B being the worst case), 
we observed distortions in the shape of the distribution, or multiple peaks and the fits with the one or two Gaussians yielded different values of the cluster parallaxes; in these cases we adopted the two-Gaussians solution.

We note that in the case of NGC~2451~B, the high field contamination
forced us to restrict the cluster radius to $0.8^\circ$ instead of $1^\circ$, and the cut in proper
motion to 2.2\,$\sigma$ rather than 3\,$\sigma$. We also mention that
the Gaussian distribution was not always the best model for the residual field
population after the loose pre-selection, but it granted an accurate positioning
of the cluster Gaussian by removing field stars in a satisfactory way after visual inspection (see the case of
NGC~6633 in Fig.~\ref{fig:6633}). 

The results are listed in Table~\ref{table:1}, where the number of
fitted Gaussians is also reported. 
The formal errors on the parallax ($\epsilon_\varpi$) were computed with the formulas by Pryor \& Meylan (1993), which take the TGAS parallax errors 
into account. The observed dispersions in the parallax distributions, $\sigma_{\varpi}$ in Table~\ref{table:1}, are significantly larger than the cluster's depth in parallax space (the latter being at most of the order of 10~$\mu$as for the closest clusters). As a result, the fitted intrinsic cluster dispersions with the adopted MLE formulation were negligible (close to zero), meaning that most of the observed spread was attributed to measurement errors. 
\subsection{Reference distances and comparison with other determinations}
Table~\ref{table:1} displays results that indicate an extremely good agreement between the
two complementary procedures, supporting the internal consistency of 
our results.
In particular, this means that false positives contaminating the selected cluster members do not significantly alter the mean parallax. 

The table also lists the parallaxes 
reported in the literature before {\it Gaia} (mainly derived from {\sc Hipparcos} data) and those derived by VL17.  Considering pre-{\it Gaia} determinations, the table shows that for five of the clusters our parallaxes agree well with previous determinations, while the agreement is less good for NGC~2516, NGC~2547, and NGC~6633. 
For the latter cluster
the parallaxes are only marginally inconsistent, given the large error of 
the {\sc Hipparcos} value, while
in the case of NGC~2516 the discrepancy is in principle
similar to the expected systematic ($\sim 0.3$ mas)
discussed by Lindegren et al. (2016). 
The disagreement for NGC~2547 is much larger; 
interestingly, in this case the distance modulus
derived from our parallax (see Table~\ref{table:2}) is consistent with that  ($m_0-M = 7.79^{+0.11}_{-0.05}$) derived by 
Naylor \& Jeffries (2006), by fitting the CMDs with isochrones.

The comparison with VL17 indicates an excellent agreement for most clusters. In particular, VL17 parallaxes for NGC~6633 and NGC~2547 are much closer to our own estimates than to those from {\sc Hipparcos}. However, a discrepancy between our results and VL17 is present for NGC~2451A and, again, NGC~2516. 
Whilst the difference for NGC~2451A (10~pc in distance) is not statistically 
significant, the discrepancy for NGC~2516 is greater than $5\sigma$ and 
equivalent to $\sim 50$~pc in distance.

In order to double check our parallax determination for NGC~2516, we ran again the procedure described in Sect. \ref{sect:method1}, but adopting
different 
initial conditions. Namely, we did not start the analysis from a list
of known literature members, but we considered the initial values $(\varpi,  \mu_\alpha^*,  \mu_\delta)_0$ from van Leeuwen (2009 --VL09), as done by VL17.
The 53 TGAS cluster members selected in this way within a radius of $1^\circ$ show a mean proper motion
$\mu_\alpha^* = -4.60 \pm  0.02~~{\rm mas yr}^{-1};~~~~    
 \mu_\delta= 11.19\pm 0.02~~{\rm mas yr}^{-1}$ and a mean parallax 
$\varpi = 2.57 \pm  0.04~~{\rm mas}$. 
The results are hence
in good agreement with the astrometric parameters reported in 
Table~\ref{table:1}, confirming the consistency of our procedure. 
We also note that
this sample includes eight bright {\sc Hipparcos} sources, whose astrometric parameters are expected to be more accurate than those of the Tycho~2 sources in TGAS. In particular, the correlation between the parallax and proper motions  are usually much smaller
for the {\sc Hipparcos} stars. 
As shown in Table \ref{hip}, all these stars have parallaxes in the range $2.2 < \varpi < 2.6$ mas, comparable to 
the mean parallax of the whole sample, although slightly smaller
(possibly due to colour and magnitude systematics). 
We hence conclude that our determination of the parallax of NGC~2516 
seems not only to be solid against initial assumptions, but also validated by the presence of the {\sc Hipparcos} stars. In 
Sect.~\ref{sec_comp} we will discuss the age and reddening estimates using VL17 parallax for this cluster.

Table~\ref{table:2} summarises the astrometric parameters of all the sample clusters resulting from the first procedure. 
We adopt these parameters as reference values in the following sections. 
%
\begin{table*}
\caption{TGAS cluster parallaxes as derived by our two independent procedures. When available, the parallaxes estimated by van Leeuwen (2009) and VL17 
are also listed for comparison.  }
\label{table:1}
\centering
\begin{tabular}{llrrrr|rrrrr|llll}
\hline\hline
   &  \multicolumn{5}{c|}{ {\it bona fide} cluster members} & \multicolumn{5}{c|}{Loosely selected members} & \multicolumn{2}{c}{\sc VL09} & \multicolumn{2}{c}{\sc VL17}\\
Cluster & 
$r$ & 
C.L. &
N & 
$\varpi$ &
$\epsilon_\varpi$ & 
N & 
$\varpi$ & 
$\epsilon_\varpi$ & 
$\sigma_\varpi$ & 
n$_{\rm{Gauss}}$ &
$\varpi$ & 
$\epsilon_{\varpi}$ & 
$\varpi$ & 
$\epsilon_{\varpi}$ \\
 & (deg) &$\sigma$ &  & (mas) & (mas) &  & (mas) & (mas) & (mas) & & (mas) & (mas) & (mas) & (mas) \\
\hline
NGC~2451A & $2$ & 2.0 & $19$ & $5.293$ & $0.069$ & 25 & 5.266 & 0.061 & 0.504 & 1 & $5.45$ & $0.11$ & 5.59 & 0.11\\
NGC~2451B  & $1$ & 1.5  & $14$ & $2.773$ & $0.089$ & 46 & 2.763 & 0.041 & 0.354 & 2 & $2.72^a$ & $0.44^a$ & -- & --\\
NGC~2516  & $0.7$ & 1.5 & $22$ & $2.586$ & $0.063$ & 33 & 2.588 & 0.052 & 0.467 & 1 & $2.92$ & $0.10$ & 2.99 & 0.08\\
NGC~2547  & $1.5$ & 1.5 & $10$ & $2.746$ & $0.091$ & 17 & 2.720 & 0.071 &  0.224 & 1 & $2.11$ & $0.17$ & 2.79 & 0.06\\
IC~2391  & $2$  & 2.0 & $17$ & $6.850$ & $0.064$ & 24 & 6.829 & 0.055 & 0.216 &  2 & $6.90$ & $0.12$ & 7.01 & 0.11\\
IC~2602  & 1.5 & 2.0 & $17$ & $6.757$ & $0.069$ & 30 & 6.749 & 0.051 & 0.198  & 1 & $6.73$ & $0.09$ & 6.74 & 0.05\\
IC~4665 & $1$ & 2.0 & $10$ & $2.732$ & $0.091$ & 15 & 2.784 & 0.073 & 0.124 &  2 & $2.81$ & $0.27$ & 2.83 & 0.05\\
NGC~6633 & $0.7$ & 1.5 & $17$ & $2.432$ & $0.066$ & 72 & 2.438 & 0.032 & 0.212 & 2 & $2.67$ & $0.32$ & 2.37 & 0.03\\
\hline
\end{tabular}
\tablefoot{$a$: from Carrier et al. (1999)}
\end{table*}
\begin{table}
\caption{{\sc Hipparcos} stars, selected as {\it bona fide} members of NGC 2516.}
\label{hip}
\centering 
\begin{tabular}{c c c c} 
\hline\hline
HIP ID & $G$ & $\varpi$ & $\sigma_\varpi$  \\ 
        &    mag  &   mas    &   mas     \\
\hline 
  38759 &  8.874 & 2.37  &   0.22 \\
  38739 &  8.163 & 2.41  &   0.24 \\
  38906 &  6.859 & 2.44  &   0.24 \\
  38966 &  7.138 & 2.20  &   0.24 \\
  39438 &  8.300 & 2.50  &   0.24 \\
  39562 &  7.683 & 2.62  &   0.23 \\
 120401 &  9.170 & 2.27  &   0.28 \\
 120404 &  7.615 & 2.28  &   0.24 \\
\hline 
\end{tabular}
\end{table}
\begin{table*}
\caption{TGAS astrometric parameters: mean cluster parallaxes, $\varpi$, and proper motions, $\mu_{\alpha^*}$, $\mu_\delta$, estimated from the selected {\it bona fide} cluster members as weighted averages.  
The derived spatial distance, $d=1/\varpi$, and the distance modulus, $DM=m_0-M=-5\log \varpi -5$, are also reported. The systematic errors  correspond to the nominal TGAS systematic uncertainty, $\epsilon_\varpi^S = 0.3$ mas. These 
dominate the error budget.
  }
\label{table:2}
\centering
\begin{tabular}{lcrrll}
\hline\hline
Cluster & 
$\varpi\pm \epsilon_\varpi$ & 
${\mu_{\alpha^*}}{\pm \epsilon_{\mu_{\alpha^*}}}$ & 
${\mu_\delta}{\pm \epsilon_{\mu_{\delta}}}  $ & 
$d\pm\epsilon_d^R\pm\epsilon_d^S$ & 
$DM\pm \epsilon_{DM}^R\pm \epsilon_{DM}^S$\\
& (mas) & $\rm{(mas~yr^{-1})}$ & $\rm{(mas~yr^{-1})}$ & (pc) &  (mag)  \\
\hline
NGC~2451A  & $5.293\pm 0.069$ & $-20.994\pm 0.186$ & $15.267\pm 0.168$ & $189\,^{+02}_{-02}$  $\,^{+11}_{-10}$ & $6.38\,^{+0.03}_{-0.03}$  $\,^{+0.13}_{-0.12}$\\
 &  &  &  &  & \\
NGC~2451B  & $2.773\pm 0.089$ & $-9.876\pm 0.305$ & $4.675\pm 0.252$ & $361\,^{+12}_{-11}$  $\,^{+44}_{-35}$ & $7.79\,^{+0.07}_{-0.07}$  $\,^{+0.25}_{-0.22}$\\
&  &  &  &  & \\
NGC~2516   & $2.586\pm 0.063$ & $-4.662\pm 0.026$ & $11.066\pm 0.027$ & $387\,^{+10}_{-09}$  $\,^{+51}_{-40}$ & $7.94\,^{+
0.05}_{-0.05}$  $\,^{+0.27}_{-0.24}$\\
&  &  &  &  & \\
NGC~2547   & $2.746\pm 0.091$ & $-8.935\pm 0.025$ & $4.388\pm 0.025$ & $364\,^{+13}_{-12}$  $\,^{+45}_{-36}$ & $7.81\,^{+0.07}_{-0.07}$  $\,^{+0.25}_{-0.23}$ \\&
&  &  &  & \\
IC~2391    & $6.850\pm 0.064$ & $-25.255\pm 0.029$ & $23.337\pm 0.029$ & $146\,^{+01}_{-01}$  $\,^{+07}_{-06}$ & $5.82\,^{+0.02}_{-0.02}$  $\,^{+0.10}_{-0.09}$\\
&  &  &  &  & \\
IC~2602    & $6.757\pm 0.069$ & $-17.213\pm 0.015$ & $10.544\pm 0.014$ & $148\,^{+02}_{-01}$  $\,^{+07}_{-06}$ & $5.85\,^{+0.02}_{-0.02}$  $\,^{+0.10}_{-0.09}$\\
&  &  &  &  & \\
IC~4665    & $2.732\pm 0.091$ & $-0.825\pm 0.023$ & $-8.565\pm 0.018$ & $366\,^{+13}_{-12}$ $\,^{+45}_{-36}$ & $7.82\,^{+0.07}_{-0.07}$  $\,^{+0.25}_{-0.23}$\\
&  &  &  &  & \\
NGC~6633   & $2.432\pm 0.066$ & $1.325\pm 0.038$ & $-1.698\pm 0.029$ & $411\,^{+11}_{-11}$  $\,^{+58}_{-45}$ & $8.07\,^{+0.06}_{-0.06}$  $\,^{+0.29}_{-0.25}$\\ 
&  &  &  &  & \\
\hline
\end{tabular}
\tablefoot{$R$: Random; $S$: Systematic}
\end{table*}
\section{Comparison between theory and observations}
\label{sec:comparison}
As mentioned in the introduction, 
the comparison between HRDs/CMDs and theoretical isochrones 
represents a key tool for
testing models of stellar evolution and to derive the cluster ages. 
Two of the main uncertainties 
affecting the comparison
of this procedure are the cluster distance and chemical composition; 
in the present case, GES analysis provides the metallicity and abundance ratios,
while the TGAS parallax measurements provide a direct cluster distance determination, enabling already a step forward. In the near future we expect that the Gaia parallaxes will be of much higher precision and uncertainties in the distance will become much less important, allowing even more detailed comparisons with the models.
Nevertheless, as we will show in the following sections, the TGAS data already allow us to test the methodology, to identify main sources of errors, and to achieve new results.
\subsection{The evolutionary models}
\label{sec:models}
In this paper we make use of and compare the results of some recent, updated, and widely used grids of evolutionary models. Namely, we chose the MIST database \citep[the MESA isochrone set, see][]{paxton2011,paxton2013,paxton2015} in the \citet{choi2016} and \citet{dotter2016} version\footnote{available at the link http://waps.cfa.harvard.edu/MIST/}, the PARSEC \citep[1.2S version, see][]{chen2014,bressan2012}\footnote{available at the 
link http://stev.oapd.inaf.it/cgi-bin/cmd}, and the PROSECCO models. We note that we only considered models that
cover the whole observed mass range of the sample clusters and include standard physics only.
Also, the PROSECCO evolutionary tracks were calculated specifically for this work by means of the Pisa version of the FRANEC evolutionary code \citep[][]{tognelli2011,dellomodarme2012,tognelli2015a,tognelli2015b}. 
For the comparison we adopted stellar models with [Fe/H]=$0$ and [$\alpha$/Fe]=0, compatibly with the results given in Table~\ref{tab_clu_abufe}.
All the grids of evolutionary models adopt updated input physics and estimates of the solar abundance mixture, summarised in  Table \ref{tab:inputmodelli}. We refer to the cited papers for a detailed description. 

The precise analysis of the effects of the different choices in the input physics is beyond the scope of the present work. Hence, we only briefly present in Appendix.~\ref{sec:models_conf} a comparison of the predictions of the selected sets of models to highlight the main differences among them, referring the interested reader to previous detailed studies on this topic \citep[i.e.][]{chabrier1997,siess2001,montalban2004,tognelli2011}. 
\begin{table*}
\caption{\label{tab:inputmodelli}Summary of the main characteristics of the selected grids of models. 
All the codes use a solar calibrated convection efficiency.} 
\centering
\begin{tiny}
\begin{tabular}{c|ccccccc}    
\hline\hline
\\
\textbf{Code: } & \textbf{EOS} & \textbf{Radiative Opacity}  & \textbf{Boundary Conditions} & \textbf{Convection} &  \textbf{BC} & \textbf{Y, Z}\\
 \hline
  & OPAL06 & OPAL & non-grey, $\tau_\mathrm{bc}=10$ & MLT & BT-Settl CFIST11\\
 \textsc{PROSECCO} & SCVH95 & F05 & BT-Settl AHF11  & \ml=2.00 & CK03 (T$_\mathrm{eff} \ge 10^4$K) & Y=0.274, Z= 0.013 \\
  & & (AS09) & CK03 (T$_\mathrm{eff} \ge 10^4$K) \\
\\
 & OPAL06 & OPAL  & non-grey, $\tau_\mathrm{bc}=100$ & MLT & ATLAS12/SYNTHE & \\
 \textsc{MIST} & SCVH95 & F05 & ATLAS12/SYNTHE & \ml=1.82 &  & Y=0.270, Z= 0.014 \\
 & & (AS09) & \citet{kurucz1993} \\
 \\
 & & OPAL & non-grey, $\tau_\mathrm{bc}=2/3$ & MLT & BT-Settl AHF11 \\
 \textsc{PARSEC}& FreeEOS & M09 &  BT-Settl AHF11 & \ml=1.7  & &  Y=0.274, Z= 0.013   \\
 & & (C11)\\
\hline  
\multicolumn{7}{l}{ }\\
\end{tabular}
\begin{flushleft}
\footnotesize
The columns provide the adopted: equation of state (EOS), radiative opacity (the heavy elements mixture is also specified), boundary condition characteristics, superadiabatic convection treatment, bolometric corrections (BC) and model chemical composition (the fractional abundance in mass of helium and metals, Y, Z). \\
References: AS09: \citet{asplund2009} solar mixture. C11: \citet{caffau2011} solar mixture.\\
BT-Settl AHF11: \citet{allard2011} with the AS09 solar mixture. BT-Settl CIFIST11: as the BT-Settl AHF11 but with the C11 solar mixture. 
CK03: \citet{castelli03}. ATLAS12/SYNTHE: \citet{castelli2005}.\\
F05: \citet{ferguson05} radiative opacity. M09: Marigo \& Aringer (2009) radiative opacity. OPAL: \citet{iglesias96}, updated in 2005. \\
FreeEOS: \citet{irwin08} EOS. OPAL06: \citet{rogers2002} EOS, updated in 2006. SCVH95: \citet{saumon1995} EOS.
\normalsize
\end{flushleft}
\end{tiny} 
\end{table*} 
\subsection{Bayesian analysis}
\label{sec:agemethod}
Our analysis was performed using a Bayesian maximum likelihood method based on a star by star comparison of observed data with isochrones, to simultaneously derive the reddening and the age for each cluster. The TGAS distances with their estimated uncertainties have been adopted. 
The maximum likelihood analysis is carried out using the 2MASS $J$, $H$ and $K_\mathrm{s}$ magnitudes that are available for the largest number of stars and are assumed to be independent from each other. After the analysis, as an additional check, the results are shown also in the ($V-K_\mathrm{s}$, $V$) and ($\log T_\mathrm{eff}$, $K_\mathrm{s}$) planes. We also discuss the effect of 
including the $V$ magnitude in the analysis.

The theoretical bases for the adopted recovery method are described in various papers in the literature; in particular we have followed the same formalism as \citet[][JL05]{jorgensen2005} and \citet[][GPT12]{gennaro2012}, to which we refer for the details of the methods. 
 
The total likelihood for each star of a given cluster has been defined as in
Eq.~\ref{eq:like_tot}.
The single or binary star likelihood for the i-th observed star is given by
\begin{equation}
\mathcal{L}_i = \prod_j^{N} \frac{1}{(2\pi)^{1/2} \sigma_{i,j}} \exp{-\frac{[q_j(\textbf{p})-\hat{q}_{i,j}]^2}{2\sigma_{i,j}^2}}
,\end{equation} where $\hat{q}_{i,j}$ and $\sigma_{i,j}$ are the j-th observed quantity and its related uncertainty for the i-th star and $q(\textbf{p})$ is the models prediction that depends on the set of parameters $\textbf{p}$.
The difference between the two contributions is that in the first term (single-star likelihood) the models are extracted from the standard isochrone, while, to evaluate the binary contribution ($\mathcal{L}_\mathrm{bin}$) to $\mathcal{L}$, the models are extracted from a binary sequence generated starting from the single star isochrone. We recall that to generate a binary sequence, one must also specify the secondary-to-primary component mass ratio, namely $q_\mathrm{bin}\equiv M_2/M_1$, which, in our cases, is not known a priori. We will discuss below the effect of adopting different values of $f_\mathrm{bin}$ and $q_\mathrm{bin}$. 

Similarly to JL05 and GPT12, we used the marginalised distribution to get the most probable value of each of the analysed parameters, which, in our case, are the age $\tau$ and the reddening $\varrho\equiv \rm{E}(B-V)$. To be more precise, the likelihood of each star depends on the stellar age, mass, and on the adopted reddening. We are not interested in deriving stellar masses; thus we integrated (marginalised) over it obtaining a likelihood that depends only on $\tau$ and $\varrho$. We do not use any prior on the mass distribution (i.e. flat prior). This procedure has been adopted for each star of a given cluster; then, the total likelihood of the whole cluster - which we used to recover the age and the reddening - is given by the product of each star likelihood marginalised on the stellar mass.

To derive the most probable value of one parameter (i.e. $\tau$ or $\varrho$), we integrated (marginalised) the likelihood over the other (i.e. to get the age we integrated over all the possible reddening values).
We defined the best value as the median of the marginalised distribution, differently from JL05 and GPT12 who adopted the mode. However, we verified that, in each analysed case, the distance between the mode and the median is much smaller than the final uncertainty on the derived parameter. We also performed
a test adopting a different definition of the best value and considering the mode of the total bi-dimensional likelihood without any marginalision process.
This corresponds to obtaining from the recovery the model with the most probable reddening and age, simultaneously;
in the case of the marginalised distributions, $\tau$ and $\varrho$ are 
instead calculated separately and they might 
not be consistent with the values that actually maximise the bi-dimensional likelihood. To better
investigate the issue, we simultaneously selected the ($\tau$, $\varrho$) 
vector corresponding to the maximum of the total likelihood and verified that 
these values are equal to the ones obtained using the marginalised distributions. In some cases there is a small difference between these two sets of values, but the difference is much smaller than the uncertainty on the estimated age and reddening. However, we preferred to use the marginalised distributions and the median to get the best value, as the total bi-dimensional likelihood is (in some cases) not very smooth and it presents spurious peaks close to its maximum.
\subsubsection{Uncertainties}\label{sec:errors}
The observables are affected by uncertainties, which in our case arise from magnitude and distance errors. The errors on $J$, $H$, $K_\mathrm{s}$ magnitudes are generally of the order of a few percent. Concerning the distance, the uncertainty slightly changes from one cluster to another (see Table~6), but it is smaller than about $15\%$ in the worst cases, which means an uncertainty on the distance modulus of less than 0.3 magnitudes. To properly evaluate the cumulative uncertainty on the estimated parameters due to the magnitude and distance errors, we adopted a Monte Carlo method similar to that used in Valle et al. (2016, 2017).
Namely, we assumed a Gaussian distribution around the central value for magnitudes and distance, with an
uncertainty (1 $\sigma$) obtained by averaging the positive and negative errors.
We discuss the effect of a non-Gaussian distribution of the error on the distance modulus at the end of this Section.
The observations were randomly perturbed within the Gaussian error for each star (stars are treated independently) and for each photometric band. Distance was also perturbed within its Gaussian error; the perturbation is the same for all the cluster stars, which are supposed to be at the same distance. 
The procedure was iterated to obtain 100 different datasets and thus 100 different estimated values of age and reddening. Among the 100 perturbed datasets we also included the unperturbed case. The most probable values of $\tau$ and $\varrho$ were selected as the median of the ordered sample of the 100 derived $\tau$ ($\varrho$), while the confidence intervals were assumed as the values corresponding to the 16 and 84 percentiles of the same ordered sample.
We verified, using the PROSECCO models in the ($J, H, K_\mathrm{s}$), plane that the recovered ages and reddening values do not change if 1000 datasets (more time consuming) were used instead of 100.

This procedure gives the cumulative uncertainty on the estimated parameters (age and reddening) due to both the photometric and distance errors. We repeated such an analysis keeping the magnitudes fixed to their central values and perturbing only the distance, to get the contribution to the parameter uncertainty due to the distance error only. As a general result, we found that the distance error has a significant impact in determining the total uncertainty on both the age and the reddening, in particular in those clusters with a relatively small
number of stars and/or scattered data. The sole distance error accounts for about 15-30\% (up to 60\% in the worst cases) of the total age uncertainty and about 30-60\% of the total reddening uncertainty. As discussed in more detail in Sect.~\ref{sec_comp}, there is obvious
room for a further increase of the precision of the results thanks to the  expected improvement in the {\it Gaia} parallaxes and photometry in subsequent data releases.

Table~\ref{table:2} shows that the errors on distance are actually
not symmetric, while, those on the distance modulus (that we actually used in the recovery) are, with a good approximation, symmetric. Small deviations are, however, present at the level of 0.02-0.04 mag. 
We then checked that the Gaussian approximation produced reliable results. 
To do this, we derived the age and reddening values considering the sole errors on the distance and adopting the central value of the DM plus or minus the errors listed in Table~\ref{table:2}. The uncertainties in the recovered parameters have been estimated by taking the difference between the age (reddening) obtained using the central value of DM and the two extremes derived using the maximum/minimum DM. We performed this analysis for all the clusters in our sample and 
found that this procedure yields
a confidence interval for age and reddening in excellent agreement with those
obtained using the Monte Carlo method: the differences between the values obtained adopting the two procedures are very small (at the level of a few percent) and
become negligible when the uncertainty on the photometry is included in the analysis. 
\subsubsection{Recovery of ages and reddening values}\label{sec:age}
The Bayesian analysis has been applied on sample clusters
using the PROSECCO isochrones.  
The first step of the analysis implied the construction of the grid of models to be employed in the parameters recovery. We started from a set of isochrones at [Fe/H]$=0$ and $[\alpha$/Fe]$=0$, as measured for the sample clusters, in the range 10-700~Myr. Where not explicitly stated, the models have been computed adopting a core overshooting parameter $\beta_\mathrm{ov} =0.150$, as recently obtained by comparing the Pisa models with the TZ Fornacis eclipsing binary system \citep{valle17}. 

The isochrone grid has been calculated with an age spacing suitable to obtain a good resolution for the age determination, namely,  $\delta\tau = 0.2$~Myr for $10\le \tau(\mathrm{Myr})\lt 25$, 0.5~Myr for $25\le \tau(\mathrm{Myr})\lt 40$, 1~Myr for $40\le \tau(\mathrm{Myr})\lt 100$, 5~Myr for $100\le \tau(\mathrm{Myr})\lt 300$ and 10~Myr for $\tau \ge 300$~Myr. For E$(B-V)$ we used a spacing of 0.0001. Concerning the prior distribution, we used a box function for both $\tau$ and E$(B-V)$. Reddening is allowed to vary with a flat probability distribution in the range $0 \le \rm{E}(B-V) \le 0.3$ to cover all the possible reddening values found in the literature for the selected clusters. For the age, the box width depends on the expected age.  More specifically, we adopted a conservative flat prior on the age for the three age intervals: $10-100$~Myr, $100-300$~Myr, and $300-700$~Myr, respectively. 

After the construction of the grid, for all the selected sets of models and all the ages, we generated the unresolved binary sequence starting from the single star isochrone. We adopted a fraction of binaries $f_\mathrm{bin}=0.3$ and $q_\mathrm{bin} = 0.8$, as reference. However, as an additional check, we also used different values for $f_\mathrm{bin}$ and $q_\mathrm{bin}$ (namely $f_\mathrm{bin}=0.1$ and 0.5; $q_\mathrm{bin}=0.5$ and 1.0), finding that both $f_\mathrm{bin}$ and $q_\mathrm{bin}$ affect the obtained age and reddening at a level smaller than the uncertainty on the estimated parameters. Finally, absolute magnitudes from the models
were converted into observed magnitudes using the DM
from the TGAS analysis and the extinction, estimated by adopting the linear 
relation between the extinction in a given band $A_{\overline{\lambda}}$ and E$(B-V)$ in terms of \citet{cardelli1989} coefficients $(A_{\overline{\lambda}}/{A_V})_\mathrm{C89}$. We used $R_V=3.1$.
\begin{figure*}
        \includegraphics[width=0.5\linewidth]{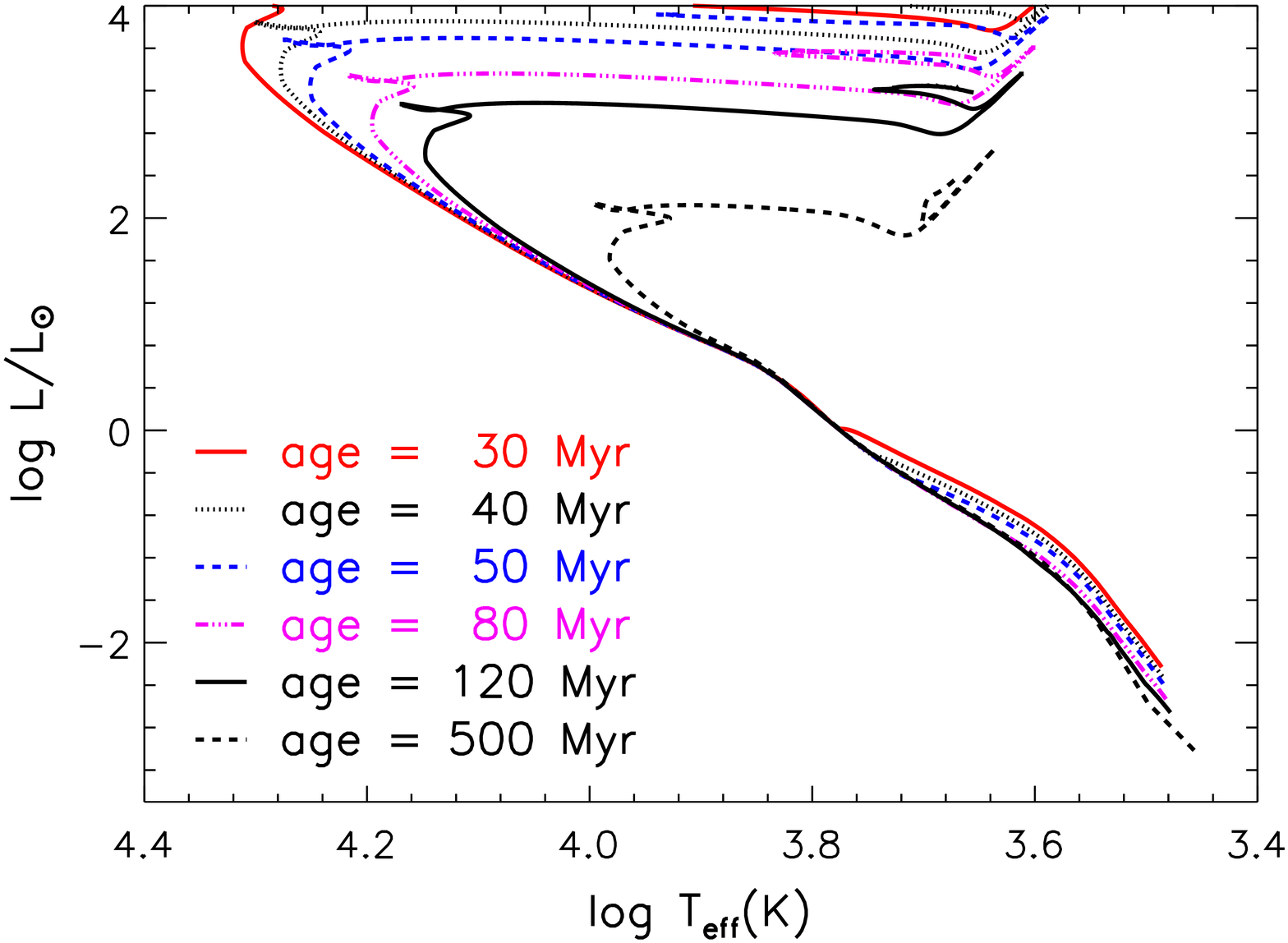}
        \includegraphics[width=0.5\linewidth]{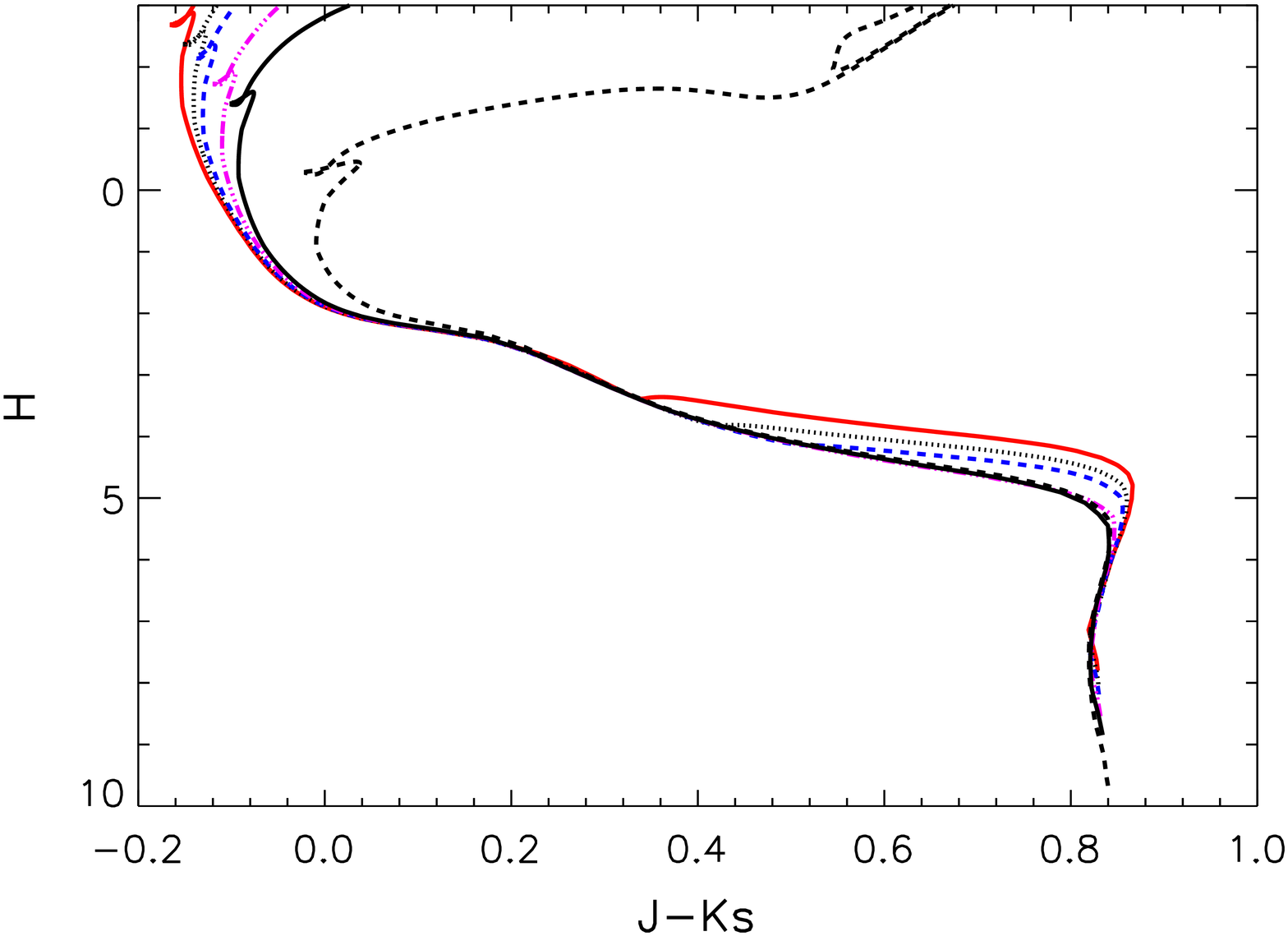}
        \includegraphics[width=0.5\linewidth]{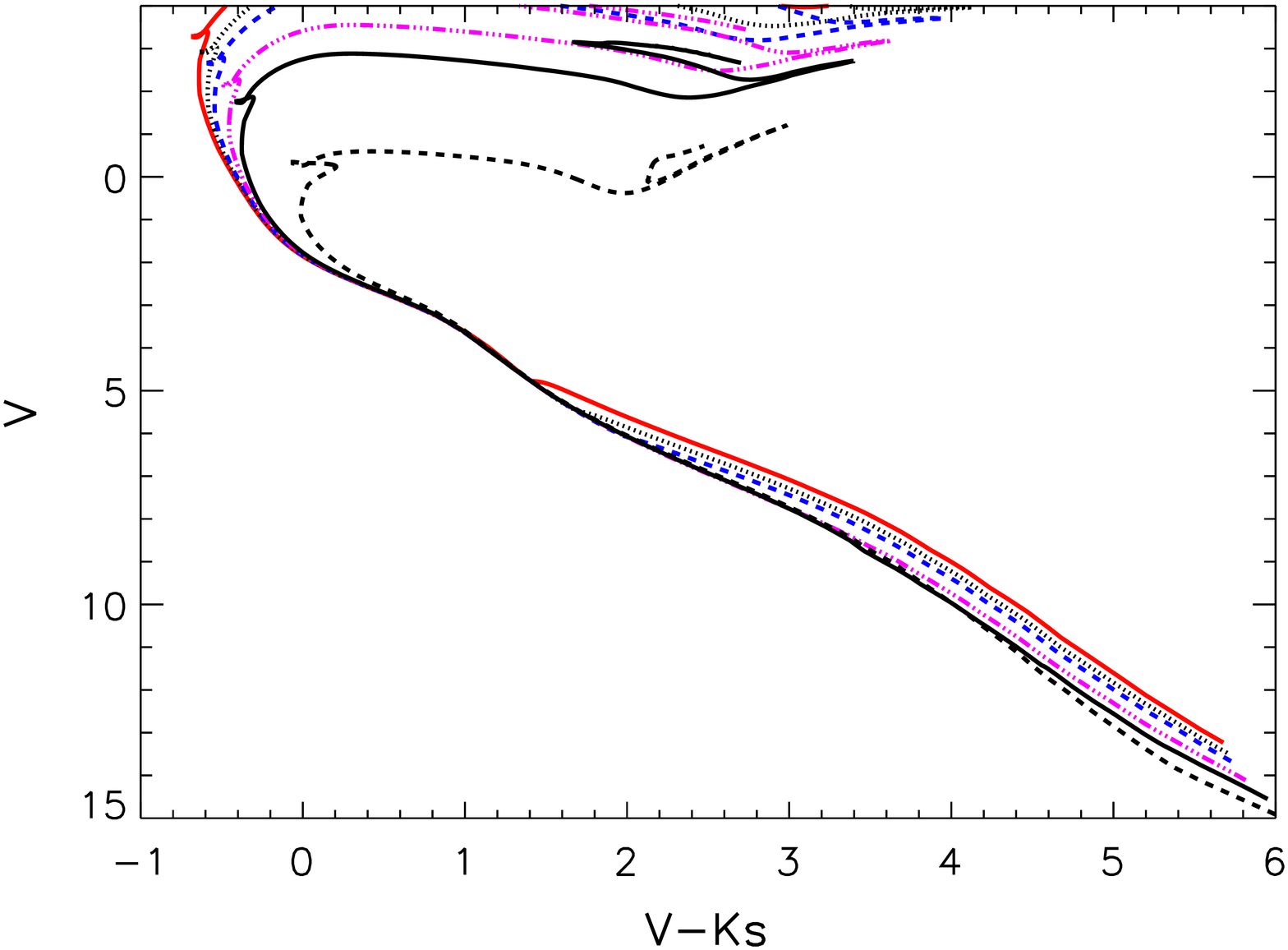}
        \caption{PROSECCO isochrones at different ages in the HR diagram, and 
the $H$ vs. $J-K_\mathrm{s}$ and $V$ vs. $V-K_\mathrm{s}$ planes.}
        \label{fig:age_cl}
\end{figure*}
The precision of the recovery procedure was tested by building synthetic clusters with artificial stars sampled from the same PROSECCO isochrone grid used for the recovery procedure. A Gaussian noise with a standard deviation equal to the assumed observational uncertainties was added to the magnitude and colour values for each generated star. This numerical experiment showed the capability of the method to recover the simulated age and reddening in the ideal case where the adopted stellar models are in perfect agreement with real stars and the observational sample is populated with sufficient number of points.

Figure~\ref{fig:age_cl} shows an example of PROSECCO isochrones of ages in the range of interest (30-500~Myr) in the different planes selected for the comparison with real clusters (as discussed in the following Section). Depending on the age and on the selected plane, different parts of the diagram are the main age indicators. In clusters younger than about 80-100~Myr low-mass stars are still sensitive to an age variation and are used to infer the age. For older clusters, low-mass stars are on the ZAMS and the age determination mainly relies on the brightest stars of the sample (if present).
 
We finally note that some of the sample clusters are characterised by very clean sequences, while others are more scattered, in spite of the {\it Gaia} and GES conservative membership analysis, suggesting that some contamination and/or large errors in photometry are still present. For each cluster, we tried to reduce the actual number of outliers using the cleaning procedure discussed in Appendix.~\ref{sec:cleaning}. 
\section{Results and discussion}
\label{sec:discussion}
\subsection{Recovered ages and reddening values}\label{sec:fits}
\begin{figure*}
        \centering
        \includegraphics[width=0.325\linewidth]{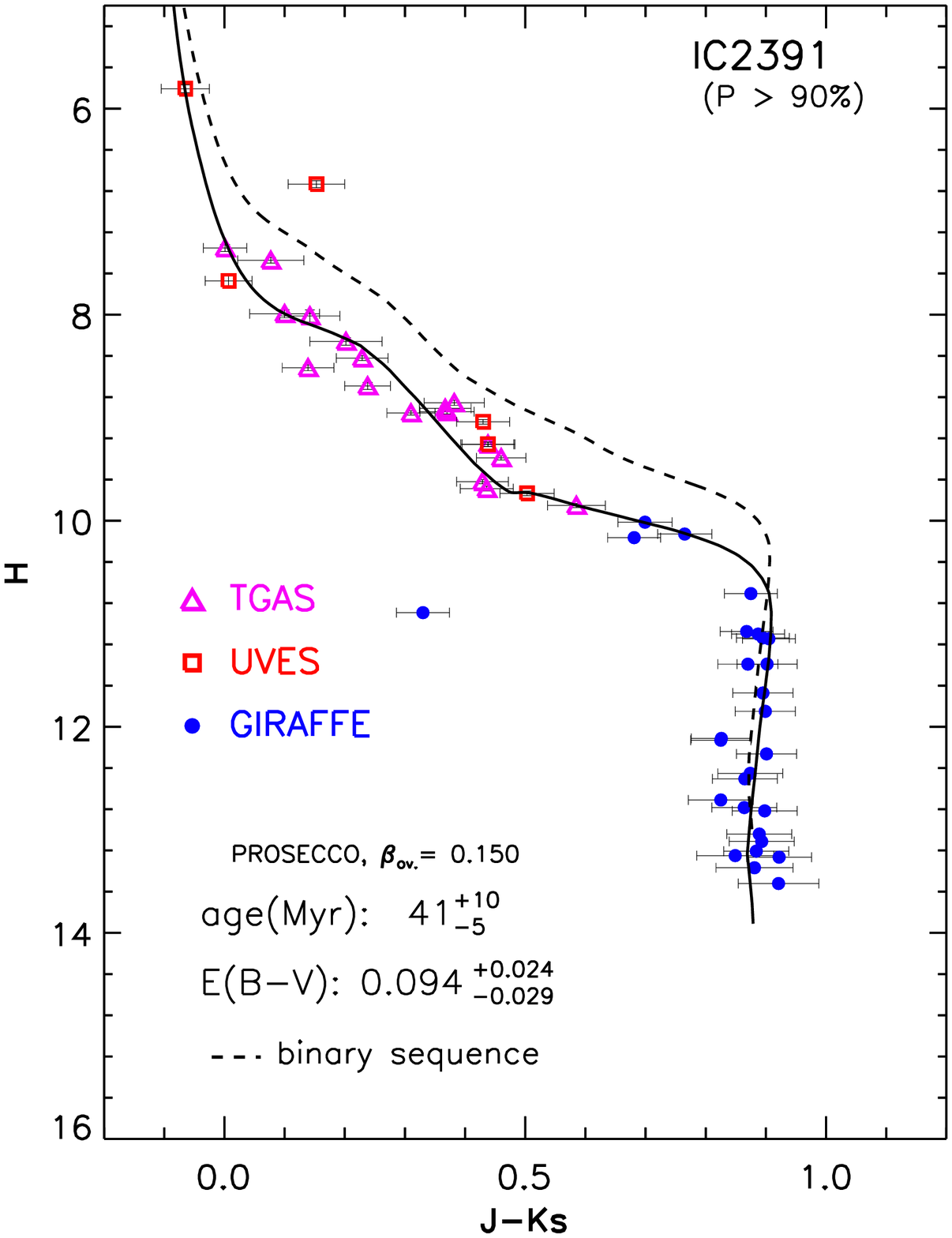}
        \includegraphics[width=0.325\linewidth]{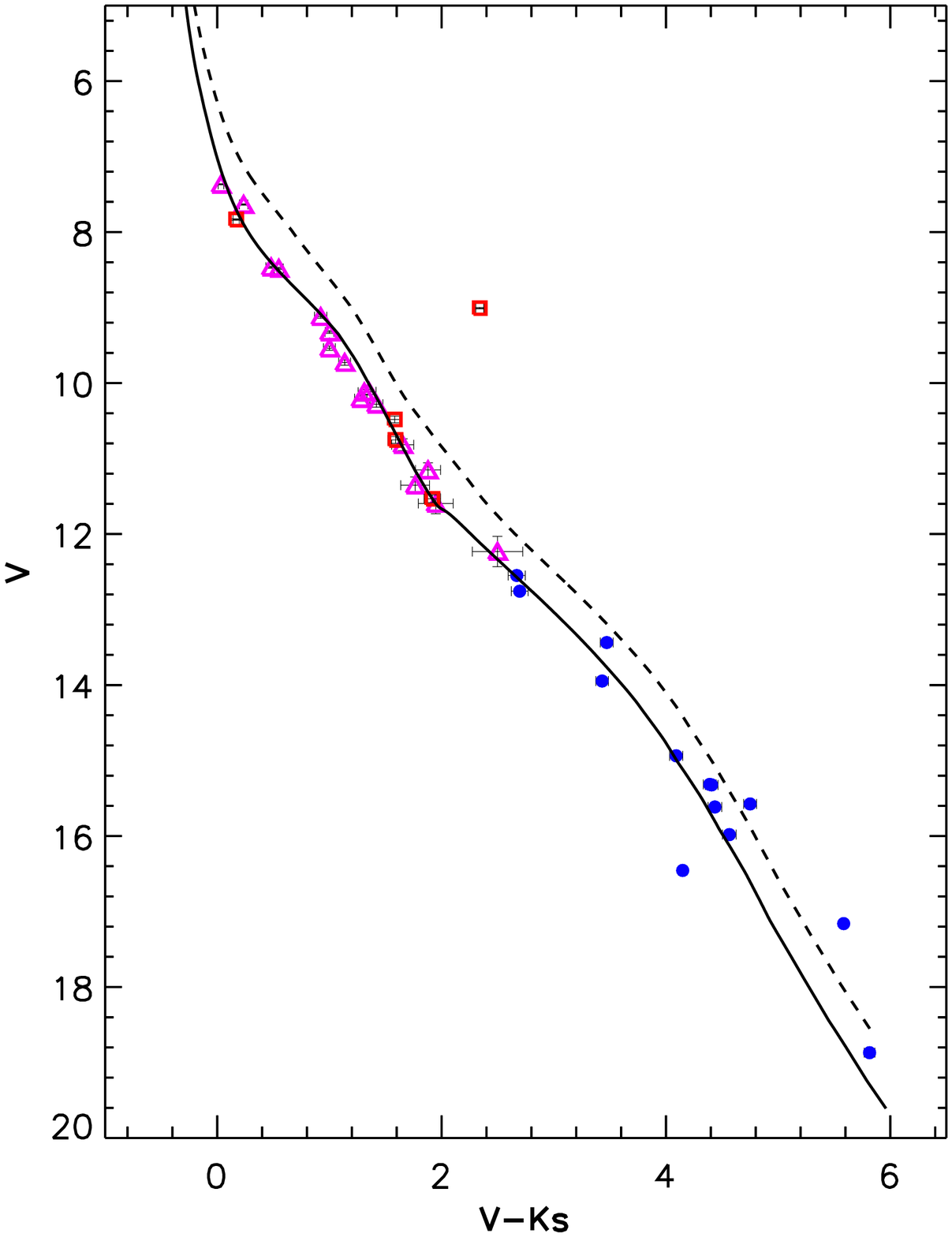}
        \includegraphics[width=0.325\linewidth]{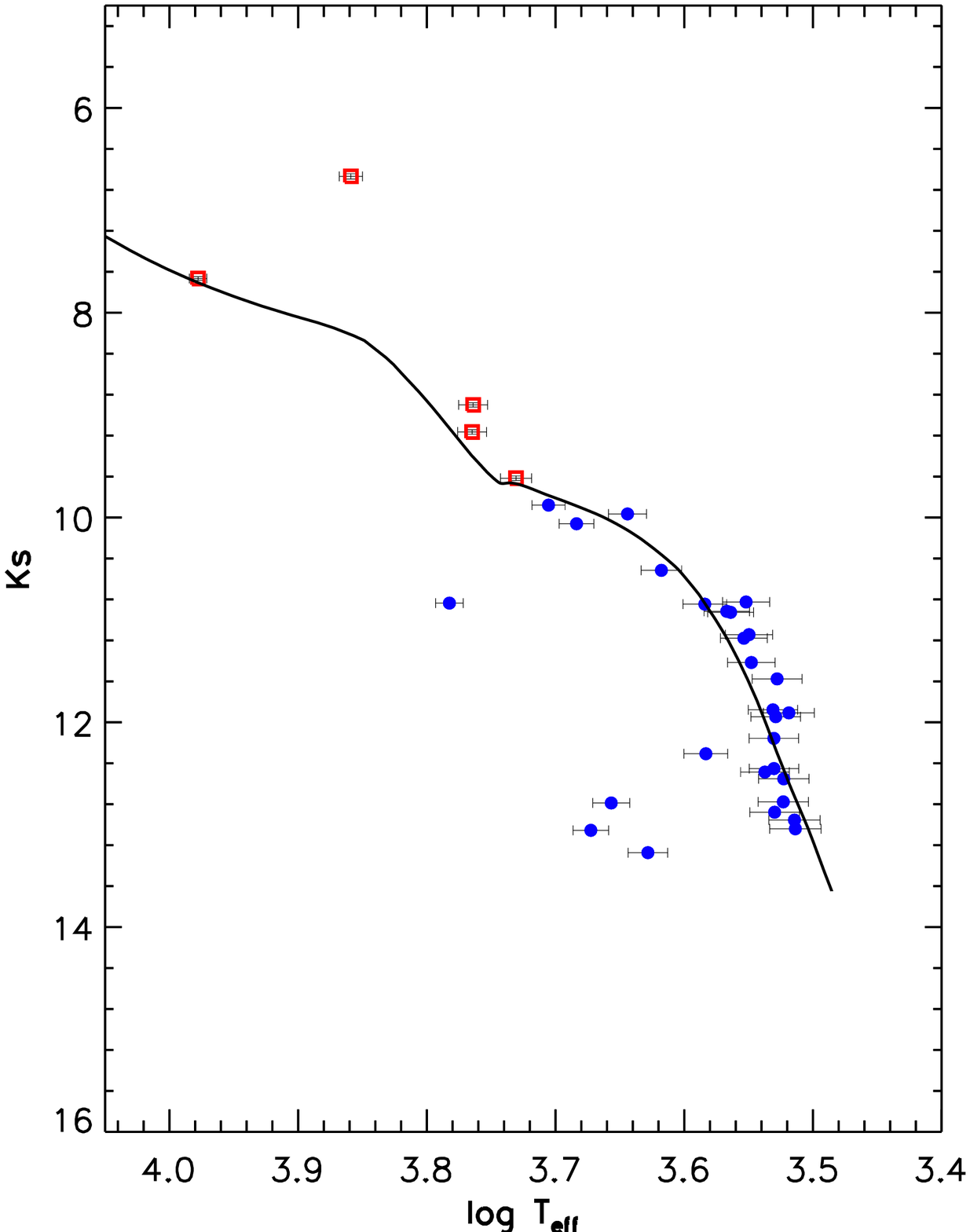}\\
        \includegraphics[width=0.325\linewidth]{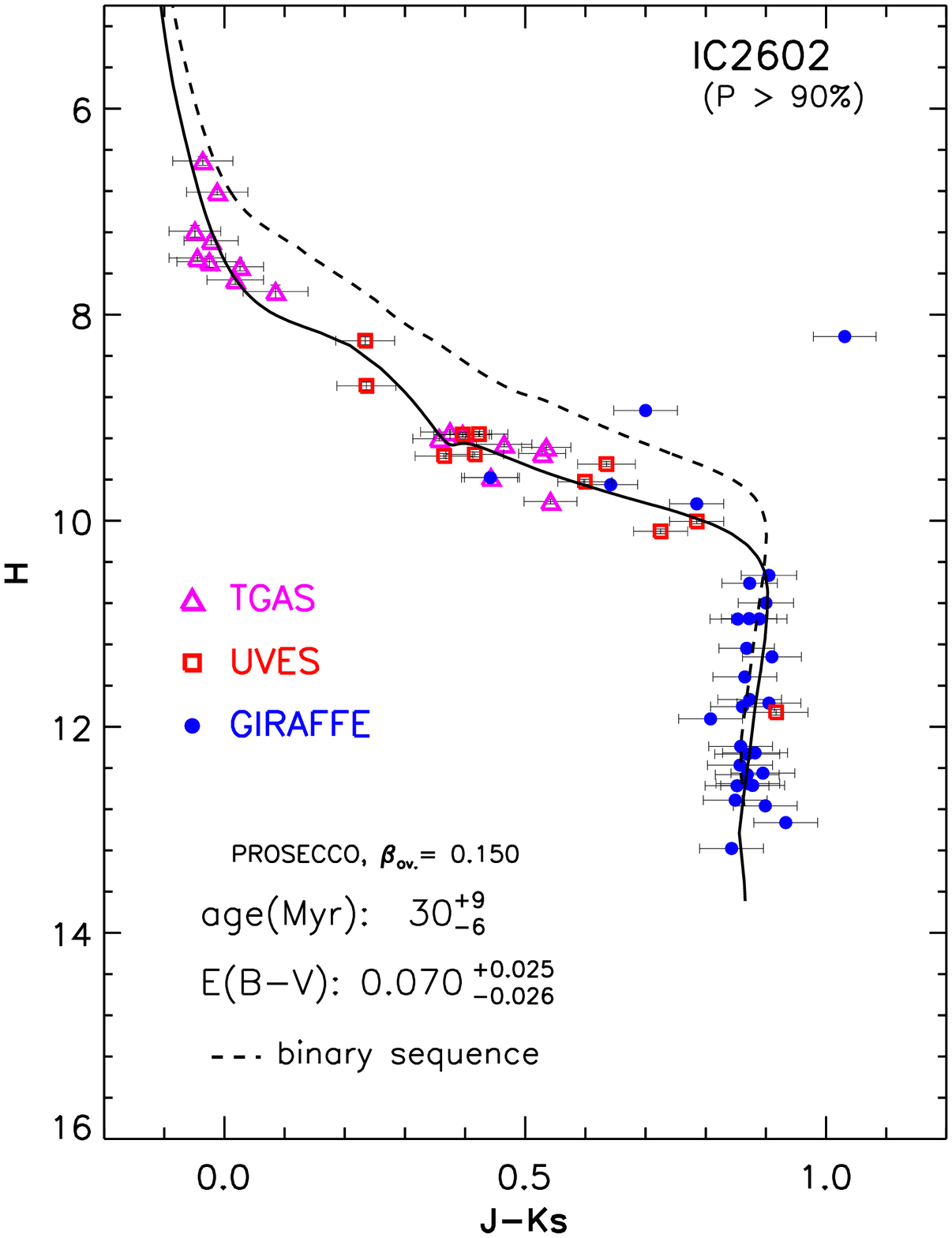}
        \includegraphics[width=0.325\linewidth]{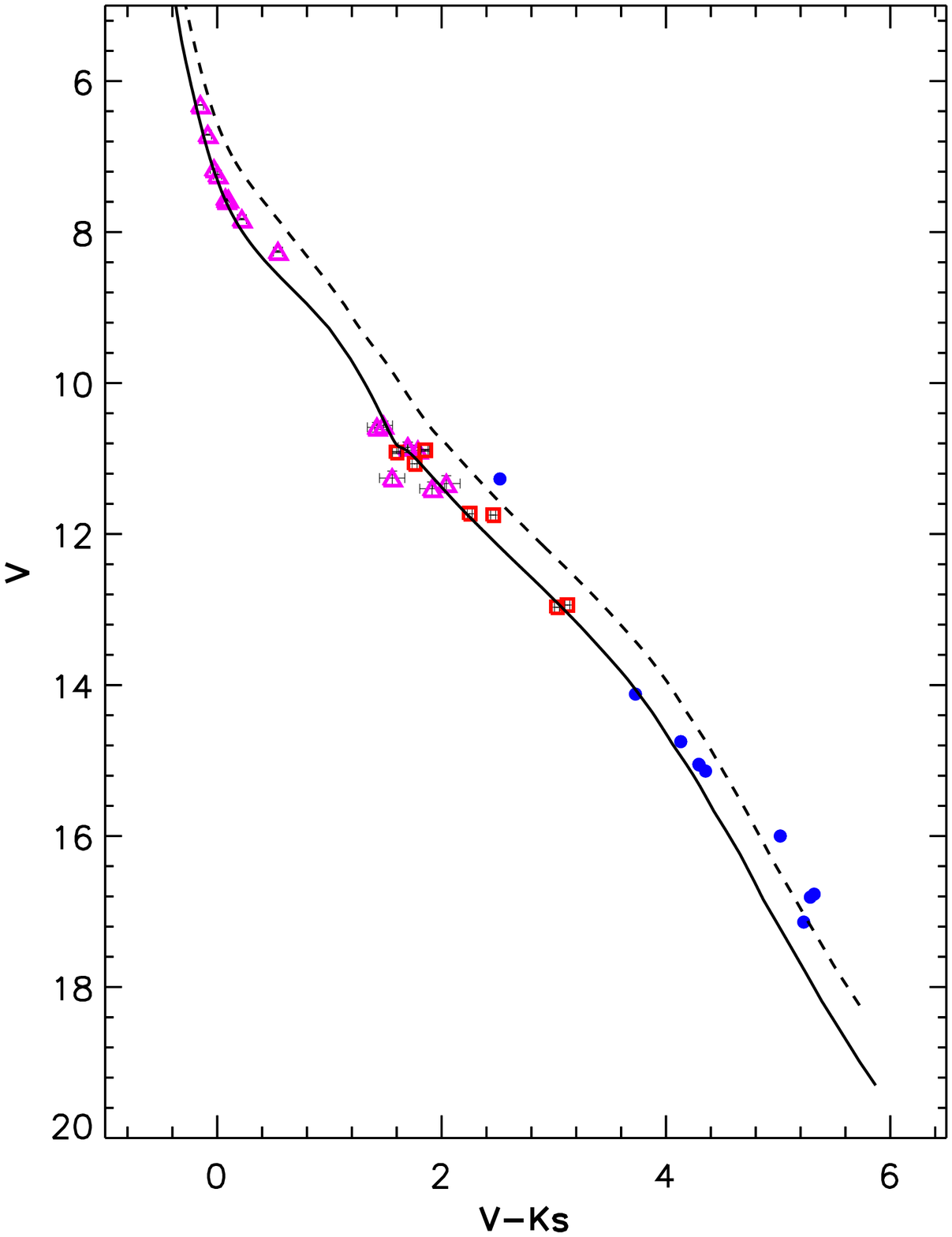}
        \includegraphics[width=0.325\linewidth]{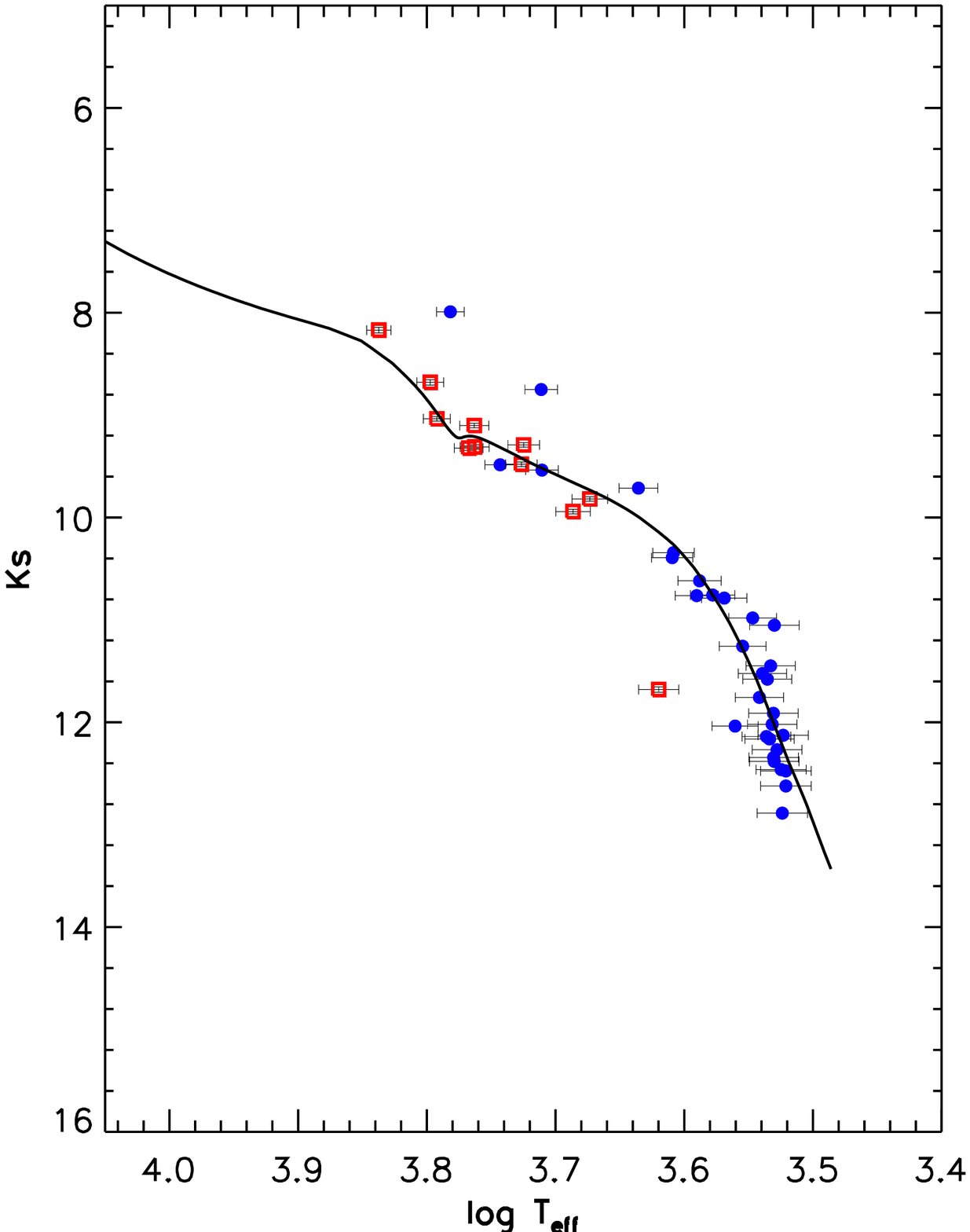}
\caption{Comparison between isochrones and observational data in the ($J-K_{\rm s}$,$H$), ($V-K_{\rm s}$,$V$), and ($\log T_\mathrm{eff}$, $K_{\rm s}$) diagrams for IC~2391 and IC~2602. Members from GES are indicated as blue full circles (GIRAFFE) and red squares (UVES), while TGAS members are plotted as magenta open triangles. Photometric data are taken from the 2MASS catalogue, AAVSO Photometric All Sky Survey (APASS) DR9 (Henden et al. 2016) and ASCC-2.5, 3rd version compilation (Kharchenko \& Roeser 2009). GES iDR4 effective temperatures are adopted for the $K_\mathrm{s}$ versus T$_\mathrm{eff}$ diagrams. Temperatures are instead not available for the TGAS members, which are not shown in this diagram. The estimated reddening and age values 
are indicated in the left-hand panels - see text for more details.
For both clusters, the unresolved binary sequence (with a constant mass $q_\mathrm{bin}=0.8$) is shown as a dashed line (see text).}
\label{fig:CMDPROSECCO_1}
\end{figure*}

\begin{figure*}
        \centering
        \includegraphics[width=0.325\linewidth]{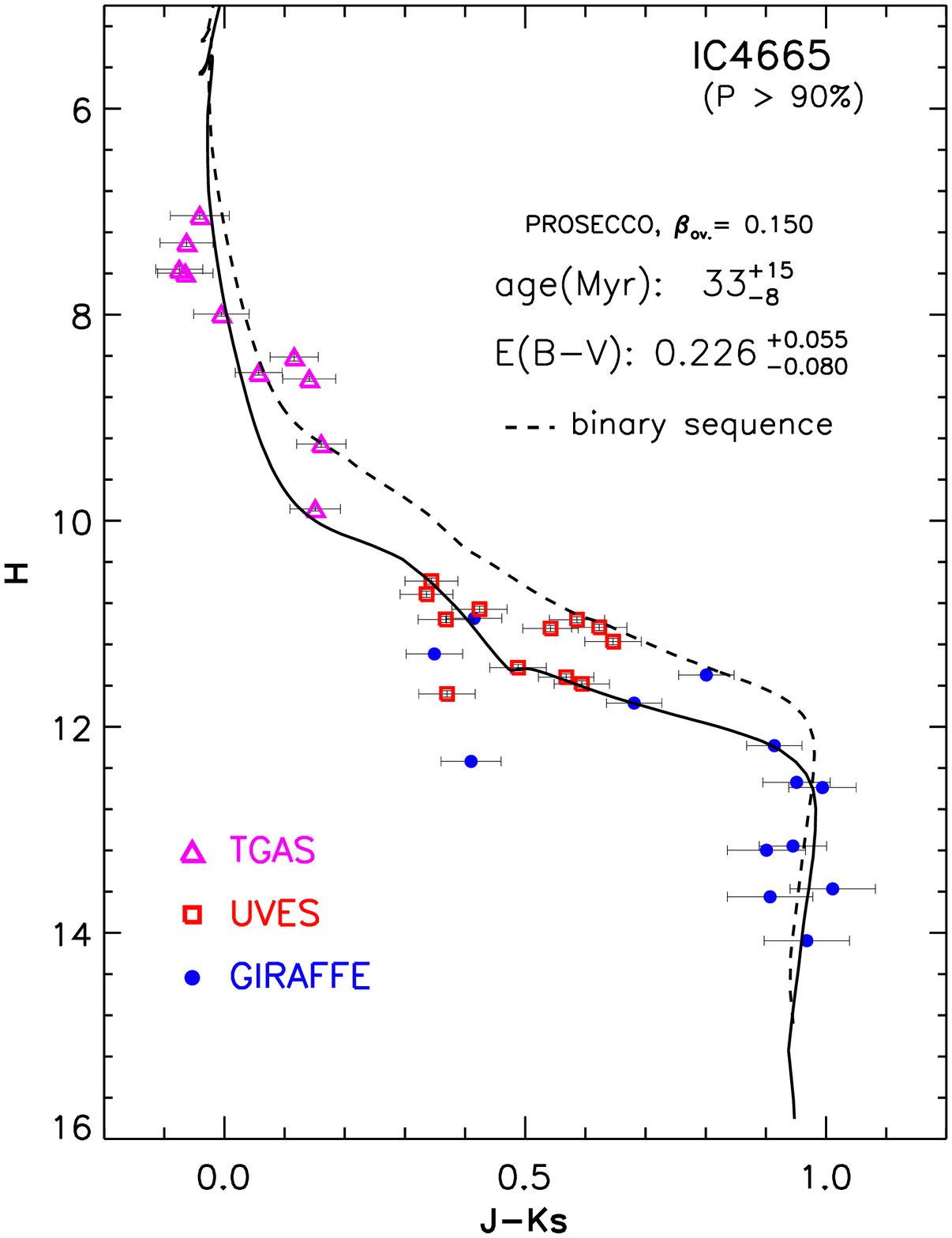}
        \includegraphics[width=0.325\linewidth]{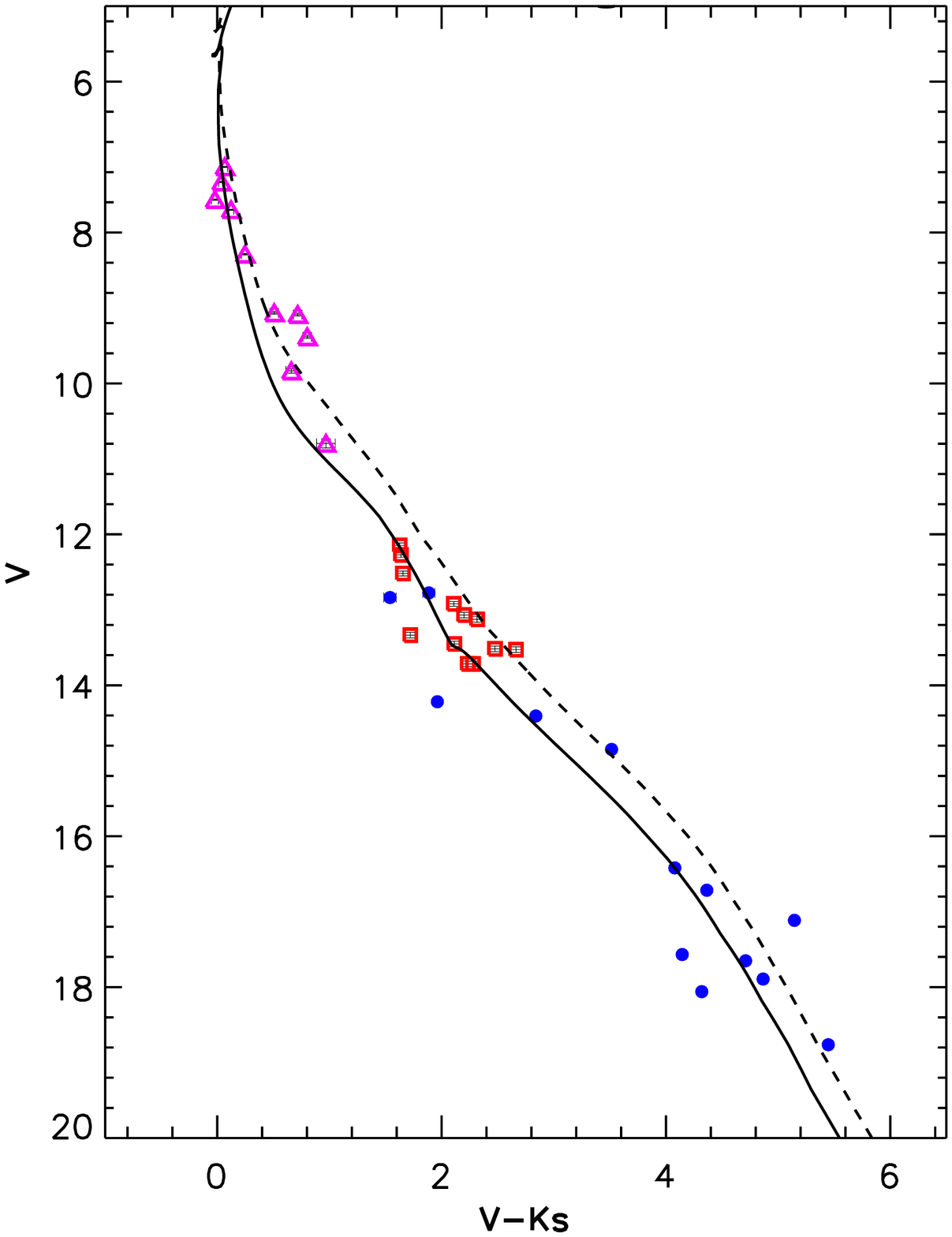}
        \includegraphics[width=0.325\linewidth]{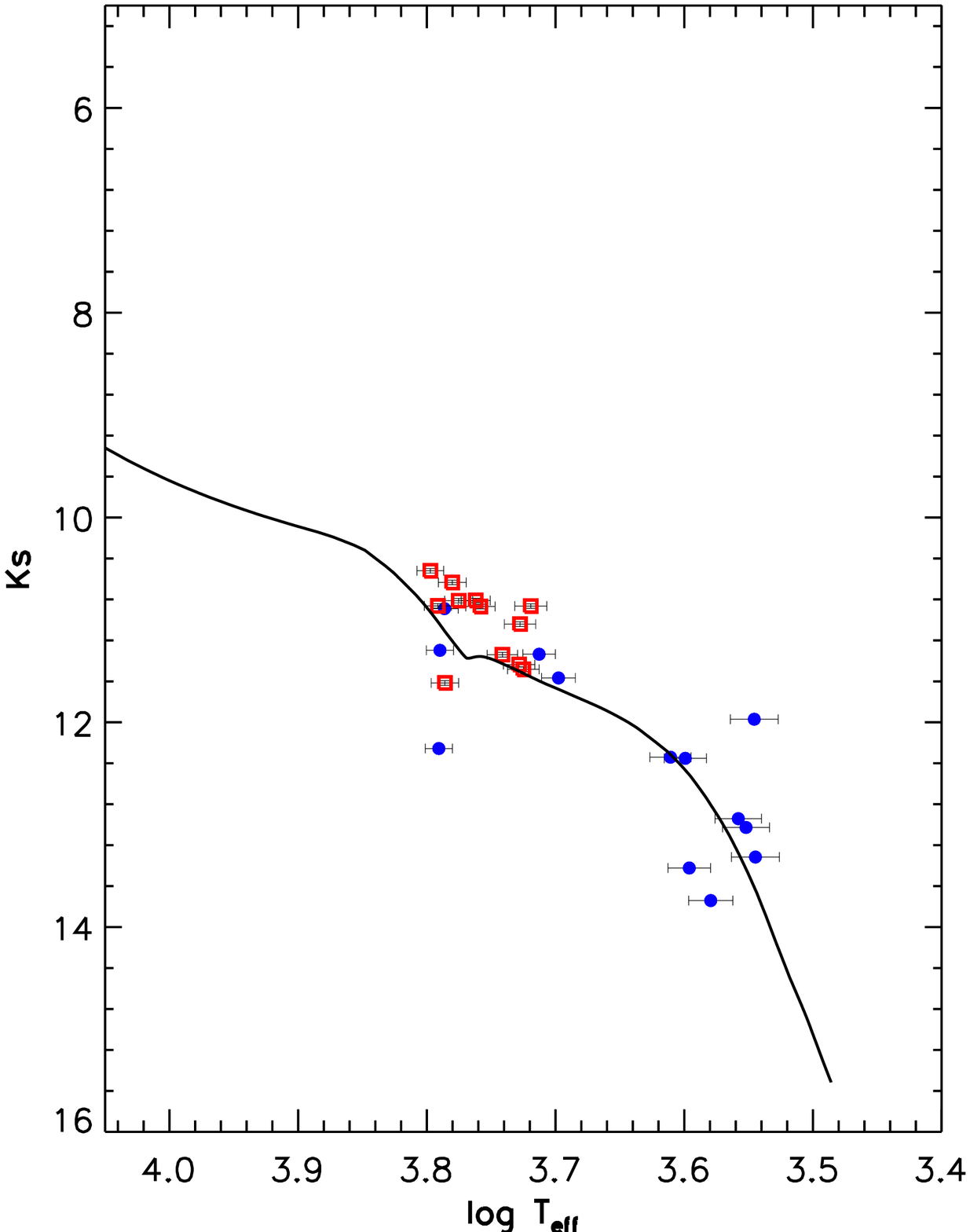}\\
        \includegraphics[width=0.325\linewidth]{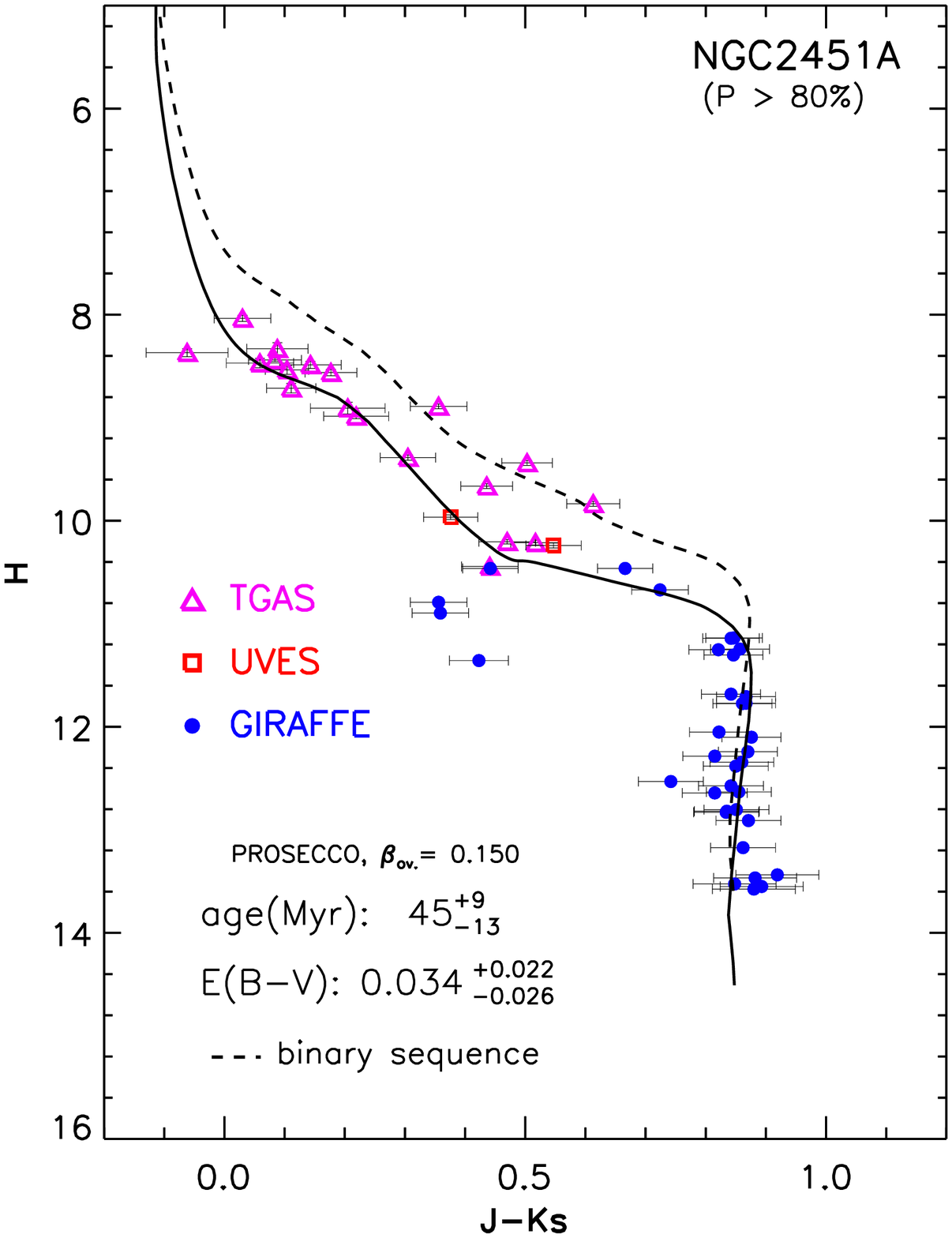}
        \includegraphics[width=0.325\linewidth]{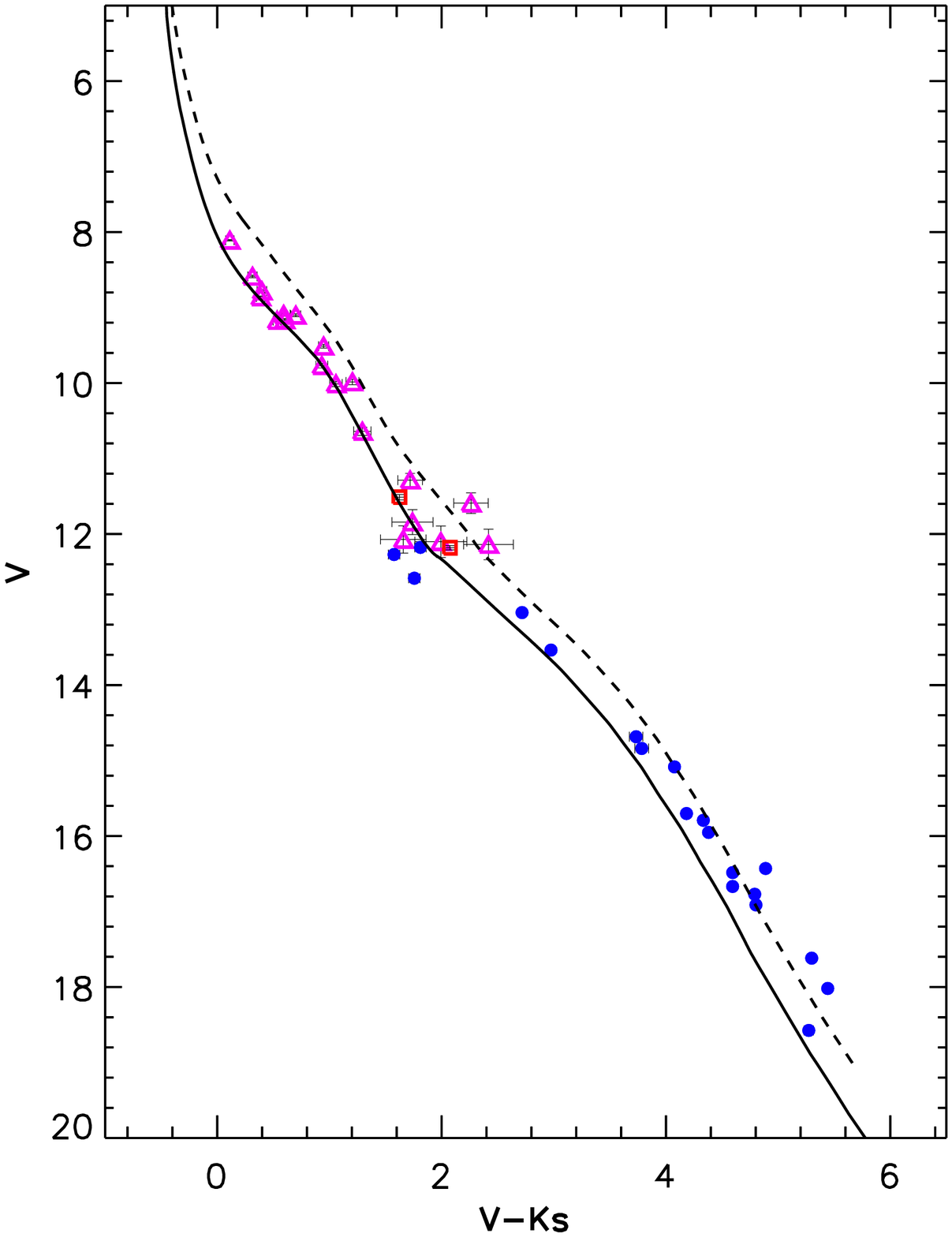}
        \includegraphics[width=0.325\linewidth]{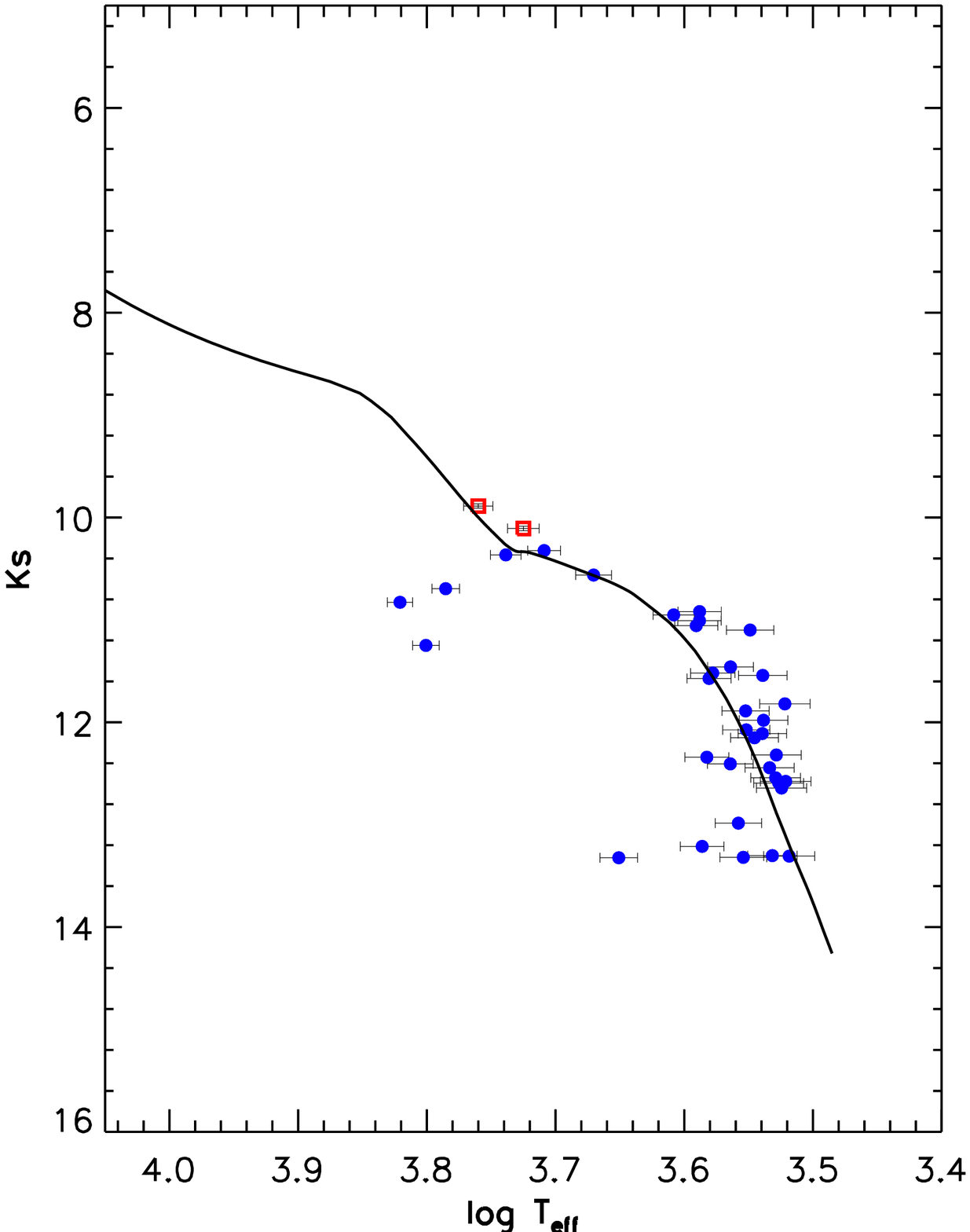}
        \caption{As in Fig.~\ref{fig:CMDPROSECCO_1} but for the clusters IC~4665 (top panel) and NGC~2451A (bottom panel). 
}
        \label{fig:CMDPROSECCO_2}
\end{figure*}

\begin{figure*}
        \centering      
        \includegraphics[width=0.325\linewidth]{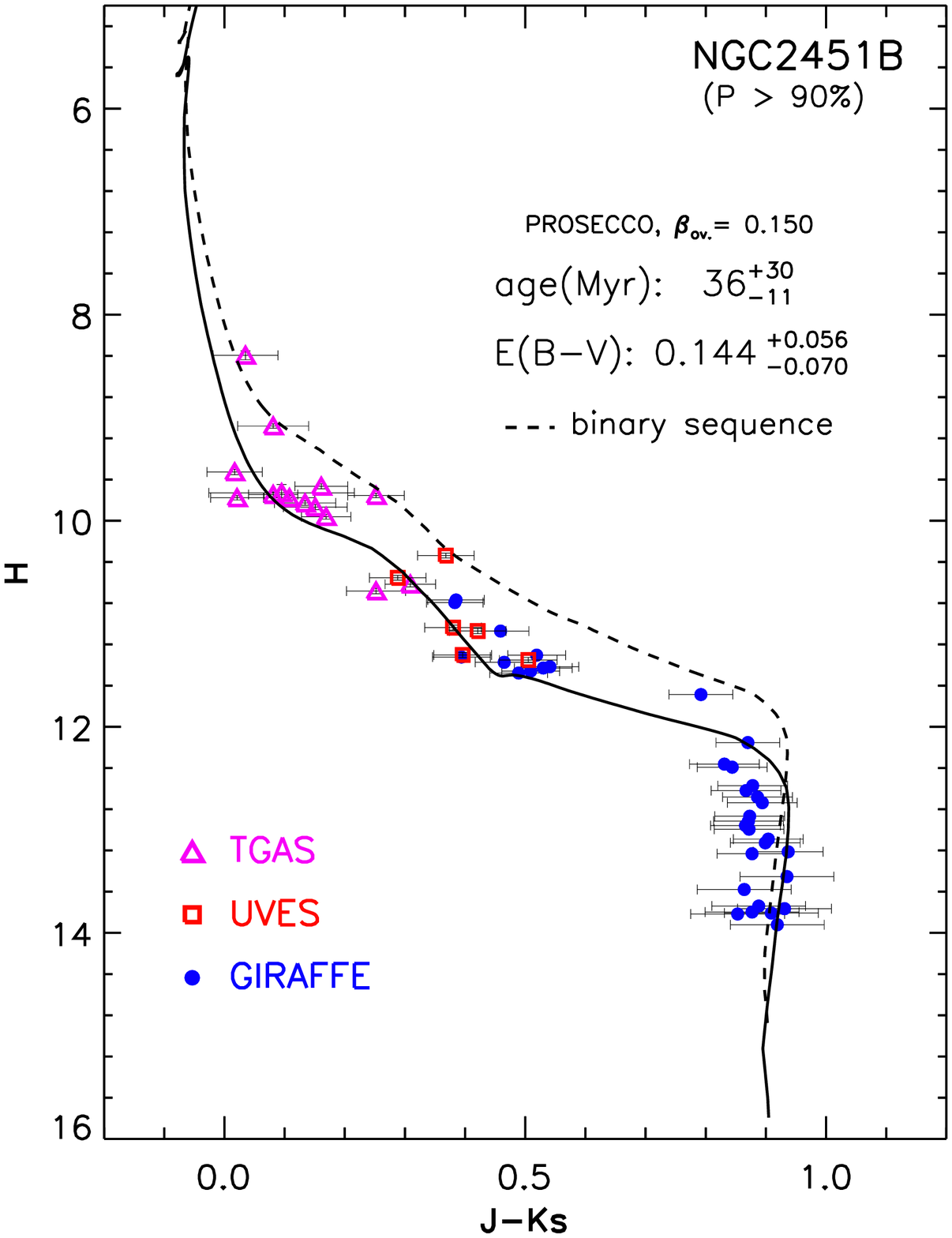}
        \includegraphics[width=0.325\linewidth]{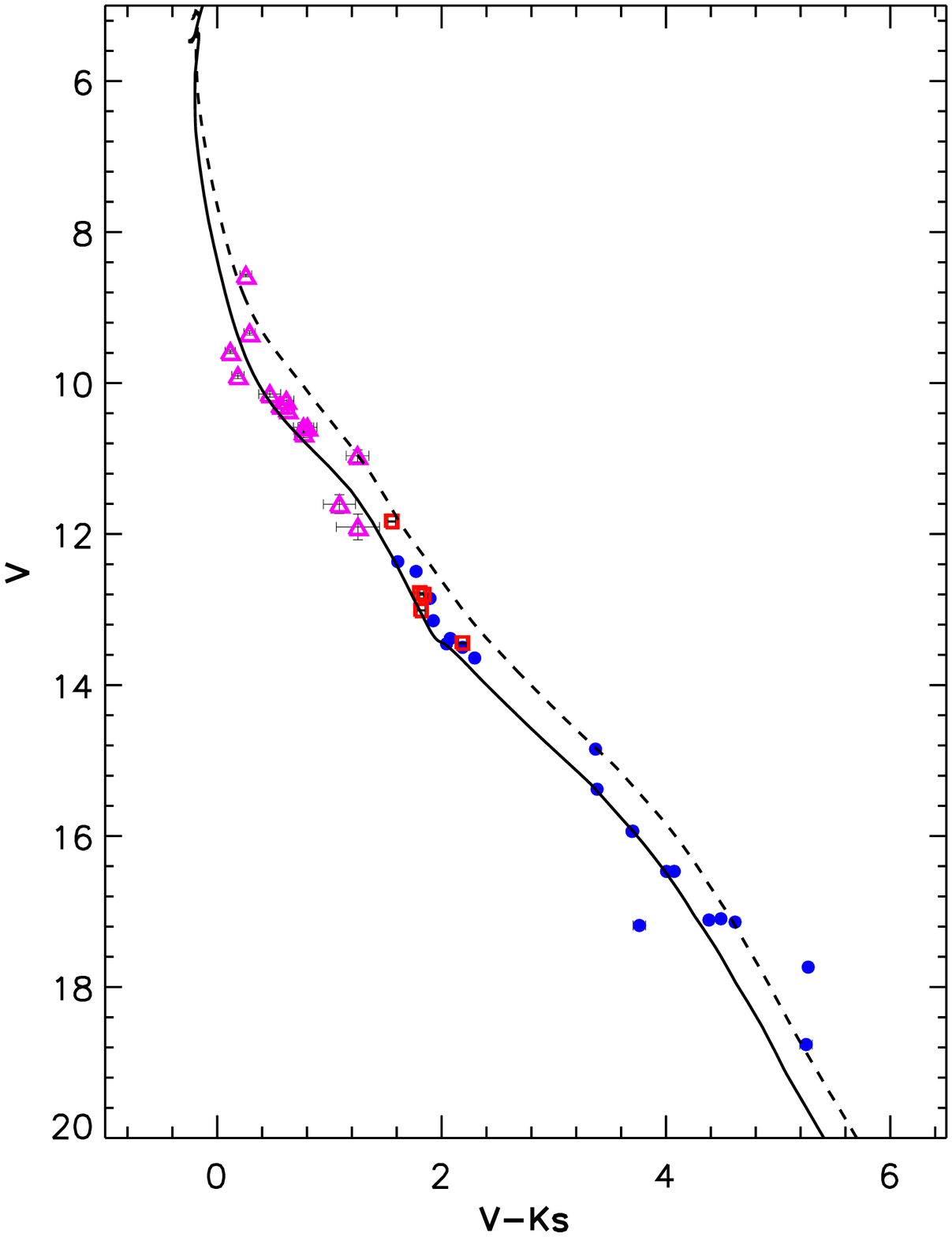}
        \includegraphics[width=0.325\linewidth]{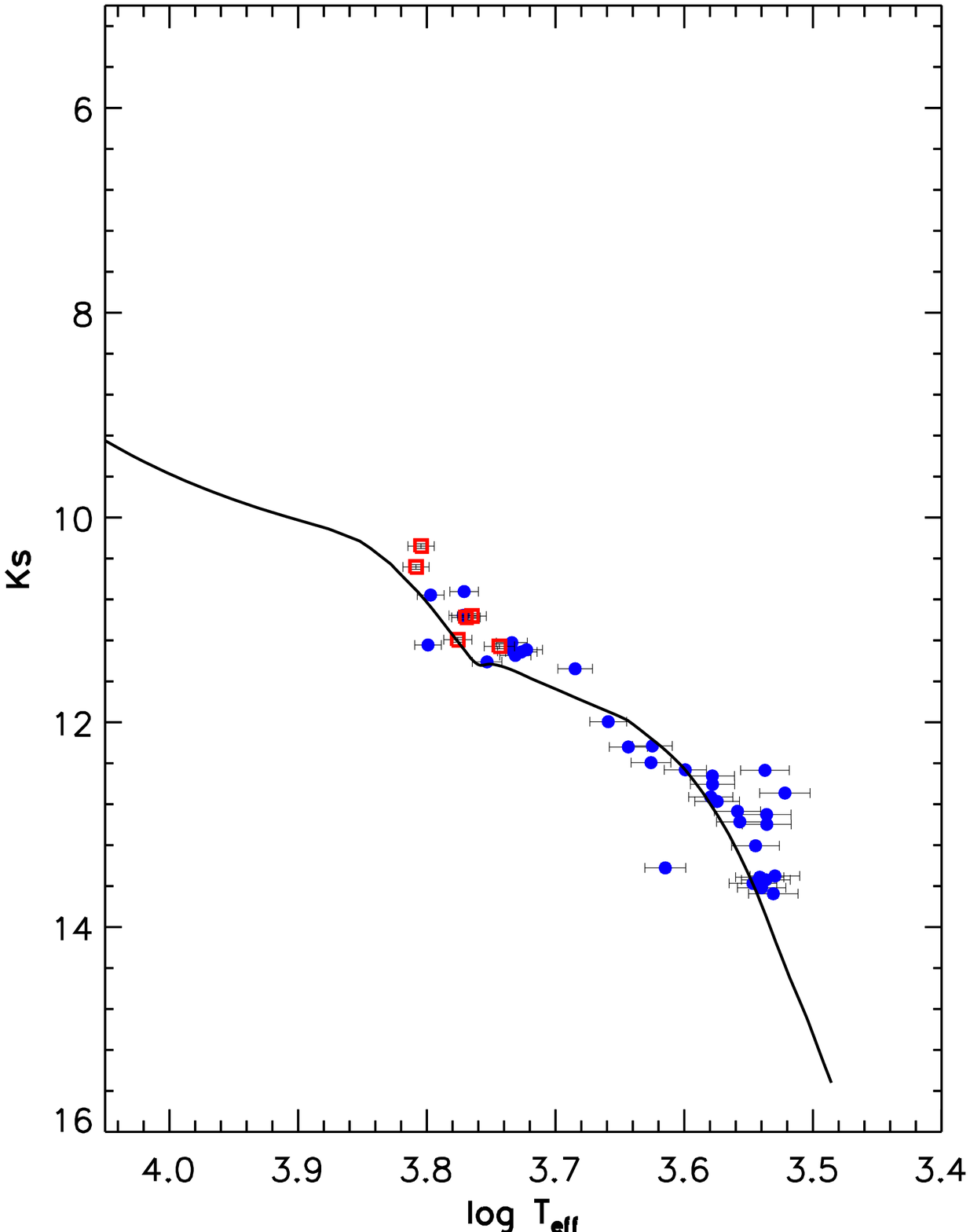}\\
        \includegraphics[width=0.325\linewidth]{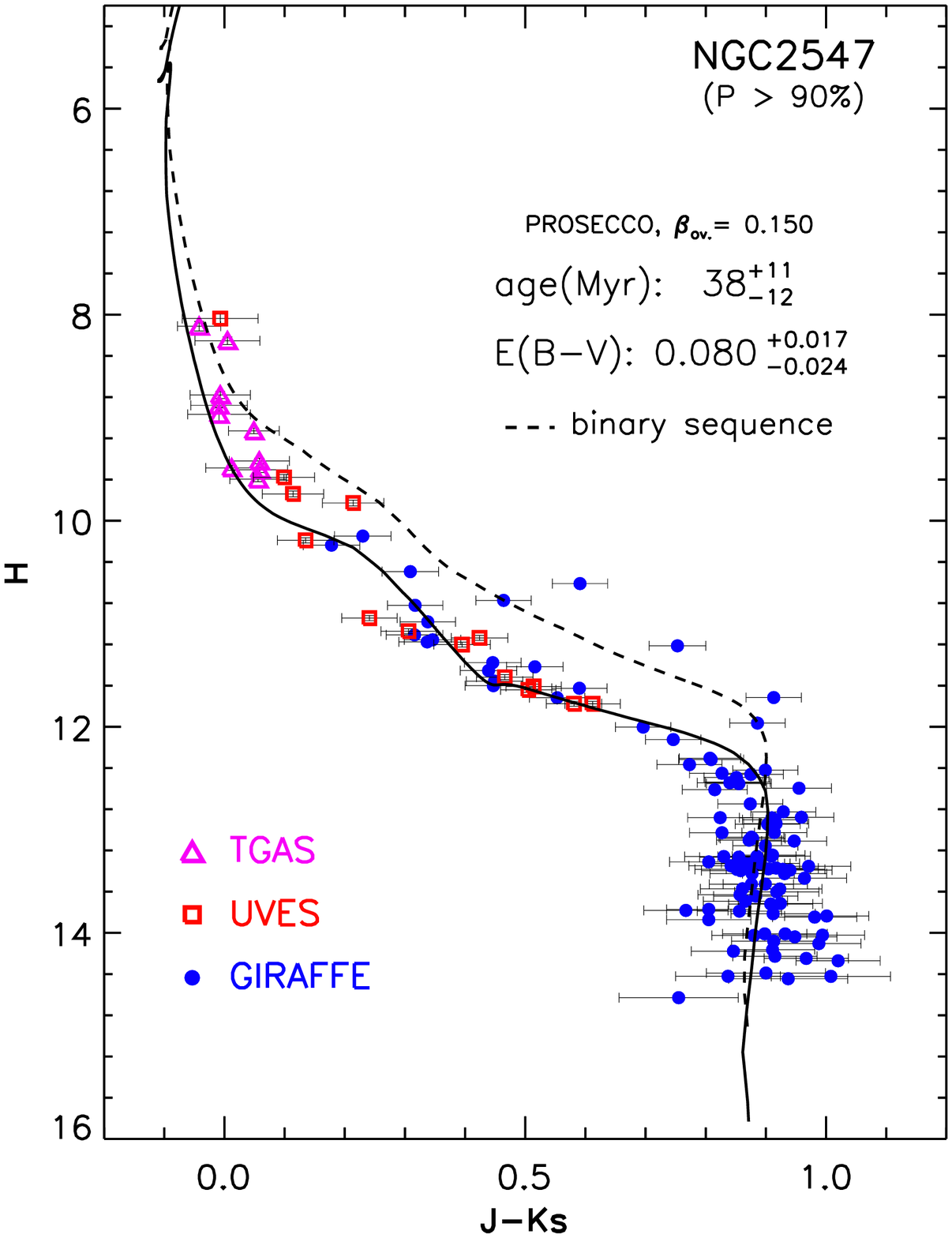}
        \includegraphics[width=0.325\linewidth]{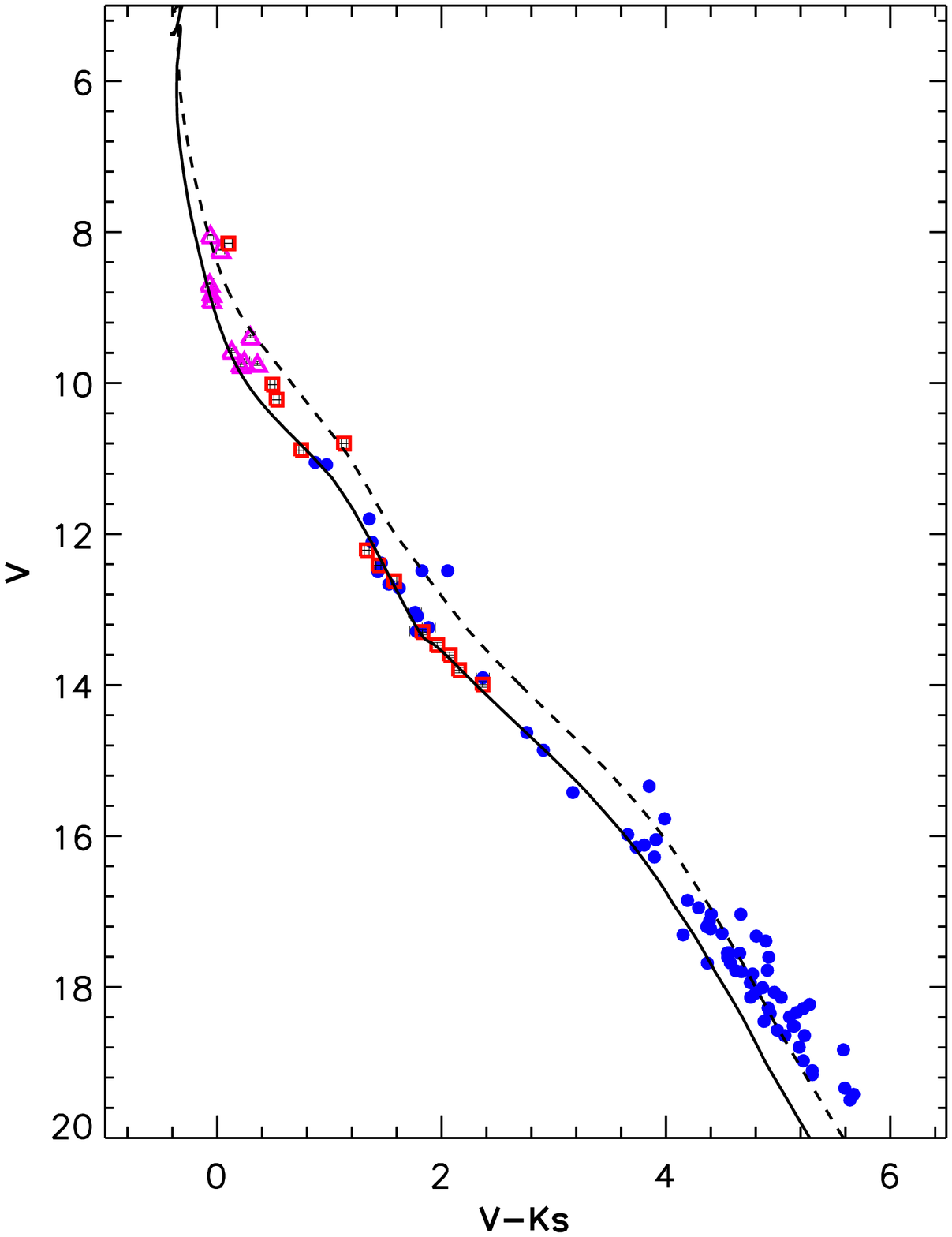}
        \includegraphics[width=0.325\linewidth]{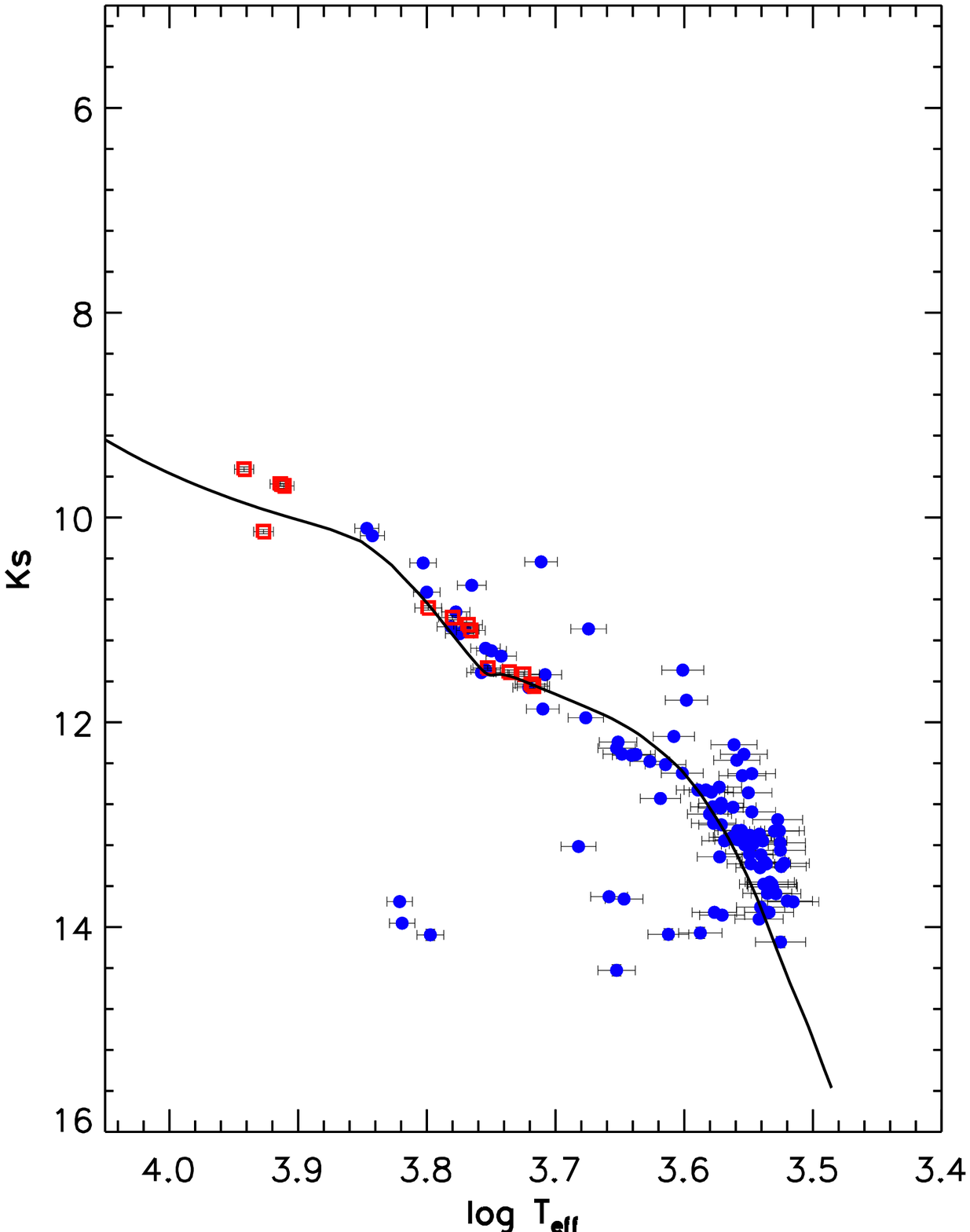}
        \caption{As in Fig.~\ref{fig:CMDPROSECCO_1} but for the clusters NGC~2451B (top panel) and NGC~2547 (bottom panel).}
        \label{fig:CMDPROSECCO_3}
\end{figure*}

\begin{figure*}
        \centering
        \includegraphics[width=0.325\linewidth]{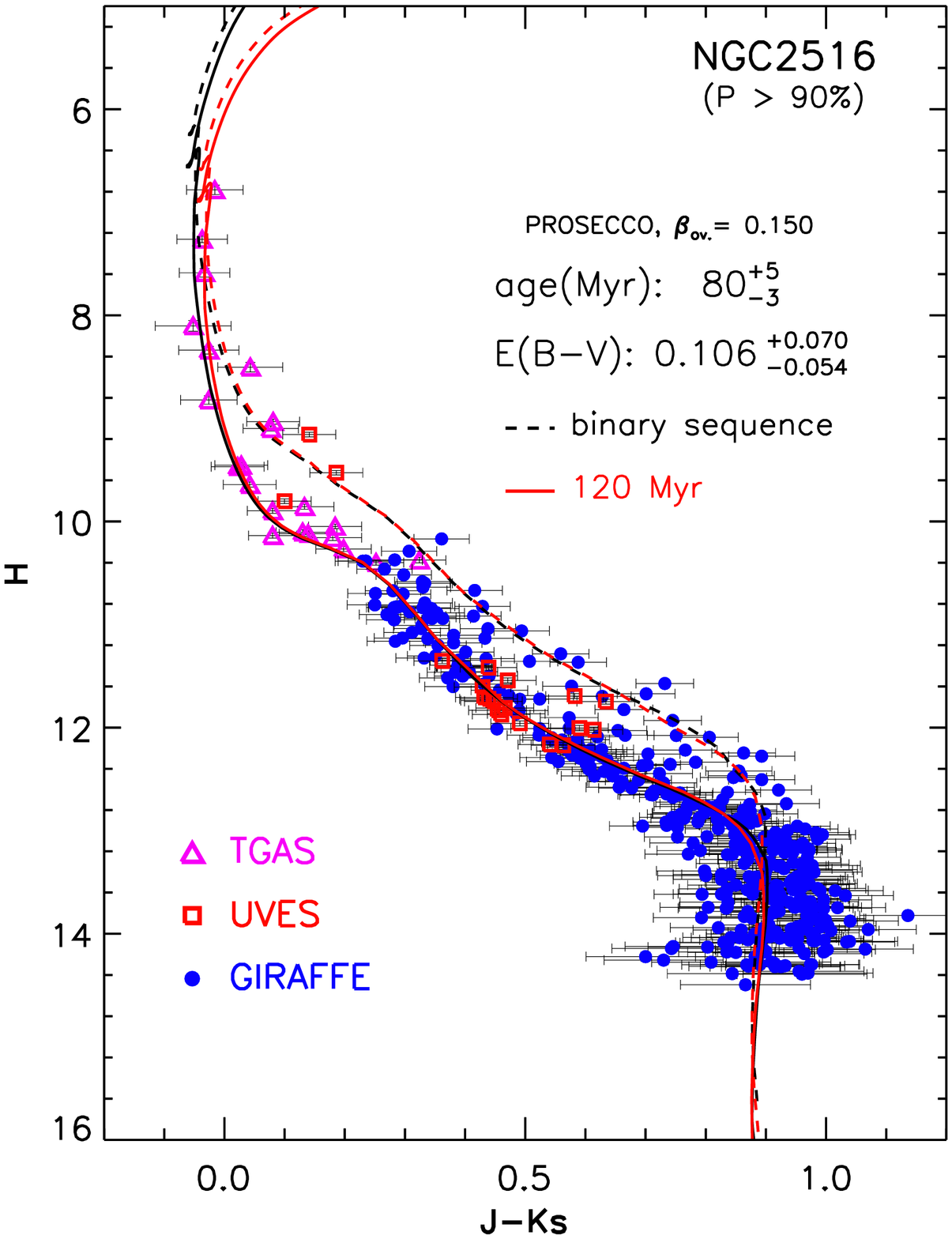}
        \includegraphics[width=0.325\linewidth]{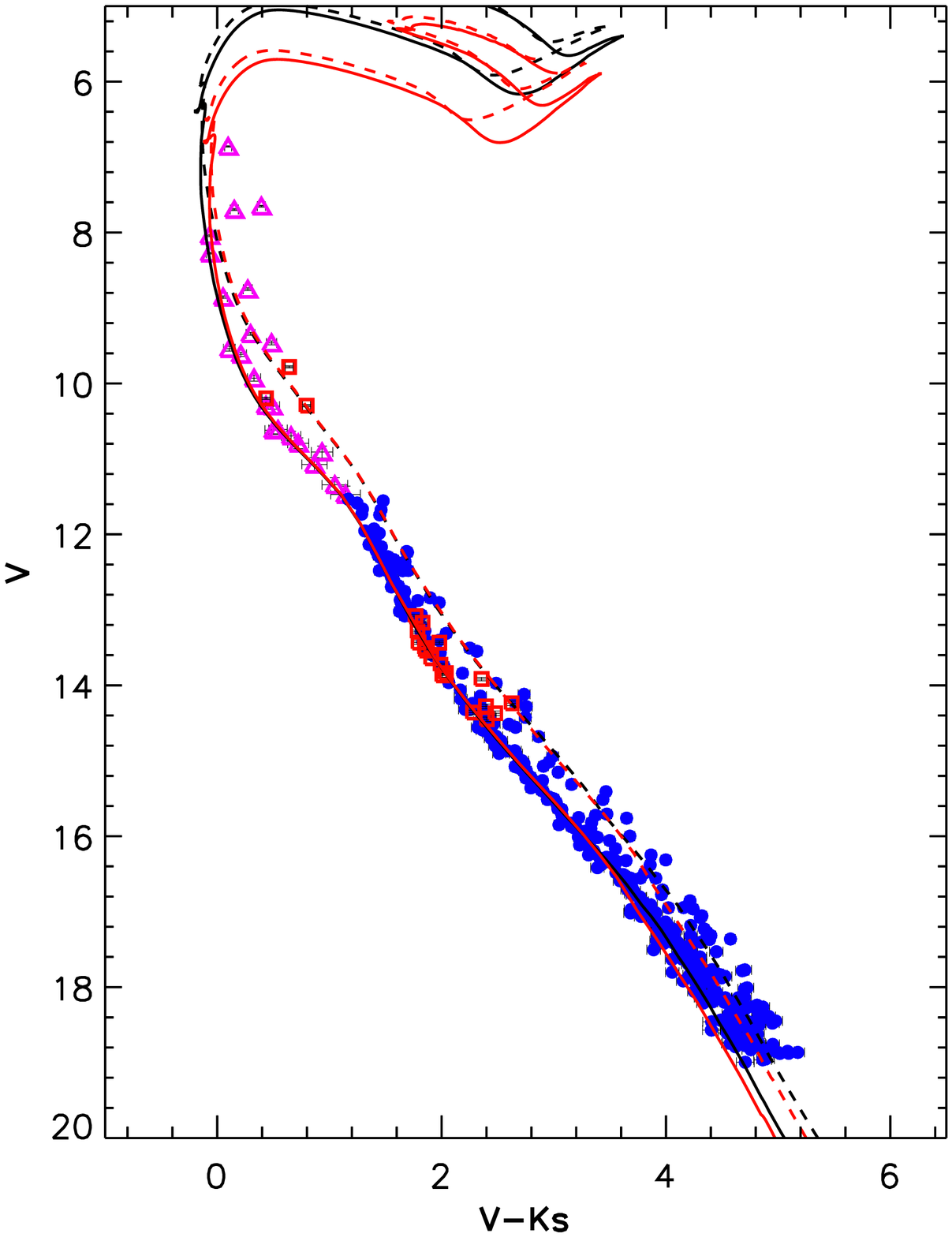}
        \includegraphics[width=0.325\linewidth]{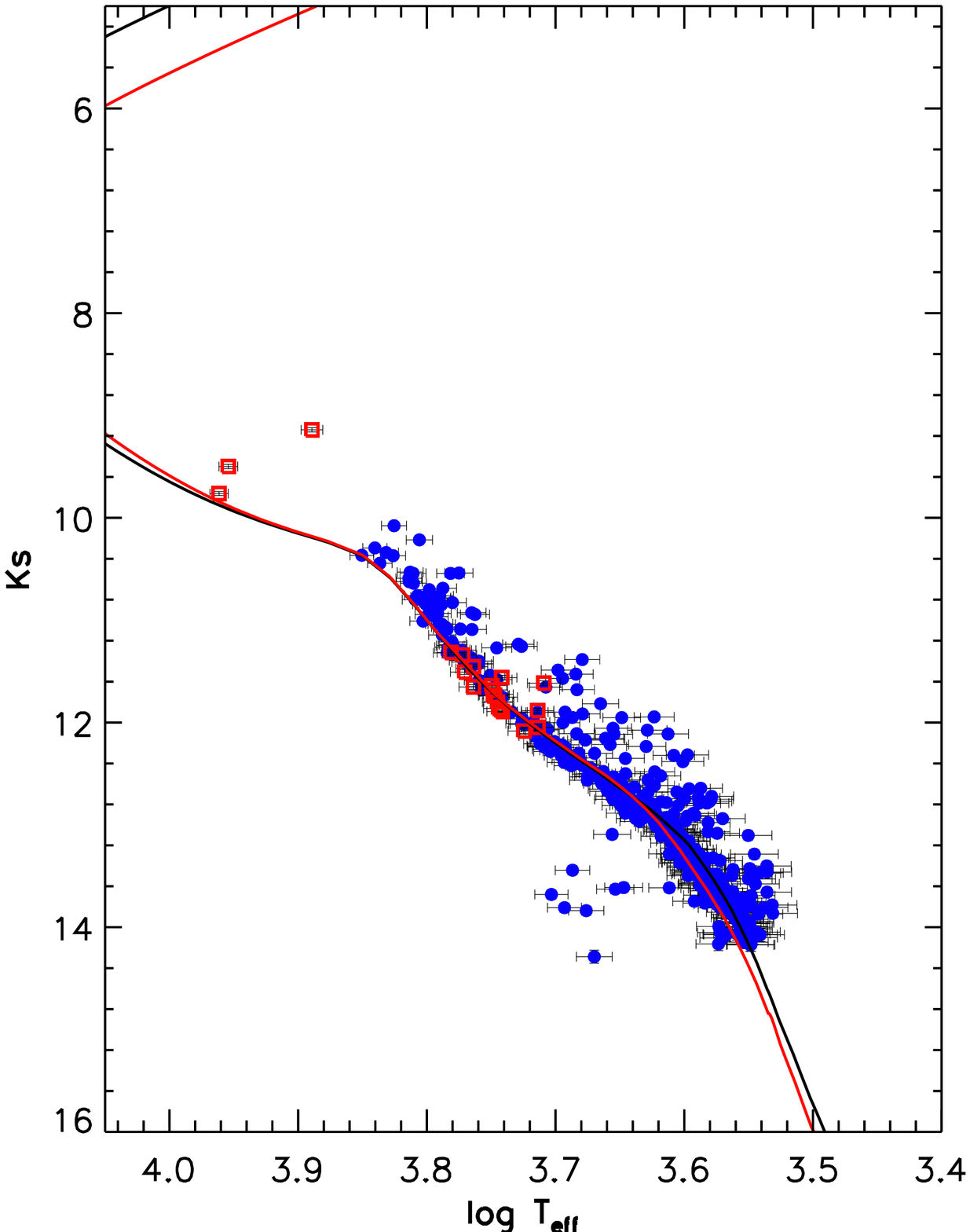}\\
        \includegraphics[width=0.325\linewidth]{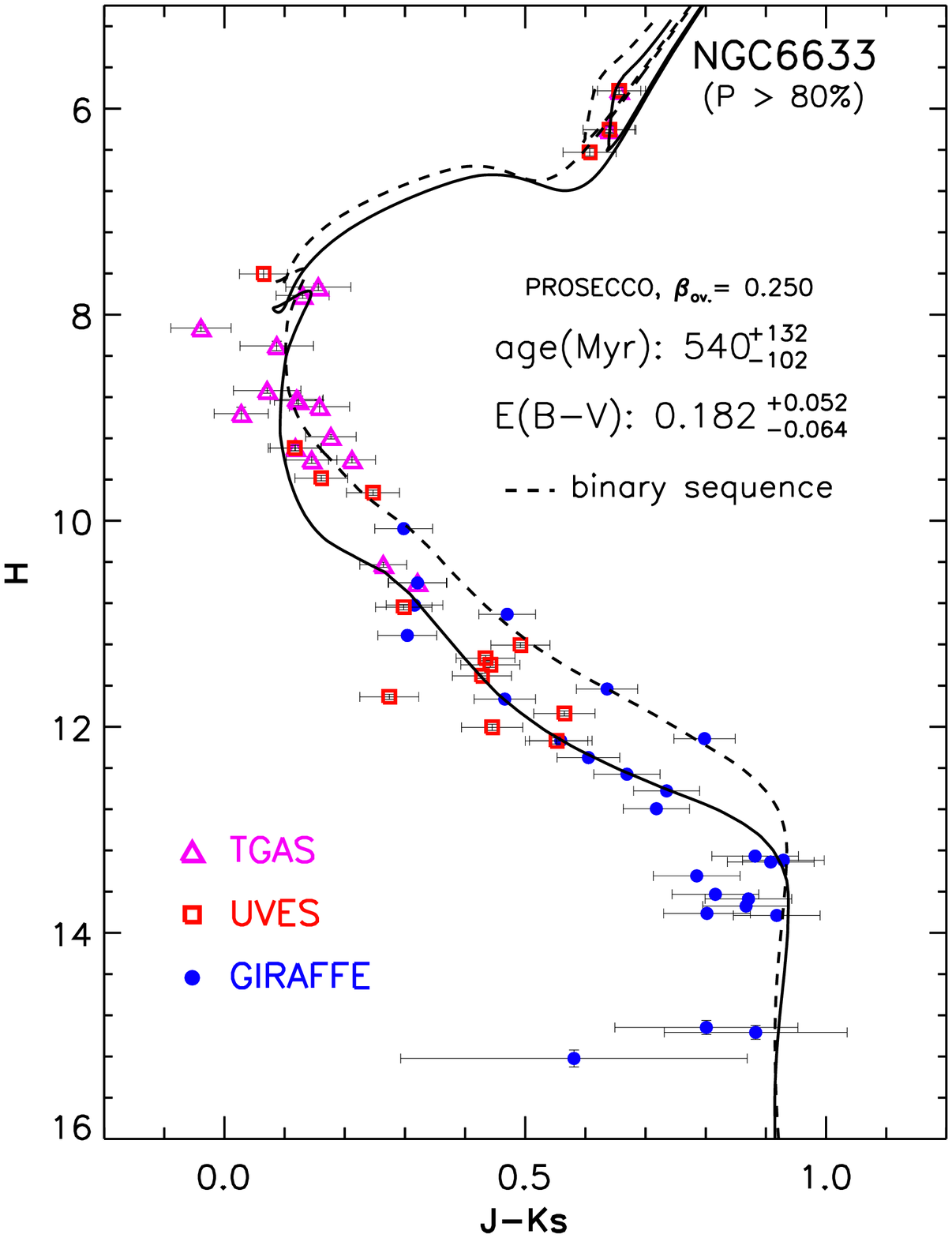}
        \includegraphics[width=0.325\linewidth]{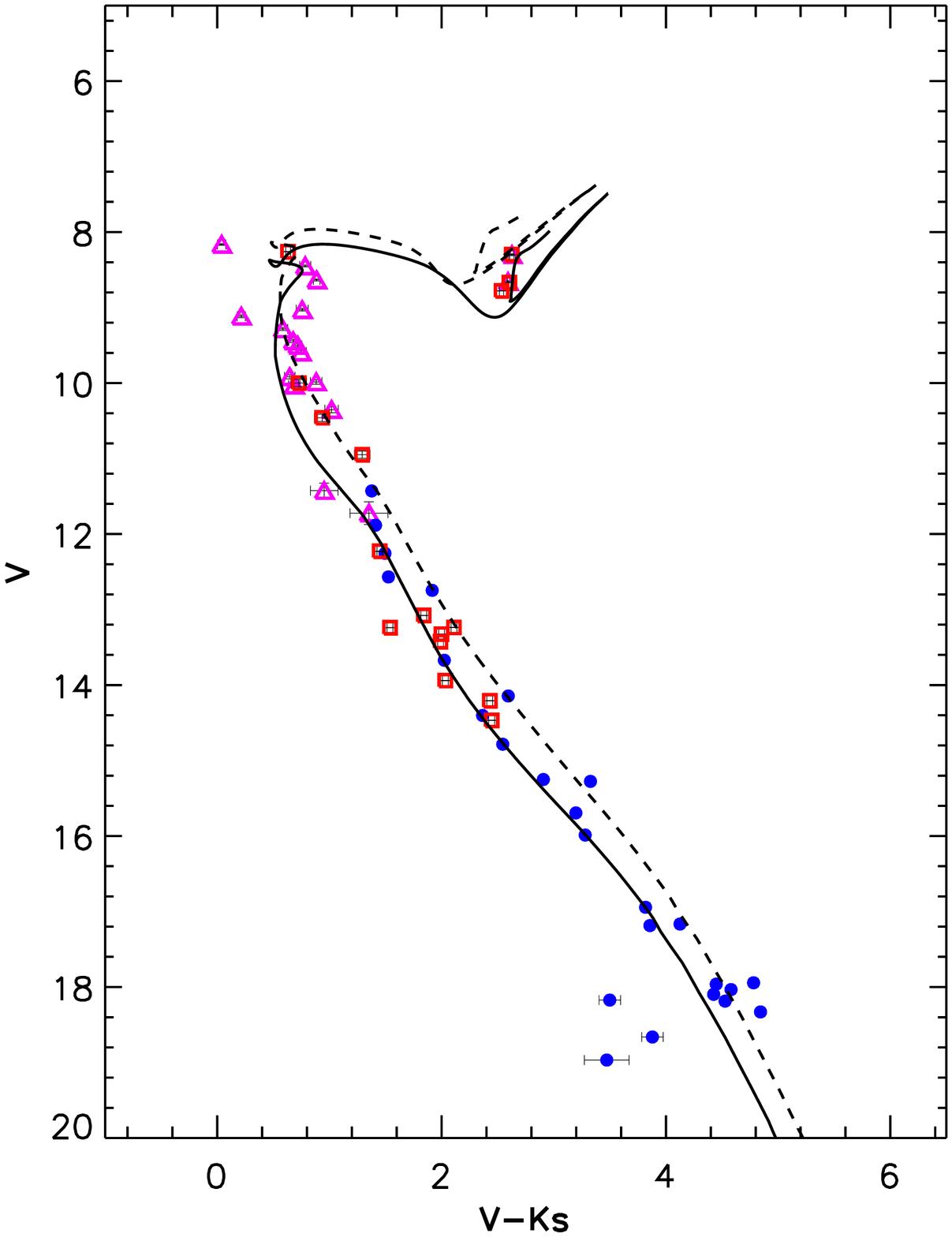}
        \includegraphics[width=0.325\linewidth]{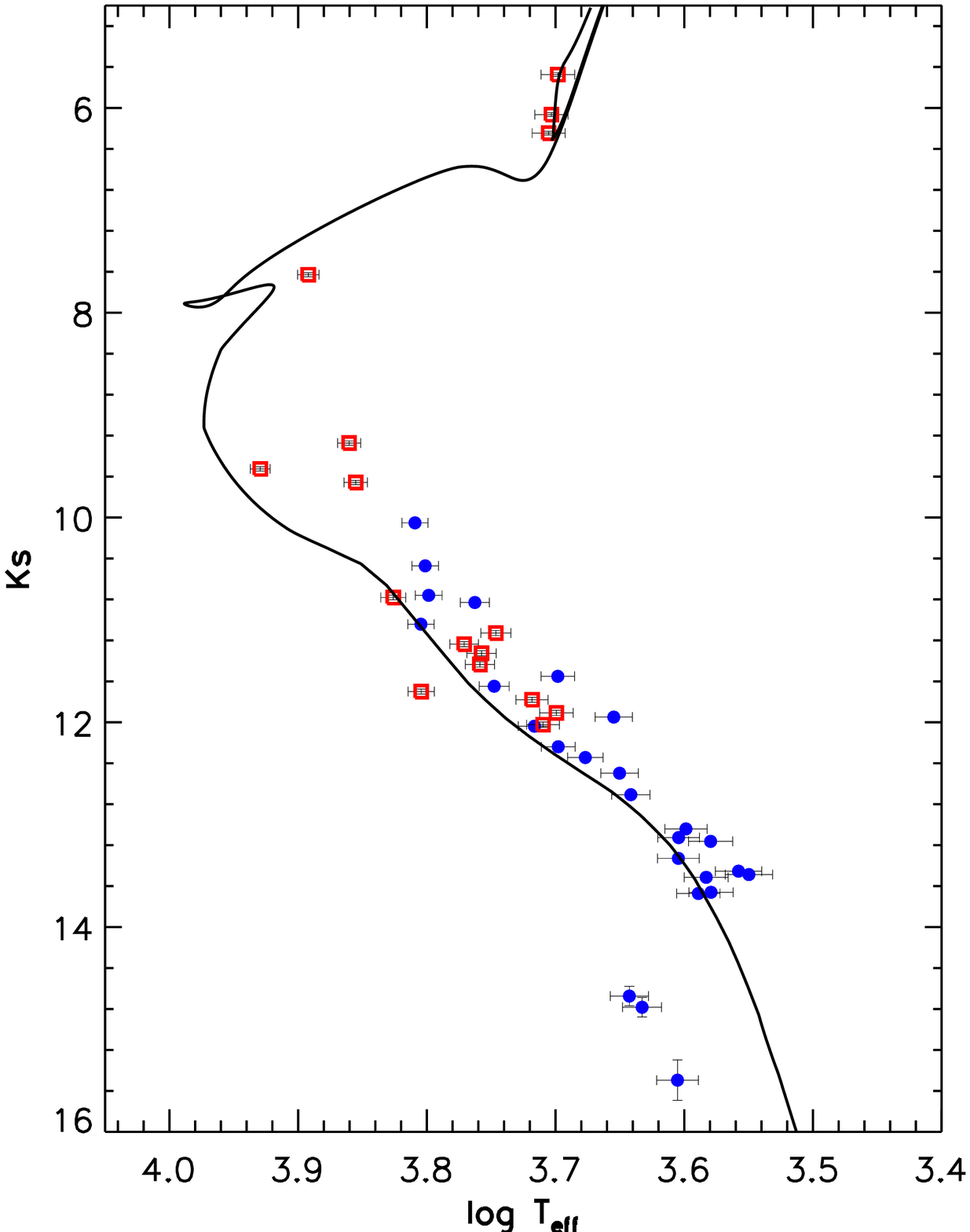}
        \caption{As in Fig.~\ref{fig:CMDPROSECCO_1} but for the clusters NGC~2516 (top panel) and NGC~6633 (bottom panel). For NGC~2516, the 120~Myr isochrone
is also shown. For NGC~6633 a core overshooting parameter $\beta_\mathrm{ov}=0.250$ has been used.}
        \label{fig:CMDPROSECCO_4}
\end{figure*}

The most probable age and reddening values derived through our analysis are listed in Table~\ref{tab:clusters_prop} and are adopted to show the comparison between models and data in the ($J-K_{\rm s}$,$H$), ($V-K_{\rm s}$,$V$), and ($\log T_\mathrm{eff}$, $K_{\rm s}$) planes. Specifically, Figs.~\ref{fig:CMDPROSECCO_1}, \ref{fig:CMDPROSECCO_2}, \ref{fig:CMDPROSECCO_3} and \ref{fig:CMDPROSECCO_4} show the CMDs and temperature-magnitude diagrams of selected relevant cases. 
The recovery has been performed for all the clusters with and without the inclusion of
unresolved binaries and the derived ages and reddening in the two cases are consistent within the estimated uncertainties (for more details see Appendix~\ref{sec:cleaning}).
We first note that, as shown by Table~\ref{tab:clusters_prop}, 
our procedure allowed us to recover ages with cumulative uncertainties between 5 and 30 \%, with the exception of NGC~2451B, which is characterised by a much larger uncertainty, 
due to the combination of relatively large error
in distance and small number of members. Also, the homogeneous analysis has enabled the establishment of an age ranking, which does not seem to depend on which model dataset was used, although there is some dependency on the magnitudes used for the recovery (see Sect.~\ref
{sect:othermodel} below). More generally, 
the Figures show a good,  global agreement between theory - isochrones with the most probable ages and reddening values - and observations for all the clusters. Noticeably, this is true for the different selected CMDs (for the entire
sequences) and even 
in the K$_{\rm s}$ versus T$_{\rm eff}$ diagrams. Some of the Figures suggest a tendency of the low-mass models to be somewhat bluer than the observations in the ($V$, $V-K_\mathrm{s}$) plane. Such a behaviour is not present if the $J-K_s$ colour or T$_\mathrm{eff}$ are considered, indicating that at least part of the problem could be related to the adopted colour transformations, as already discussed in the literature \citep[see e.g. ][]{kucinskas2005,casagrande2014}.

The two older clusters, NGC~2516 and NGC~6633, deserve some further comments. NGC~2516 (top panel of Fig.~\ref{fig:CMDPROSECCO_4}) is the cluster with the largest number of members among those analysed in the present work. Both the sequence of single stars and that of binaries of equal mass are clearly visible. In this cluster, low-mass stars are very close to their ZAMS position, weakly sensitive to age, which in principle should be mainly determined from stars 
close to the turn-off.
Unfortunately, given the bright limit of both GES and available photometric 
surveys, relatively few members in that position ($V < 9$) 
are present in the current
sample. Also, while we identified 22 stars from TGAS, only a few of them are close to the turn-off; a few more may come from the stars identified in VL17, once these are confirmed to be members.

Finally, NGC~6633 (bottom panel of Fig.~\ref{fig:CMDPROSECCO_4}) is the only cluster in our sample with a few stars in the central helium burning phase, sensitive to the extension of the convective core during the MS phase. Thus, we derived age and reddening using two values of the core overshooting parameter, namely $\beta_\mathrm{ov} =0.15$ and 0.25. Due again to the small sample of stars close to the 
turn-off region and in the He-burning phase, the two models with different values of core overshooting are almost indistinguishable, both in terms of ages and reddening, although the case with $\beta_\mathrm{ov} =0.25$ seems to be favoured by a slightly better fit quality.

We evaluated the impact of a slightly different [Fe/H] on the age and reddening determination by running the analysis considering [Fe/H]$=-0.1$ for all the clusters. This value corresponds to the lowest metallicity listed in Table~2; that for NGC\,2516. 
Adopting a different metallicity does result in changed estimates for the age and reddening. On the one hand, however, the metallicity of these clusters has been homogeneously measured by GES with good precision, and on the other hand, in the case of NGC~2516, with [Fe/H]$=-0.1$, we obtained a reduction in age of 1-2\% and an increase in E$(B-V)$ of 0.02-0.03 mag. Similar numbers are found for the other clusters. This is a small effect, smaller than the currently estimated uncertainties, but will become more important when the systematic uncertainties in the Gaia parallaxes are removed.  
\begin{table*}
\centering
\caption{Ages and reddening values for the sample clusters, as obtained by adopting the PROSECCO, PARSEC and MIST isochrones in the ($J$, $H$, $K_\mathrm{s}$) and in the ($J$, $H$, $K_\mathrm{s}$, $V$) planes - first and second set
of values, respectively.}
\label{tab:clusters_prop}
\begin{tabular}{l|rr|rr|rr}
\hline\hline \\
         & \multicolumn{2}{c}{PROSECCO} & \multicolumn{2}{c}{PARSEC} & \multicolumn{2}{c}{MIST}\\
        Cluster  & $\tau$(Myr) & E$(B-V)$ & $\tau$(Myr) & E$(B-V)$ & $\tau$(Myr) & E$(B-V)$ \\
        \hline
& & & & & & \\
NGC~2451A & $45^{+9}_{-13}$ & $0.034^{+0.022}_{-0.026}$ & $52^{+26}_{-20}$ & $0.042^{+0.025}_{-0.028}$ &  $48^{+8}_{-14}$ &  $0.006^{+0.011}_{-0.006}$ \\
& & & & & & \\
NGC~2451B & $36^{+30}_{-11}$ & $0.144^{+0.056}_{-0.070}$ & $34^{+52}_{-13}$ & $0.174^{+0.051}_{-0.072}$ & $44^{+32}_{-16}$ & $0.126^{+0.092}_{-0.079}$\\
& & & & & & \\
NGC~2516 & $80^{+5}_{-3}$ & $0.106^{+0.070}_{-0.054}$ & $113^{+15}_{-6}$ & $0.148^{+0.072}_{-0.056} $ & $77^{+3}_{-3}$ &  $0.084^{+0.060}_{-0.048}$\\
& & & & & & \\
NGC~2547 & $38^{+11}_{-12}$ & $0.080^{+0.017}_{-0.024}$ & $33^{+11}_{-10}$ & $0.110^{+0.026}_{-0.031}$ & $40^{+13}_{-15}$ & $0.046^{+0.017}_{-0.026}$\\
& & & & & & \\
IC~2391 & $41^{+10}_{-5}$ & $0.094^{+0.024}_{-0.029}$ & $43^{+15}_{-7}$ & $0.100^{+0.021}_{-0.027}$ & $45^{+10}_{-6}$ &  $0.048^{+0.026}_{-0.028}$\\
& & & & & & \\
IC\,2602 & $30^{+9}_{-6}$ & $0.070^{+0.025}_{-0.026}$ & $29^{+10}_{-7}$ & $0.092^{+0.028}_{-0.025}$ & $30^{+9}_{-7}$ & $0.039^{+0.026}_{-0.027}$ \\
& & & & & & \\
IC~4665 & $33^{+15}_{-8}$ & $0.226^{+0.055}_{-0.080}$ & $31^{+10}_{-13}$ & $0.268^{+0.027}_{-0.087}$ & $34^{+19}_{-8}$ &  $0.204^{+0.066}_{-0.088}$ \\
& & & & & & \\
NGC~6633$^{a}$ & $540^{+132}_{-102}$ & $0.182^{+0.052}_{-0.064}$ & $552^{+114}_{-116}$ & $0.218^{+0.049}_{-0.062}$ &  $490^{+132}_{-94}$ & $0.166^{+0.056}_{-0.076}$\\
& & & & & & \\
\hline
& & & & & & \\
NGC~2451A & $32^{+4}_{-4}$ & $0.114^{+0.030}_{-0.021}$ & $44^{+2}_{-3}$ & $0.078^{+0.057}_{-0.039}$ & $36^{+4}_{-5}$ & $0.080^{+0.034}_{-0.036}$\\
 & & & & & & \\
NGC~2451B & $32^{+8}_{-6}$ & $0.168^{+0.051}_{-0.051}$ & $40^{+18}_{-13}$ & $0.177^{+0.033}_{-0.059}$ & $33^{+9}_{-6}$ & $0.184^{+0.057}_{-0.060}$\\
 & & & & & & \\
NGC~2516 & $79^{+5}_{-9}$ & $0.111^{+0.065}_{-0.047}$ & $118^{+9}_{-8}$ & $0.103^{+0.064}_{-0.056}$ & $69^{+4}_{-8}$ & $0.123^{+0.061}_{-0.058}$\\
 & & & & & & \\
NGC~2547 & $37^{+8}_{-11}$ & $0.095^{+0.042}_{-0.050}$ & $39^{+11}_{-12}$ & $0.084^{+0.034}_{-0.033}$ & $36^{+8}_{-11}$ & $0.110^{+0.029}_{-0.047}$\\
 & & & & & & \\
IC~2391 & $40^{+9}_{-10}$ & $0.082^{+0.022}_{-0.020}$ & $42^{+10}_{-14}$ & $0.100^{+0.029}_{-0.031}$ & $44^{+3}_{-4}$ & $0.063^{+0.031}_{-0.028}$\\
 & & & & & & \\
IC\,2602 & $31^{+6}_{-5}$  & $0.068^{+0.029}_{-0.026}$ & $30^{+7}_{-8}$ & $0.065^{+0.032}_{-0.028}$ & $32^{+7}_{-5}$ & $0.032^{+0.029}_{-0.024}$ \\
 & & & & & & \\
IC~4665 & $31^{+13}_{-11}$ & $0.238^{+0.028}_{-0.042}$ & $31^{+21}_{-9}$ & $0.226^{+0.029}_{-0.055}$ & $32^{+14}_{-6}$ & $0.237^{+0.035}_{-0.041}$\\
 & & & & & & \\
NGC~6633$^{a}$ & $620^{+61}_{-190}$ & $0.180^{+0.076}_{-0.023}$ & $620^{+46}_{-201}$ & $0.181^{+0.074}_{-0.042}$ & $590^{+81}_{-210}$ & $0.217^{+0.062}_{-0.072}$\\
& & & & & & \\
\hline
\end{tabular}
\footnotesize
\begin{flushleft}
$^{a}$ $\beta_\mathrm{ov} = 0.250$ in the PROSECCO models.
\end{flushleft}
\normalsize
\end{table*}
\begin{figure*}
        \centering
        \includegraphics[width=0.325\linewidth]{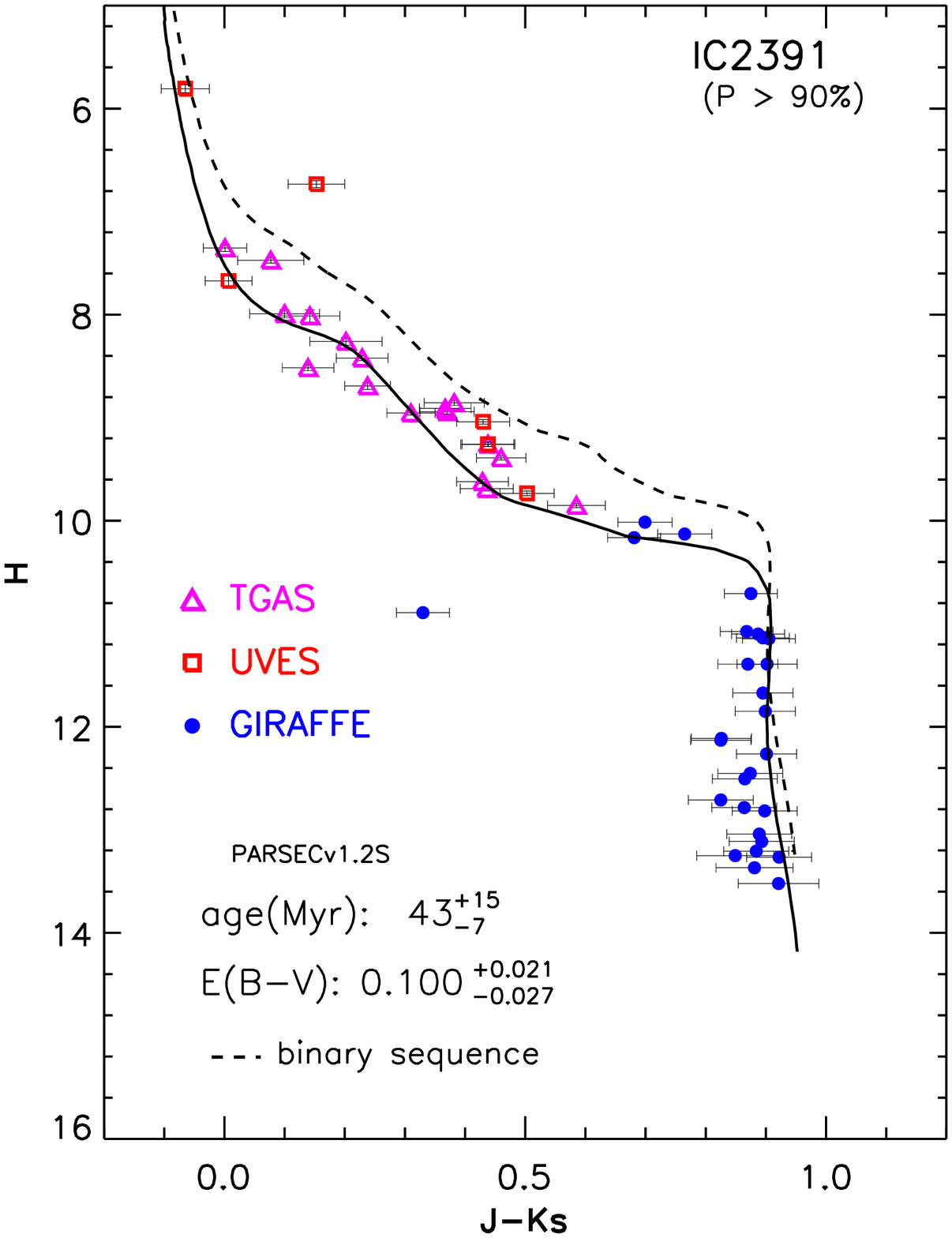}
        \includegraphics[width=0.325\linewidth]{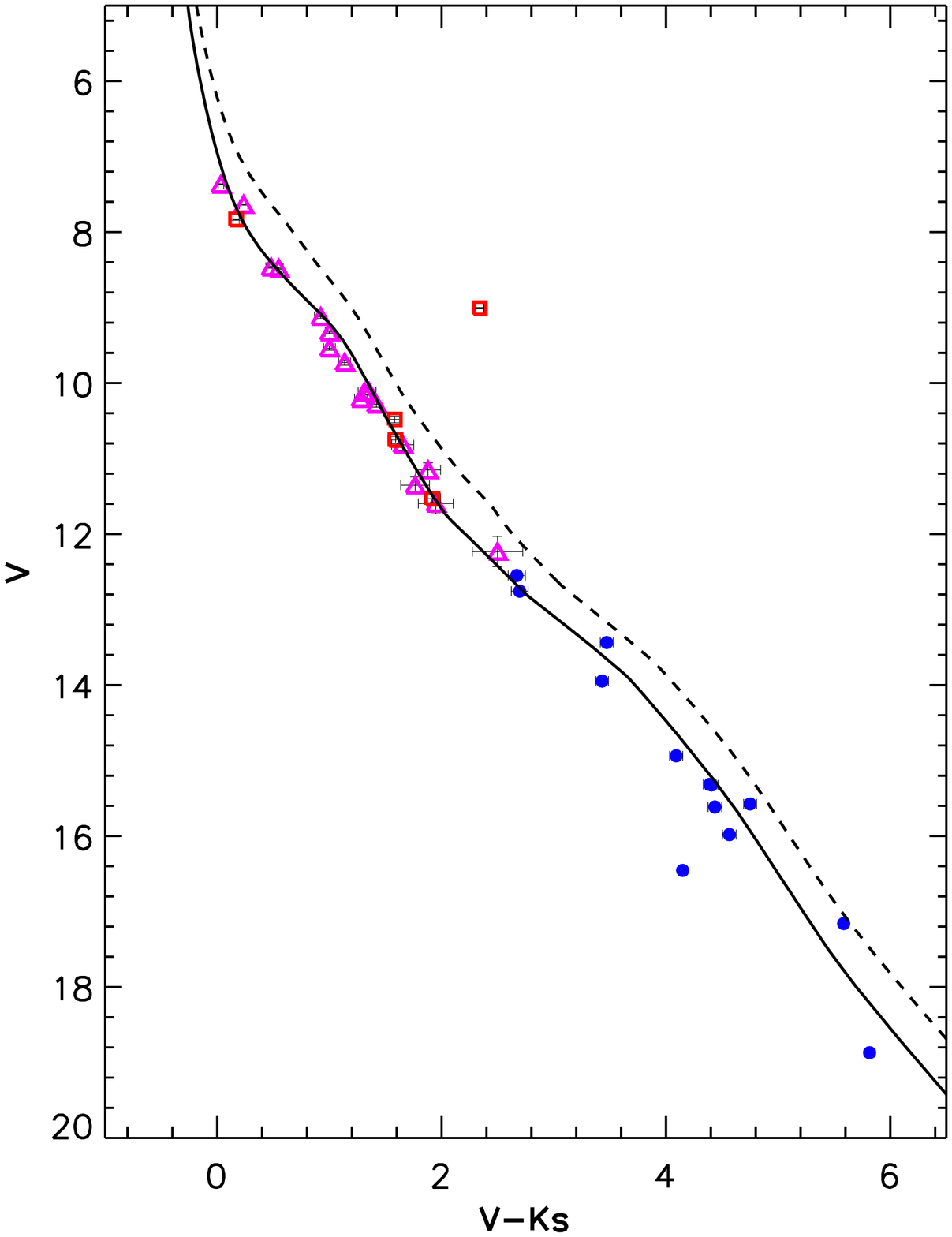}
        \includegraphics[width=0.325\linewidth]{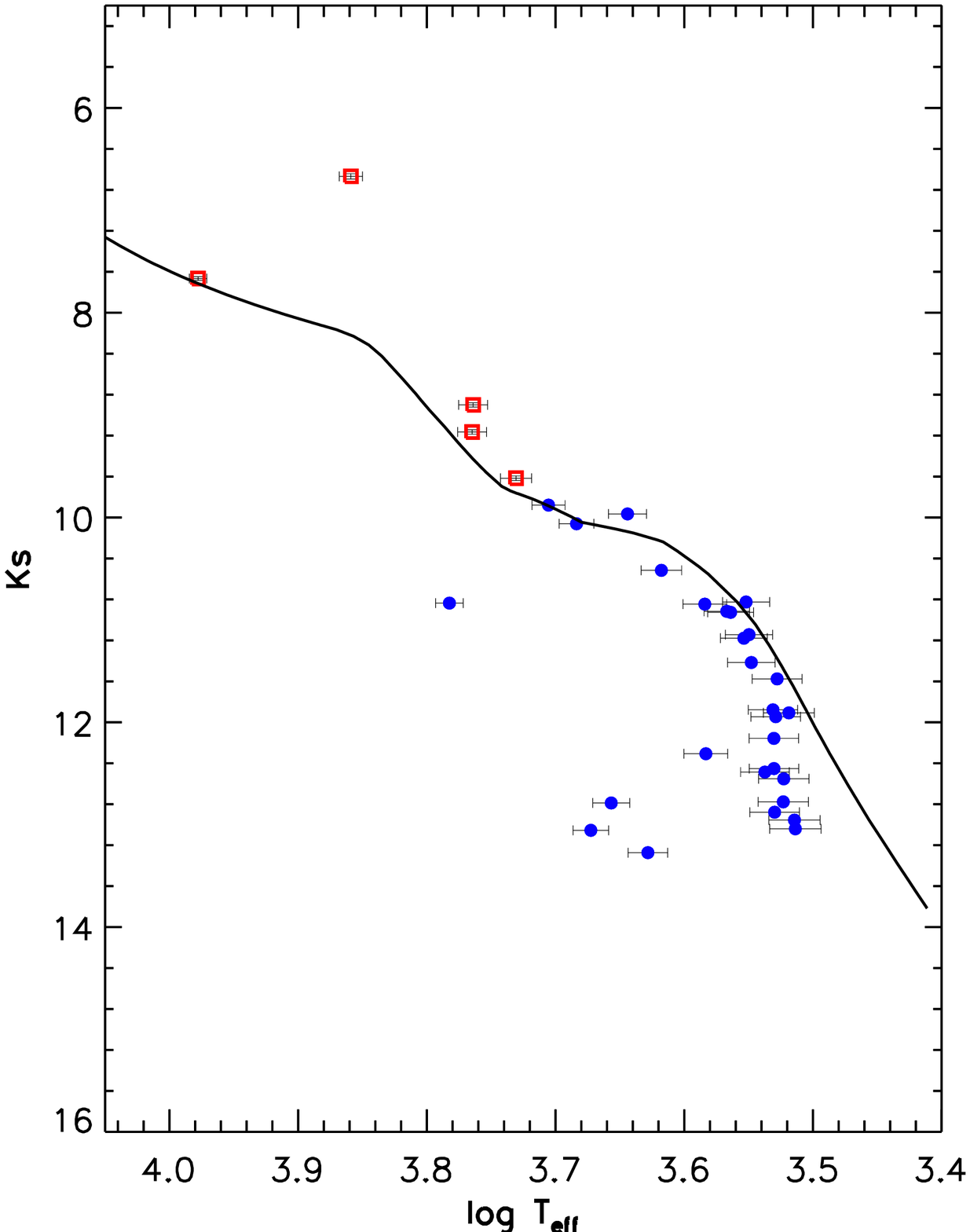}\\
        \includegraphics[width=0.325\linewidth]{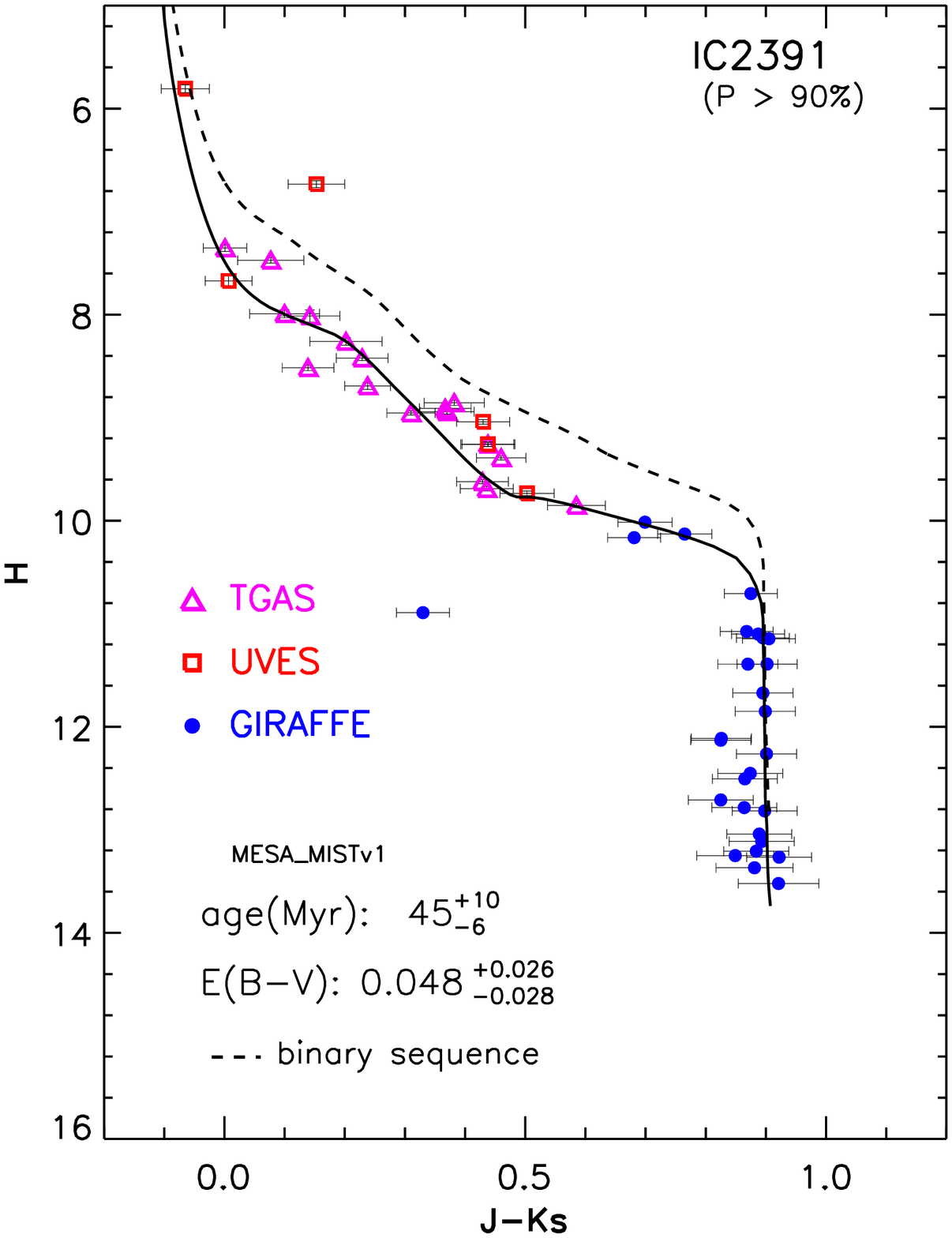}
        \includegraphics[width=0.325\linewidth]{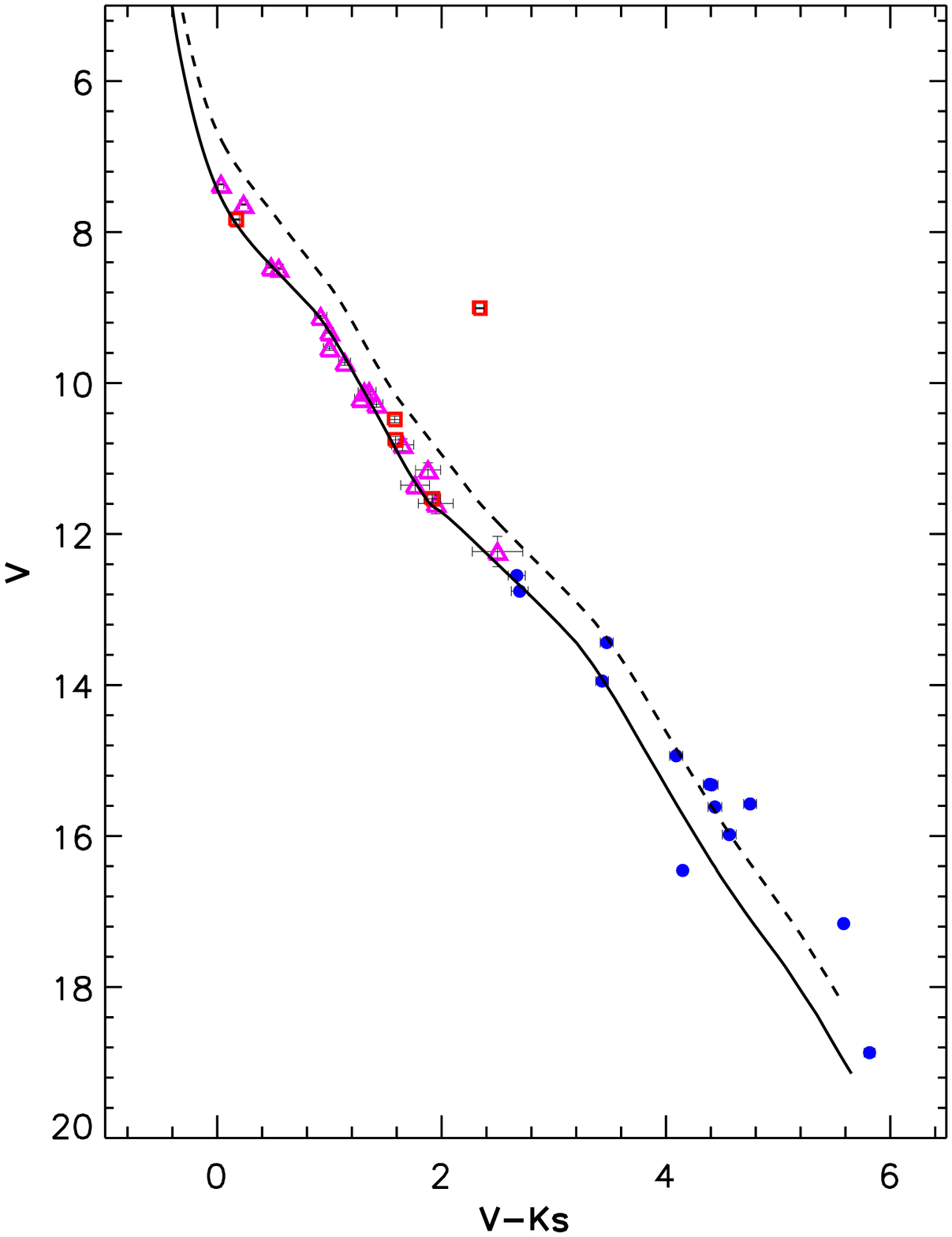}
        \includegraphics[width=0.325\linewidth]{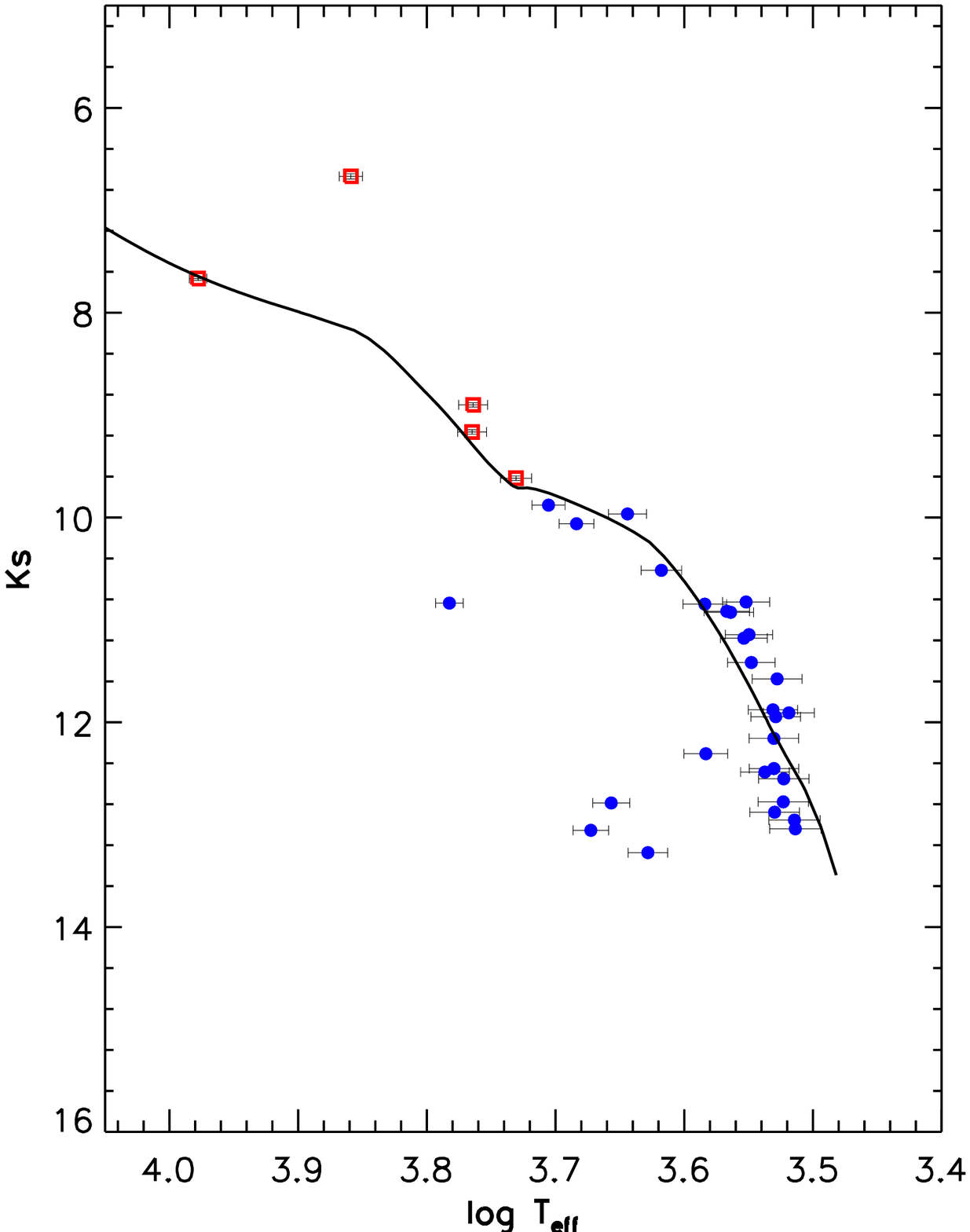}
        \caption{As in Fig.~\ref{fig:CMDPROSECCO_1} but for the cluster IC~2391 with the PARSEC (top panel) and MIST (bottom panel) isochrones.}
        \label{fig:CMDPARSEC_MESA_1}
\end{figure*}
\subsection{MIST and PARSEC models}\label{sect:othermodel}
In order to investigate the dependence of the recovered ages and reddening values on the adopted set of stellar models, we performed the same analysis using the PARSEC and MIST isochrones. 
The relevant sets of isochrones in the same age intervals used for PROSECCO were downloaded from the databases.

The age and reddening values obtained with the PARSEC and MIST models are listed in Table~\ref{tab:clusters_prop} together with the PROSECCO results. The table shows that with the exception of NGC~2516, the ages derived using the three model sets are in good agreement. The differences for the reddening are instead more significant; in particular, we note that MIST models and the analysis with the 2MASS magnitudes provide the lowest values, 
likely due to the fact that MIST isochrones for low-mass stars are slightly redder than the other two datasets in the $H$ versus $J-K_{\rm s}$ plane.

The comparison of the full set of sample clusters with the MIST and PARSEC models is shown in Appendix.~\ref{sec:CMD}; 
Fig~\ref{fig:CMDPARSEC_MESA_1} only shows the comparison between the data for IC~2391 and the PARSEC (top panel) and MIST (bottom panel) isochrones. 
The agreement between the models and observed stars in the 2MASS colour-magnitude diagrams appears good. PARSEC isochrones better reproduce the low-mass tail in the ($V-K_\mathrm{s},$ $V$) plane, while MIST models are bluer 
than data in this region. On the other hand, when the effective temperature is considered, MIST models appear in good agreement with observations, like the PROSECCO ones,
while PARSEC isochrones
show large deviations for $\log \rm T_\mathrm{eff} \la 3.6$. This is probably due to the $T-\tau$ modification introduced in the PARSEC models, discussed in Appendix~\ref{sec:models_conf}. The same deviation in the T$_\mathrm{eff}$ plane found for low-mass stars when adopting PARSEC isochrones is present for all the analysed clusters. 

As already mentioned, $V-K$ colours in the models (the PROSECCO and MIST ones at least) tend to be bluer 
than observations for low-mass stars (i.e.,
for $V$ magnitudes larger than about 16 in the selected clusters), 
thus possibly leading to a systematic effect in the derived age and reddening. 
In order to check the robustness of our procedure, we hence also made a test including in the recovery the Johnson $V$ magnitude together with the 2MASS bands. The results are shown in the bottom part of
Table~\ref{tab:clusters_prop} for the three selected model sets. The inclusion of the $V$ band in the recovery procedure yields ages that are generally in agreement within their estimated uncertainties with those derived using only the 2MASS bands; formal errors in ages are generally smaller, because of the inclusion of an additional band (although with fewer points).  

Concerning the recovered reddening, the E$(B-V)$ value is also sensitive to the use of the $V$ band in the recovery. However, generally, the derived E$(B-V)$ is compatible with that obtained in the 2MASS case, with the exception of NGC~2451A when the MIST and PROSECCO models are used. In this case, the derived reddening is much larger than that found using the 2MASS bands. 
\begin{figure*}
\includegraphics[width=0.5\linewidth]{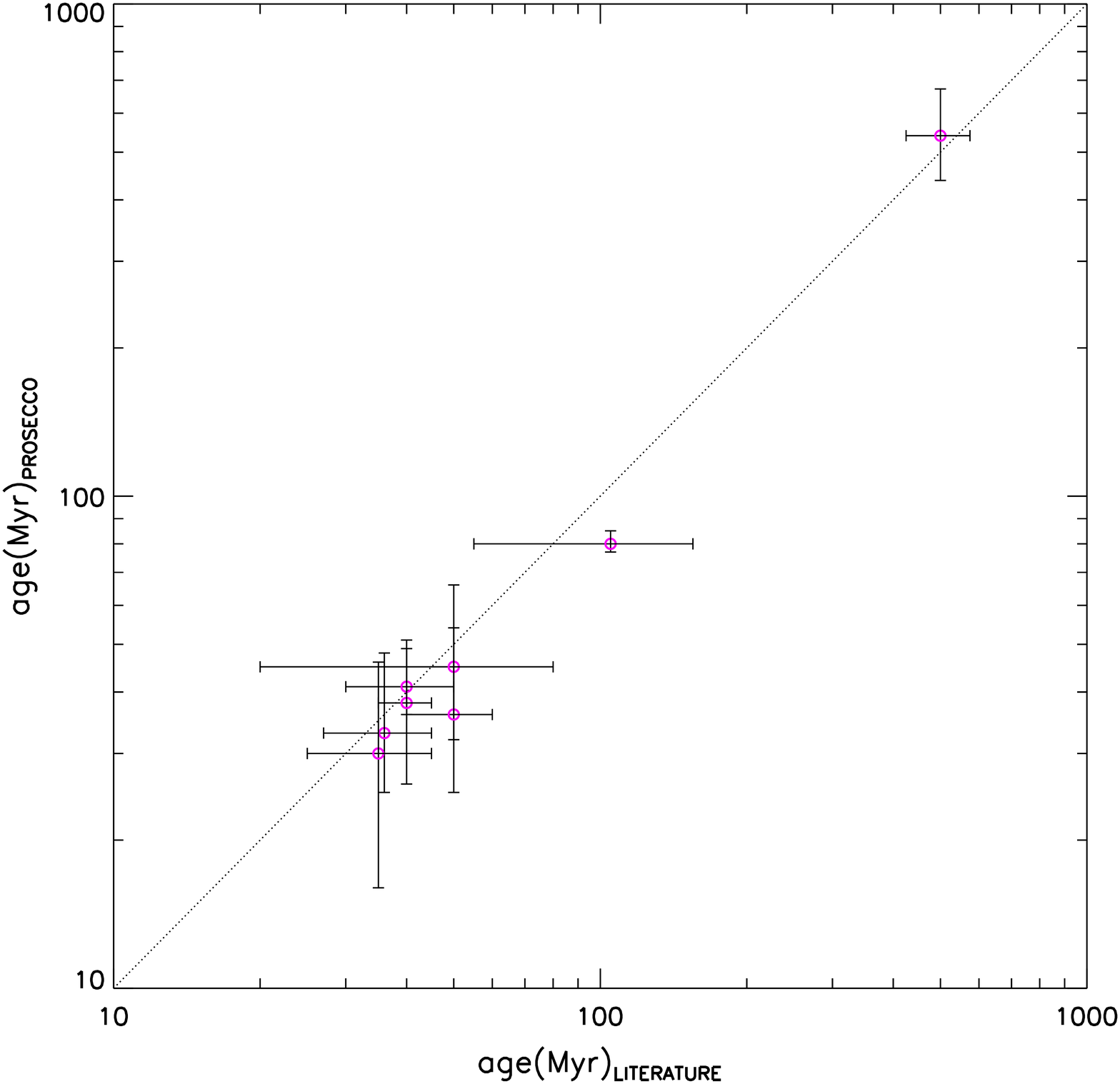}
\includegraphics[width=0.5\linewidth]{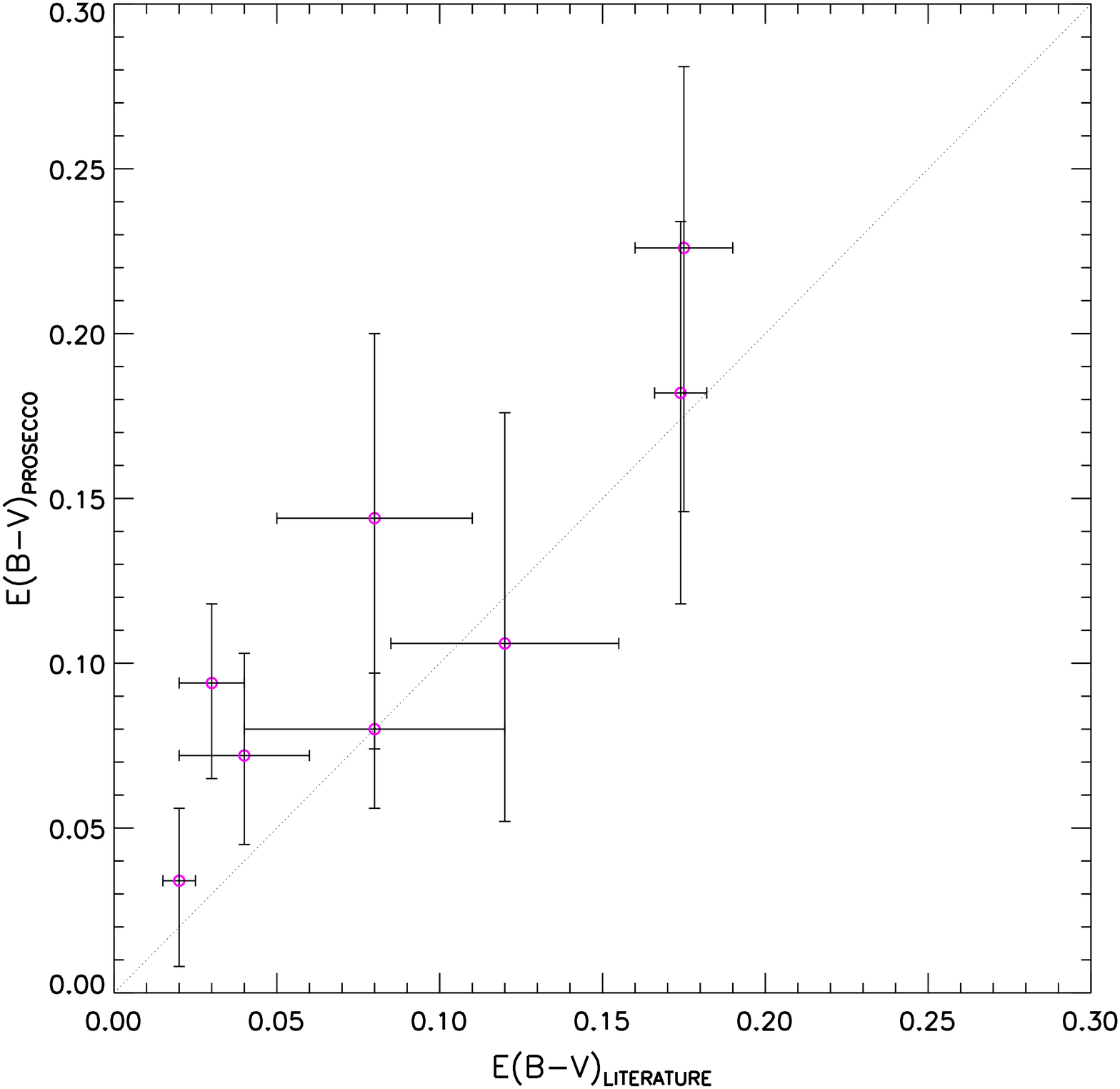}
\caption{Ages (left-hand panel) and reddening (right-hand panel) derived in this work using PROSECCO models vs. values from the literature (see Table~1).}\label{fig:comp}
\end{figure*}
\subsection{Comparison with the literature}\label{sec_comp} 
In Fig.~\ref{fig:comp} we plot our determination of ages and 
reddening obtained with the PROSECCO models and 2MASS magnitudes against literature values (see Table~\ref{cluster_sample}). For each cluster we show the average of the literature values and adopt the difference between the maximum and minimum values as error bars.
The Figure suggests a general good agreement for the majority of the clusters, although the values derived by us are somewhat smaller than literature determinations. Independently from the absolute values of the ages, the relevant result here is that all the sample clusters have been analysed in a homogeneous way and are now on the same age scale. Our rigorous statistical approach also indicates that cluster ages are generally very solid against the chosen models; this represents an interesting result, which was not obvious a priori. 
As already noted, Table~\ref{tab:clusters_prop} shows that for all clusters, with the exception of NGC~2516 (which we discuss below), the ages recovered with the three sets of models are consistent within the errors, which have a significant contribution coming from the uncertainties in the cluster distances.

For some of the clusters
the final cumulative errors on the ages are still larger than or comparable to those achieved in the literature using photometry only and leaving the distance as a free parameter (e.g. Cargile \& James (2010) for IC~4665; Naylor \& Jeffries (2006) for NGC~2547).
As mentioned, the TGAS systematic error in parallax is several times larger than the random errors on the mean cluster parallaxes, it dominates the cumulative error in distance, and thus the age uncertainty.
\begin{table}
\centering
\caption{Comparison of the age confidence intervals obtained for all the 
clusters adopting a systematic uncertainty of 0.3~mas (reference --ref.) 
and 0.1~mas (reduced --red.) on TGAS parallaxes.
}
\label{tab:uncertainties_phot_dist_sys0p1mas}
\begin{tabular}{l|rr}
Cluster & $\tau_\mathrm{err,ref}$ & $\tau_\mathrm{err,red}$ \\
        & (Myr) & (Myr) \\
\hline
\hline
NGC2451A & $  +9,-13$  & $ +6, -7$  \\
NGC2451B & $ +30,-11$  & $+28, -7$  \\
NGC2516  & $  +5, -3$  & $ +4, -2$  \\
NGC2547  & $ +11,-12$  & $ +3, -5$  \\
IC2391   & $ +10, -5$  & $ +5, -3$  \\
IC2602   & $ +9,  -6$  & $ +2, -2$ \\
IC4665   & $ +15, -8$  & $ +4, -3$  \\
NGC6633  & $+132,-102$  & $+42,-48$ \\
\hline
\end{tabular}
\end{table}
The next {\it Gaia} DR2 (April 2018) will include a consistent astrometric solution based on {\it Gaia} observations only, and the small-scale spatial variations of the parallax systematics are expected to be significantly reduced. In order to test how this will affect the uncertainty of the age recovery, we analysed a case in which we assume that the parallax/distance is known with better accuracy. Namely, if we reduce the adopted systematic uncertainty in the mean cluster parallax to only 0.1~mas, instead of 0.3~mas, we find that the systematic distance uncertainty becomes comparable to the current statistical uncertainty (or in the worst cases, still twice as large). We ran the parameter recovery process using this smaller total distance uncertainty for all the clusters.
Table~\ref{tab:uncertainties_phot_dist_sys0p1mas} shows the comparison between the original age uncertainties, determined when the 0.3 mas systematic distance error was considered (along with the random part and the photometric one), and the values obtained when a systematic error of 0.1 mas was adopted. The Table shows that for all the clusters, the age confidence intervals are significantly reduced, with the exception of NGC\,2451B for which it is still not satisfactory. This test clearly confirms that distance uncertainty has a large impact on the final errors on ages; therefore, there is obvious room for a significant improvement of the results with the anticipated increase in the quality of \emph{Gaia} parallaxes in subsequent data releases. In addition, the fainter magnitude limit of {\it Gaia} DR2 (astrometric parameters available down to $G=20$ mag instead of $G=13$ mag as for TGAS) will greatly increase the number of cluster members, which will help to significantly reduce the random errors and estimate more accurately the local systematic errors. Finally, precise homogeneous photometry available in {\it Gaia} DR2 will also contribute a significant improvement in the precision of the age estimates.

Focusing again on NGC~2516, the recovered ages using the PROSECCO and MIST models are younger by about 30\% than commonly quoted values in the literature, while the ages from the PARSEC models are in better agreement with previous estimates. 
Figure~\ref{fig:CMDPROSECCO_4} shows that the 120~Myr PROSECCO isochrone provides in principle a 
very good fit to the few stars at the turn-off.
Indeed, we believe that the recovered age of $\sim 80$~Myr is driven by the stars close to the knee; whereas, as mentioned, these stars are less sensitive
to age than stars close to the turn-off; they are much more 
numerous, hence amplifying the small age dependence. For the PROSECCO and MIST models, the 80~Myr isochrone is close in colour to the barycentre  of those stars at the knee, while the same happens with the 100-120~Myr PARSEC isochrone.
To better constrain the age for this cluster, we require either more stars at the turn-off region or more precise photometry
for stars close to the knee.

We finally note that we performed an analysis of NGC~2516 adopting the 
parallax/distance of VL17, which corresponds to a smaller distance to the 
cluster. We derive an age of 80~Myr, but a very low reddening (E($B-V)=0.032_{-0.015}^{+0.019}$); this is clearly far too small, hence
supporting the larger distance derived by us.

The right-hand panel of Fig.~\ref{fig:comp} (see also Table~\ref{tab:clusters_prop}) indicates that reddening is far less well constrained than age. Non-negligible differences are present with respect to the literature values (with our own estimates being generally larger) and our values are characterised by  relatively large errors. 
Also the agreement between the results from the three models is less good than for the ages, with 
the MIST models generally providing values that are more similar to the literature ones. Part of the discrepancy between the reddening values derived here and those found in the literature might arise from the fact that the data used for the recovery mainly relies on near-infrared bands, which are less sensitive to the reddening. We believe that future analysis using {\it Gaia} photometry in the optical bands -- which will be available for all cluster members -- will allow improvement of this aspect.

\subsection{Lithium depletion boundary ages}\label{LDB}
For the youngest clusters in our sample, a lithium depletion boundary (LDB) age determination has been performed in the literature. Given the very final goal of a consistent age calibration, which implies comparing cluster ages obtained with different methods, we again derive LDB ages here. We used our own
estimates of the cluster distances, the PROSECCO models, and bolometric corrections, as well as the reddening values derived in this paper.  LDB data have been retrieved from Jeffries \& Oliveira (2005 -NGC~2547), Barrado y Navascu\' es et al. (2004 --IC~2391), Dobbie et al. (2010 --IC~2602), and  Manzi et al. (2008 --IC~4665), respectively. We note that the LDB age errors include the TGAS distance modulus uncertainty, but also an additional 0.1 mag uncertainty for photometric calibration (in colour and magnitude) and the likely uncertainty in the LDB location.
\begin{table*}
\caption{Lithium depletion boundary ages.}
\label{tabldb}
\centering
\begin{tabular}{cccccc}
\hline\hline
Cluster & DM & E($B-V$) &  $K_{\rm LDB}$ & ($I-K$)$_{\rm LDB}$  & Age \\
          & (mag)&            &                         &                                 &  (Myr)\\
\hline
NGC~2547 & $7.81\pm 0.25$ & $0.080\pm 0.024$ & $14.86\pm 0.12$ & $2.70\pm 0.10$ & $37.7^{+5.7}_{-4.8}$\\
IC~2391  & $5.82\pm 0.10$ & $0.088\pm 0.027$ & $13.49\pm0.10$& $2.75\pm 0.10$ & $51.3^{+5.0}_{-4.5}$\\
IC~2602  &  $5.85\pm 0.10$ & $0.068\pm 0.025$ & $13.23\pm 0.11$ & $2.60\pm 0.20$ & $43.7^{+4.3}_{-3.9}$\\
IC~4665  & $7.82\pm 0.25$ & $0.226\pm 0.080$ & $13.95\pm 0.13$ & $2.60\pm 0.10 $ & $23.2^{+3.5}_{-3.1}$\\
\hline
\end{tabular}
\end{table*}
LDB ages are shown in Table~\ref{tabldb}. Comparison with the ages reported in Table~\ref{tab:clusters_prop} suggests a good agreement; more specifically, the agreement is excellent for NGC~2547 and within $\sim 1 \sigma$ for IC~2391 and IC~4665. This implies that by using the same set of models and assumptions, discrepancies between LDB and isochronal ages become less significant than previously claimed. The difference in inferred ages is somewhat larger for IC~2602, with the LDB age being more than 2$\sigma$ older than the isochronal one. In order to further check the age obtained with the Bayesian analysis, we tried to change (manually) the age of IC~2602 in the range 30-40 Myr.
An age of more than about 32-33 Myr gets progressively less and less compatible with the data in the H versus $J-K_{\rm s}$, $V$ versus $V-K_{\rm s}$ or $K_{\rm s}$ versus T$_{\rm eff}$
diagrams, and the isochrones reproduce only the lower envelope of the observed sequence.
We also recall that an age around 30 Myr seems to be very robust even when adopting PARSEC or MIST models and/or including the $V$-band. We note however that the recovery of the age for this cluster is mainly driven by the stars located in the region where there is a change of slope in the isochrones and where even a small difference in magnitude may lead to a significant difference in age. 
Whilst there are many stars at that position, they do not follow a very narrow sequence, but are relatively scattered, possibly due to the uncertainties in the 2MASS photometry. Also in this case, future {\it Gaia} releases will allow a more detailed analysis, possibly also indicating a front-to-back distance spread, given the proximity of the cluster.
\section{Summary}
\label{sec:conclusions}
We have homogeneously
combined spectroscopic information from GES and {\it Gaia}-DR1 TGAS parallaxes
for eight OCs, to derive a uniform set of cluster ages and reddening values,
 using a statistical Bayesian analysis applied to three different sets of standard evolutionary models. Parallaxes from TGAS data have been calculated and compared with the results of VL17.
The agreement is good for all but one cluster, NGC~2516, for which we derive a smaller parallax (larger distance) than VL17. Both astrometric tests and the comparison with the isochrones support our own determination.

Our results show that the proposed approach is very promising. We have not only successfully tested our methodology, which will be further exploited with {\it Gaia}-DR2, but also the eight clusters have been put onto a consistent age scale for the first time. Although this ranking is model dependent, comparison of the results using the three different sets of models generally shows very good agreement, indicating that ages are robust against the chosen models for all but one cluster, NGC~2516. For this cluster,  with two of the models (MIST and PROSECCO) we derived an age which is significantly younger than most values from the literature, while the PARSEC model provides an age in agreement with literature values. The younger age may be due to the lack of stars near the turn-off region, which prevents a robust age determination. 

The LDB ages, determined in this paper for the four youngest clusters in the sample using the same evolutionary models (the PROSECCO ones), also generally agree well with isochronal ones. It has been claimed (Soderblom et al. 2014) that the LDB ages are far less model-dependent than ages determined from isochronal fits to high- or low-mass stars. It is therefore encouraging that three quarters of the clusters for
which we have LDB and isochronal ages are in excellent agreement.
E($B-V$) values are instead less well constrained, since the Bayesian analysis has been performed using mainly near-infrared magnitudes that are not very sensitive to reddening. 

Our analysis confirms that current errors in parallaxes, which are dominated by a relatively large systematic error, are a significant source of uncertainty in the derived ages (about 15-30\%), especially in clusters with few and/or scattered photometric data. Indeed, ages and distances derived in the literature using photometry only 
can be more precise, but are not necessarily more accurate.
We have quantitatively shown, however, how age uncertainties are expected to significantly improve once better {\it Gaia} parallaxes are available. Also, at this stage we cannot yet put extremely tight constraints on the models; nevertheless, we showed already that, among the chosen model sets, the PARSEC ones do not provide a good agreement with the observed sequences in the magnitude versus temperature diagrams.

{\it Gaia}-DR2, along with the full GES cluster dataset, will allow the extension of the present analysis to many more OCs, better sampling the age interval from a few Myr to several Gyr, and covering also metallicities above and below the solar value. The combination of {\it Gaia} proper motions with spectroscopic membership indicators will yield very clean sequences from the very bright down to the very faint stars.
Extremely precise and homogeneous photometry from {\it Gaia} for virtually all cluster stars will also provide very valuable information. All this will allow more detailed comparisons with the models, including also those accounting for non-standard processes like rotation and magnetic fields, resulting in more
crucial tests on the input physics. The dataset will not only enable tighter constraints on ages of our current sample and of many additional, more-distant clusters, but will also allow us to use that age information to calibrate a variety of commonly used age tracers, which, in turn, will allow estimates of the ages of field stars.
\begin{acknowledgements}
Based on data products from observations made with ESO Telescopes at the La Silla Paranal Observatory under programme IDs 188.B-3002, 193.B.0936, and 197.B-1074. 
These data products have been processed by the Cambridge Astronomy Survey Unit (CASU) at the Institute of Astronomy, University of Cambridge, and by the FLAMES/UVES reduction team at INAF/Osservatorio Astrofisico di Arcetri. These data have been obtained from the Gaia-ESO Survey Data Archive, prepared and hosted by the Wide Field Astronomy Unit, Institute for Astronomy, University of Edinburgh, which is funded by the UK Science and Technology Facilities Council.
This work has made use of data from the European Space Agency (ESA)
mission {\it Gaia} (\url{http://www.cosmos.esa.int/gaia}), processed by
the {\it Gaia} Data Processing and Analysis Consortium (DPAC,
\url{http://www.cosmos.esa.int/web/gaia/dpac/consortium}). Funding
for the DPAC has been provided by national institutions, in particular
the institutions participating in the {\it Gaia} Multilateral Agreement. This research was made possible through the use of the AAVSO Photometric All-Sky Survey (APASS), funded by the Robert Martin Ayers Sciences Fundation.
This research has made use of the SIMBAD and VizieR databases,
operated at CDS, Strasbourg, France.
This research has made use of NASA's Astrophysics Data System. 
This work was partly supported by the European Union FP7 programme through ERC grant number 320360 and by the Leverhulme Trust through grant RPG-2012-541. We acknowledge the support from INAF and Ministero dell' Istruzione, dell' Universit\`a' e della Ricerca (MIUR) in the form of the grant "Premiale VLT 2012" and by PRIN-INAF 2014. The results presented here benefit from discussions held during the Gaia-ESO Survey workshops and conferences supported by the ESF (European Science Foundation) through the GREAT Research Network Programme.
E.T., P.G.P.M. and S.D. acknowledge PRA Universit\'a di Pisa 2016 (\emph{Stelle di piccola massa: le pietre miliari dell'archeologia galattica}, PI: S. Degl'Innocenti) and INFN (Iniziativa specifica TAsP). M.T.C acknowledge the financial support from the Spanish Ministerio de Econom\'{\i}a y Competitividad, through grant AYA2016-75931. U.H. acknowledges support from the Swedish National Space Board (SNSB/Rymdstyrelsen).
PRF and AS acknowledge useful discussions with Mario Lattanzi. We thank an anonymous referee for the very helpful and constructive comments.
\end{acknowledgements}
%
%
%
\bibliographystyle{aa} 

\begin{appendix} 
\section{Evolutionary Models} 
\label{sec:models_conf} 
The main characteristics of the selected model grids are shown in Table~\ref{tab:inputmodelli}. The 
selected stellar libraries have been computed adopting similar EOS, opacity, and
solar mixture, while the atmospheric models used to specify the outer boundary 
conditions are different. This last quantity is particularly important 
in the case 
of low and very-low mass stars, where it can be responsible for relatively 
large 
deviations among models which adopt different BCs (see 
e.g. Tognelli et al. 2011).

\begin{figure*}
        \centering 
        \includegraphics[width=0.325\linewidth]{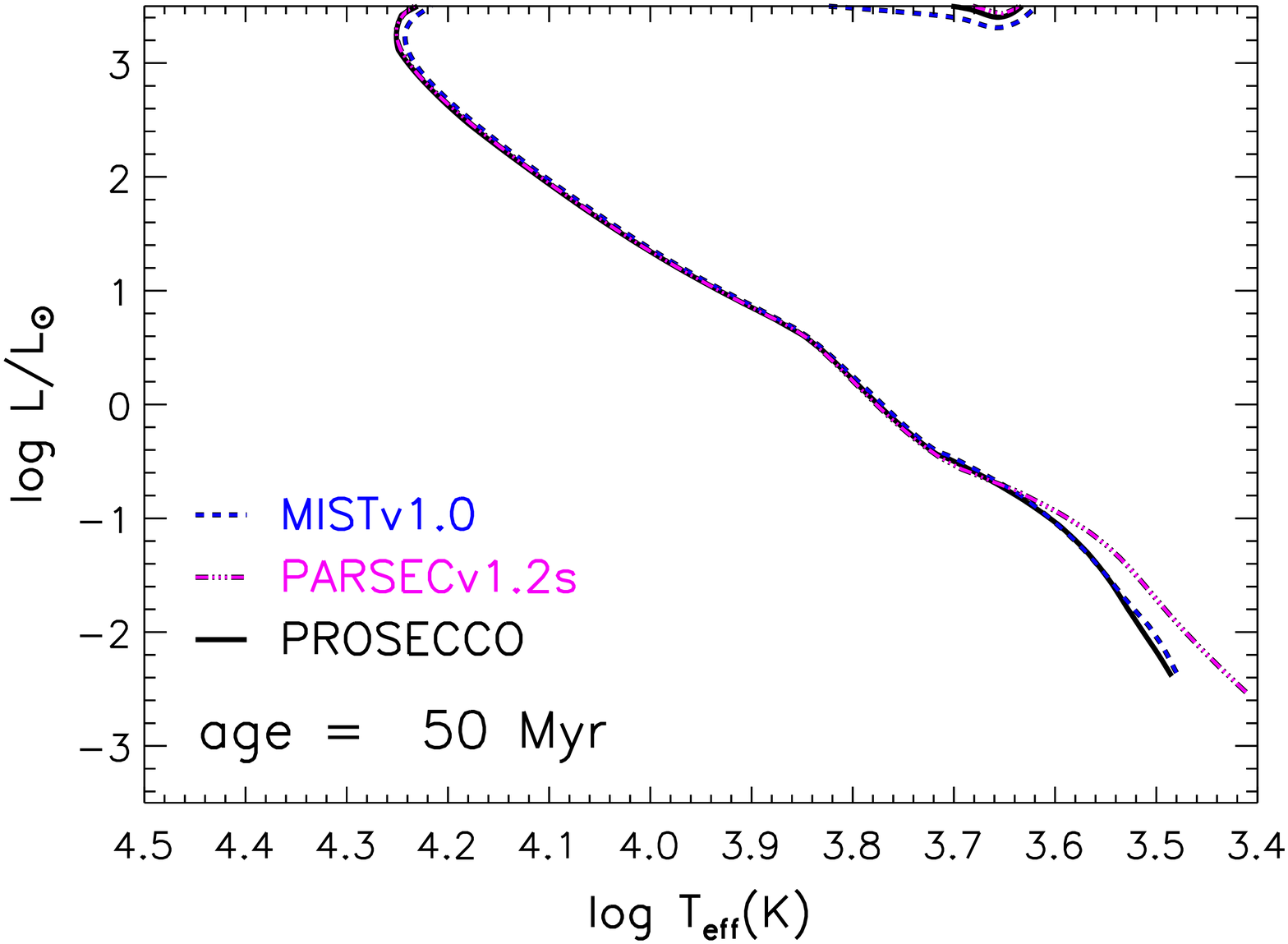} 
        \includegraphics[width=0.325\linewidth]{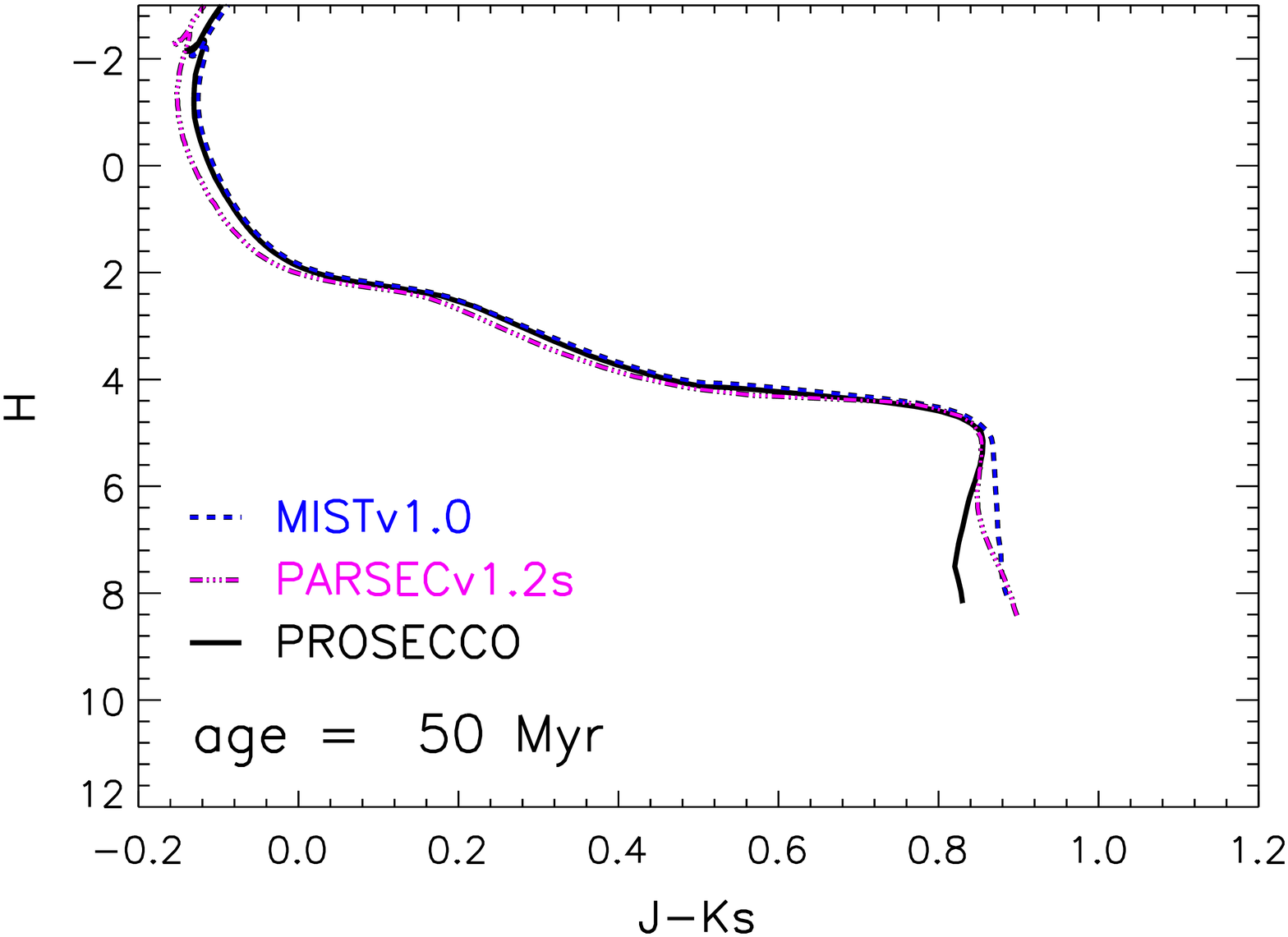}
        \includegraphics[width=0.325\linewidth]{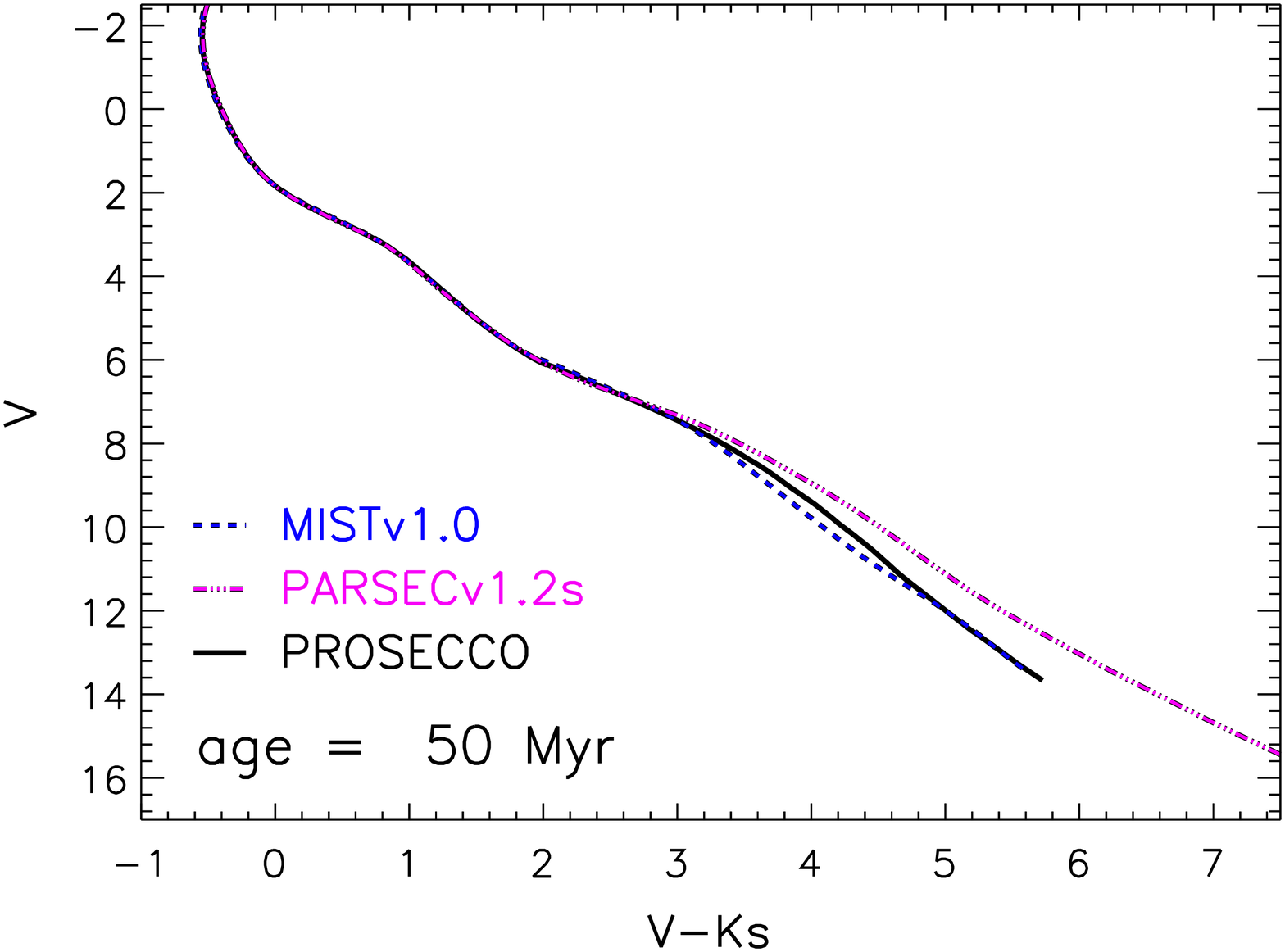}\\
        \includegraphics[width=0.325\linewidth]{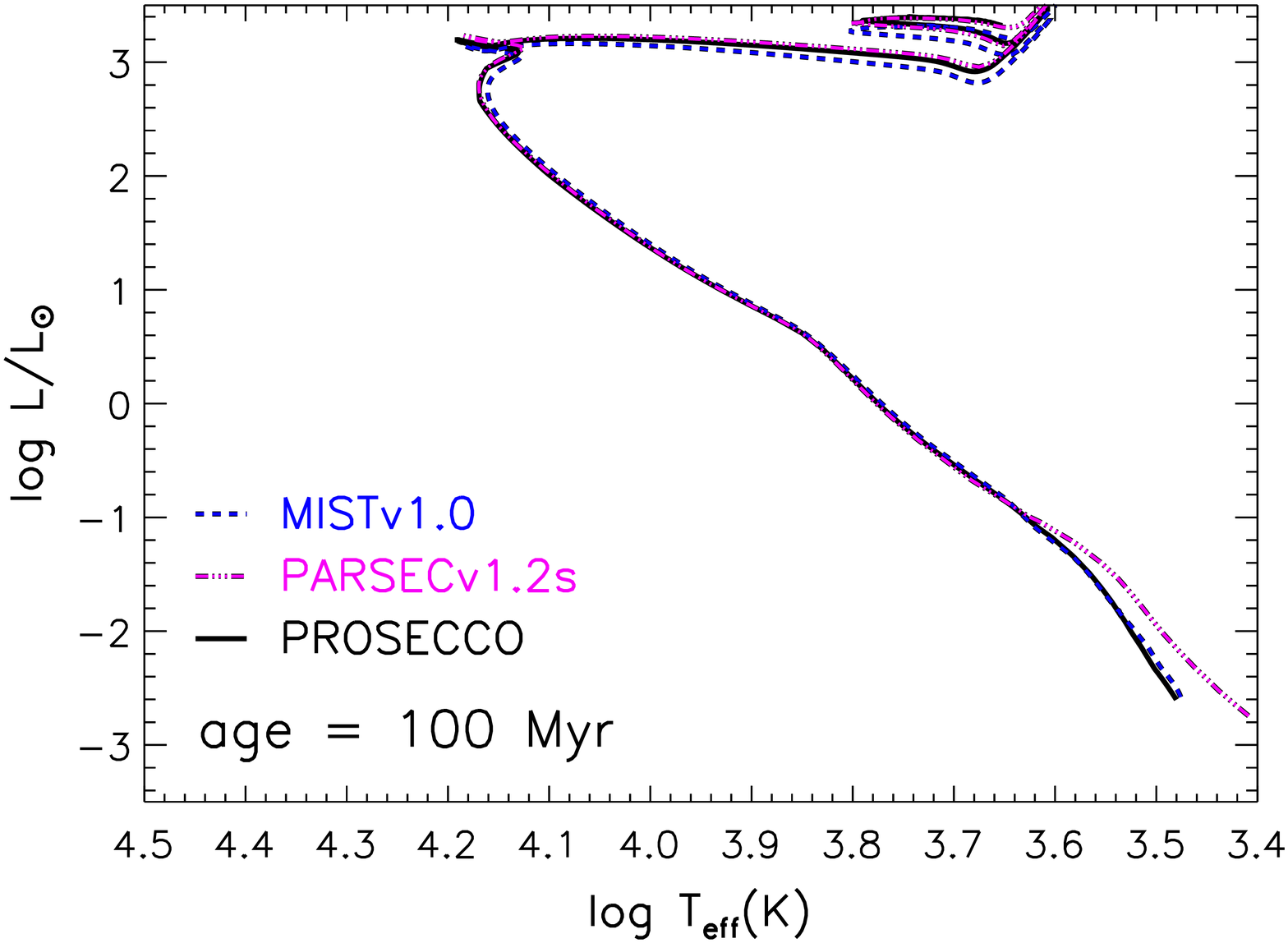} 
        \includegraphics[width=0.325\linewidth]{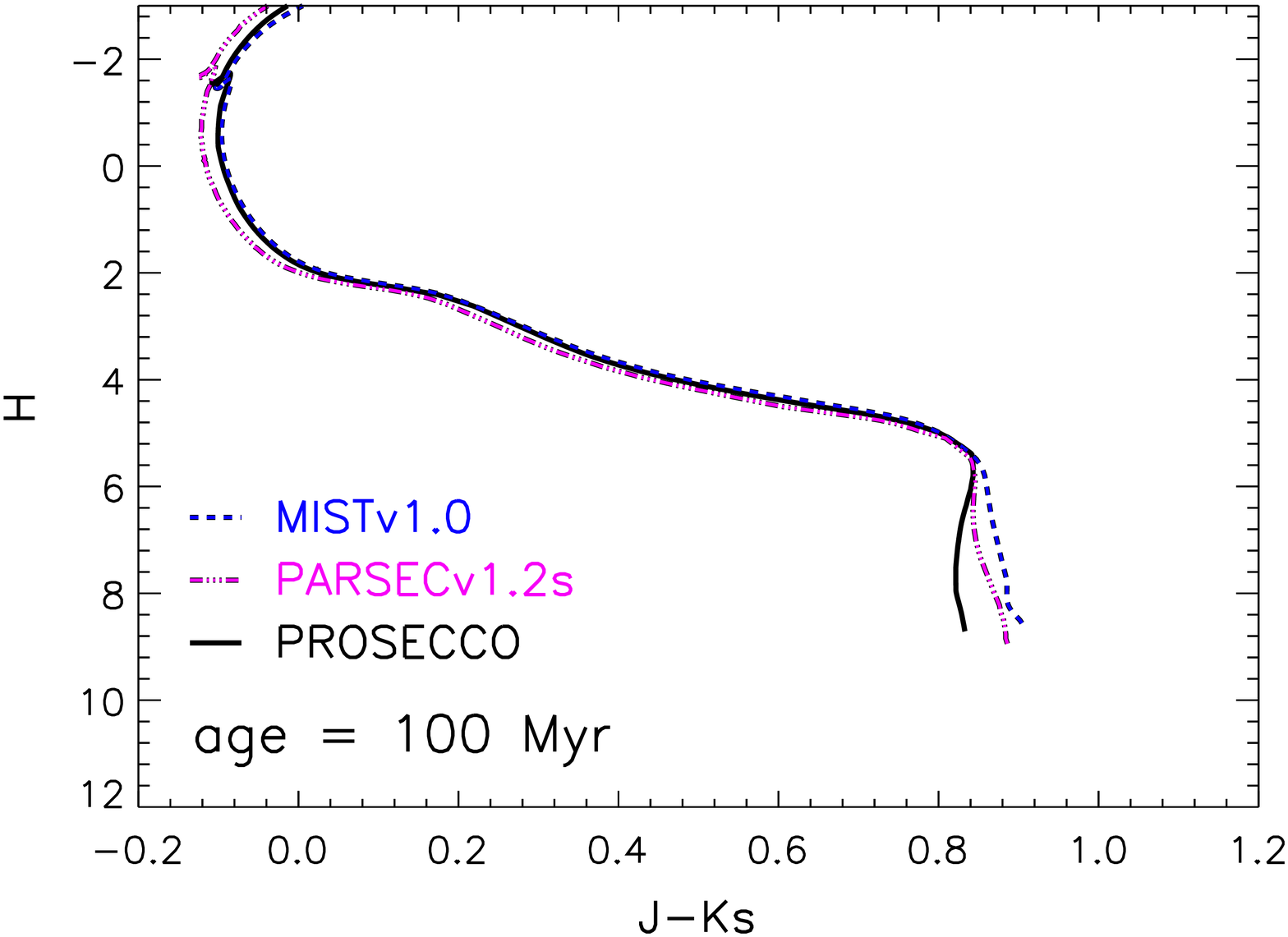}
        \includegraphics[width=0.325\linewidth]{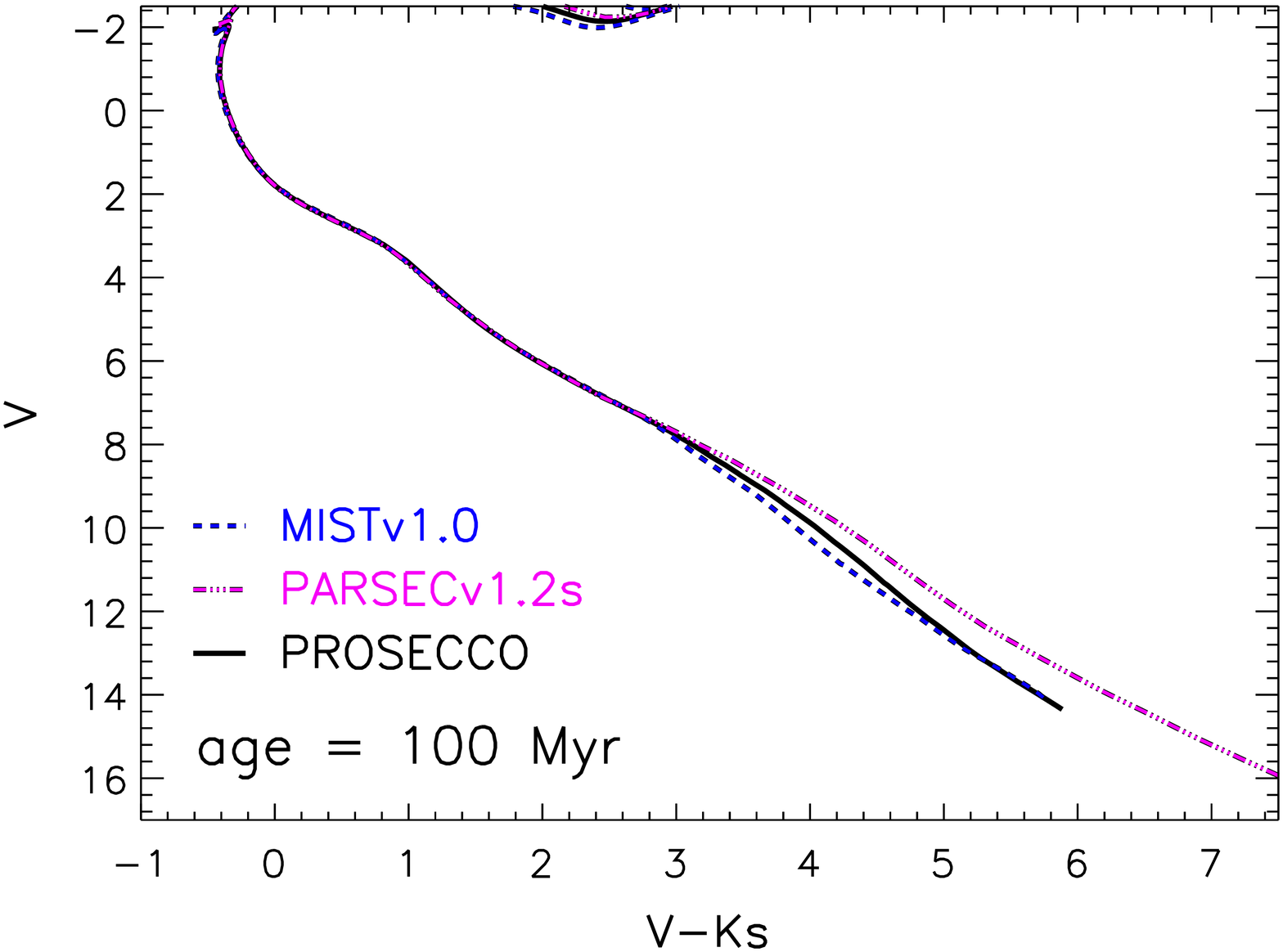}\\
        \includegraphics[width=0.325\linewidth]{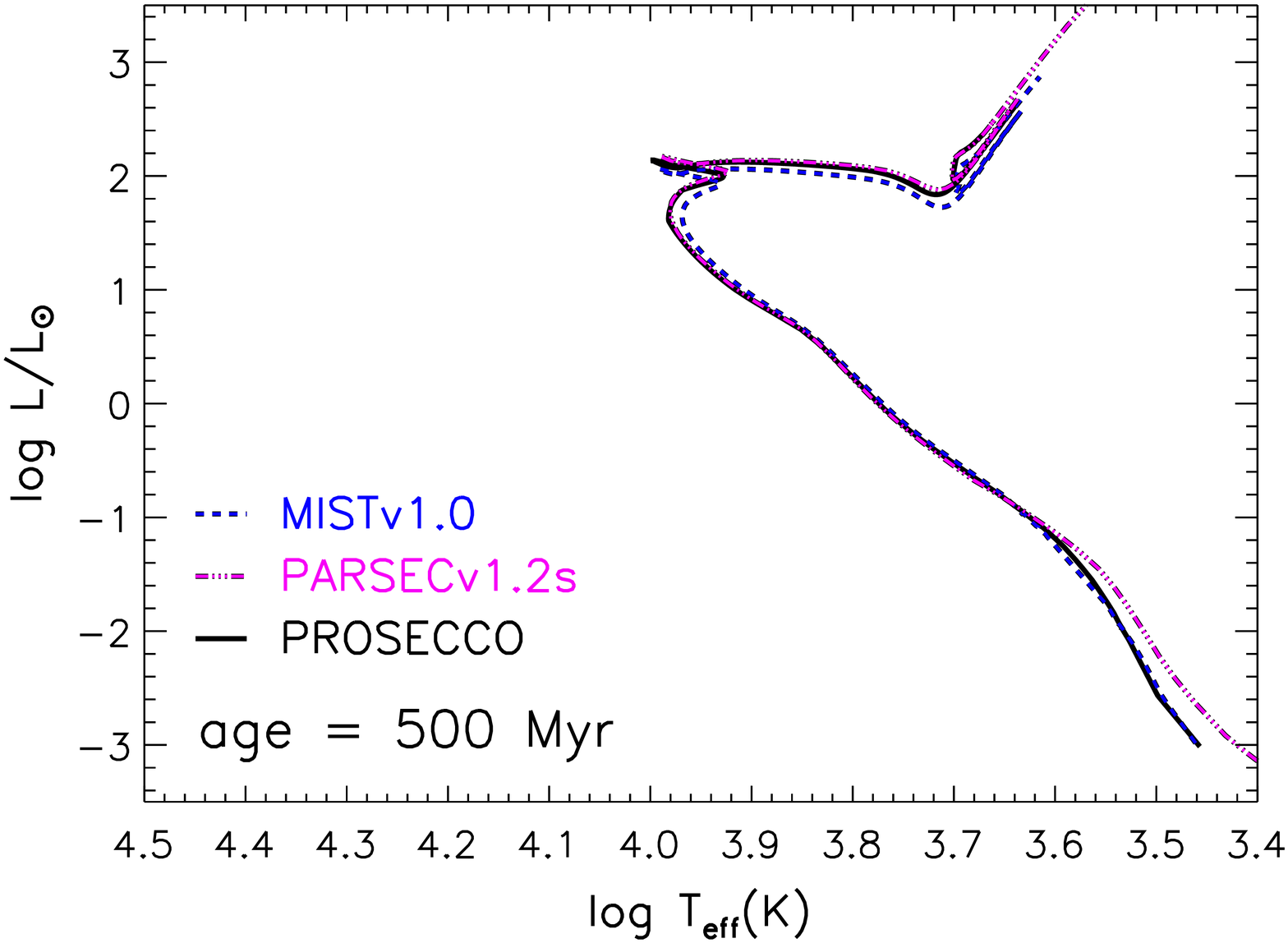} 
        \includegraphics[width=0.325\linewidth]{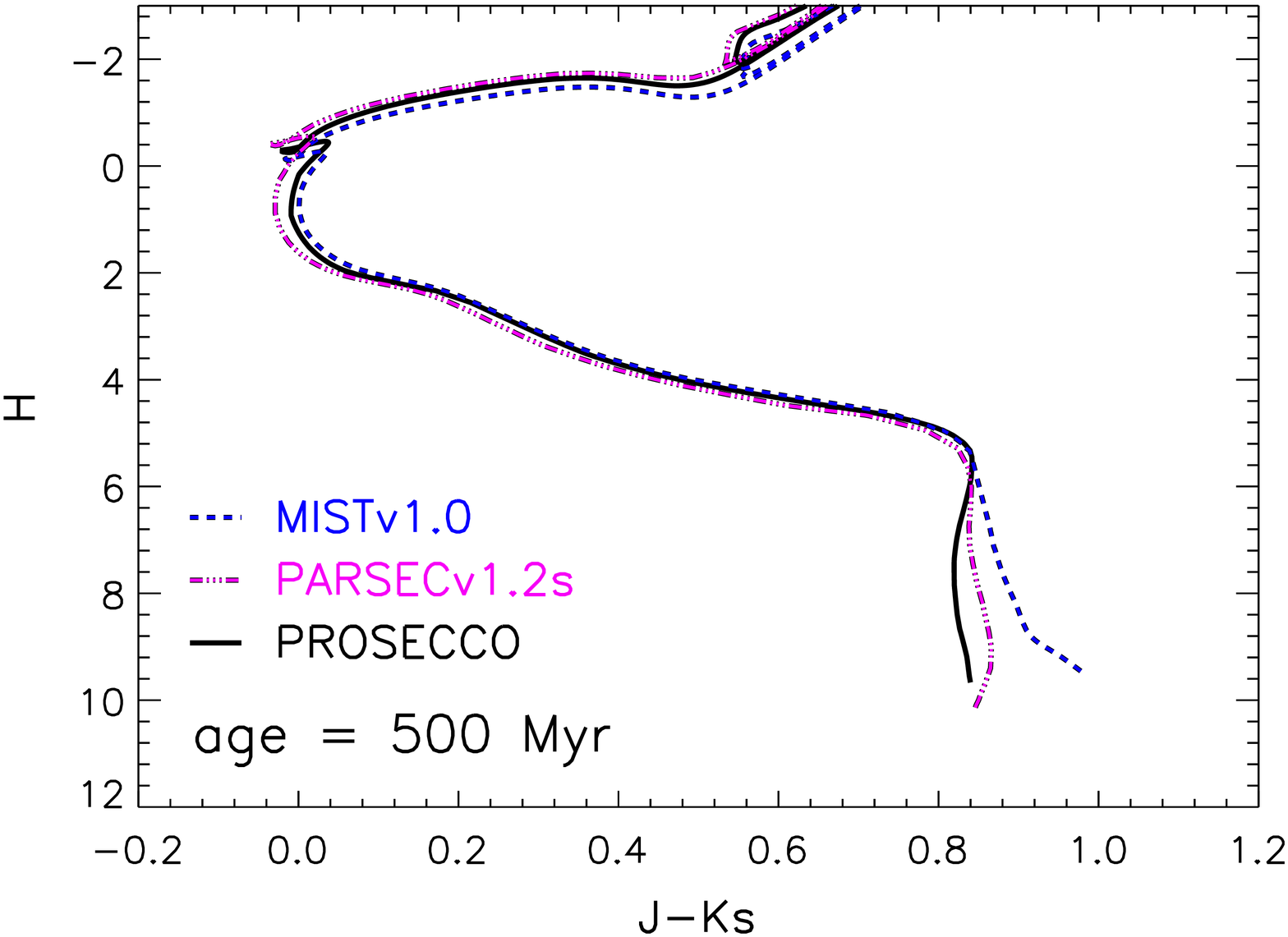}
        \includegraphics[width=0.325\linewidth]{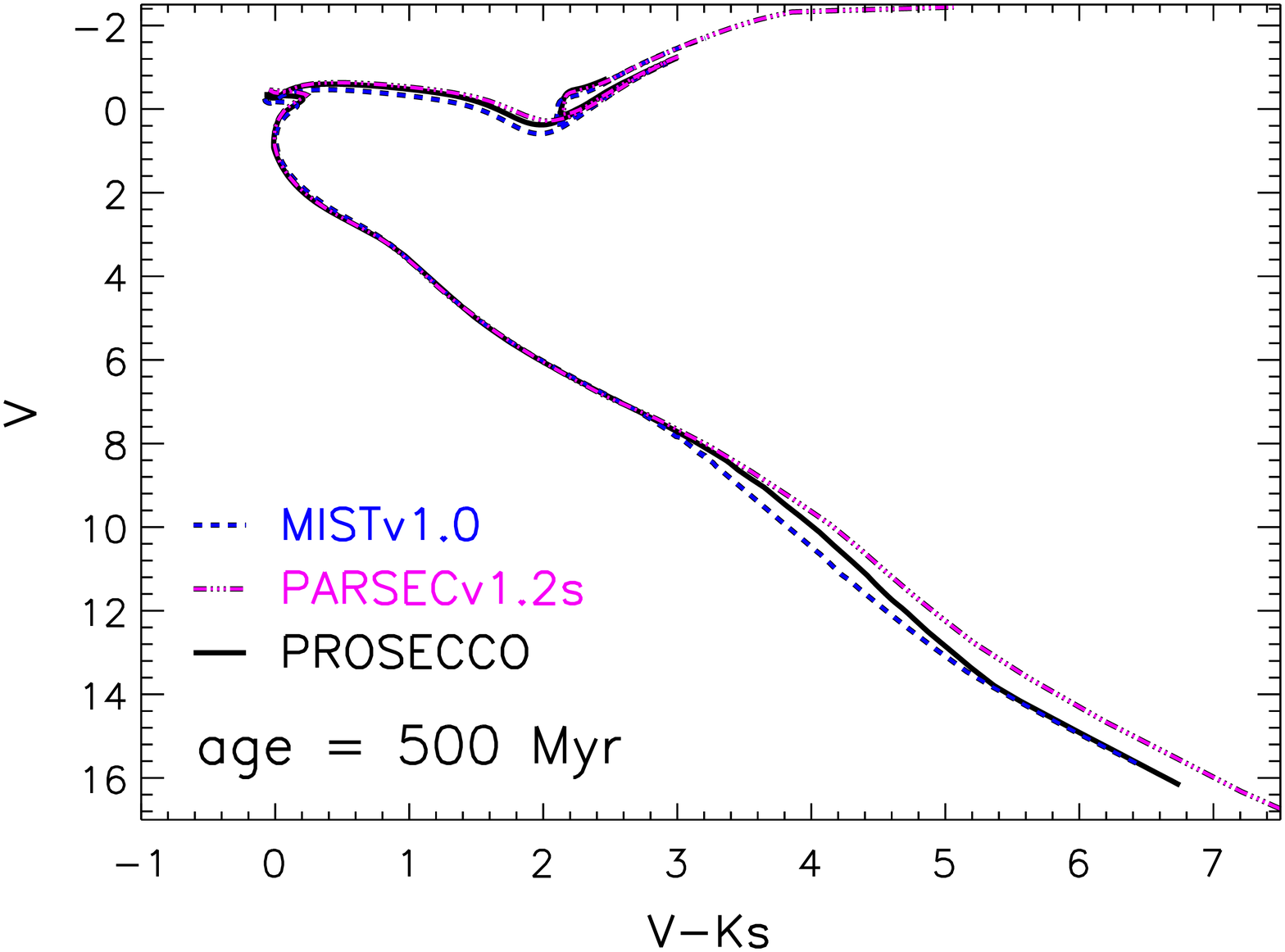} 
        \caption{Comparison among 50, 100 and 500~Myr isochrones from the selected sets in the HR diagram and in various CM diagrams. MISTv1.0= MESA isochrones by \citet{choi2016,dotter2016}, PARSECv12.s= PARSEC isochrones in the PARSEC 1.2S version \citep[][]{chen2014}, PROSECCO= isochrones specifically calculated for the present work; see text. }
        \label{fig:comparison}
\end{figure*}

Figure~\ref{fig:comparison} compares solar composition isochrones of the selected sets for three ages in the range covered by the sample of observed OCs: 50~Myr, 100~Myr and 500~Myr. The comparison is made in the HRDs and in the CMDs in the photometric bands for which observational data are available (2MASS $J$, $H$, $K_\mathrm{s}$ and Johnson $V$ bands).
The Figure shows that in the HRD the PROSECCO and MIST isochrones below the turn-off are in excellent agreement among each other at all ages, while PARSEC isochrones are different from the other sets at low luminosities. This deviation is probably caused by an empirical recalibration of the atmospheric T$-\tau$ relation adopted in the PARSEC models for T$_\mathrm{eff} \le 4730$~K ($\log T_\mathrm{eff} \approx 3.67$) to reproduce the observed mass-radius relation \citep{chen2014}. Such a recalibration seems to produce a large effect on the predicted T$_\mathrm{eff}$ which gets significantly cooler than those of the other models at the same luminosity, reaching a maximum difference of 200-400~K at $\log T_\mathrm{eff}\sim 3.5$. 

Near the turn-off region and in the central He burning phase the differences among the isochrones (mainly between PROSECCO/PARSEC and MIST) increase. However, in the age range we are dealing with, these two evolutionary phases are both sensitive also to the core overshooting efficiency during central hydrogen burning, which is different in the selected models.

The overshooting phenomenon, that is, the extension of the central mixing regions beyond the Schwarzschild border, $\ell_\mathrm{ov}$, is generally parametrised in terms of the pressure scale height $H_\mathrm{P}: \ell_\mathrm{ov} = \beta_\mathrm{ov}\times H_\mathrm{P}$, where $\beta_\mathrm{ov}$ is a free parameter. For this comparison the PROSECCO models are calculated with $\beta_\mathrm{ov}=0.25$. Note that PARSEC and MIST isochrones adopt a different treatment to estimate the core overshooting length $\ell_\mathrm{ov}$. However, both formalisms are well reproduced by standard overshooting treatment if $\beta_\mathrm{ov}=0.2$-0.25 is adopted \citep{bressan2012,choi2016}.
In our cluster sample only clusters with age of the order of or higher than 100~Myr include stars in the central hydrogen exhaustion region (turn-off)
and in more advanced evolutionary phases (Red Giant Branch, RGB or central He burning). The CMDs show additional differences among the various sets with respect to the ones present in the HR diagram. This is due to the choice of synthetic spectra adopted to compute the bolometric corrections, required to convert luminosities into magnitudes. The differences in magnitude among the selected sets of isochrones vary with the chosen photometric system and with the stellar mass. In the ($J-K_{\rm s}$,$H$) plane a maximum colour difference of about 0.04~mag (similar to typical photometric errors in the 2MASS CMDs of the selected clusters) is reached for the masses which populate the isochrone, with the exception of the low-mass stars tail, where the differences can exceed 0.1~mag. Similar differences are present in the ($V-K_{\rm s}$,$V$) plane among the PROSECCO 
and MIST models, while the PARSEC ones can reach a colour difference of about 0.4~mag in the case of low-mass stars 
\section{Cluster CMDs cleaning procedure and best models choice}
\label{sec:cleaning}
As mentioned in the text, the CMDs of some of the analysed clusters are affected by contamination (field stars, unresolved binaries) even after the member selection process. Some of these stars, that have a membership probability higher than the imposed threshold, actually lie in regions of the CM diagrams incompatible neither with either a single star isochrone sequence, or with a plausible binary sequence. The presence of such outliers does not affect the final age and reddening determinations, but contributes with a large $\chi^2$ value to the total likelihood. Hence, we preferred to remove such objects from the sample used to derive the most probable age and reddening. However, in all the sample clusters the cleaning procedure removes a maximum of 3-4 objects which are those lying clearly below/above the cluster sequences and that can easily be identified by the reader.

The adopted cleaning procedure relies on four steps.
\begin{itemize}
        \item We identified the location of the ZAMS for each cluster. This has been done by comparing PROSECCO isochrones with data using the TGAS dereddened distance modulus and the average E$(B-V)$ available in the literature. We also checked that the adopted E$(B-V)$ value is compatible with the data for all the used photometric bands. Then, we identified the stars that have minimum distance from the single star ZAMS (above or below it) larger than $10\sigma$ and simultaneously in all the adopted photometric bands. These objects have been removed from the data sample used in the age analysis.
        \item We ran the recovery procedure for the sample without the identified outliers, obtaining age and reddening values. We checked that the removed stars are outliers even adopting the new derived E$(B-V)$ (always verified). 
At this stage we did not include the unresolved binary sequence.
        \item We used the age and reddening values to obtain the fiducial sequence. We built the unresolved binary sequence from such isochrone and we identified objects above the binary sequence with a minimum distance larger than $10\sigma$ (in all the used photometric bands). These stars have been removed.
        \item We ran again the recovery with the new cleaned sample, including the unresolved binary sequence,  and checked that the outliers still verify the distance conditions with the new derived age and reddening. 
\end{itemize}

From such an analysis, we found that the derived age and reddening are not affected by the inclusion or exclusion of the outliers. The adoption of the cleaned sequence allows us to achieve a lower $\chi^2$ value (hence a larger likelihood). 
In addition, we checked that the most probable set actually reproduces the data and that, in the case of the inclusion of binaries, the fraction of stars within $1\sigma$ from the binary sequence is less than the stars within $1\sigma$ from the single star sequence. This is an empirical check to avoid that the single star sequence is less populated than the binary one, which would be unlikely.
\section{Cluster CMDs and temperature-magnitude diagrams} 
\label{sec:CMD} 
\begin{figure*}
        \centering
        \includegraphics[width=0.325\linewidth]{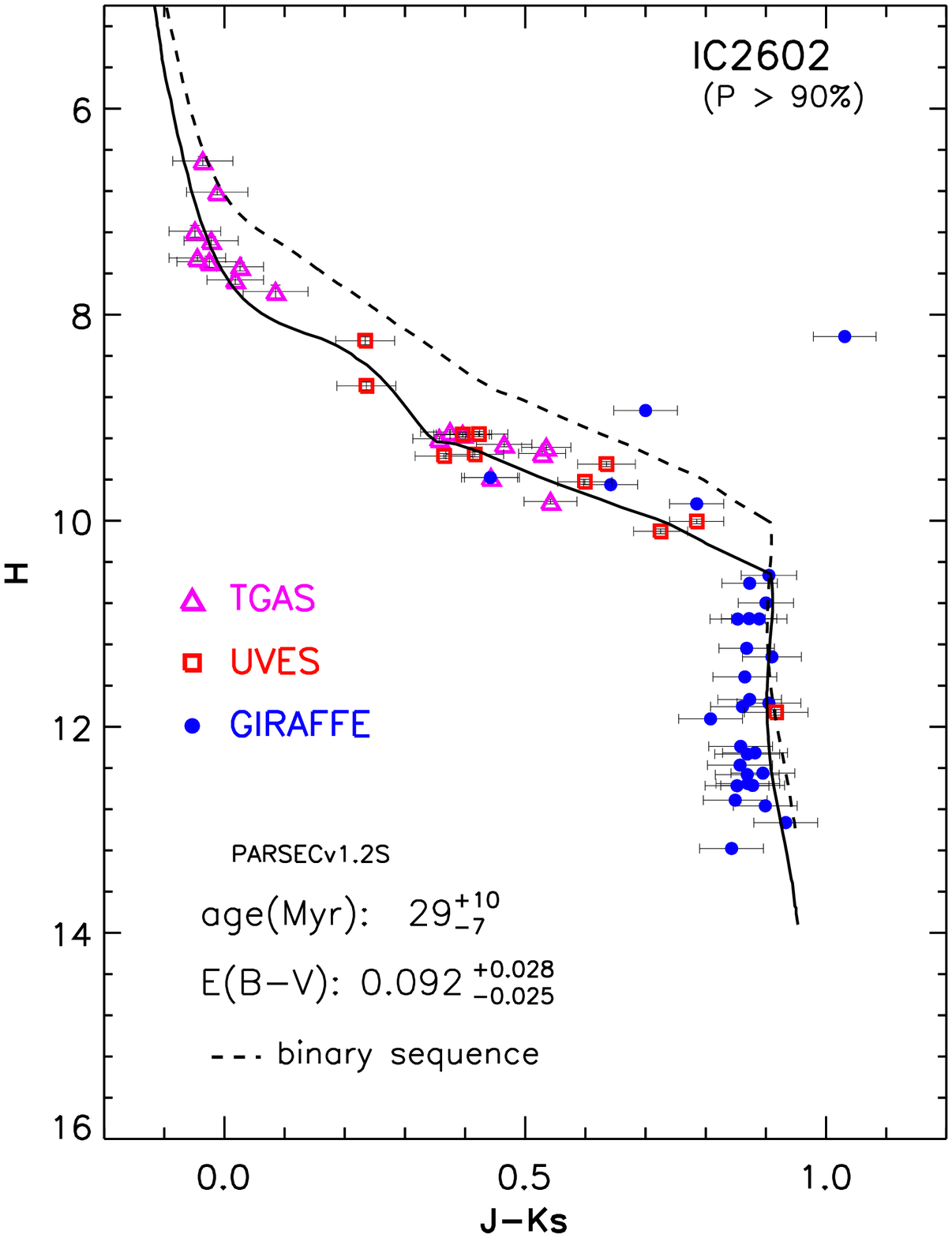}
        \includegraphics[width=0.325\linewidth]{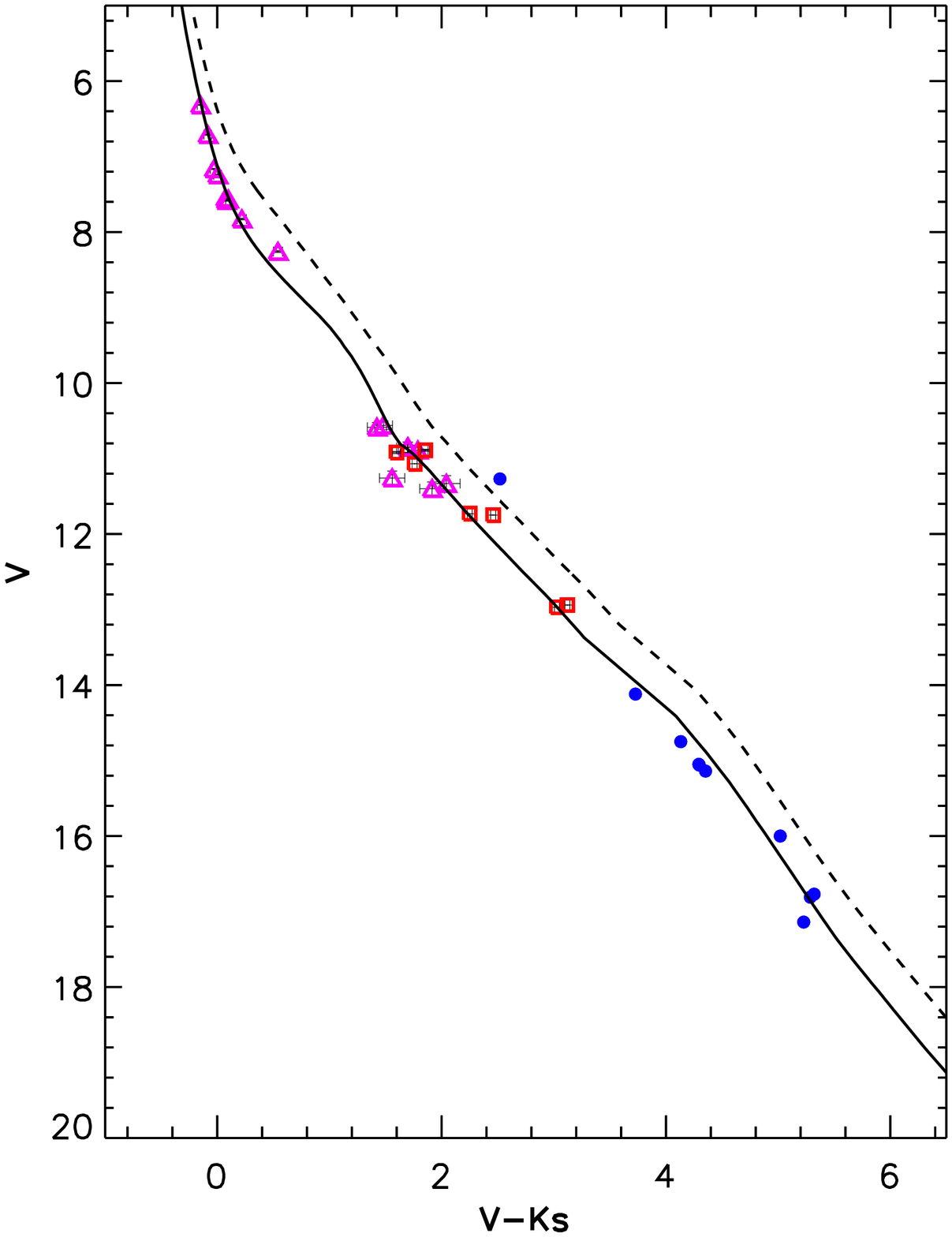}
        \includegraphics[width=0.325\linewidth]{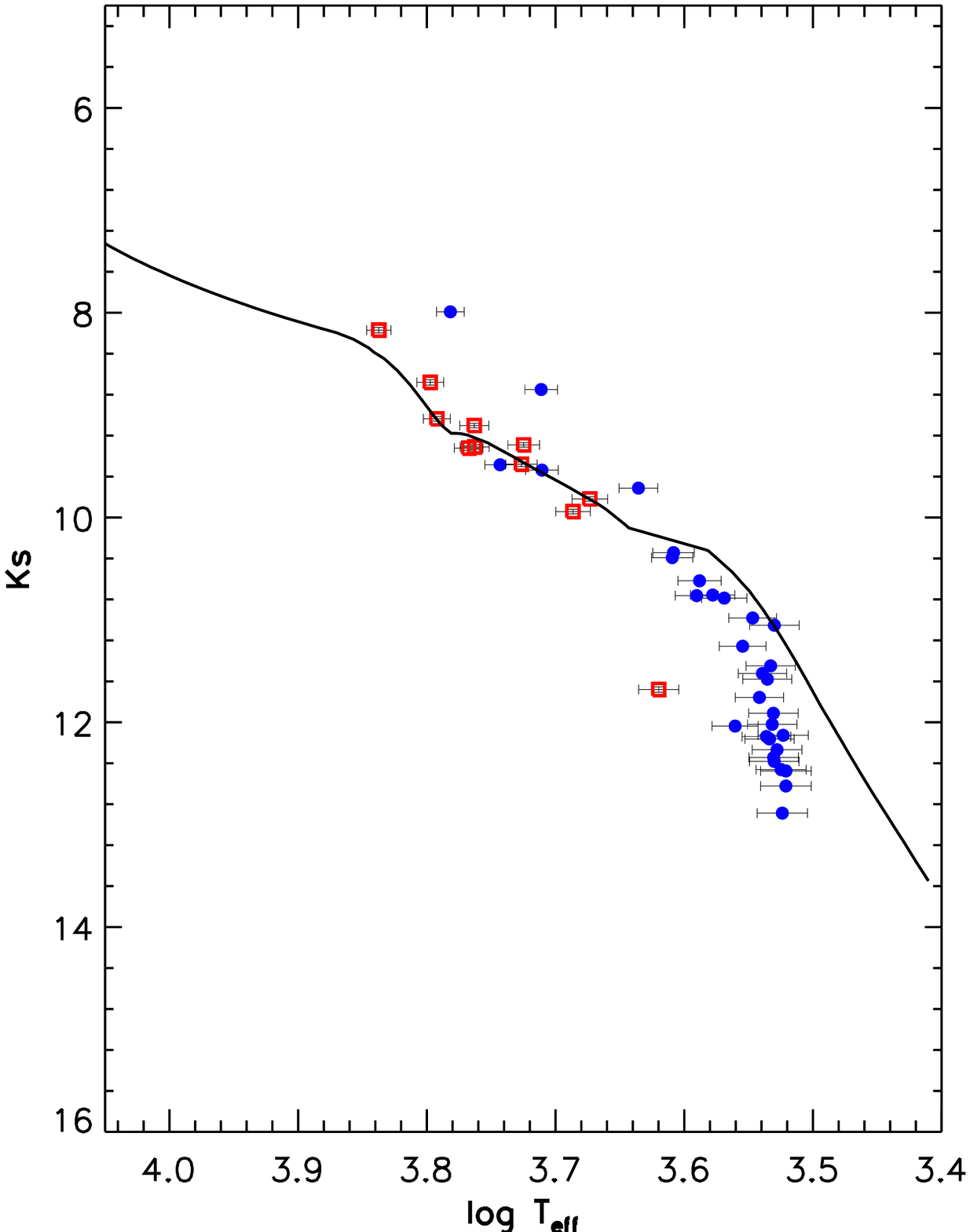}\\
        \includegraphics[width=0.325\linewidth]{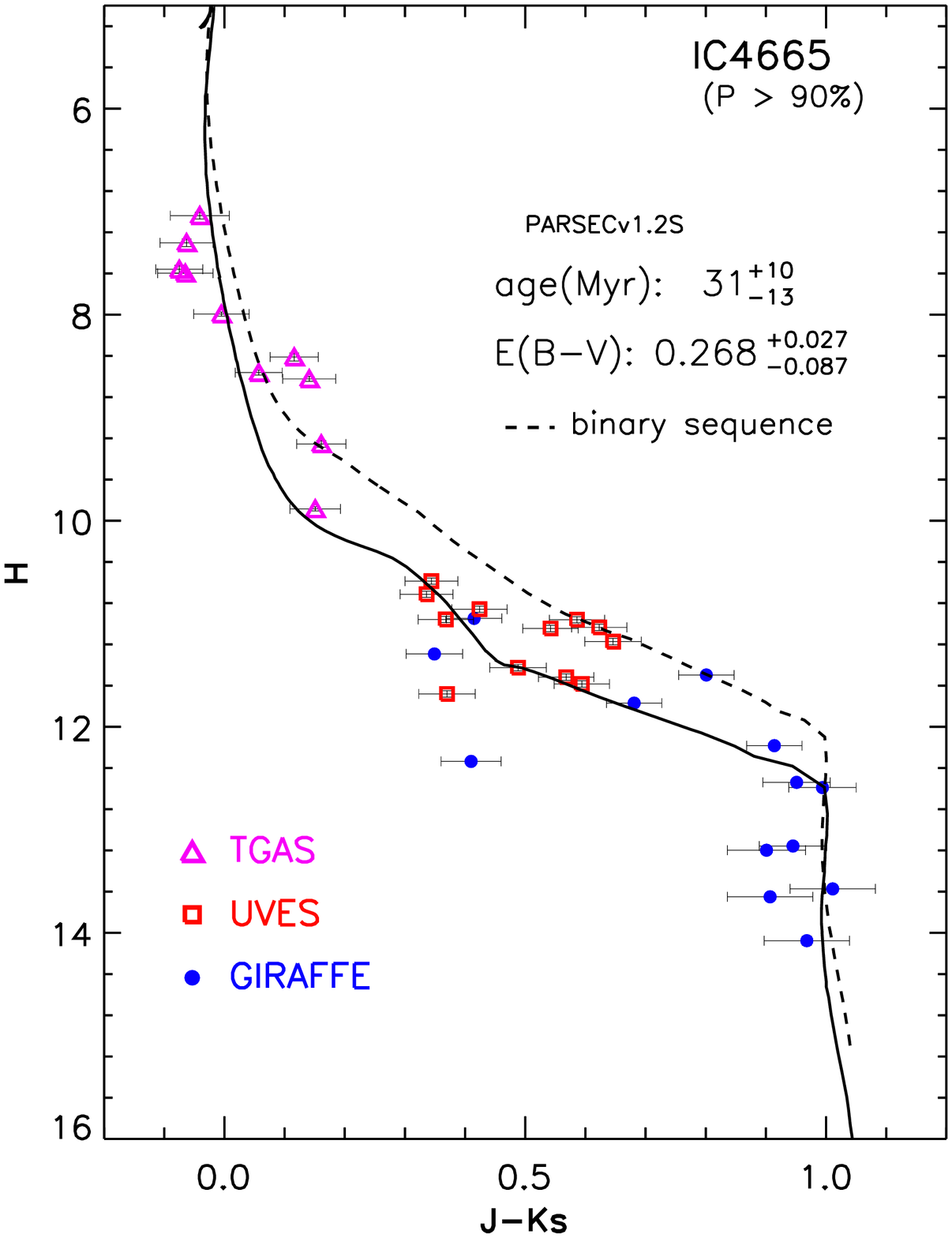}
        \includegraphics[width=0.325\linewidth]{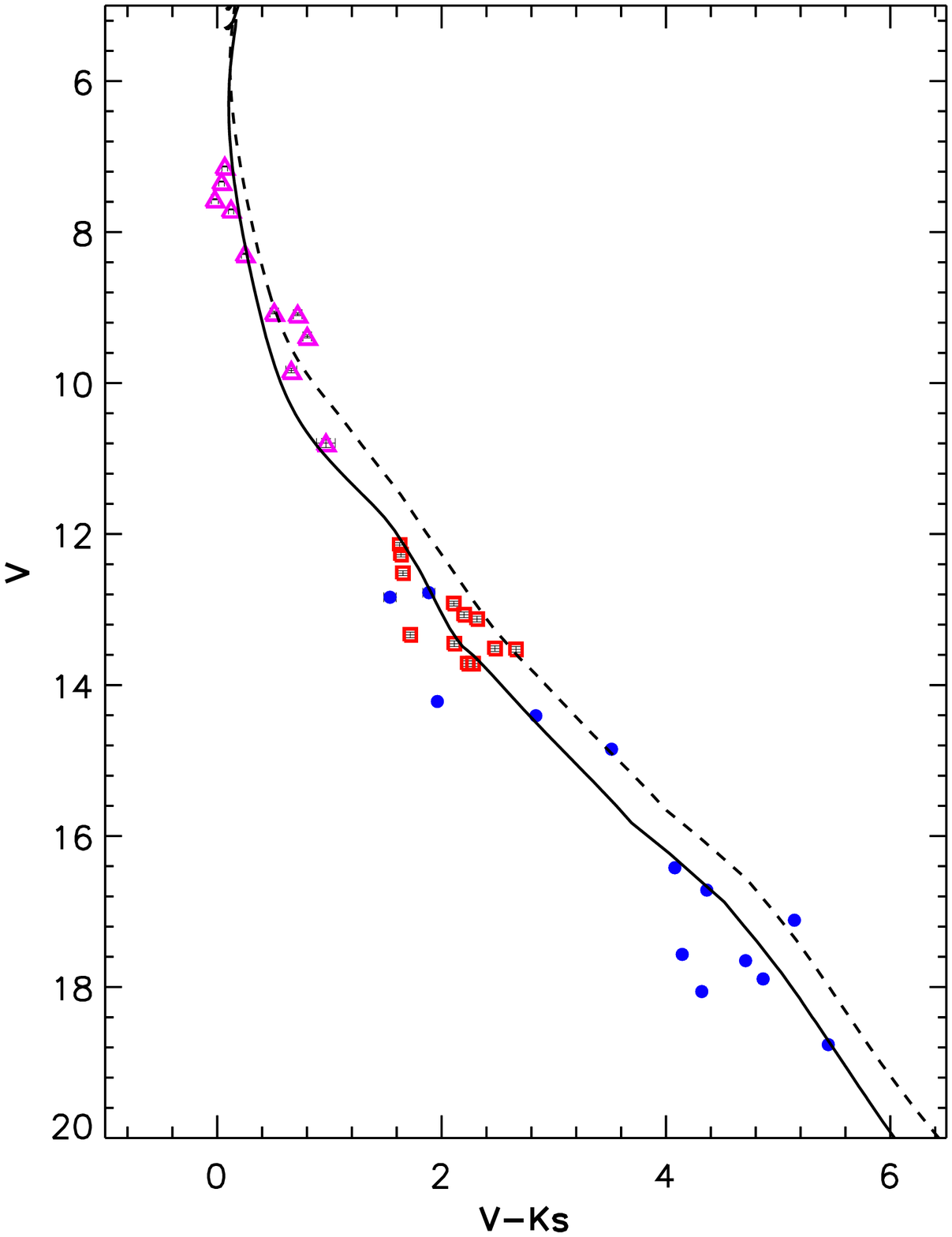}
        \includegraphics[width=0.325\linewidth]{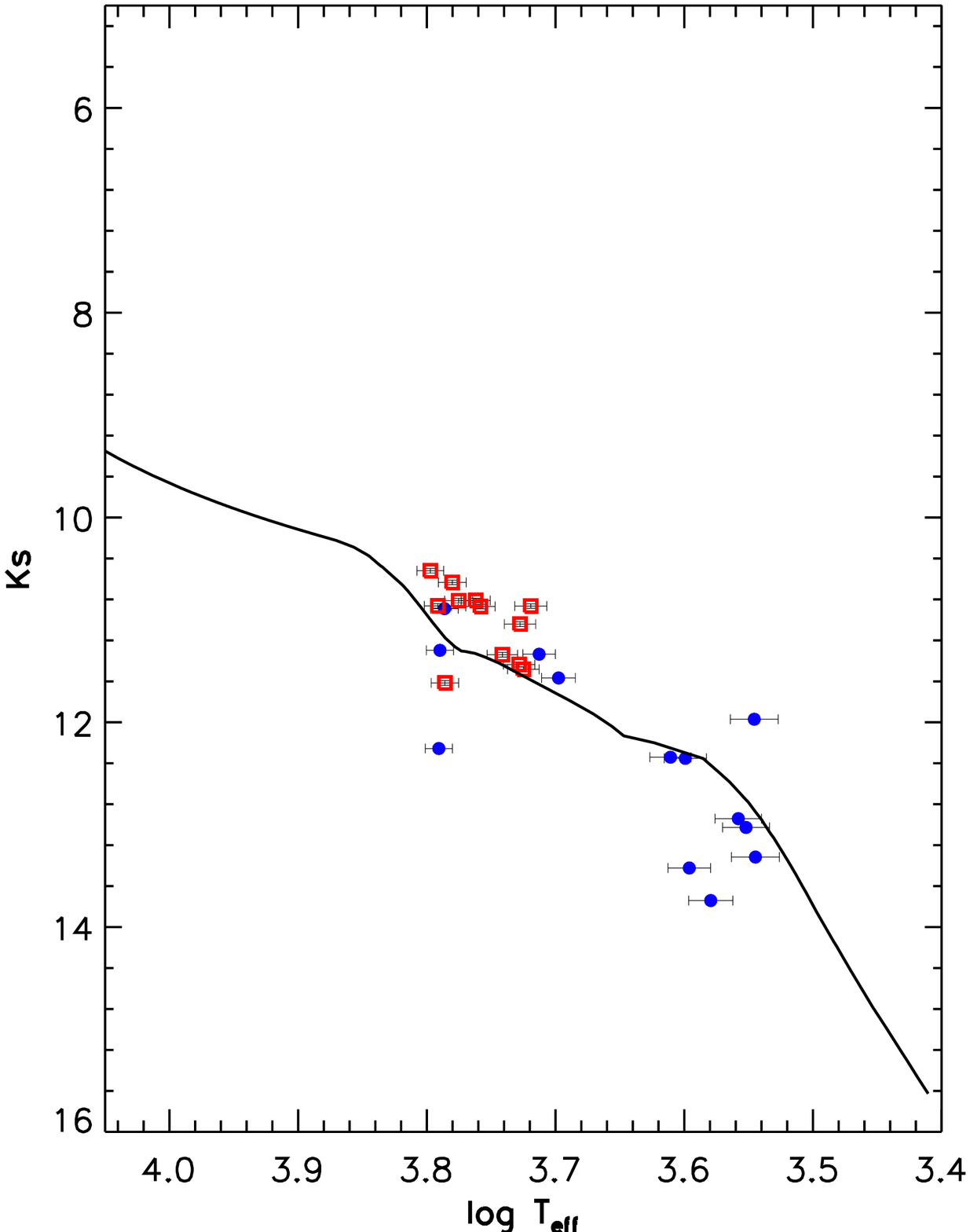}\\    
        \includegraphics[width=0.325\linewidth]{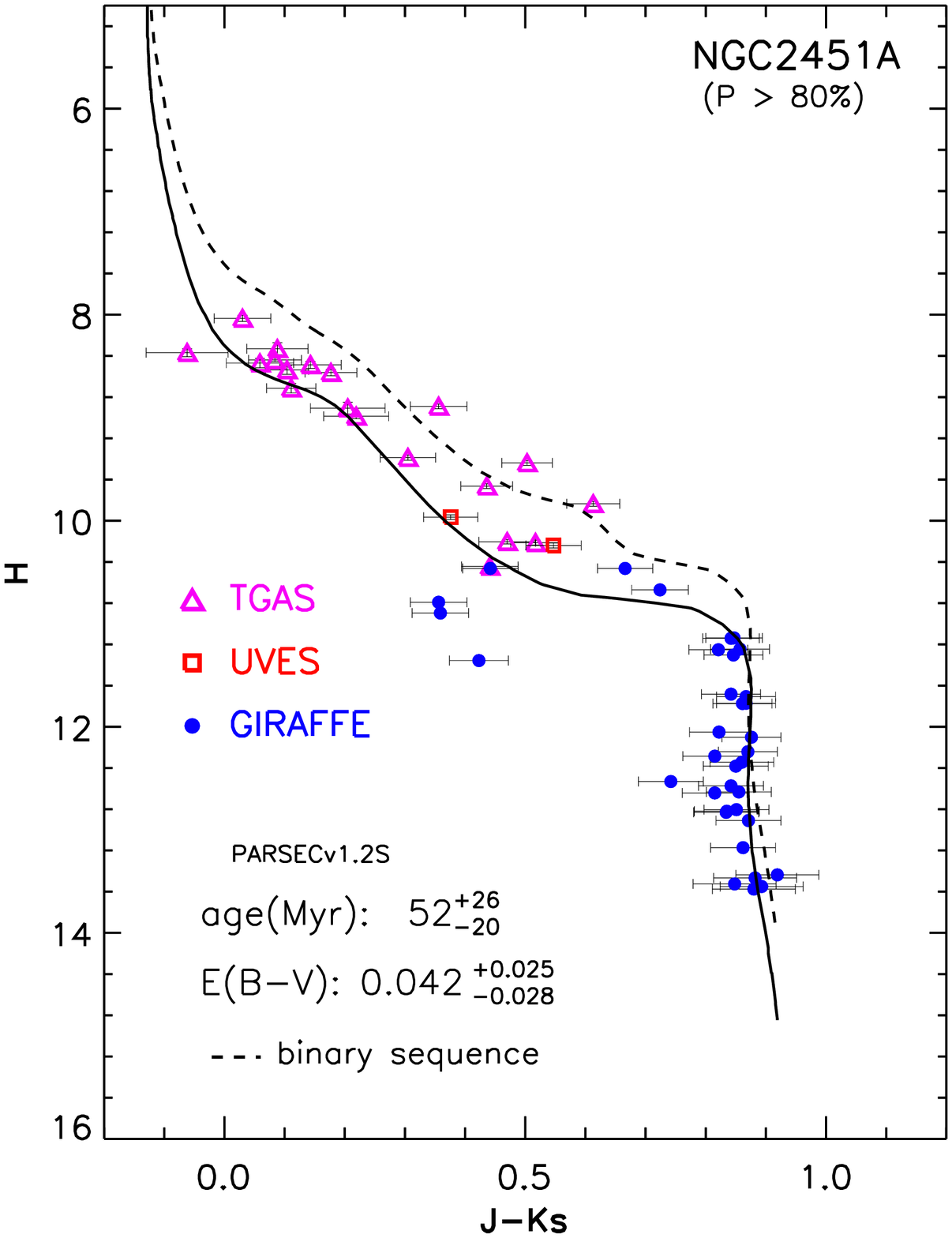}
        \includegraphics[width=0.325\linewidth]{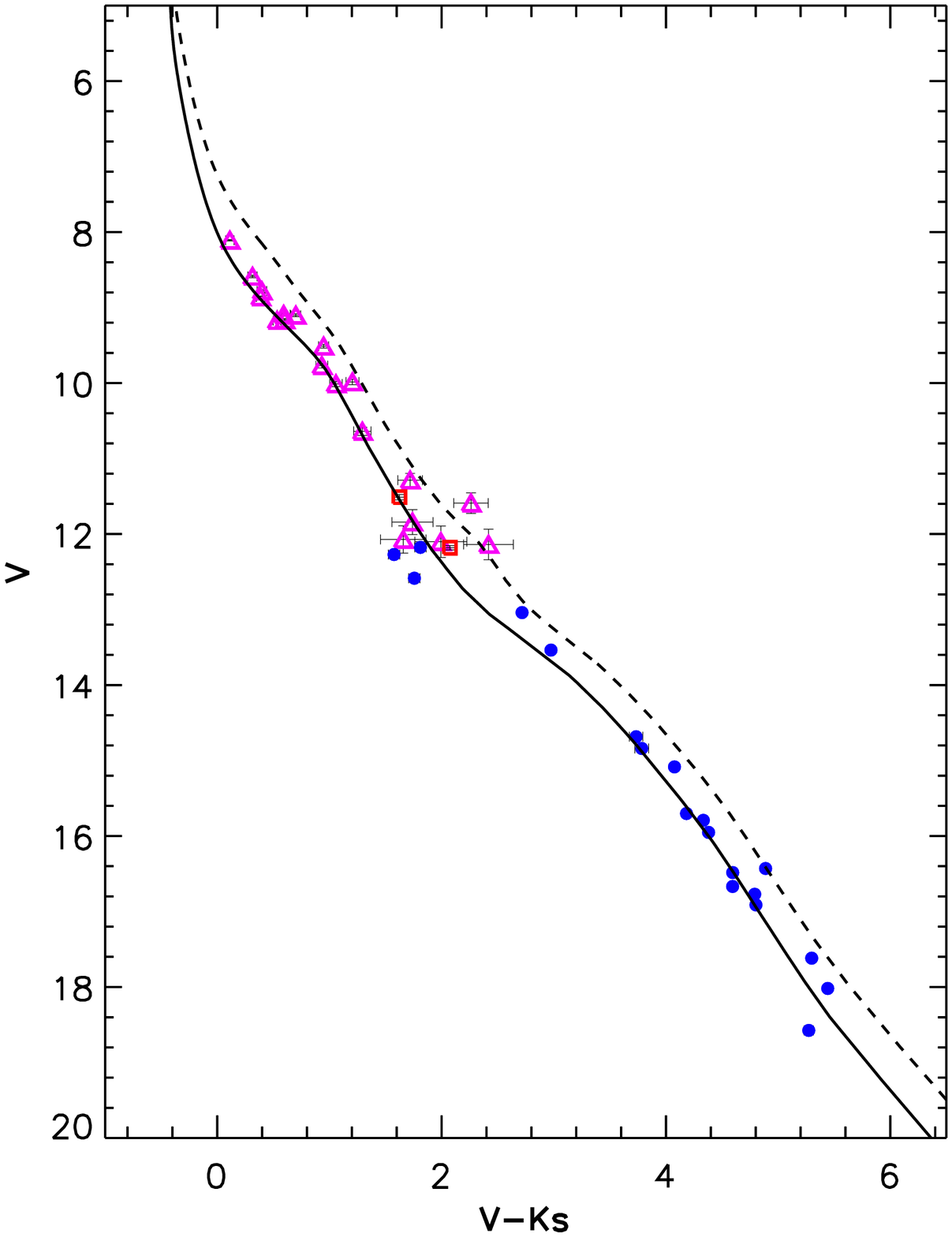}
        \includegraphics[width=0.325\linewidth]{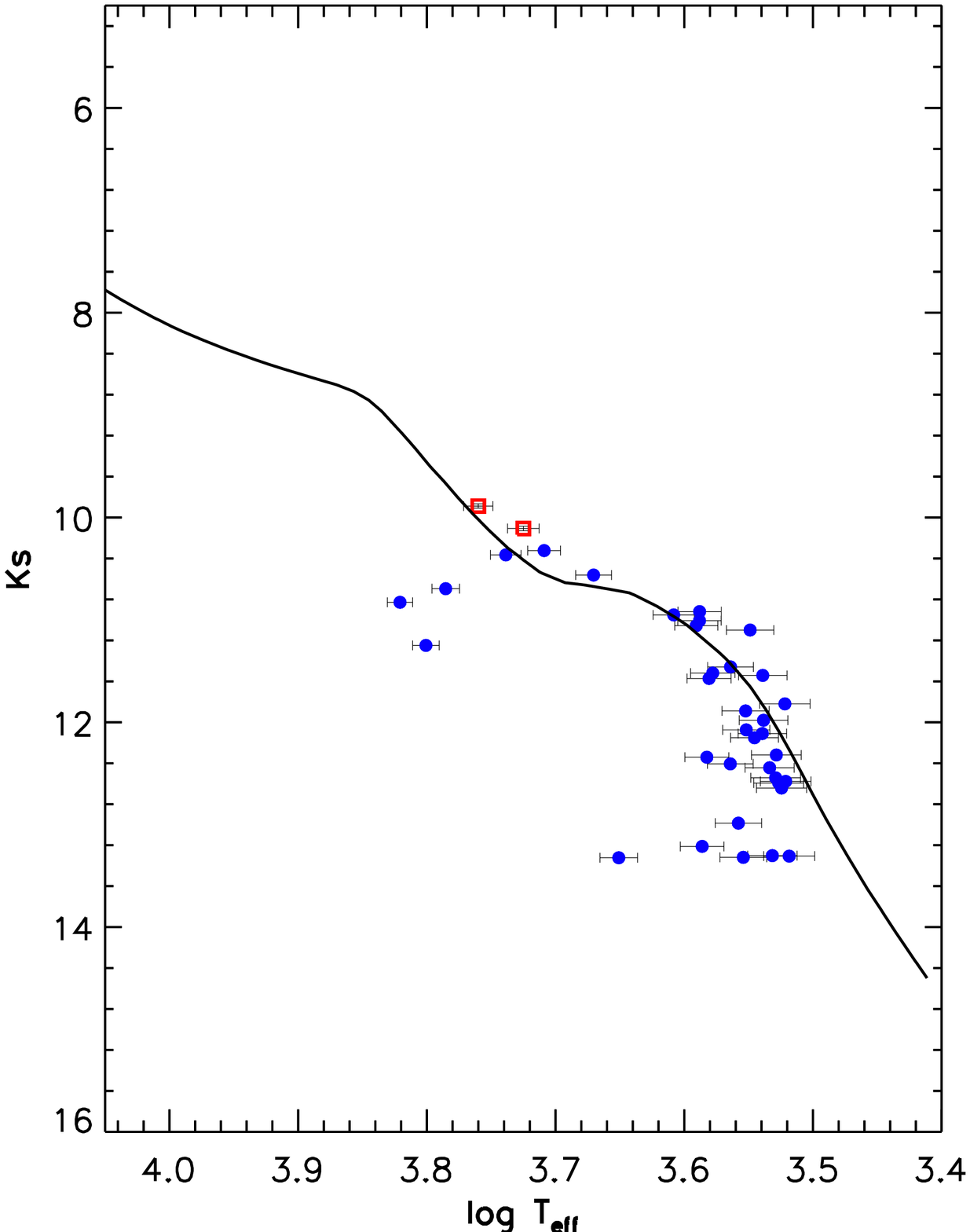}\\  
        \caption{As in Fig.~\ref{fig:CMDPROSECCO_1}, but for the clusters IC~2602, IC~4665, and NGC~2451A with the PARSEC isochrones.}
        \label{fig:CMDPARSEC_1}
\end{figure*}

\begin{figure*}
        \centering
        \includegraphics[width=0.325\linewidth]{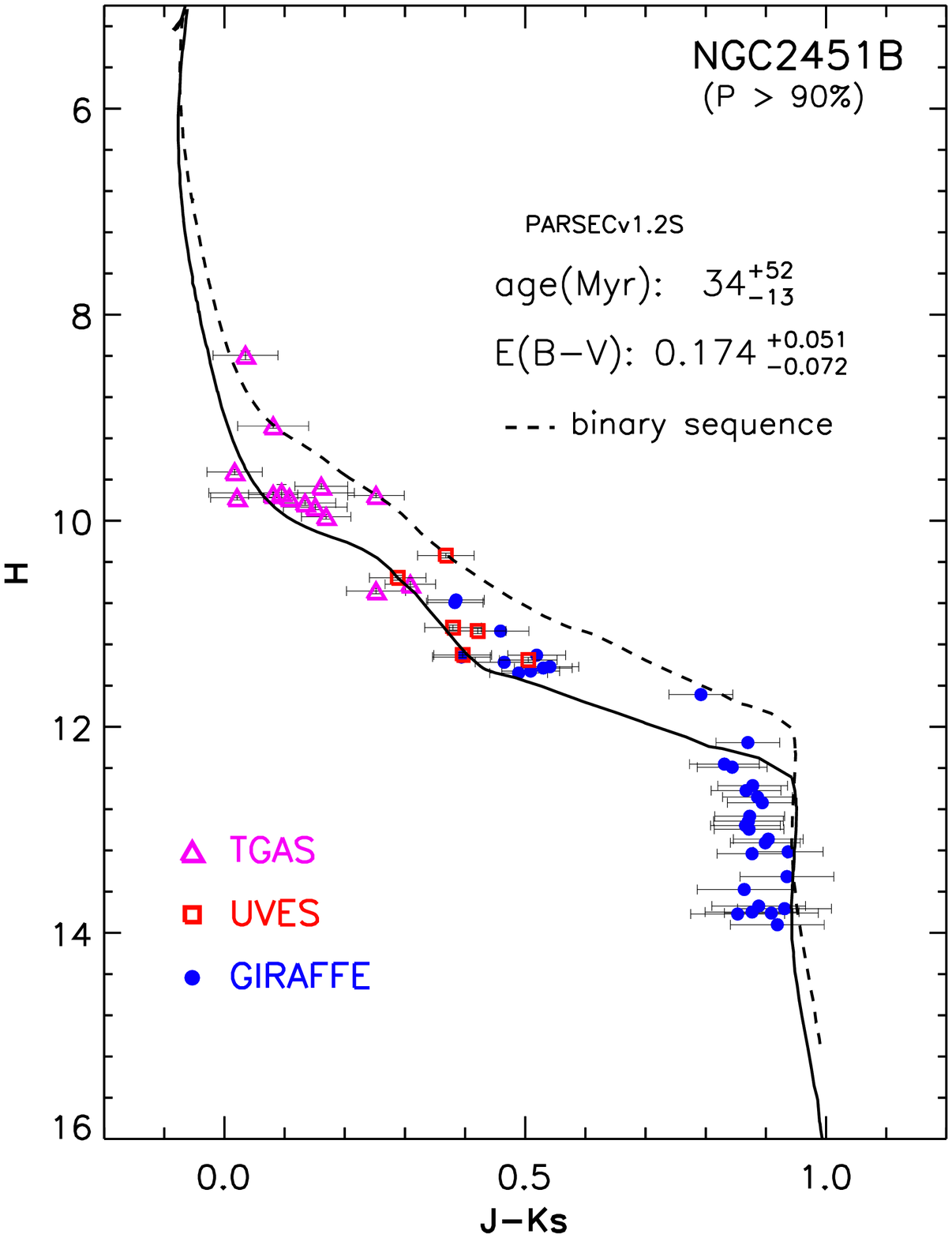}
        \includegraphics[width=0.325\linewidth]{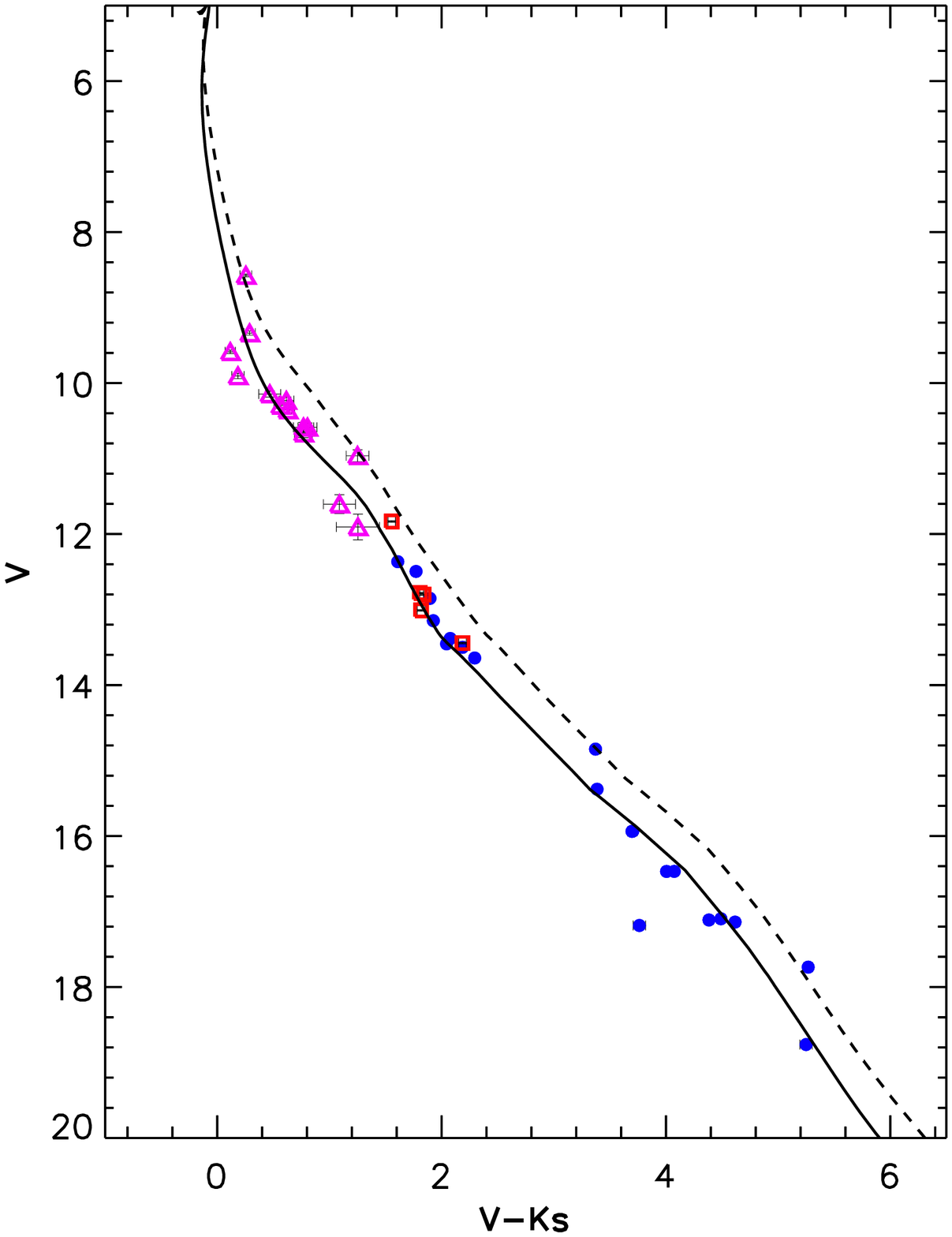}
        \includegraphics[width=0.325\linewidth]{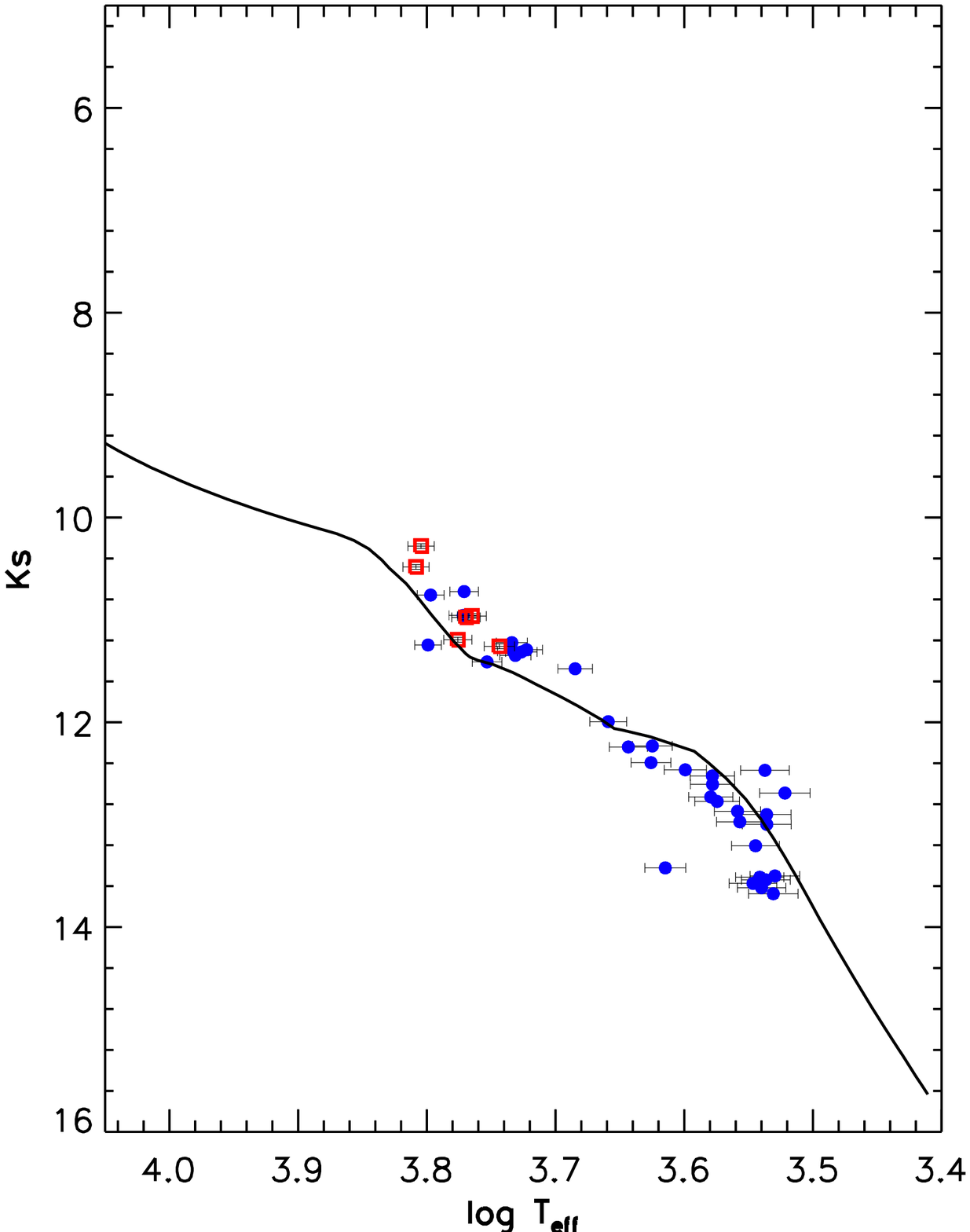}\\
        \includegraphics[width=0.325\linewidth]{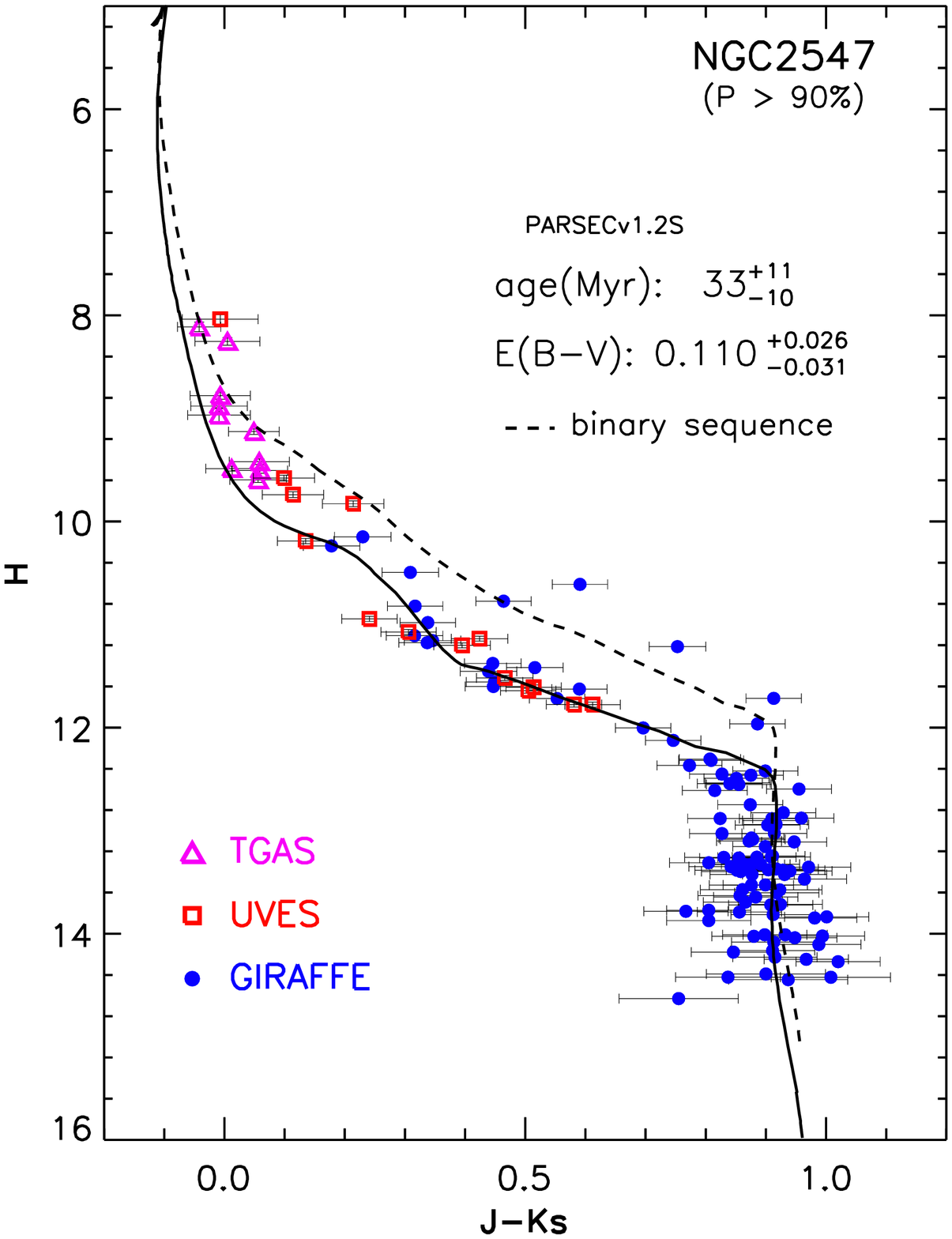}
        \includegraphics[width=0.325\linewidth]{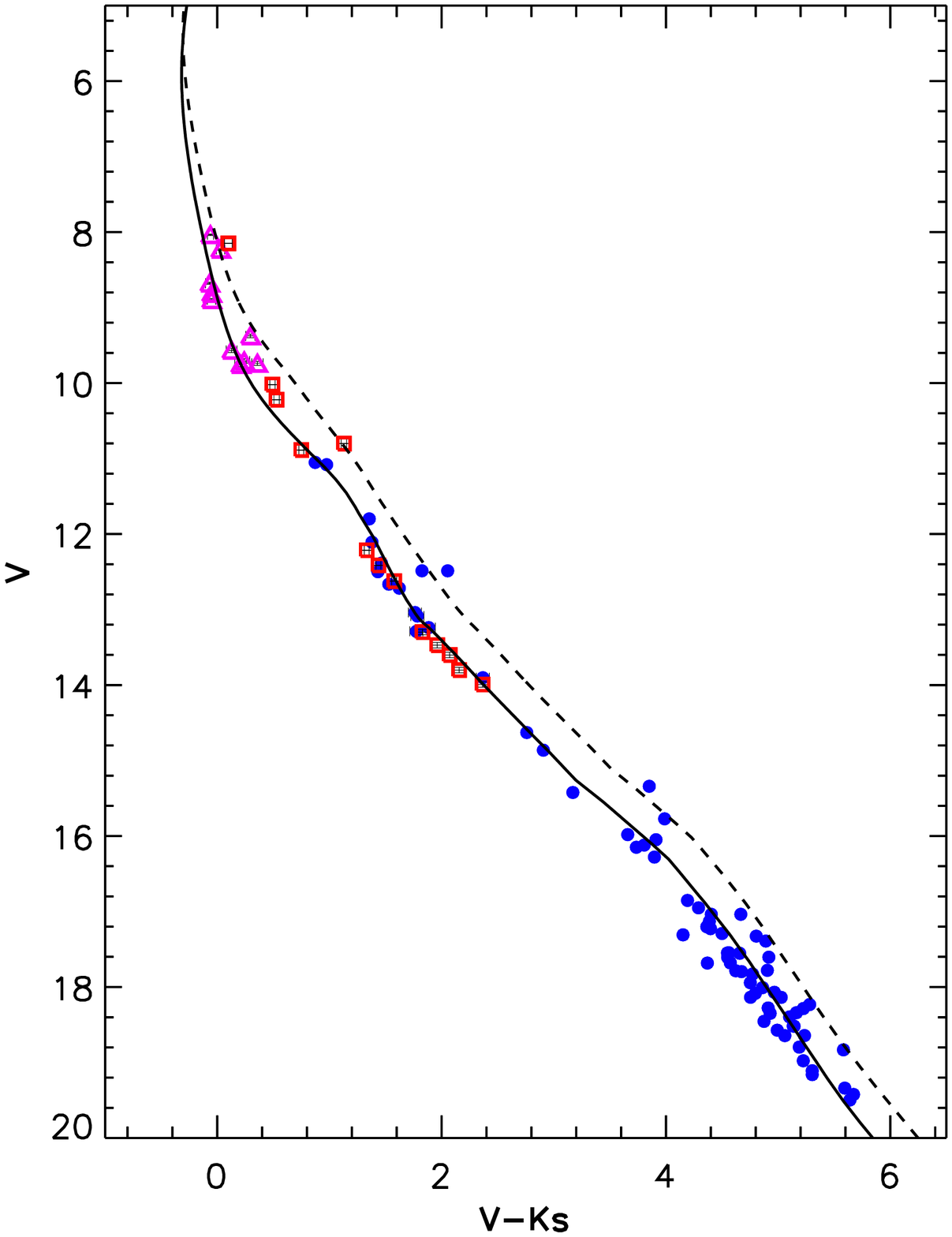}
        \includegraphics[width=0.325\linewidth]{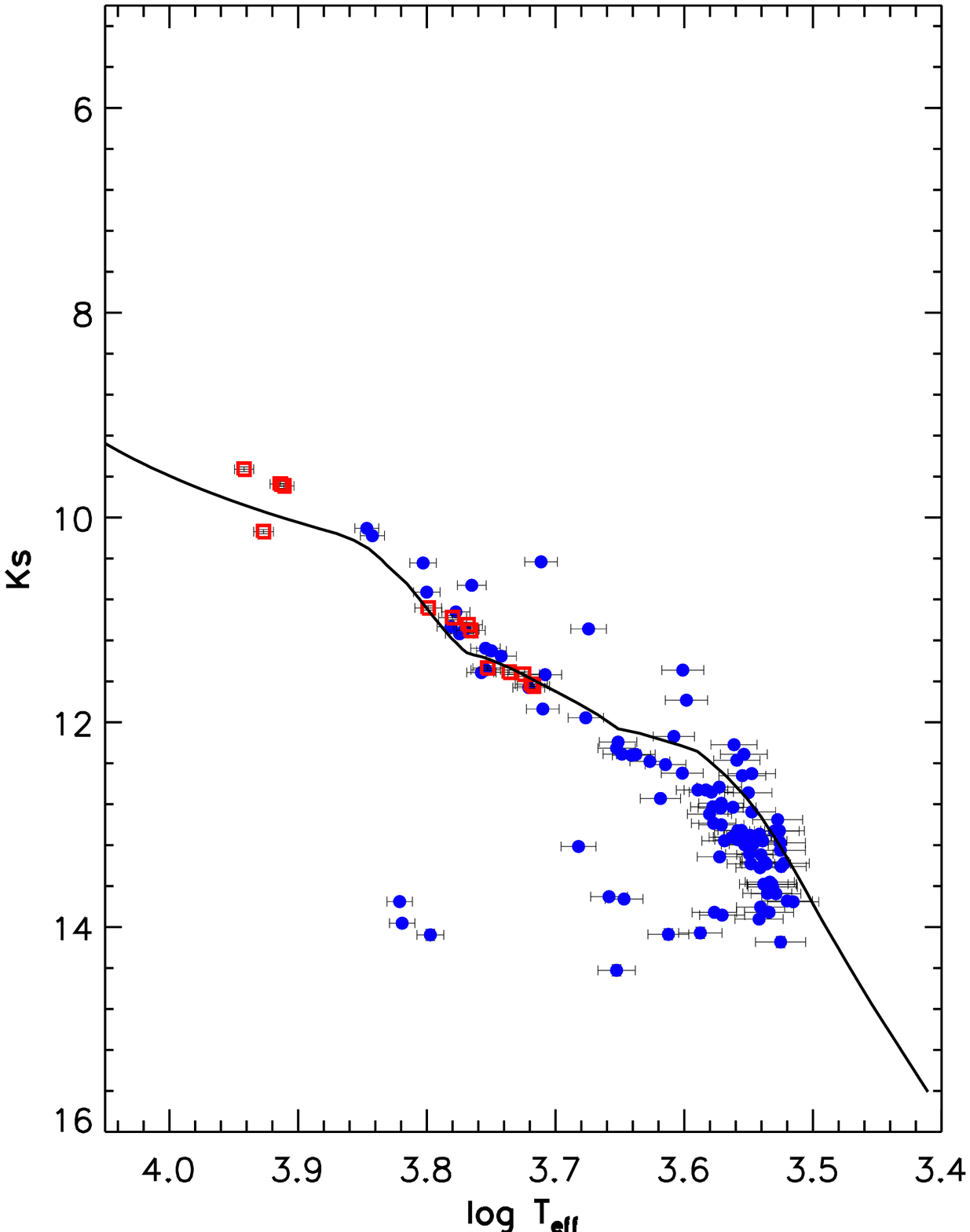}\\   
        \includegraphics[width=0.325\linewidth]{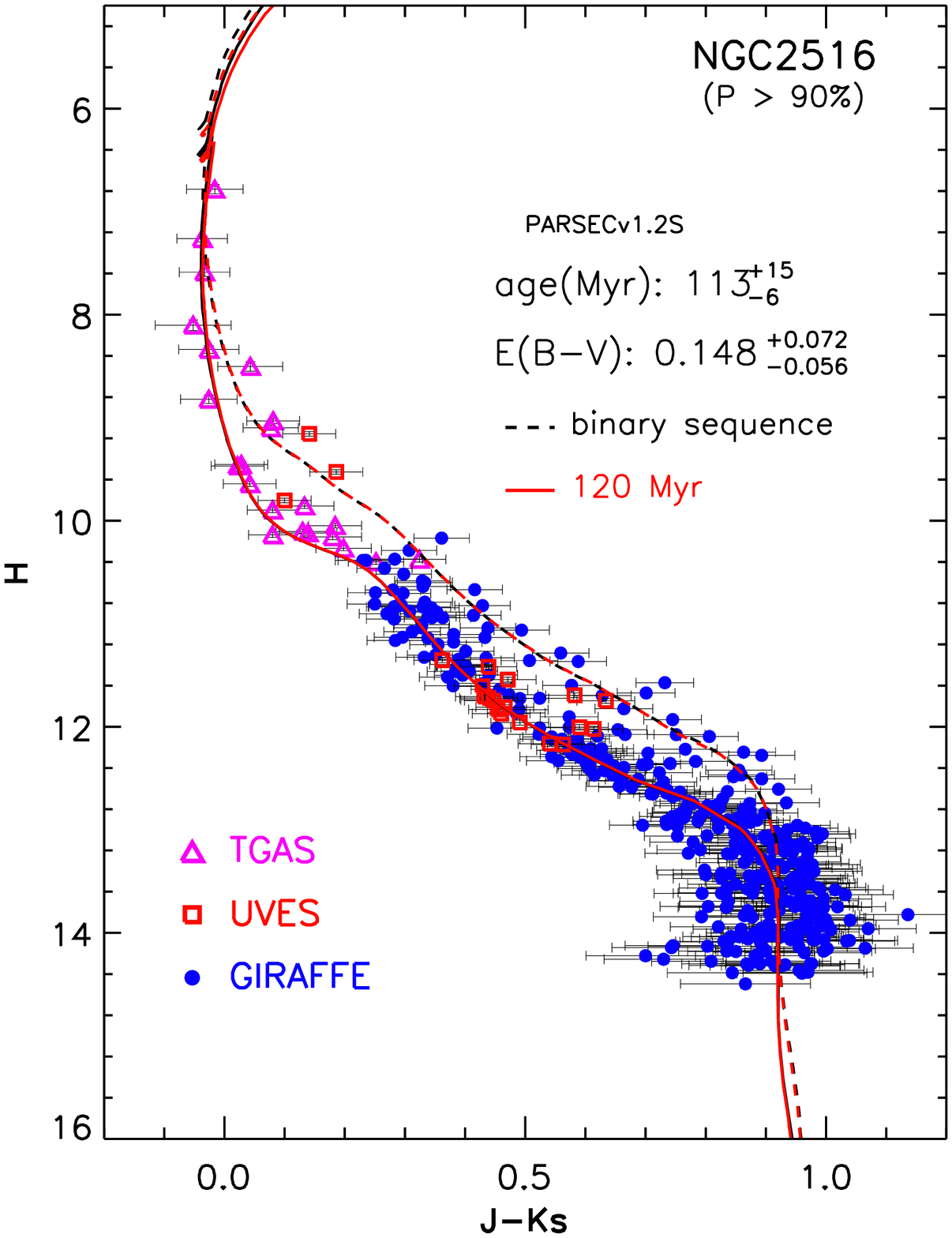}
        \includegraphics[width=0.325\linewidth]{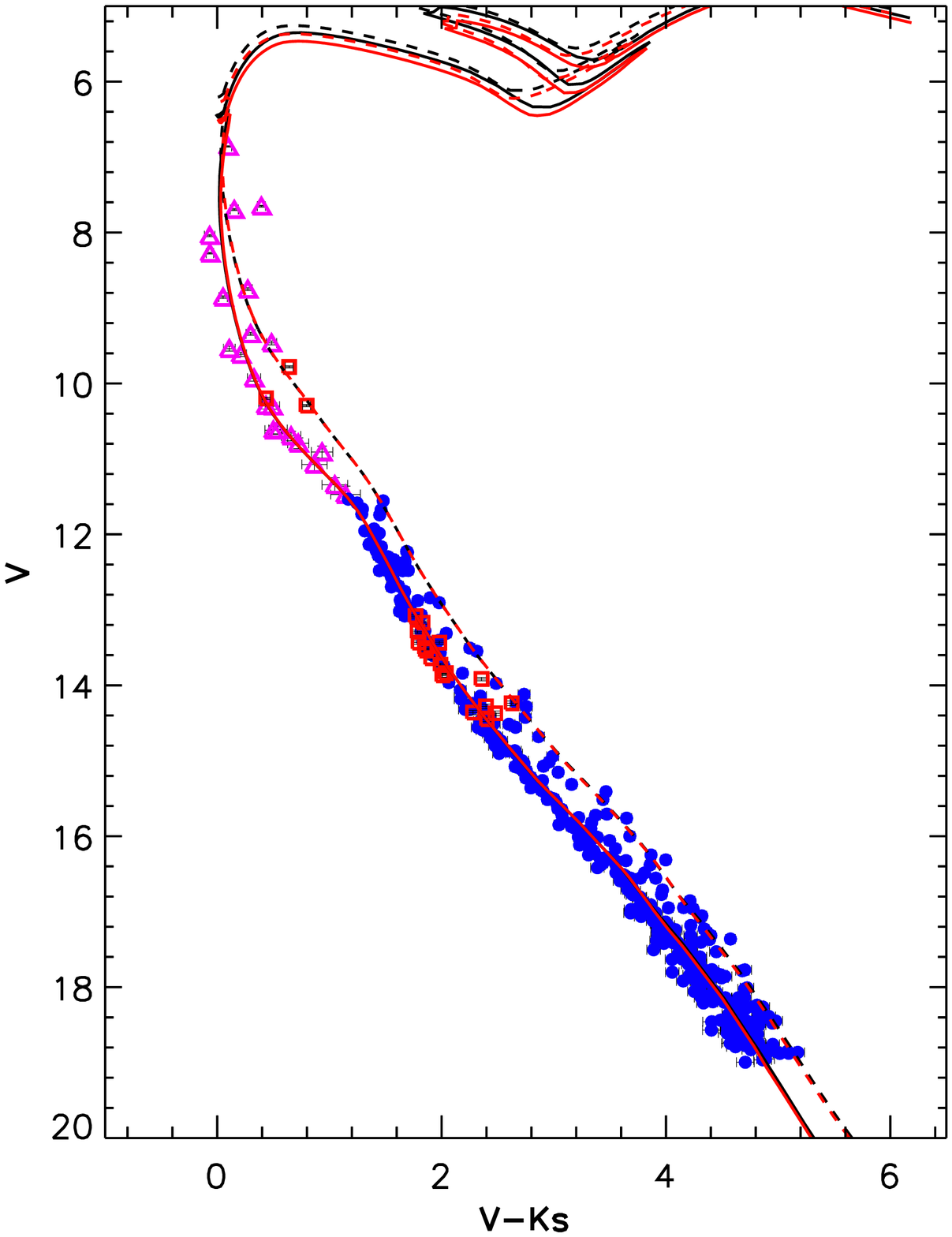}
        \includegraphics[width=0.325\linewidth]{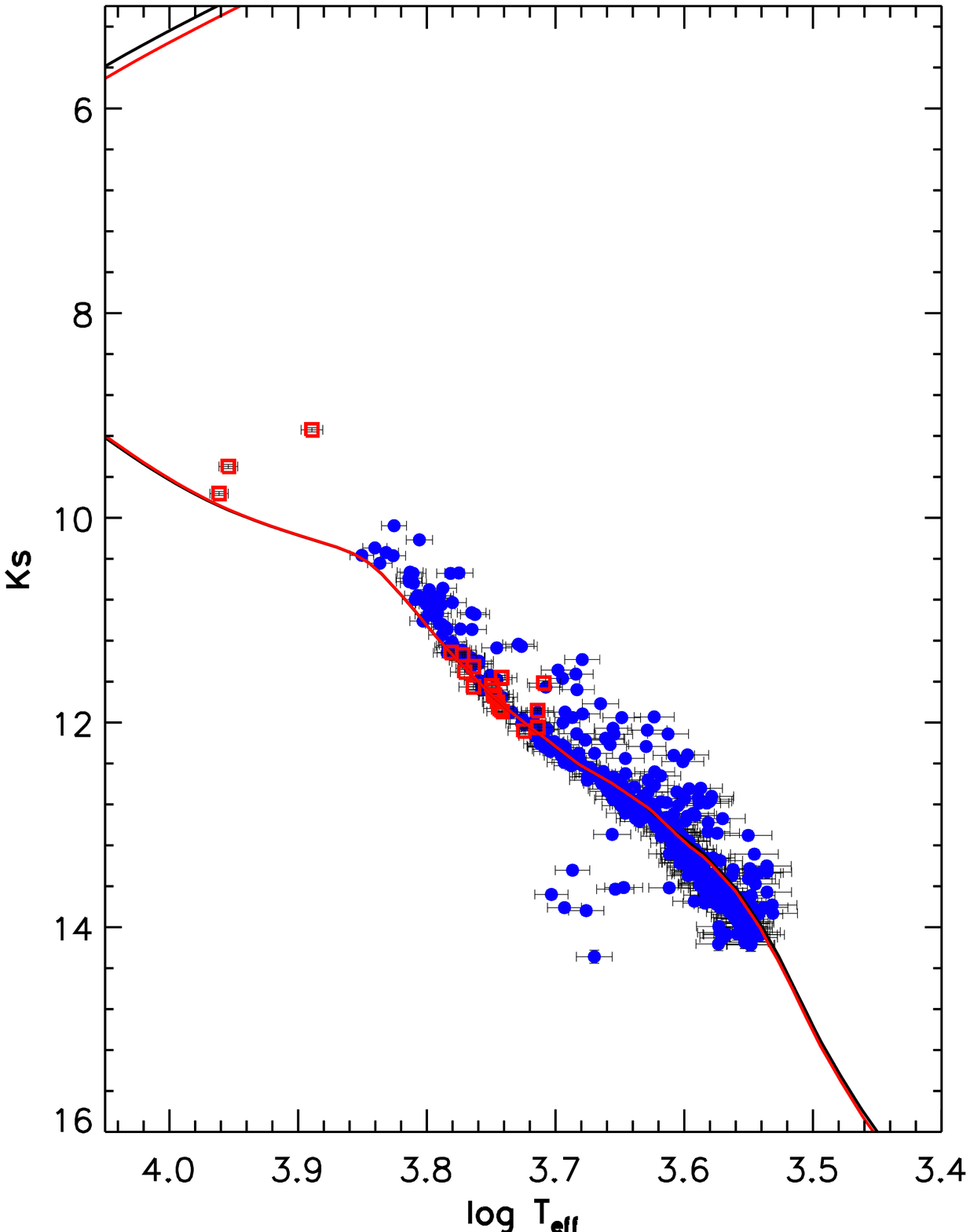}\\   
        \caption{As in Fig.~\ref{fig:CMDPROSECCO_1}, but for the clusters NGC~2451B, NGC~2547, and NGC~2516 with the PARSEC isochrones.}
        \label{fig:CMDPARSEC_2}
\end{figure*}

\begin{figure*}
        \centering
        \includegraphics[width=0.325\linewidth]{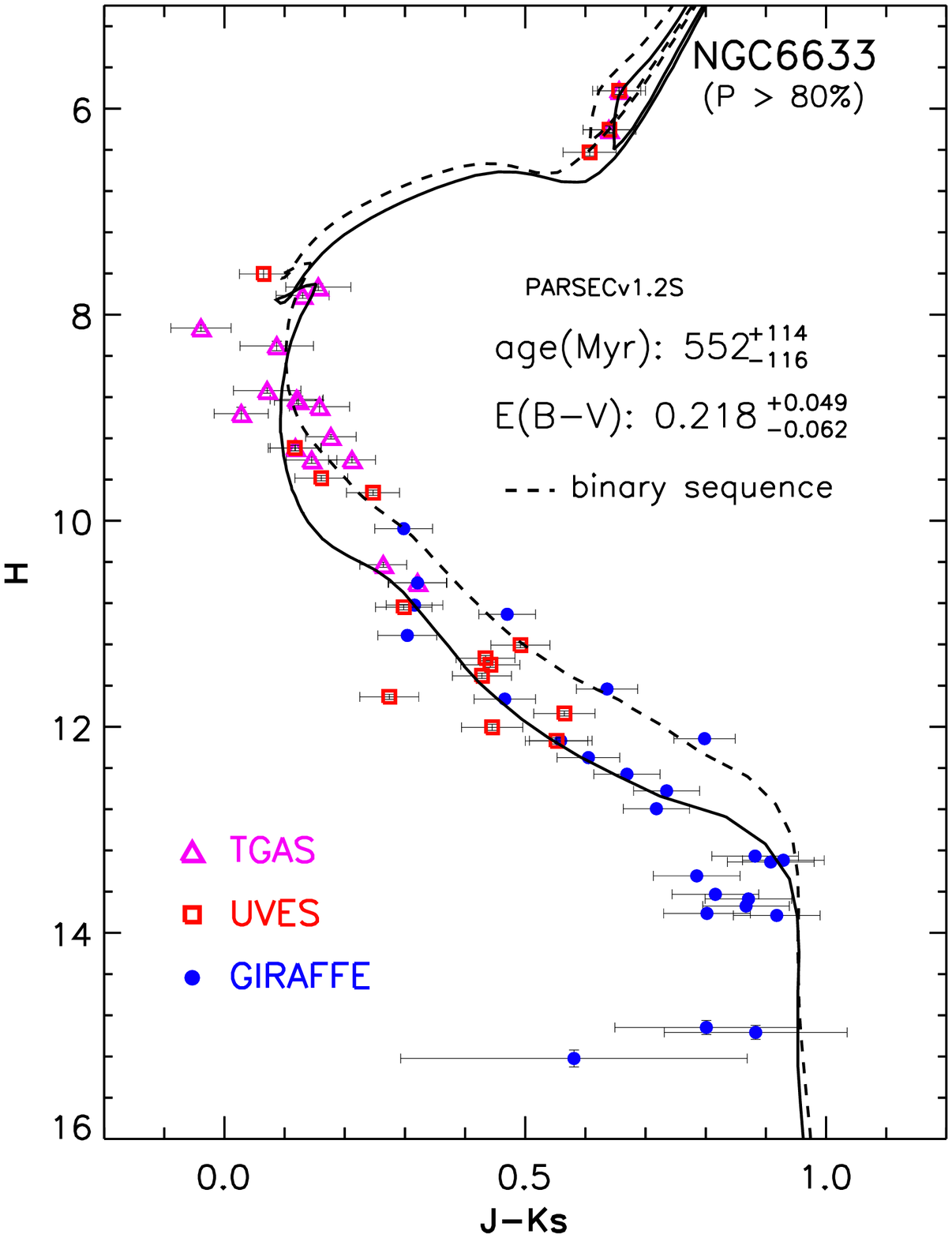}
        \includegraphics[width=0.325\linewidth]{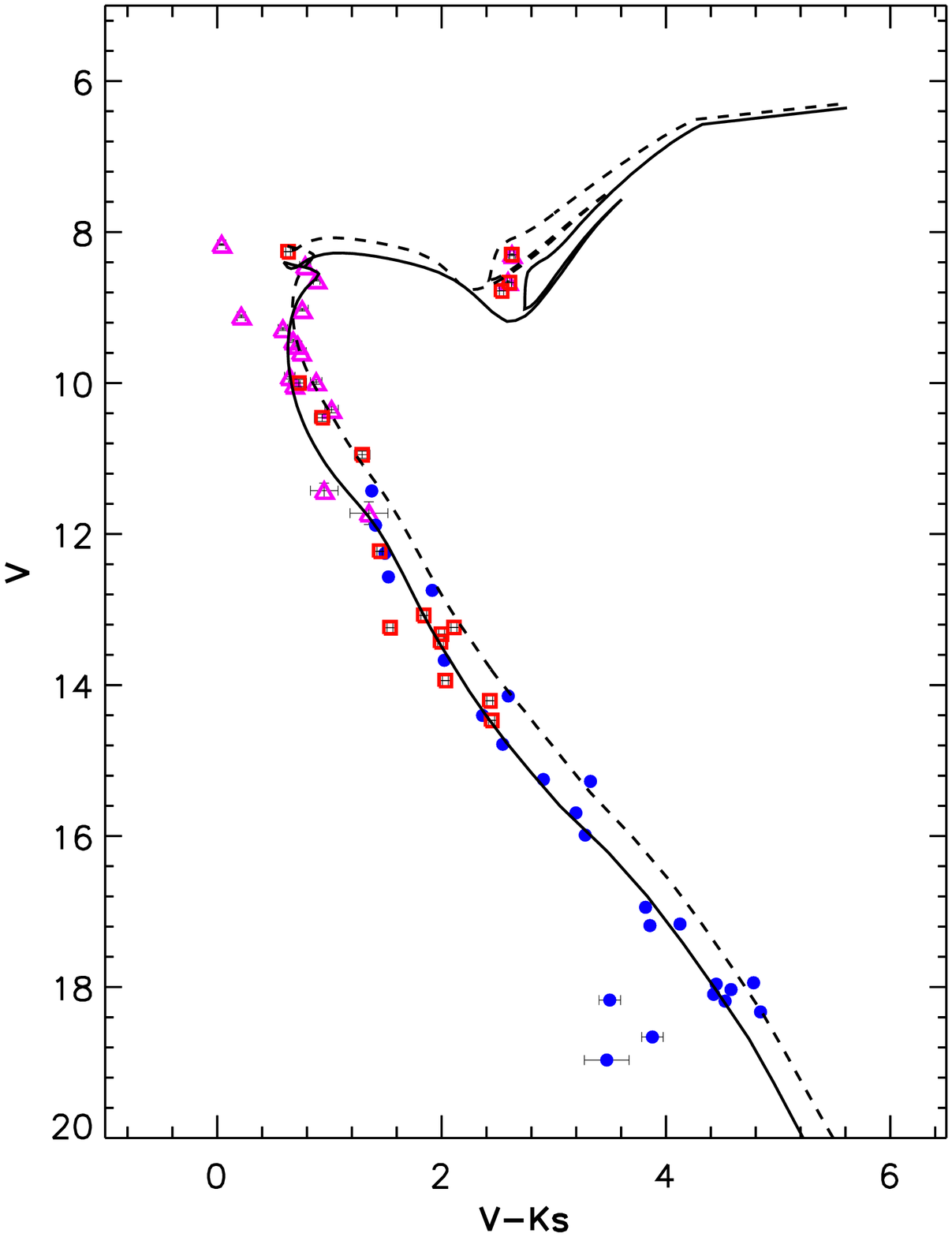}
        \includegraphics[width=0.325\linewidth]{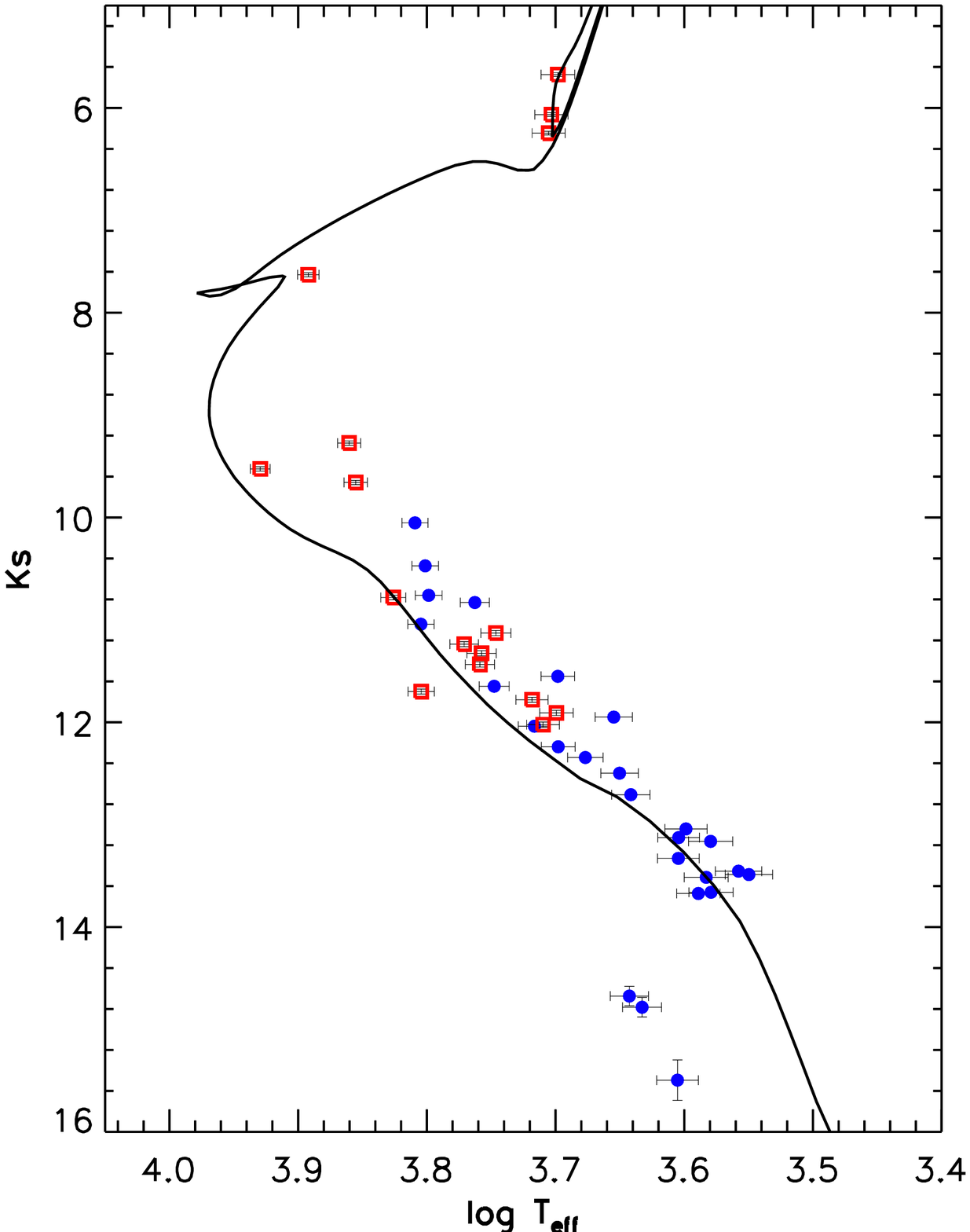}
        \caption{As in Fig.~\ref{fig:CMDPROSECCO_1}, but for the cluster NGC~6633 with the PARSEC isochrones.}
        \label{fig:CMDPARSEC_3}
\end{figure*}

\begin{figure*}
        \centering
\includegraphics[width=0.325\linewidth]{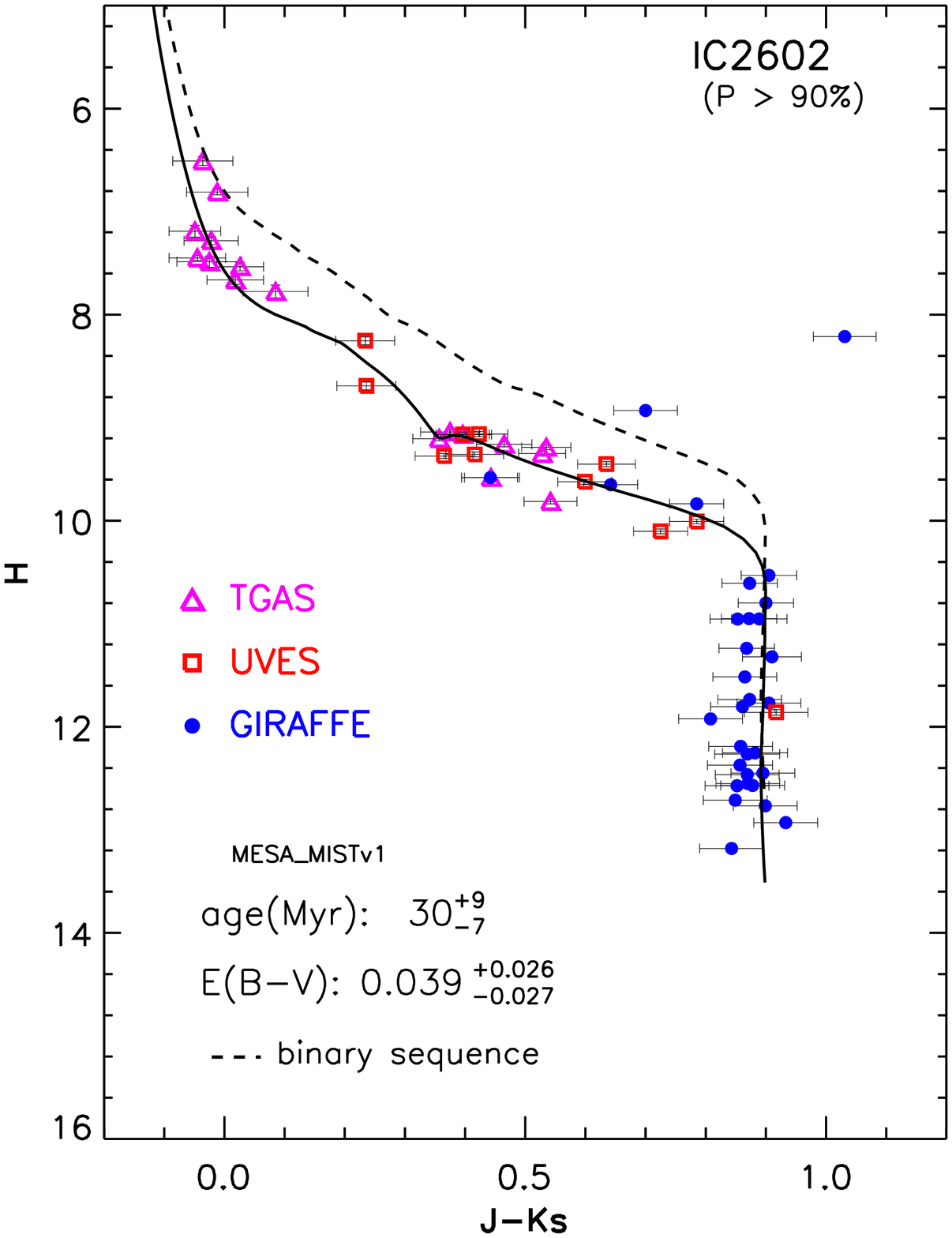}
\includegraphics[width=0.325\linewidth]{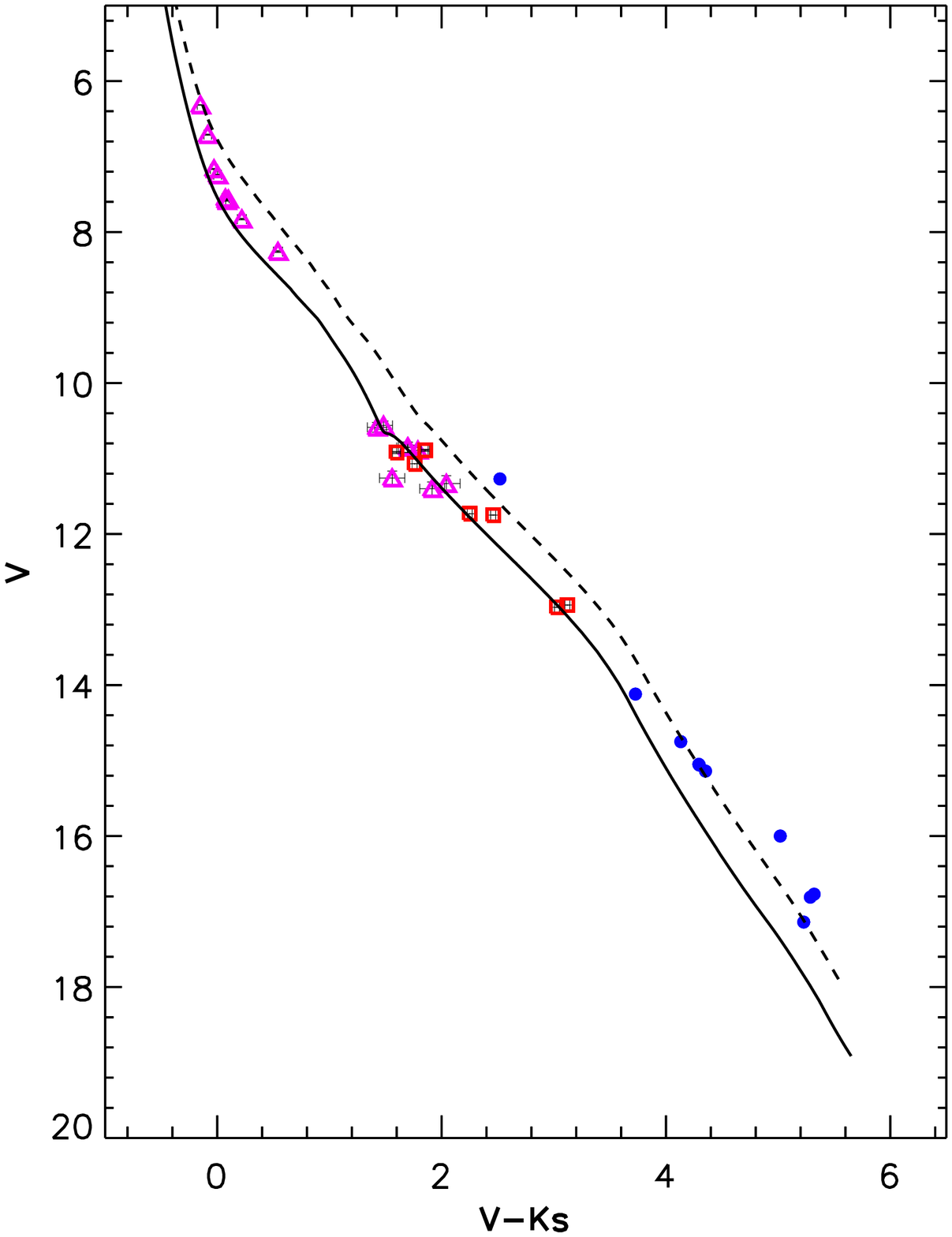}
\includegraphics[width=0.325\linewidth]{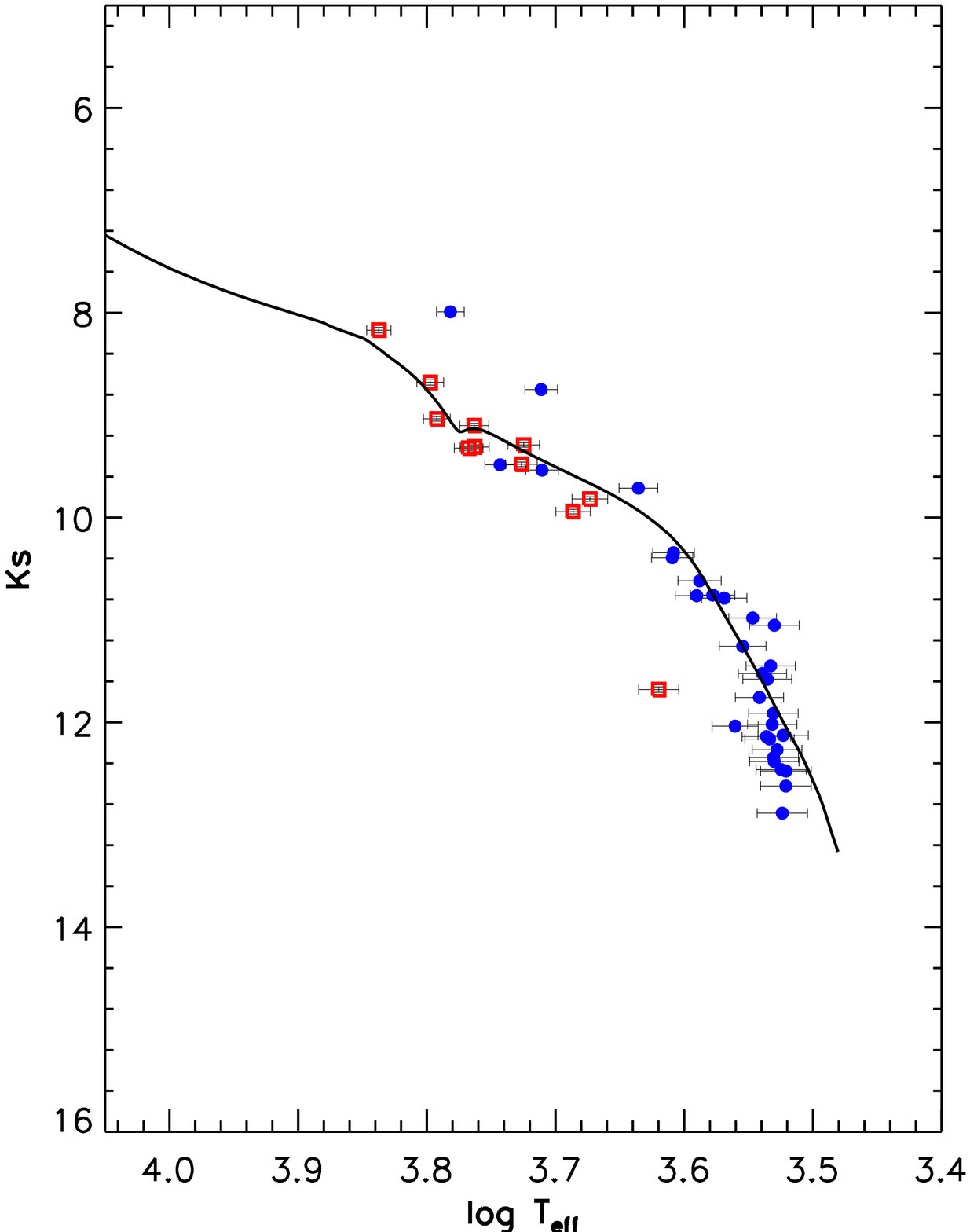}\\
\includegraphics[width=0.325\linewidth]{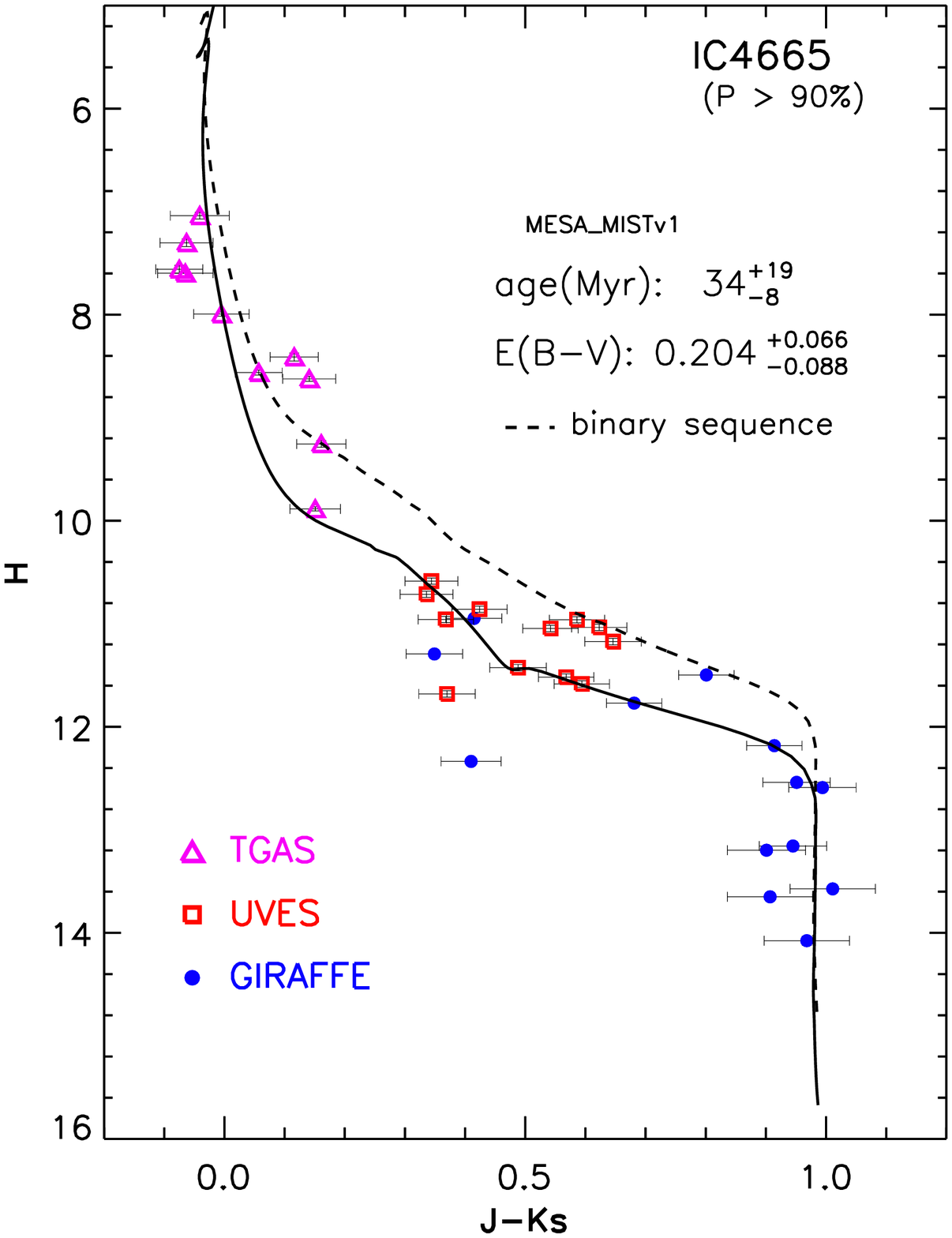}
\includegraphics[width=0.325\linewidth]{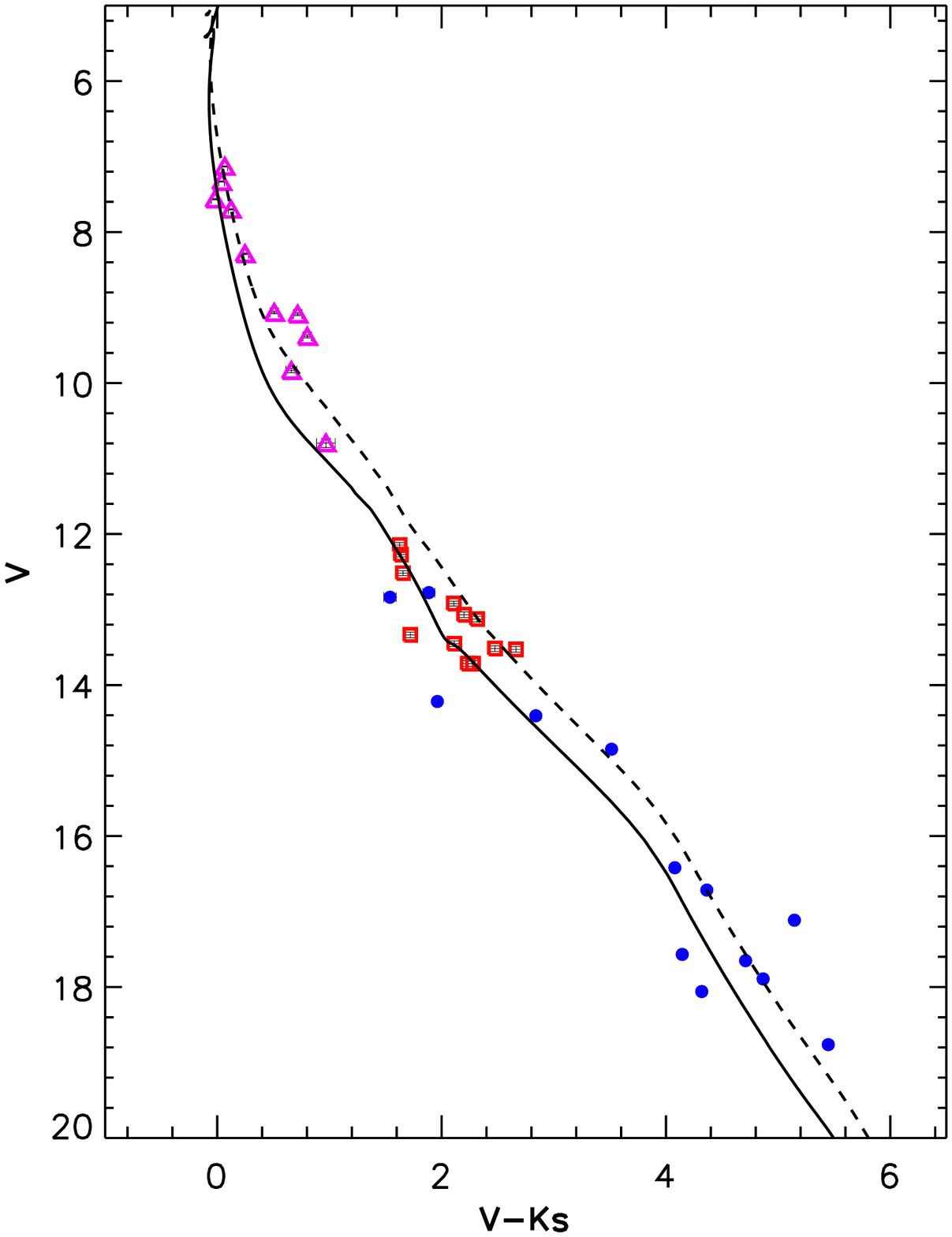}
\includegraphics[width=0.325\linewidth]{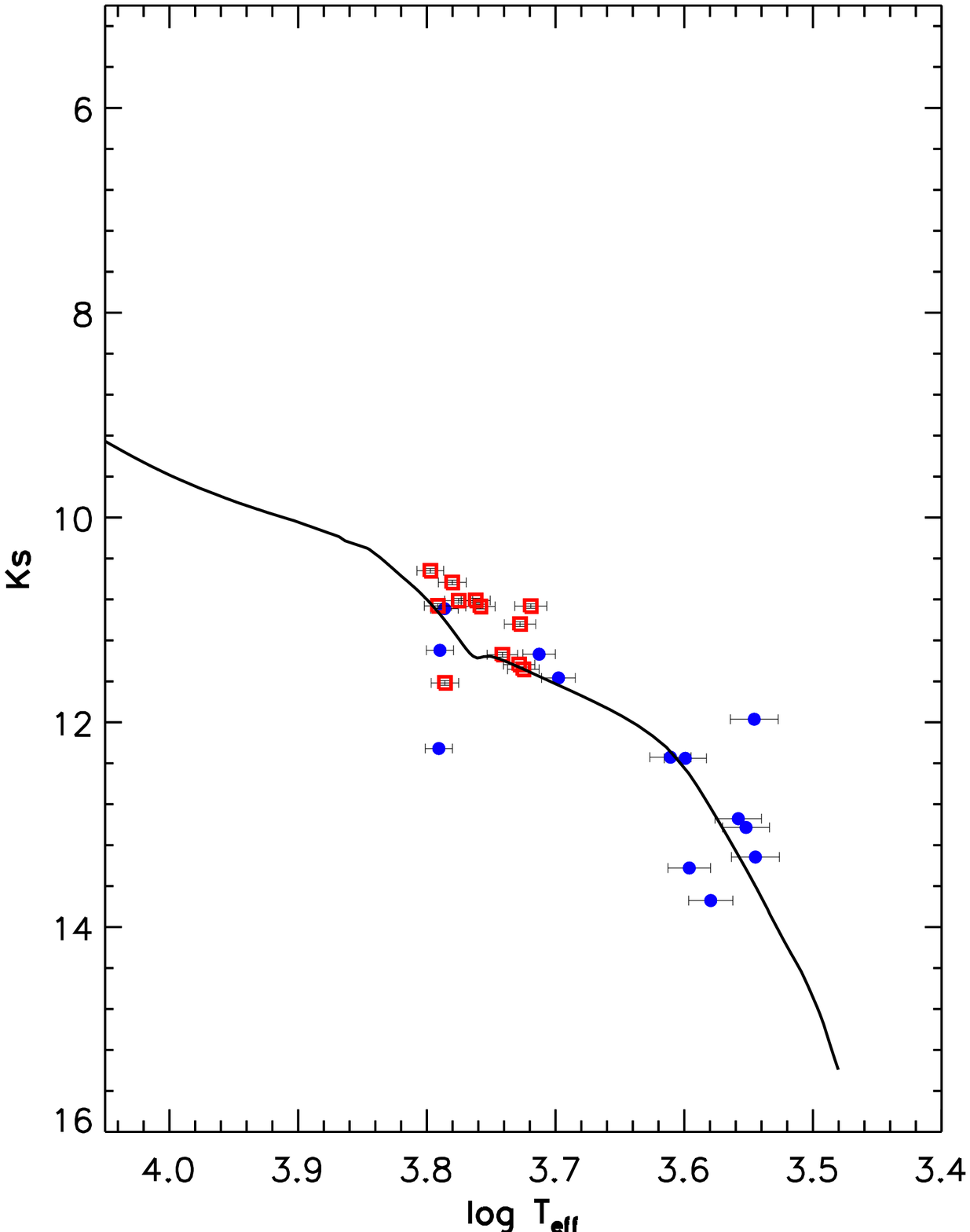}\\      
\includegraphics[width=0.325\linewidth]{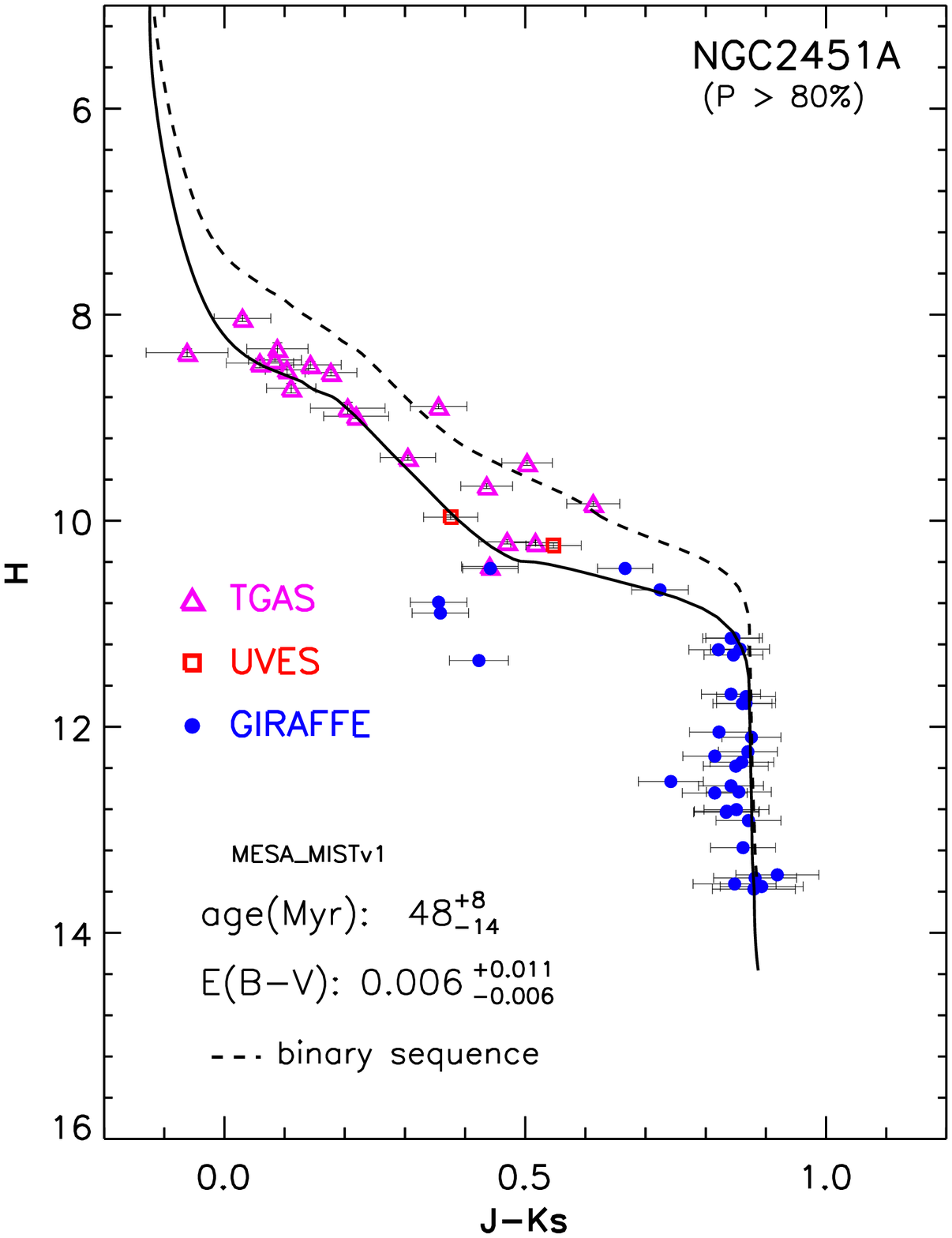}
\includegraphics[width=0.325\linewidth]{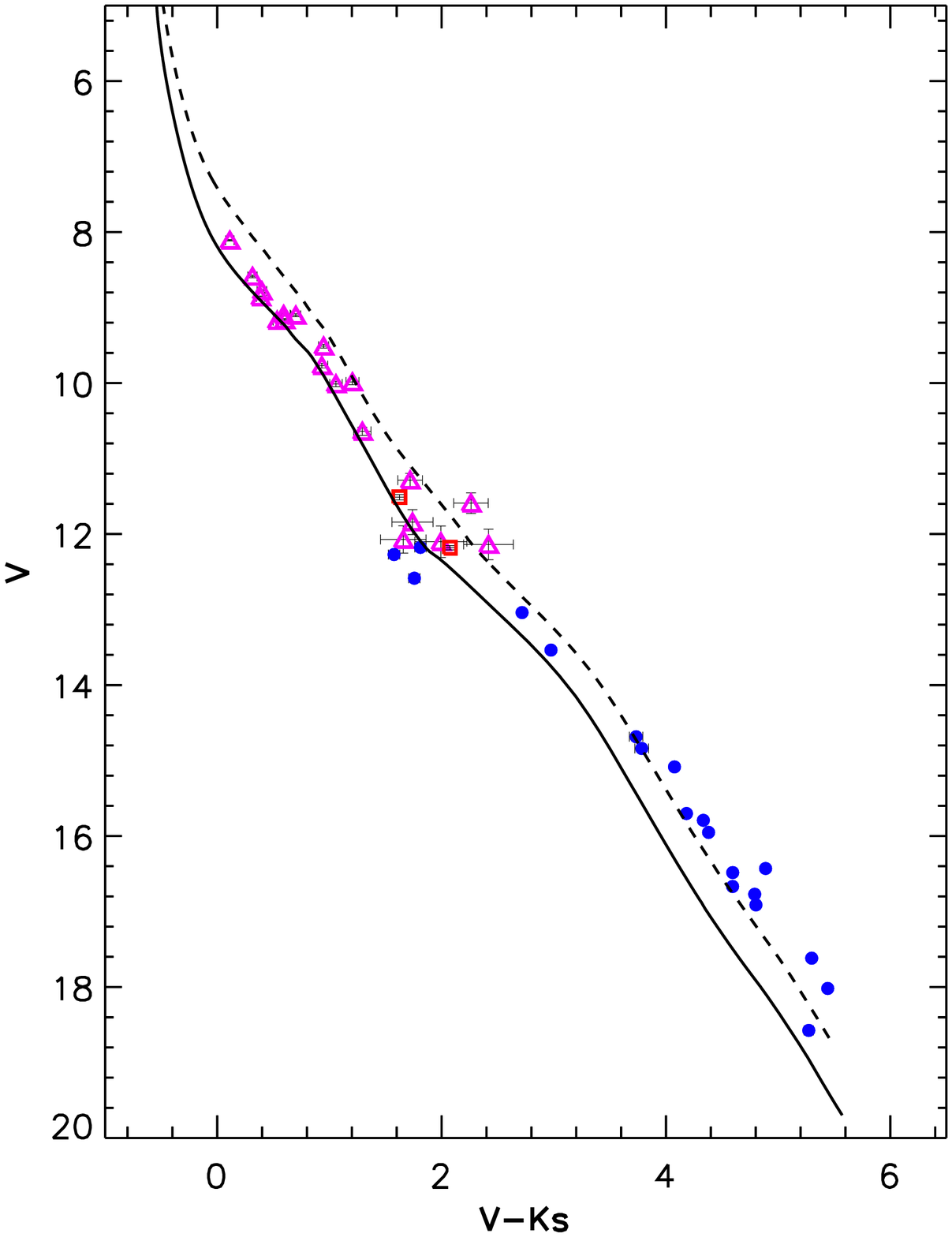}
\includegraphics[width=0.325\linewidth]{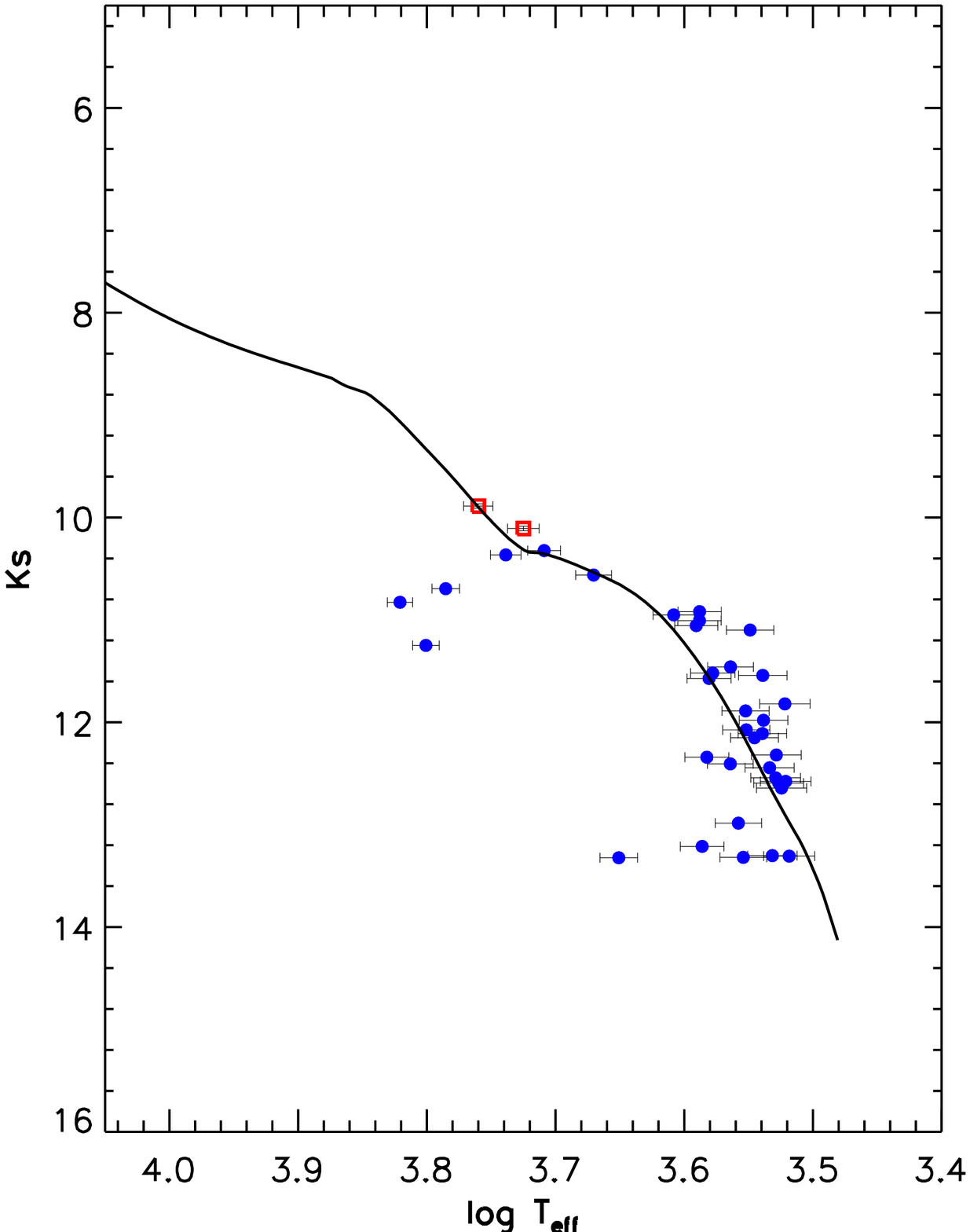}\\    
\caption{As in Fig.~\ref{fig:CMDPROSECCO_1}, but for the clusters IC~2602, IC~4665, and NGC~2451A with the MIST isochrones.}
        \label{fig:CMDMESA_1}
\end{figure*}

\begin{figure*}
        \centering
        \includegraphics[width=0.325\linewidth]{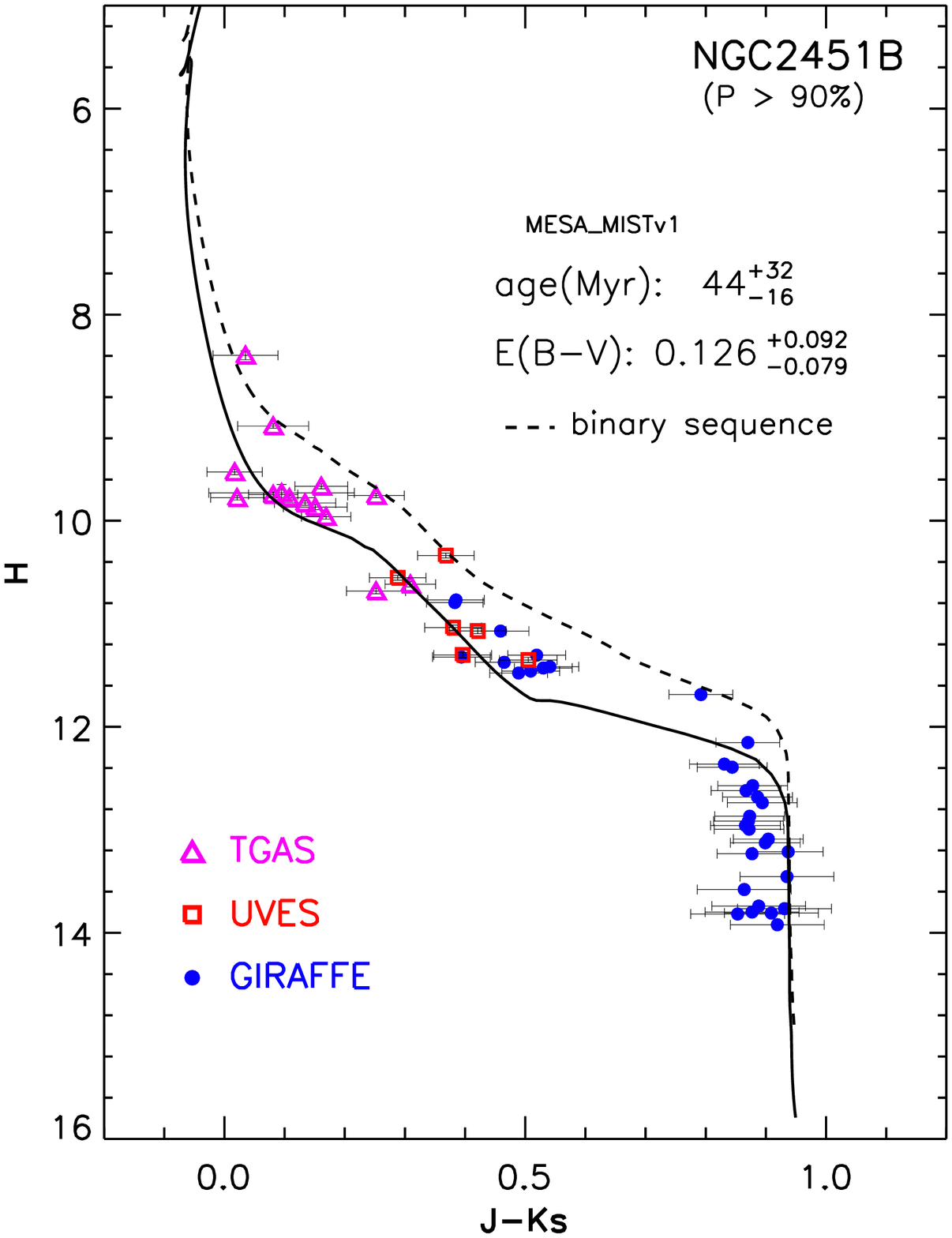}
        \includegraphics[width=0.325\linewidth]{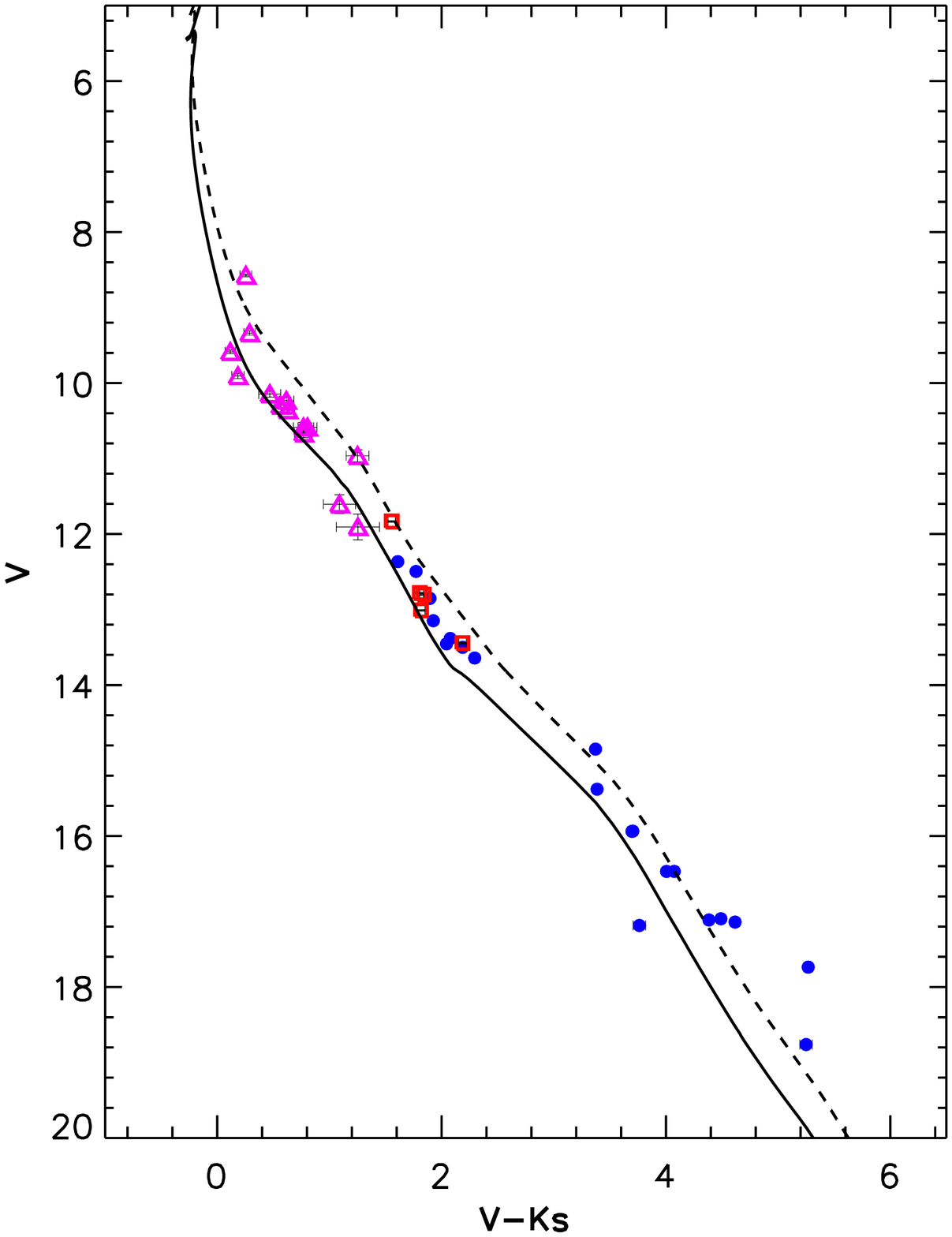}
        \includegraphics[width=0.325\linewidth]{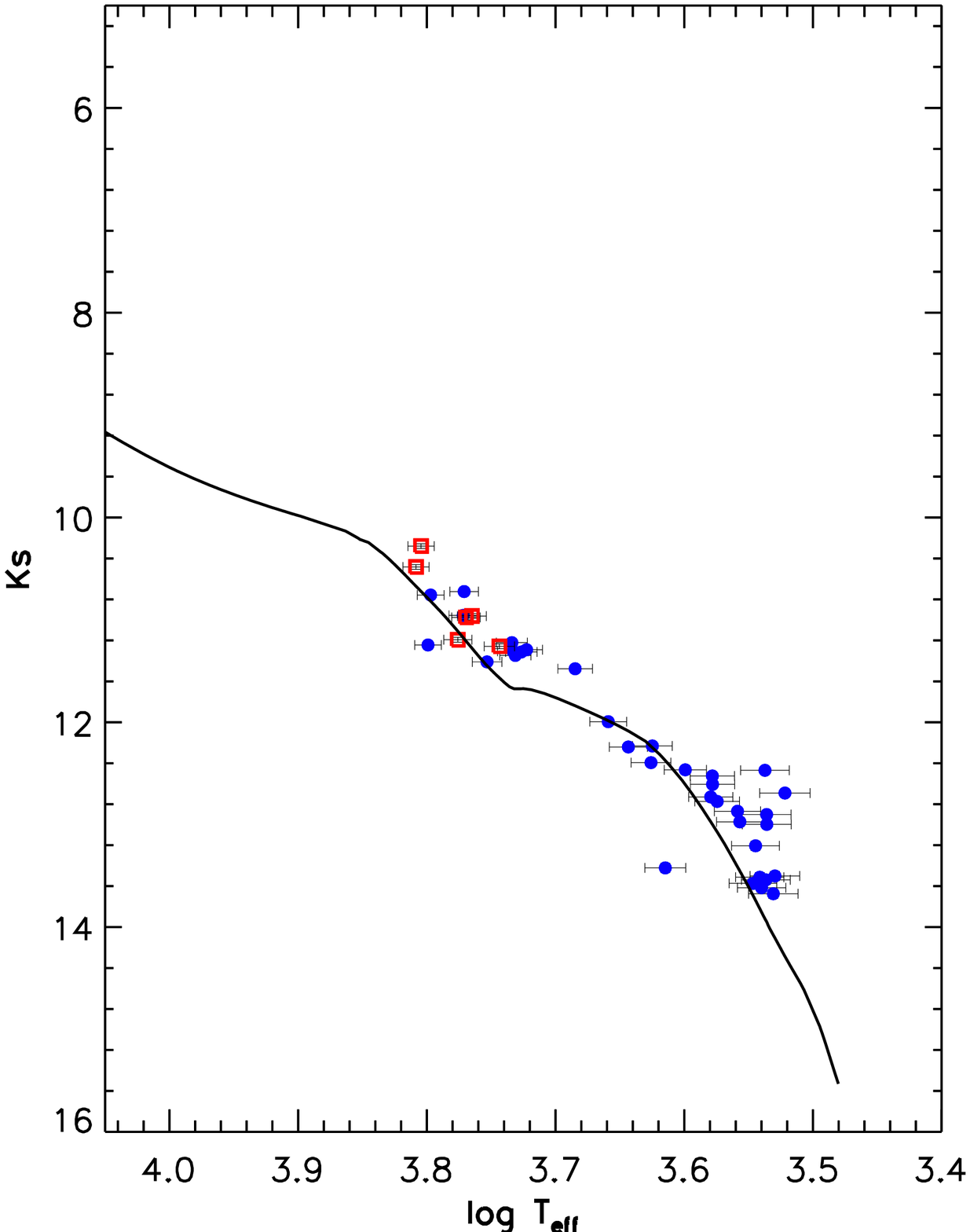}\\
        \includegraphics[width=0.325\linewidth]{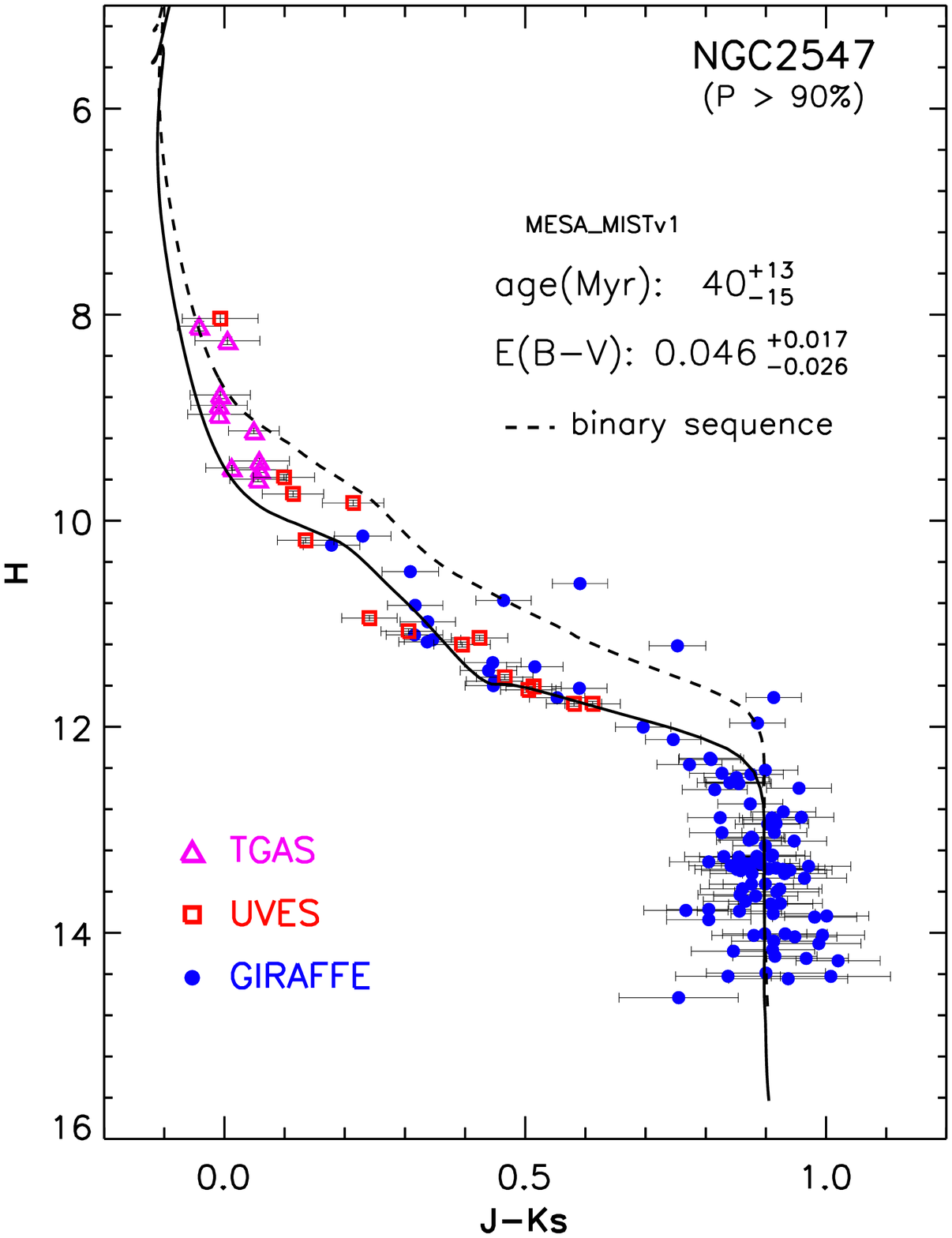}
        \includegraphics[width=0.325\linewidth]{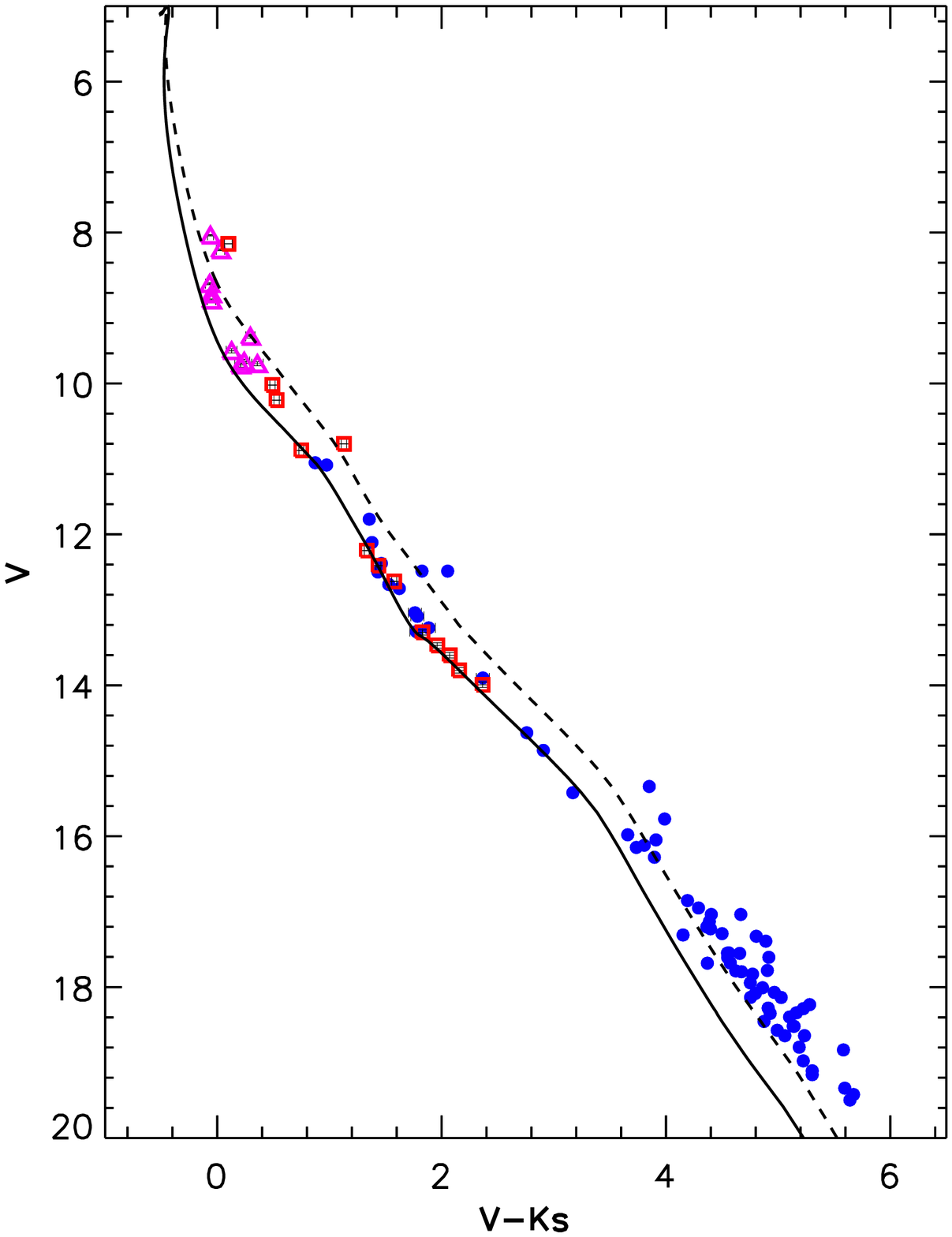}
        \includegraphics[width=0.325\linewidth]{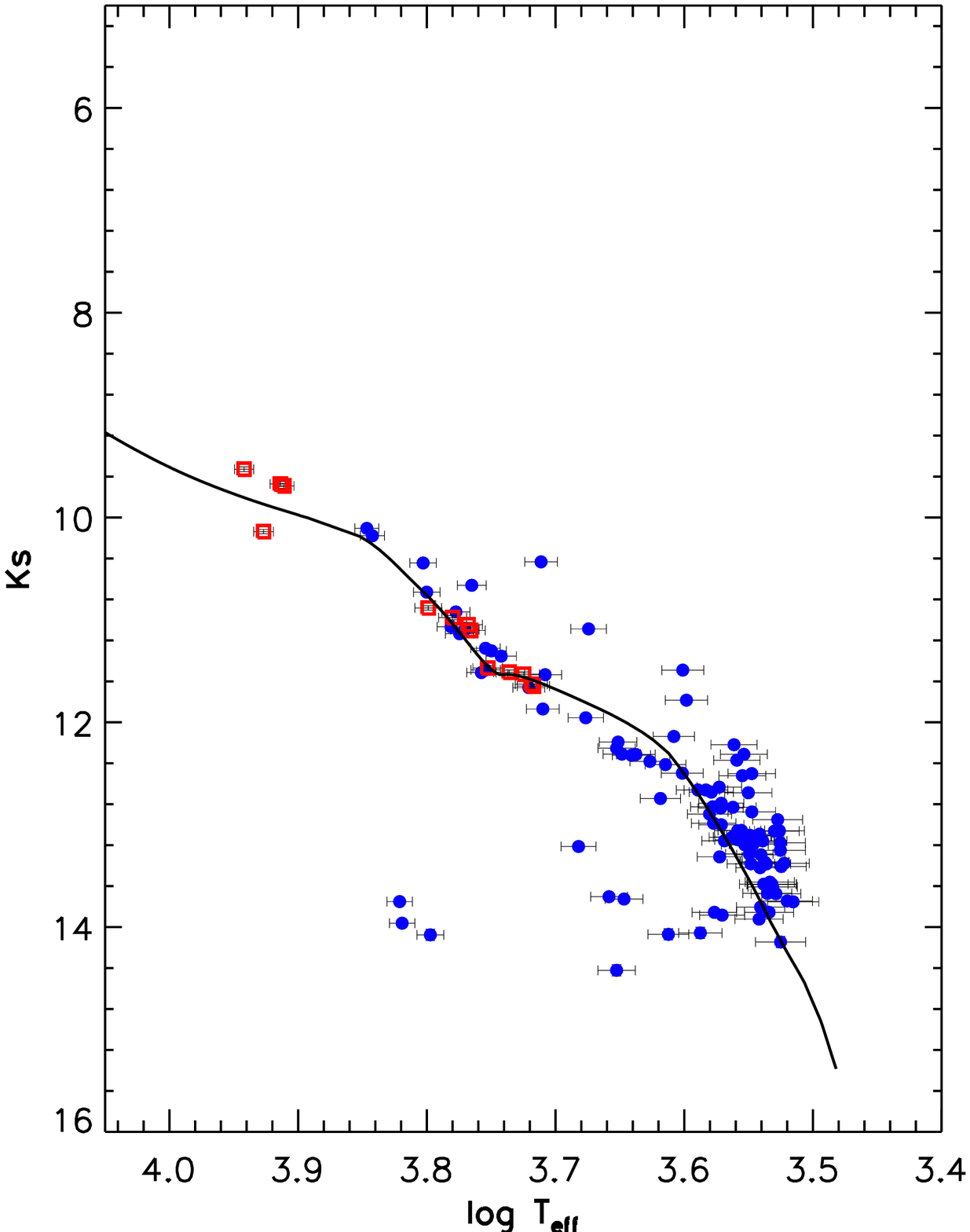}\\     
        \includegraphics[width=0.325\linewidth]{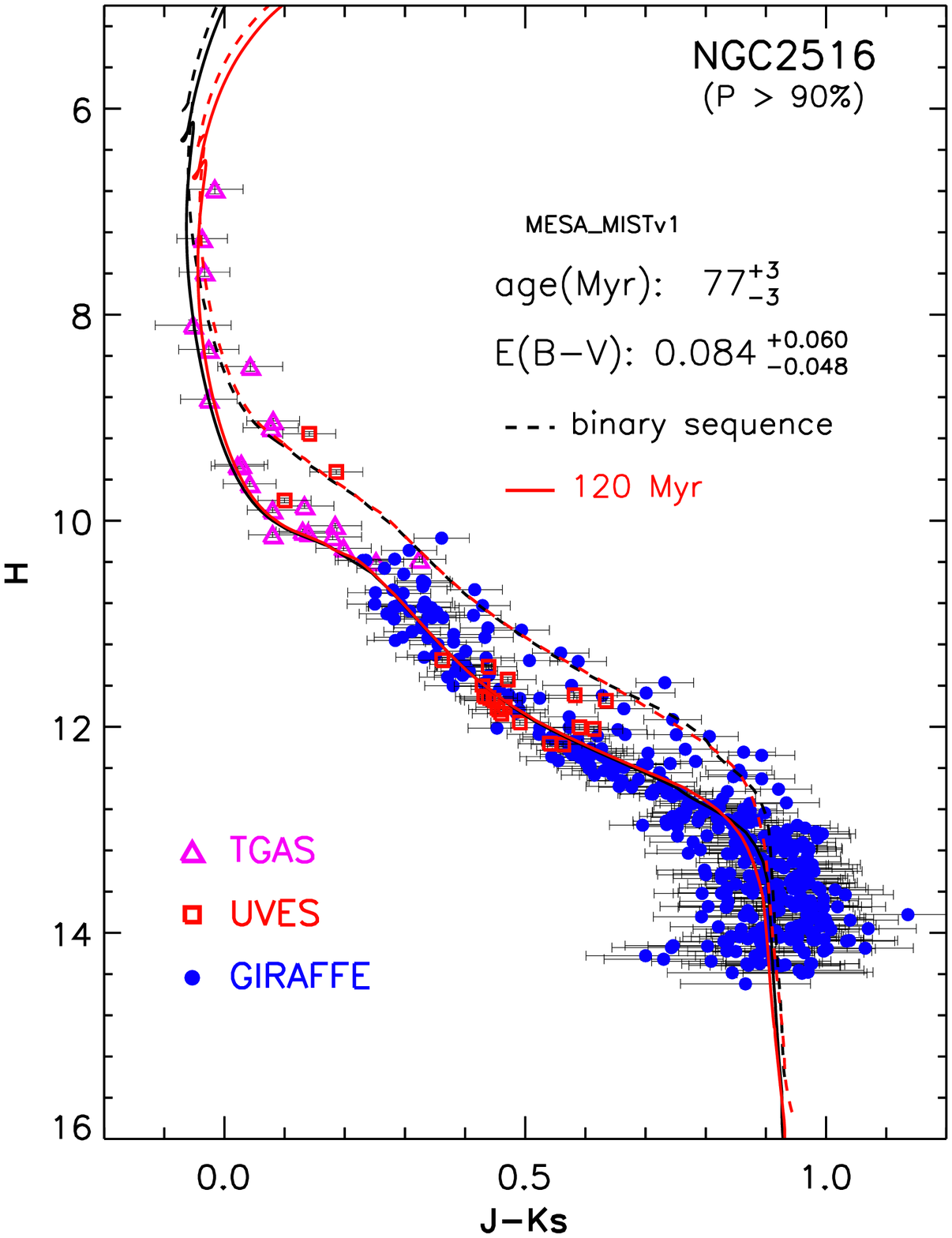}
        \includegraphics[width=0.325\linewidth]{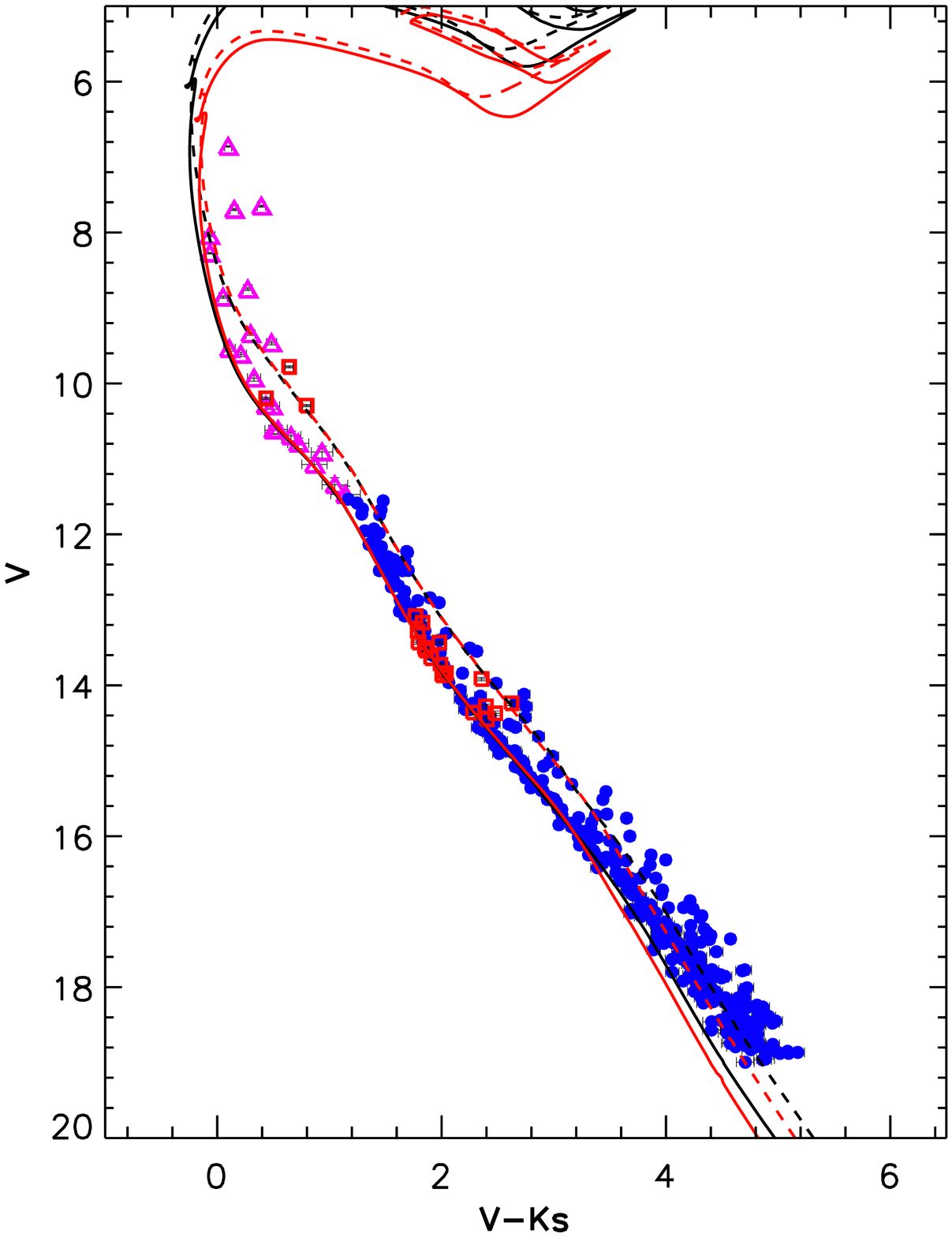}
        \includegraphics[width=0.325\linewidth]{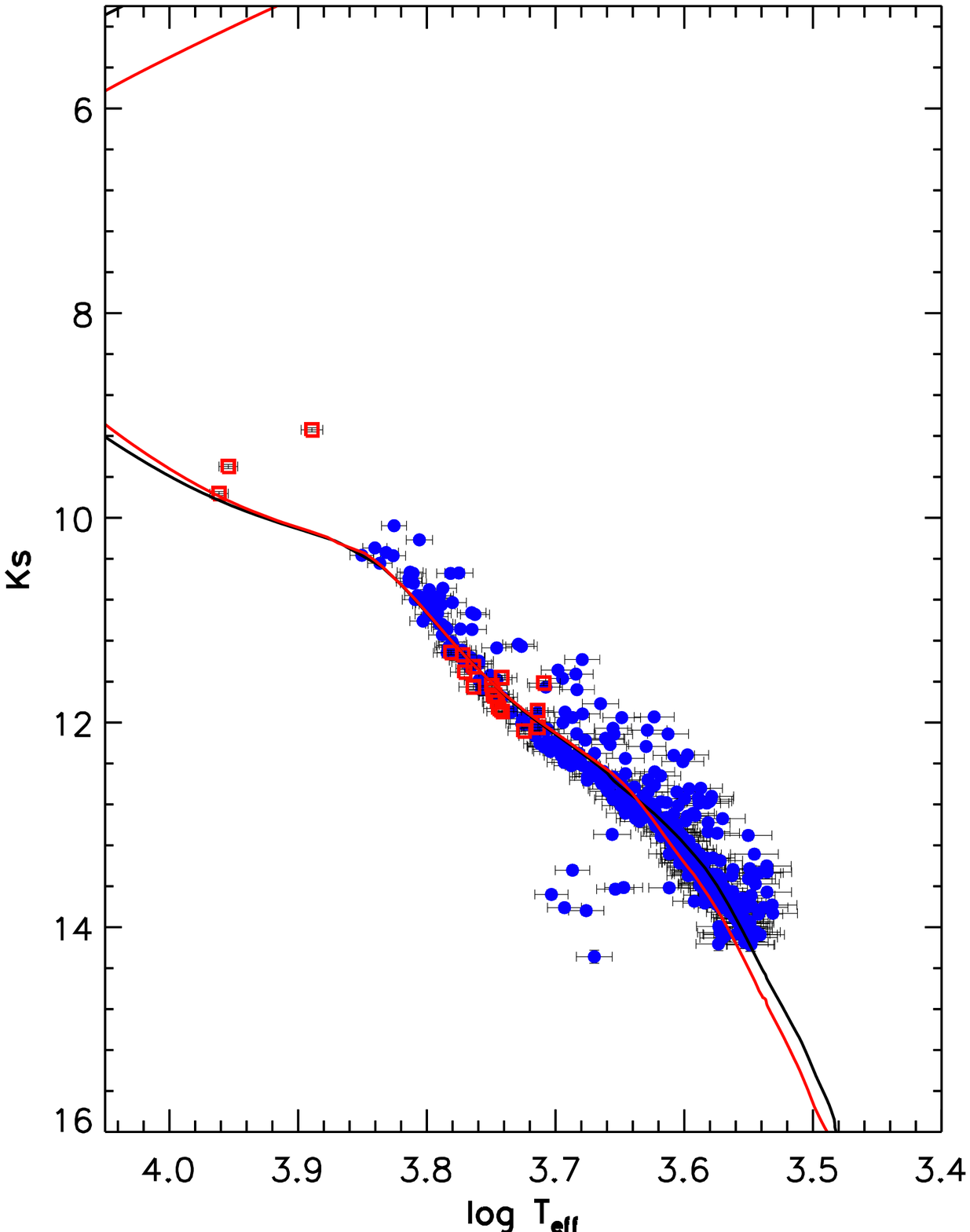}\\     
        \caption{As in Fig.~\ref{fig:CMDPROSECCO_1}, but for the clusters NGC~2451B, NGC~2547, and NGC~2516 with the MIST isochrones.}
        \label{fig:CMDMESA_2}
\end{figure*}

\begin{figure*}
        \centering
        \includegraphics[width=0.325\linewidth]{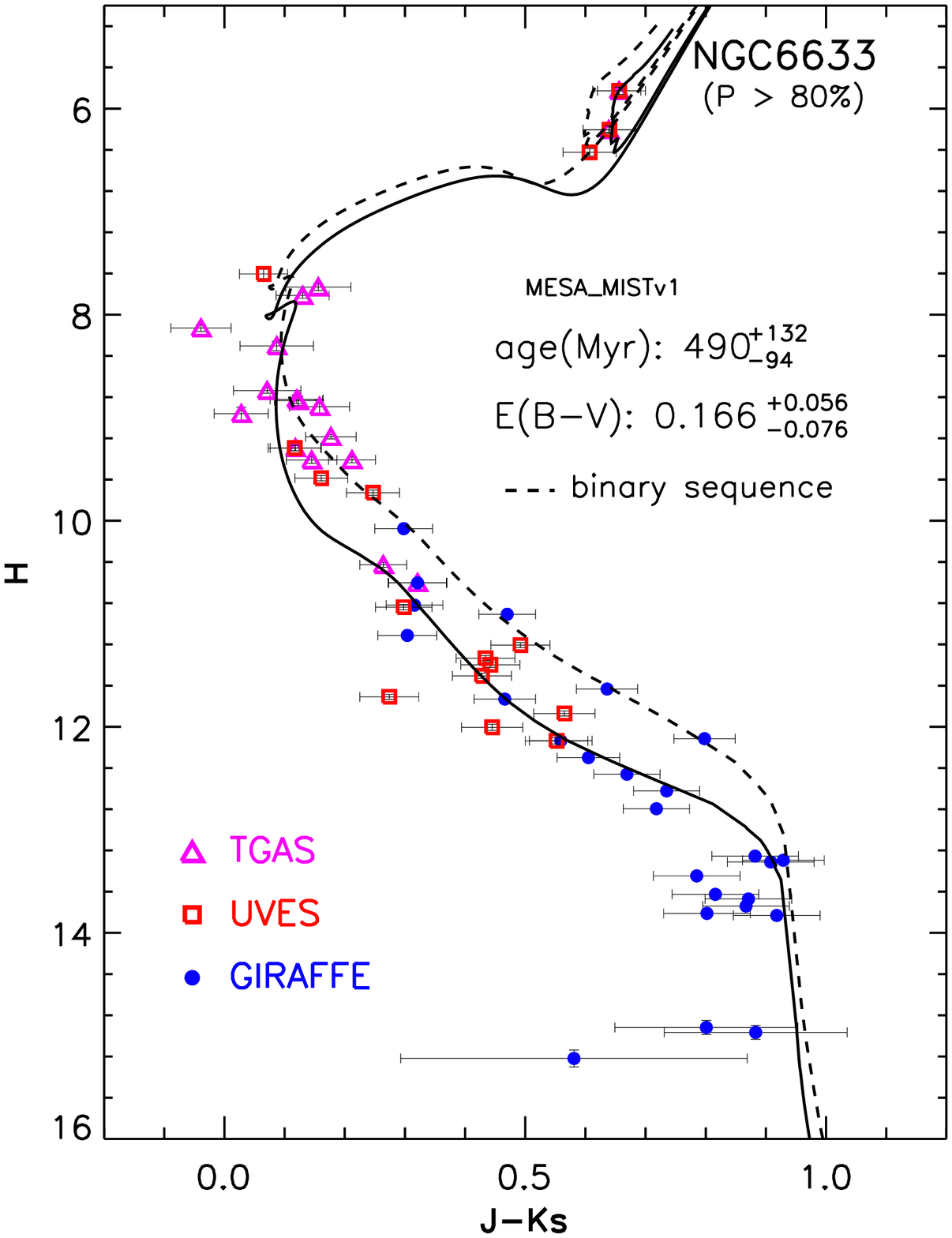}
        \includegraphics[width=0.325\linewidth]{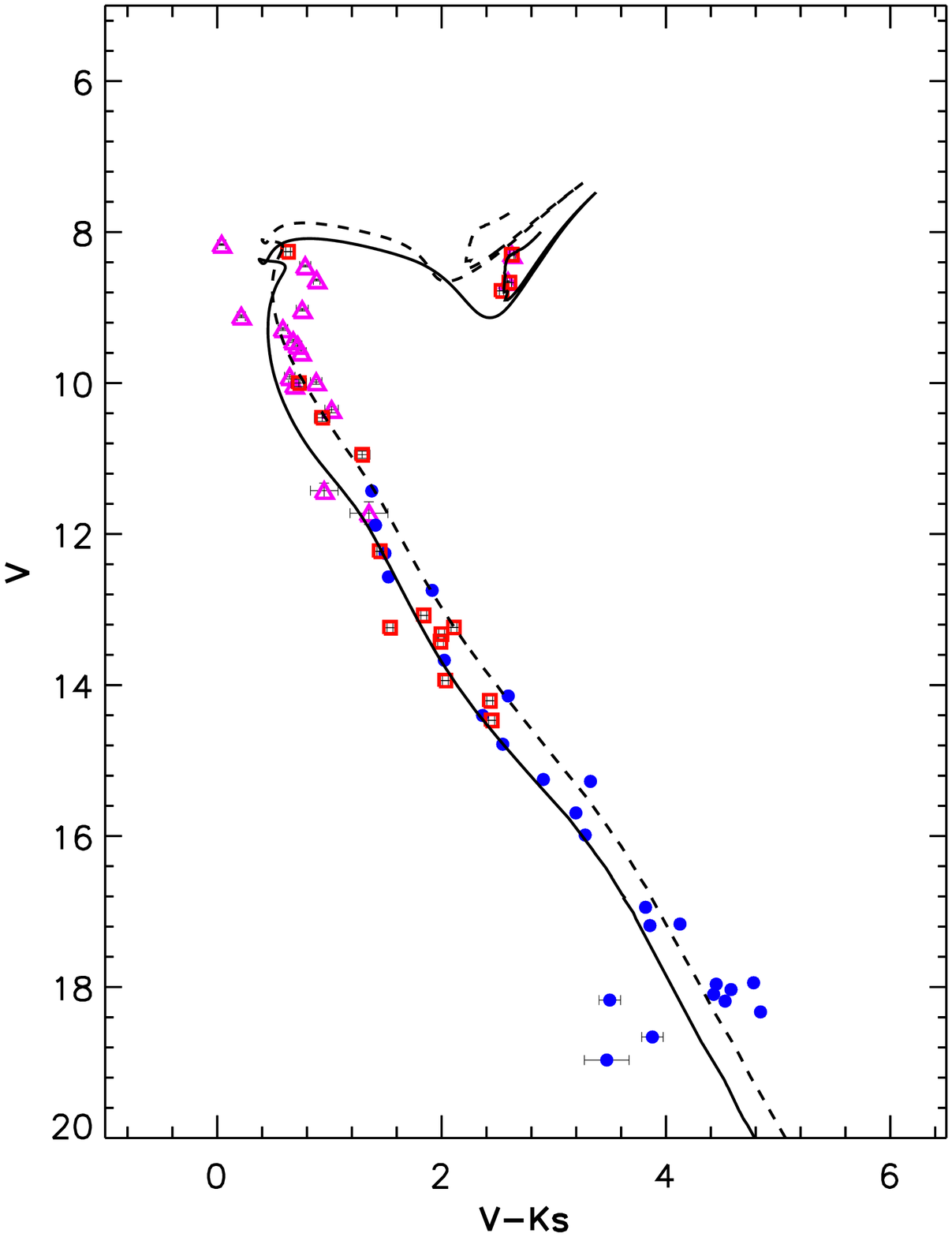}
        \includegraphics[width=0.325\linewidth]{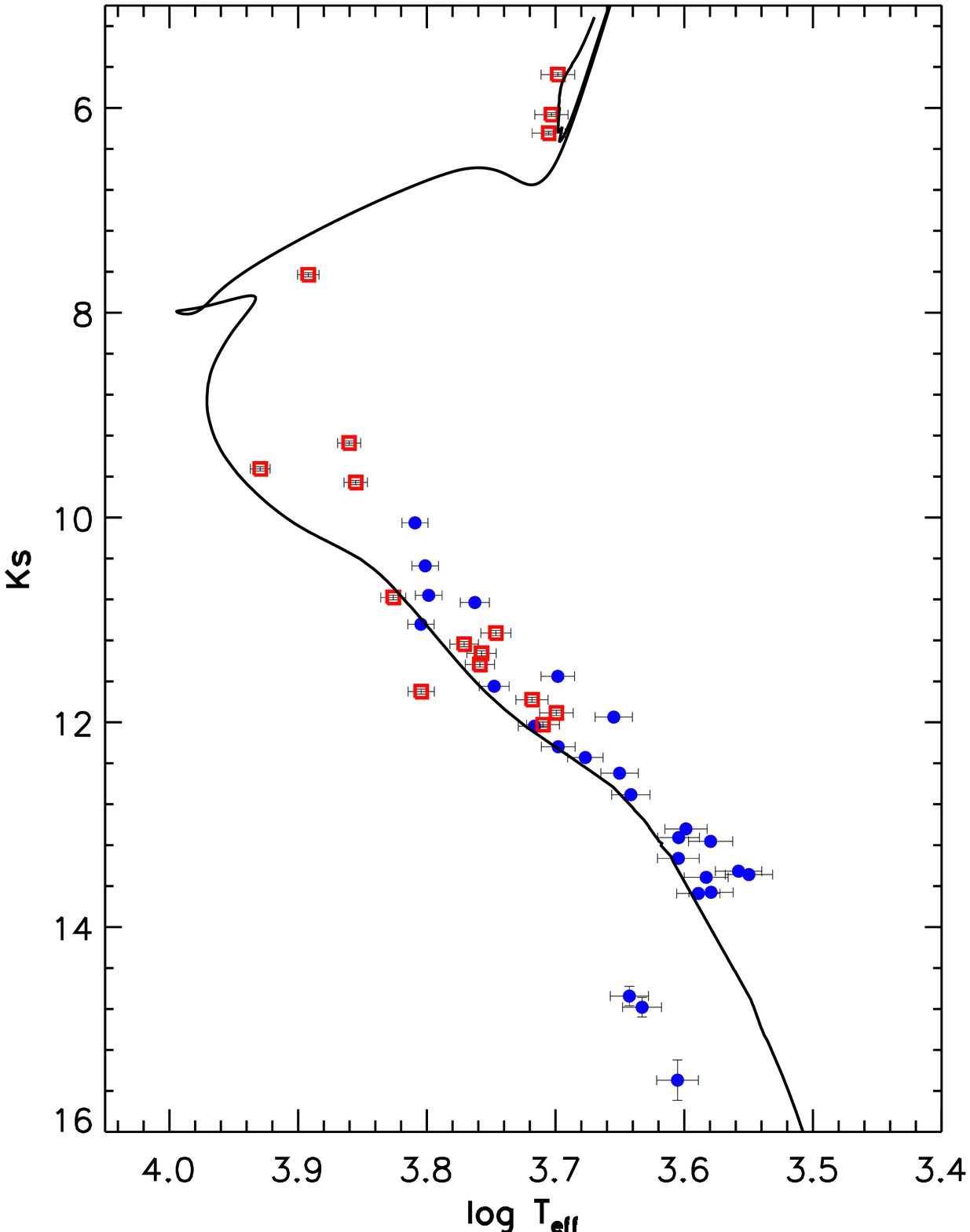}
        \caption{As in Fig.~\ref{fig:CMDPROSECCO_1} but for the cluster NGC~6633 with the MIST isochrones.}
        \label{fig:CMDMESA_3}
\end{figure*}

In this Section we show the comparison between the full sample of clusters but IC~2391 (previously shown in Sec.~\ref{sec:age}) and the PARSEC (Figs.~\ref{fig:CMDPARSEC_1}-\ref{fig:CMDPARSEC_3}) and MIST (Figs.~\ref{fig:CMDMESA_1}-\ref{fig:CMDMESA_3}) isochrones in the ($J-K_{\rm s}$, $H$), ($V-K_{\rm s}$,$V$) and (T$_\mathrm{eff}$, $K_\mathrm{s}$) diagrams.
\end{appendix}
\end{document}